\documentclass[twocolumn]{aastex701}
\usepackage[dvipsnames,x11names]{xcolor}
\usepackage{rotating}

\begin{document}


\title{\large H\,{\sc{i}} Depletion Begins Well Beyond the Virial Radius: A FAST Stacking Study of 36 Galaxy Clusters to 
$5 \times R_{200}$ 
}

\author[orcid=0000-0003-0202-0534]{Cheng Cheng}
\affiliation{Chinese Academy of Sciences South America Center for Astronomy, National Astronomical Observatories, CAS, Beijing 100101, China} 
\affiliation{Key Laboratory of Optical Astronomy, NAOC, 20A Datun Road, Chaoyang District, Beijing 100101, China}
\email[show]{chengcheng@nao.cas.cn}

\author[0000-0002-4428-3183]{Chuan-Peng Zhang}
\affiliation{National Astronomical Observatories, Chinese Academy of Sciences, Beijing 100101, People's Republic of China}
\affiliation{Guizhou Radio Astronomical Observatory, Guizhou University, Guiyang 550000, People's Republic of China}
\email[show]{cpzhang@nao.cas.cn}

\author[0009-0008-9801-2224]{Edo Ibar}
\affiliation{Instituto de F\'isica y Astronom\'ia, Universidad de Valpara\'iso, Avda. Gran Breta\~na 1111, Valpara\'iso, Chile}
\affiliation{Millennium Nucleus for Galaxies (MINGAL), Avda. Gran Breta\~na 1111, Valpara\'iso, Chile}
\email[show]{eduardo.ibar@uv.cl}

\author[0000-0001-5303-6830]{Rory Smith}
\affiliation{Universidad Tecnica Federico Santa Maria, Avenida Espa\~na 1680, Valpara\'iso, Chile}
\email[]{rorysmith274@gmail.com}

\author[0000-0002-9587-6683]{Weiwei Xu}
\affiliation{National Astronomical Observatories, Chinese Academy of Sciences, Beijing 100101, People's Republic of China}
\affiliation{School of Physics and Astronomy, Beijing Normal University, Beijing 100875, People's Republic of China}
\email[]{wwxu@nao.cas.cn}

\author[0000-0003-4032-8572]{Hyowon Kim}
\affiliation{Universidad Tecnica Federico Santa Maria, Avenida Espa\~na 1680, Valpara\'iso, Chile}
\affiliation{Korea Astronomy and Space Science Institute, Daejeon 34055, Republic of Korea}
\email[]{hyowon.kim@usm.cl}

\author[0000-0001-6511-8745]{Jia-Sheng Huang}
\affiliation{Chinese Academy of Sciences South America Center for Astronomy, National Astronomical Observatories, CAS, Beijing 100101, China}
\affiliation{Harvard-Smithsonian Center for Astrophysics, 60 Garden Street, Cambridge, MA 02138, USA}
\email{jhuang@nao.cas.cn}
\author[]{Wei Zhang}
\affiliation{National Astronomical Observatories, Chinese Academy of Sciences, Beijing 100101, People's Republic of China}
\email[]{xtwfn@bao.ac.cn}

\author[]{Pei Zuo}
\affiliation{National Astronomical Observatories, Chinese Academy of Sciences, Beijing 100101, People's Republic of China}
\email[]{peizuo@nao.cas.cn}

\author[]{Hugo M\'endez-Hern\'andez}
\affiliation{
Departamento de Astronom\'ia, Universidad de La Serena, Avda. Ra\'ul Bitr\'an 1305, La Serena, Chile}
\affiliation{Millennium Nucleus for Galaxies (MINGAL), Avda. Gran Breta\~na 1111, Valpara\'iso, Chile}
\email[]{hugo.mendez@userena.cl}

\author[0000-0002-8136-8127]{Juan Molina}
\affiliation{Instituto de F\'isica y Astronom\'ia, Universidad de Valpara\'iso, Avda. Gran Breta\~na 1111, Valpara\'iso, Chile}
\affiliation{Millennium Nucleus for Galaxies (MINGAL), Avda. Gran Breta\~na 1111, Valpara\'iso, Chile}
\email[]{juan.molinato@uv.cl}

\author[0000-0001-6083-956X]{Ming Zhu}
\affiliation{National Astronomical Observatories, Chinese Academy of Sciences, Beijing 100101, People's Republic of China}
\affiliation{Guizhou Radio Astronomical Observatory, Guizhou University, Guiyang 550000, People's Republic of China}
\email{mz@nao.cas.cn}

\author[]{Qian Yu}
\affiliation{National Astronomical Observatories, Chinese Academy of Sciences, Beijing 100101, People's Republic of China}
\affiliation{Department of Physics and Astronomy, University of Missouri, Columbia, MO 65211, USA}
\email[]{YUQIAN@ARIZONA.EDU}

\begin{abstract}

We present a stacking study of the neutral atomic hydrogen (H\,{\sc{i}}) content in and around 36 local galaxy clusters at $z<0.07$, using a combination of the  FAST all sky H\,{\sc{i}} survey (FASHI) and the extensive spectroscopic catalog mainly from the Dark Energy Spectroscopic Instrument (DESI). We employ spectral stacking techniques to probe the average H\,{\sc{i}} mass and HI-to-stellar mass ratio ($M_{\rm HI}/M_*$) for member galaxies down to stellar masses of $M_*\sim 10^9M_\odot$, spanning a projected cluster-centric distance to $5R_{200}$. Our analysis reveals a pronounced environmental effect: both $M_{\rm HI}$ and $M_{\rm HI}/M_*$ decrease steadily towards the cluster center, dropping by $\sim 0.5$ dex on average from the outskirts to the core. Crucially, we find that $M_{\rm HI}/M_*$ of galaxies remain lower than the field galaxies even at the $5R_{200}$. This provides direct, statistical evidence for substantial gas stripping and pre-processing in the cluster outskirts, likely occurring in infalling groups and large-scale filaments. By further splitting the sample by $g-r$ color, we show that the H\,{\sc{i}} deficiency persists at fixed galaxy color: even the bluest cluster members exhibit $\sim 0.5$~dex lower $M_{\rm HI}/M_*$ than field galaxies of similar color, reflecting environmental effects on the cold gas reservoir prior to full optical transformation. The total H\,{\sc{i}} mass within clusters and their outskirts agrees broadly with predictions from cosmological simulation. Our results underscore the critical role of the extended cluster environment in quenching galaxies by depleting their cold gas reservoirs well before they enter the dense cluster core.
\end{abstract}



\section{Introduction} 

Galaxy clusters reside in the most massive dark matter halos, continuously accreting cold gas from the cosmic web, processing it through star formation and feedback, and heating it into the hot intracluster medium \citep[ICM, e.g., ][]{1991ApJ...379...52W}. Typically, local galaxy clusters host a significant population of early-type galaxies but are deficient in cold gas \citep[e.g., ][]{1984AJ.....89..758H, 2009MNRAS.400.1962K, 2018A&A...618A.126O}. The diffuse, hot intracluster gas is believed to originate from gas infall and shock heating \citep{2006MNRAS.368....2D}. High-resolution X-ray observations can provide critical insights into the temperature, density, and dynamical structures of this hot gas, revealing shocks, tails, and gas bridges \citep{2004ApJ...610L..81H}. However, the detailed interactions between hot and cold gas remain poorly understood-particularly how galaxy clusters acquire cold gas or how star-forming galaxies migrate from the field into clusters.

Understanding the links between hot and cold gas is crucial for studying the final stages of massive galaxy evolution. Neutral atomic hydrogen (HI) 21-cm emission as the best tracer of the cold atomic gas is sensitive to environmental effects and galaxy interactions because of the diffuse nature. Previous observations show that for the X-ray detected galaxy clusters, about 40\% of HI-rich galaxies within $R_{200}$ are affected by ram pressure stripping from the ICM \citep{2017MNRAS.466.1275B, 2020ApJ...903..103W}. Beyond the cluster cores, a significant fraction of galaxy transformation occurs in the surrounding large-scale structures, extending out to at least 5$\times R_{200}$. Observations reveal a suppressed star formation rate and a lower fraction of star-forming galaxies at 2–5 $\times R_{200}$ compared to field galaxies \citep{2015ApJ...806..101H, 2024MNRAS.527L..19L}. This suggests that galaxies may already undergo environmental quenching in galaxy groups within the broader cluster-feeding filaments \citep{2004PASJ...56...29F, 2018MNRAS.477.4931H}, highlighting the key role of the large-scale environment in galaxy evolution and large scale structure formation \citep{2017MNRAS.466.4692K, 2018ApJ...852..142C, 2024MNRAS.529.2595B}. To understand the environment quenching in galaxy clusters, it is necessary to map the H\,{\sc{i}} emission to $5\times \rm R_{200}$ of the massive halos to cover the pre-process region of the clusters.

While environmental quenching dominates the outskirts of galaxy clusters \citep{2015ApJ...806..101H, 2021PASA...38...35C, 2022ApJS..262...31L, 2024MNRAS.528..919P}, recent studies reveal a countervailing process: star-forming galaxies are also found inflowing into massive halos via cosmic filaments \citep[e.g., ][for the Virgo cluster, where $R_{200}\simeq $ 5 degrees]{2011MNRAS.414.2101U, 2016ApJ...833..207K, 2021ApJ...923..235C}. This duality highlights the critical need to map clusters to several $\times \rm R_{200}$ -- probing both gas depletion and replenishment mechanisms. However, previous studies of gas properties in galaxy clusters have mainly focused on the central regions \citep[Arecibo mapping to Virgo cluster with $10\times 2\rm \, deg^2$,][]{2012MNRAS.423..787T}, while the vast physical extent of local clusters \citep[e.g., $5\times \rm R_{200} \sim 25$ degrees for the Virgo cluster][]{2025arXiv251023260C} poses challenges in balancing survey depth and coverage. Consequently, previous studies about H\,{\sc{i}} in clusters are focused on the H\,{\sc{i}} properties up to about $\sim 2 R_{200}$, such as the ASKAP observations of the Hydra cluster 
\citep{2021MNRAS.505.1891R, 2021ApJ...915...70W}, 
Blind Ultra Deep H I Environmental Survey \citep[BUDHIES,][]{2013MNRAS.431.2111J}, 
Westerbork Coma Survey \citep{2022A&A...659A..94M}.
However, statistical studies of gas abundance within $5\times \rm R_{200}$ remain scarce.

Recent studies have shown that H\,{\sc{i}} depletion in galaxies is not only confined to dense cluster cores \citep{2017MNRAS.466.1275B}. Using WSRT observations toward the SDSS South Galactic Cap, \citet{2021MNRAS.507.5580H} applied spectral stacking to group members identified from the SDSS group catalog \citep{2007ApJ...671..153Y, 2012ApJ...752...41Y}. They found a clear decline of about 0.5 dex in H\,{\sc{i}} mass toward group centers, with the trend extending beyond the virial radius to more than $3R_{180}$, suggesting significant pre-processing before cluster infall. However, the WSRT survey area covers only 35 pointings ($35'$ primary beam each), and the stacking bins contain only 20–100 galaxies, which leaves potential risks of random errors or systematic biases; deeper H\,{\sc{i}} data cubes with dense spectroscopic coverage are still needed for robust constraints on gas abundance. 

A complementary approach was taken by \citet{2013MNRAS.429.2191Z}, who make use of the H\,{\sc{i}} mass scaling relation and found that galaxies in clusters or high-density environments are more gas-poor, with gas depletion strongly depending on structural properties, especially stellar mass surface density.

Together, these studies suggest that both large-scale environment and intrinsic galaxy structure play critical roles in regulating the cold gas reservoir. Motivated by these findings, we aim to investigate H\,{\sc{i}} abundance in and around clusters out to $5\times R_{200}$ by combining the wide-field The Five-hundred-meter Aperture Spherical radio Telescope (FAST) survey with DESI spectroscopy, using stacking techniques to probe the evolution characteristics below the individual detection limit.

The FAST all sky H\,{\sc{i}} survey (FASHI) is one of the largest H\,{\sc{i}} blind survey projects \citep{2024SCPMA..6719511Z}. It covered about 4000 deg$^2$ to the depth about 0.5 mJy/beam (where the beam is 3 arcmin at 1.5 GHz), and the DR1 catalog includes more than 40,000 H\,{\sc{i}} bright targets from local Universe up to redshift 0.08. The survey is sensitive to H\,{\sc{i}} masses as low as $\sim10^{8}\,M_\odot$ at redshift 0.01 \citep{2025ApJS..281...66C}. Meanwhile, the recent released DESI dr1 spec-$z$ provides us with a reliable spec-$z$ catalog to about r band 20 AB mag, which is deep enough to detect the targets with $M_*\sim 10^9 M_\odot$ at redshift 0.05. Therefore, although the depth of the blind survey is not deep enough to detect all the members in galaxy clusters, the combination of the DESI spec-$z$ and the FASHI coverage provide us with a unique chance to explore the H\,{\sc{i}} abundance by stacking.

In this work, we present the stacking H\,{\sc{i}} properties of galaxy clusters from the AXES cluster catalog \citep{2024A&A...690A..52D}. Section 2 describes the sample selection, and Section 3 and 4 present the data reduction and results. We discuss our results in Section 5 and summary in Section 6. Throughout this paper, we adopt the $\Lambda$CDM cosmology with $H_0=70\, \rm km\, s^{-1}\, Mpc^{-1}$, $\Omega_{\rm M} = 0.3$, and $\Omega_{\rm \Lambda} = 0.7$. All the magnitudes are in the AB magnitude system \citep{1983ApJ...266..713O}.

\begin{deluxetable*}{lccccccccc}
\tablewidth{0pt}
\tablecaption{Catalog of galaxy clusters for H\,{\sc i} stacking\label{tab1}}
\tablehead{
\colhead{AXES ID} & \colhead{RA (J2000, deg)} & \colhead{Dec (J2000, deg)} & 
\colhead{$z_{\rm spec}$} & \colhead{$R_{200}$ (arcmin)} & \colhead{$R_{200}$ (Mpc)} & \colhead{$\log (M_{200}/M_\odot$)} & \colhead{Flag} & \colhead{Stage ($R<R_{200}$)}
}
\startdata
AXES\_E\_468    &  236.21292   &   36.118090 &   0.0663038  &  18.56 &  1.42  &  14.5365	&E & Disturbed\\
AXES\_E\_828    &  153.45190   & -0.91429000 &   0.0460370  &  22.63 &  1.23  &  14.3422	&E & Relaxed\\
AXES\_E\_2416   &   240.52026  &   16.048090 &   0.0356283  &  35.95 &  1.53  &  14.6236	&E & Disturbed\\
AXES\_E\_2496   &   216.77689  &   16.725600 &   0.0545393  &  20.06 &  1.28  &  14.3965	&E & Disturbed\\
AXES\_E\_3220   &   186.25158  &   32.068650 &   0.0615888  &  17.63 &  1.26  &  14.3787	&E & Disturbed\\
AXES\_E\_5441   &   248.00373  &   13.636990 &   0.0527733  &  19.89 &  1.23  &  14.3446	&E & Disturbed\\
AXES\_E\_5596   &   229.17347  &   7.0143000 &   0.0353297  &  38.50 &  1.62  &  14.7022	&E & Disturbed\\
AXES\_E\_5686   &   208.54612  &   33.131380 &   0.0506048  &  23.05 &  1.37  &  14.4843	&E & Relaxed\\
AXES\_E\_7498   &   170.36292  &   2.8231600 &   0.0502842  &  20.20 &  1.19  &  14.3044	&E & Disturbed\\
AXES\_E\_8887   &   177.06272  &   54.608240 &   0.0603254  &  17.74 &  1.24  &  14.3615	&E & Disturbed\\
AXES\_E\_15574  &   159.85551  &   5.1859000 &   0.0699293  &  16.81 &  1.35  &  14.4727	&E & Disturbed\\
AXES\_E\_17458  &   180.04817  &   56.220460 &   0.0654503  &  17.24 &  1.30  &  14.4249	&E & Disturbed\\
AXES\_E\_20316  &   118.34494  &   29.384770 &   0.0613446  &  18.12 &  1.29  &  14.4096	&E & Relaxed\\
AXES\_C\_513    &   248.25393  &   11.743030 &   0.0518593  &  19.87 &  1.21  &  14.3213	&C & Relaxed\\
AXES\_C\_672    &   214.38044  &   8.1914200 &   0.0587548  &  19.74 &  1.35  &  14.4679	&C & Relaxed\\
AXES\_C\_1484   &   255.89479  &   36.063130 &   0.0629381  &  16.27 &  1.18  &  14.3011	&C & Disturbed\\
AXES\_C\_2236   &   207.92897  &   46.368900 &   0.0632097  &  17.63 &  1.29  &  14.4107	&C & Disturbed\\
AXES\_C\_2383   &   242.78157  &   36.927210 &   0.0675788  &  15.61 &  1.21  &  14.3342	&C & Disturbed\\
AXES\_C\_2723   &   181.10861  &   1.8965800 &   0.0212001  &  56.59 &  1.46  &  14.5551	&C & Relaxed\\
AXES\_C\_3307   &   246.76723  &   14.360540 &   0.0502313  &  20.47 &  1.21  &  14.3207	&C & Disturbed\\
AXES\_C\_3978   &   230.78113  &   8.6018700 &   0.0351075  &  45.11 &  1.89  &  14.9006	&C & Relaxed\\
AXES\_C\_4398   &   254.06732  &   39.280440 &   0.0621263  &  16.52 &  1.19  &  14.3051	&C & Relaxed\\
AXES\_C\_5255   &   216.20935  &   2.7114700 &   0.0548892  &  18.72 &  1.20  &  14.3144	&C & Disturbed\\
AXES\_C\_5567   &   230.45953  &   7.7015100 &   0.0453269  &  41.22 &  2.20  &  15.1043	&C & Disturbed\\
AXES\_C\_5835   &   156.25400  &   47.830690 &   0.0634621  &  16.42 &  1.20  &  14.3233	&C & Disturbed\\
AXES\_C\_5886   &   233.33449  &   31.160180 &   0.0668029  &  17.39 &  1.34  &  14.4609	&C & Disturbed\\
AXES\_C\_6372   &   220.16284  &   3.4727300 &   0.0276671  &  38.51 &  1.28  &  14.3928	&C & Relaxed\\
AXES\_C\_12349  &   200.05824  &   33.140390 &   0.0372790  &  27.40 &  1.22  &  14.3268	&C & Disturbed\\
AXES\_C\_13894  &   242.41146  &   5.8749600 &   0.0647107  &  19.19 &  1.43  &  14.5501	&C & Disturbed\\
AXES\_C\_20685  &   172.49668  &   36.638020 &   0.0628619  &  16.52 &  1.20  &  14.3197	&C & Disturbed\\
AXES\_C\_25593  &   224.60960  &   48.550270 &   0.0373847  &  32.41 &  1.44  &  14.5489	&C & Disturbed\\
 AXES\_C\_6155  &   161.16041  &   38.755780 &   0.0339169  &  29.77 &  1.21  &  14.3154	   &C & --- \\
 AXES\_C\_39639 &   127.90746  &   19.391180 &   0.0395083  &  37.25 &  1.75  &  14.7999	   &C & --- \\
 AXES\_C\_55129 &   131.72855  &   25.380090 &   0.0668807  &  25.37 &  1.95  &  14.9544	   &C & --- \\
 AXES\_C\_60247 &   255.82569  &   37.626500 &   0.0330494  &  33.61 &  1.33  &  14.4410	   &C & --- \\
 AXES\_C\_82283 &   249.21096  &   26.525720 &   0.0440580  &  27.90 &  1.45  &  14.5602	   &C & --- \\
\enddata
\tablecomments{E: extended sample and C: compact sample from AXES catalog. This is the subset of AXES clusters that are used for this study.}
\end{deluxetable*}

\section{Sample Selection and Data Reduction}

\subsection{Data preparation}
\subsubsection{Archival Galaxy Cluster sample}

The galaxy cluster sample used in our study is drawn from the All-sky X-ray Extended Sources - Sloan Digital Sky Survey (AXES-SDSS) cluster catalog \citep{2024A&A...690A..52D}. This catalog systematically identifies and confirms galaxy groups and clusters with extended X-ray emission by combining X-ray data from the ROSAT All-Sky Survey (RASS) and optical data from SDSS. Using a screening method based on optical properties, AXES effectively removes contaminants, resulting in a sample with high purity, reaching up to 95\%. The catalog covers systems spanning a wide mass range, from group scales to rich clusters, and is optimized for the detection of extended X-ray sources, significantly reducing scatter in both optical and X-ray scaling relations \citep{2024A&A...690A..52D}. The SDSS spectroscopic data further improves the identification of BCGs and provides reliable cluster redshifts.  Additionally, AXES accounts for the influence of the large-scale density field and demonstrates that removing systems with low optical luminosity located in dense regions of the cosmic web further reduces scatter. These characteristics make AXES a statistically robust and clean sample of X-ray clusters, well-suited for investigating the relationship between gas and galaxy evolution in cluster environments.

AXES-SDSS provides two classes of galaxy clusters: extended clusters and compact clusters. Since the two types do not show significantly different properties \citep{2024A&A...690A..52D}, we combine the compact cluster sample with the extended cluster sample to form the parent sample used in this work. We remove the compact sources that overlap with the extended sample, because for clusters appearing in both samples, the extended targets have a larger $R_{200}$.

\subsubsection{Archival Spectroscopic Redshift Catalog}

The Dark Energy Spectroscopic Instrument (DESI) is a state-of-the-art spectroscopic survey designed to investigate fundamental questions in cosmology, galaxy formation and evolution, and in particular about the nature and evolution of dark energy \citep{2022AJ....164..207D}. Equipped with 5,000 robotically positioned fibers, DESI is capable of efficiently obtaining high-quality spectroscopic redshifts (spec-z) for millions of galaxies and quasars. The first data release (DR1\footnote{\url{https://fastspecfit.readthedocs.io/en/latest/iron.html}}) of DESI currently covers approximately 9000 deg$^2$ of sky at declinations greater than -23.5 degrees, reaching a depth of around $r\sim 20$ AB mag \citep{2025arXiv250314745D}. At this magnitude limit, DESI can effectively detect galaxies down to a stellar mass ($M_*$) of approximately $10^{8-9.5} M_\odot$ at low to intermediate redshifts ($z < 0.1$), enabling accurate identification of the galaxy cluster members, and statistically robust studies of galaxy properties.

Besides the DESI data, we incorporate the Siena Galaxy Atlas 2020 (SGA) catalog \citep{2023ApJS..269....3M} to complement the spectroscopic coverage, particularly for bright galaxies. While DESI is highly efficient in observing targets faint to $r\sim 20$ mag, its spectroscopic completeness for bright galaxies in DR1 can be incomplete. The SGA catalog addresses this gap by aggregating and homogenizing archival spectroscopic redshifts from multiple sources, including NED, SIMBAD and HyperLEDA. It also provides consistent photometric measurements across the surveyed area. In our study of galaxy clusters, the Brightest Cluster Galaxies (BCGs) are often too bright and may not have released spectroscopic data in DESI DR1. The SGA catalog therefore serves as a crucial supplement, ensuring a more complete census of spec-$z$s for bright sources within our cluster fields.

We note that the DESI DR1 spectroscopic sample is not fully complete, and the incompleteness varies with both stellar mass and cluster-centric radius. The primary limitation is the survey magnitude limit ($r \sim 20$~AB mag), corresponding to a stellar-mass completeness limit of $M_* \sim 10^{9.5}\, M_\odot$ at $z\sim 0.07$ and $M_* \sim 10^{8}\,M_\odot$ at $z\sim 0.02$ (Figure~\ref{masszspec}, left panel). Below these limits, the sample becomes increasingly incomplete, particularly for dwarf galaxies. The stellar mass distributions at different $R/R_{200}$ are, however, broadly similar (Figure~\ref{masszspec}, right panel), suggesting that this magnitude-dependent incompleteness affects all radial bins in a comparable manner.

An additional source of incompleteness arises from fiber collisions in the DESI focal plane, which most strongly affect the dense cluster cores ($R \lesssim 0.5\,R_{200}$). The galaxies most likely to be missed are quenched, low-mass satellites with weak emission lines, which are typically HI-poor. Their omission would, if anything, bias the stacked H\,{\sc{i}} mass slightly upward at small radii, making our measured H\,{\sc{i}} deficiency a conservative estimate of the true environmental quenching effect. We also note that our analysis is based on stacking the H\,{\sc{i}} spectra of galaxies with spectroscopic redshifts, rather than on absolute number counts or luminosity functions. The incompleteness therefore primarily reduces the sample size per bin rather than introducing a systematic bias in the average properties, provided the incompleteness is not strongly correlated with H\,{\sc{i}} content in a way that differs across radial bins. 

Finally, the combination of incomplete spectroscopy and the large FAST beam (3\arcmin) introduces a further systematic: the 6\arcmin\ extraction aperture at the position of a known member may also capture H\,{\sc{i}} emission from nearby galaxies that lack DESI redshifts. This flux is currently attributed entirely to the cataloged galaxy, potentially overestimating the per-galaxy $\langle M_{\rm HI}\rangle$. Future, more complete spectroscopic catalogs will help separate these contributions. A more detailed discussion of uncertainties is presented in Section~\ref{sec:uncertainty}.

\subsubsection{The FAST all sky H\,{\sc{i}} survey (FASHI)}
FASHI \citep{2024SCPMA..6719511Z, 2026arXiv260631539Z} provides an exceptional HI-selected sample for studying gas content in galaxy clusters. Covering $\sim$ 19,500 deg$^2$ with an rms sensitivity of about 0.56 mJy Beam$^{-1}$ at a velocity resolution of 6.4 km s$^{-1}$ and the beam size of $3'$. Its second data release (DR2) offers over 156 400 HI-detected sources within $z<0.09$. The on-going FASHI project is observing the sky of north with $ -14< \rm dec < 66 $ deg. 

FASHI observation setup is drift scan, and reduced by the FAST pipeline HIFAST \citep{2024SCPMA..6759514J}, including RFI mask, stand wave and radio continuum subtraction, calibration and re-gridding the data into the data cube with $1'\times 1'$ pixel size. The pipeline is run after each observing session and processes all available raw data in the field, including both the new and previous observations.
We make use of all the FASHI current coverage in this work.

The high sensitivity of FASHI enables the detection of galaxies with lower H\,{\sc{i}} masses at comparable redshifts compared to previous surveys like ALFALFA, making it particularly powerful for probing gas-deficient and gas-rich systems in diverse environments. Moreover, the blind survey provides large-area H\,{\sc{i}} datacubes that are crucially important for both H\,{\sc{i}} stacking and detailed statistical studies of the cold gas content in and around galaxy clusters.

\begin{figure*}
    \centering
    \includegraphics[width=0.99\linewidth]{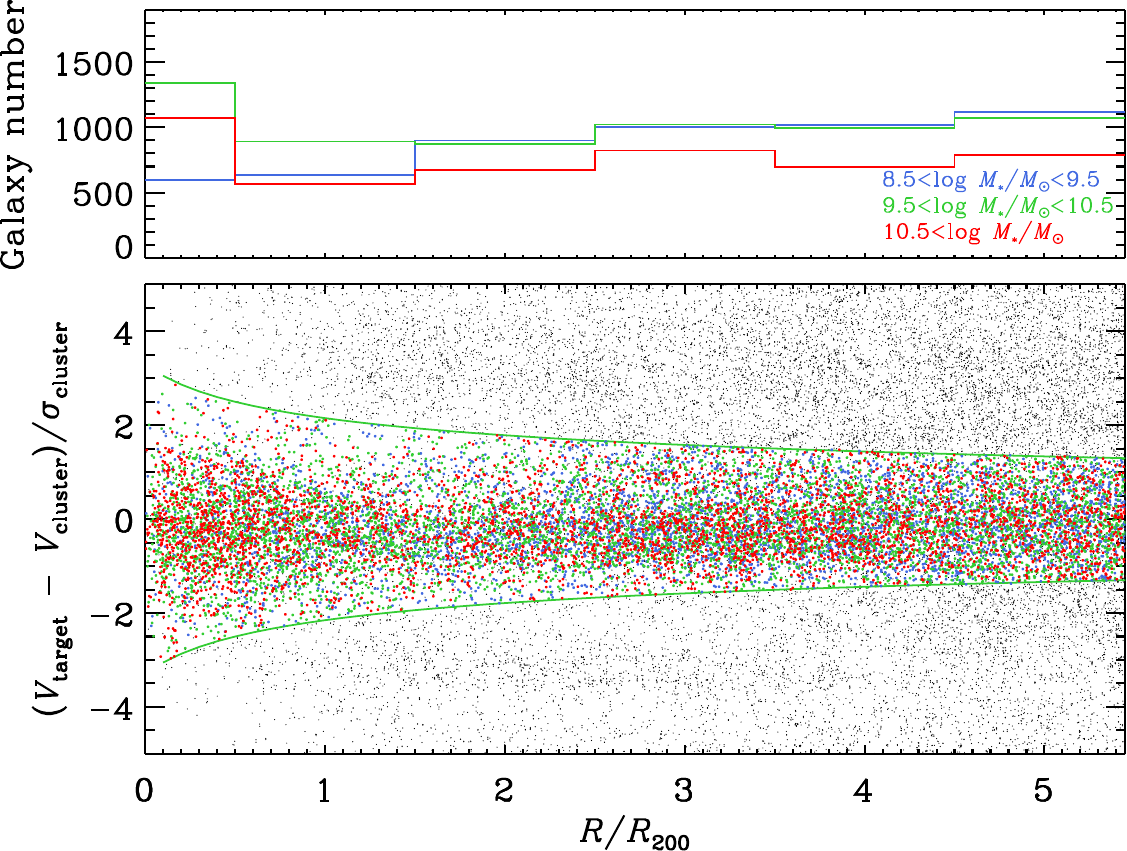}
    \caption{{\bf Lower panel:} The velocity–distance distribution for all sample clusters. The green lines mark the boundaries defined by Equation \ref{eqbound}. The blue, red and green dots are cluster members in low mass bin ($8.5<\log(M_*/M_\odot)<9.5$), medium mass bin ($9.5<\log(M_*/M_\odot)<10.5$) and high mass bin ($10.5<\log(M_*/M_\odot)$). The black dots are the targets close to the clusters, yet not bounded by the cluster. {\bf Upper panel:} Galaxy number histogram of the cluster member galaxies with spec-z for each mass bin. Number of galaxies in virialized system should decrease steady within $R/R_{200}$. The flat distribution of the low mass bin is caused by the incompleteness of the spec-z dwarf sample.
    }
    \label{phaseplot}
\end{figure*}

\subsection{Galaxy cluster sample selection}

To construct a reliable cluster sample, we utilize all systems from the AXES catalog irrespective of their original compact or extended classification, as this distinction is primarily influenced by the specific target detection kernel used during the X-ray source identification process and does not reflect intrinsic physical properties of the clusters \citep{2024A&A...690A..52D}.

We then combine the sky coverage of DESI and FASHI and apply the following selection criteria to the AXES clusters:

\begin{itemize}
    \item Cluster redshift $z_{\rm cluster} < 0.07$ to maintain high sensitivity for H\,{\sc{i}} observations with FAST 
    and to ensure that the corresponding H\,{\sc{i}} frequencies are covered by the FASHI survey. We emphasize that this selection does not require individual H\,{\sc{i}} detections for member galaxies. The $z_{\rm cluster}$ is adopted from AXES-SDSS catalog;
    \item Declination $\rm dec > -2^\circ$ to ensure sufficient sensitivity from FAST and the coverage from DESI observations;
    \item A focus on the most massive clusters $M_{200}>2\times 10^{14}M_\odot$, which generally possess the largest $R_{200}$ values. The large size of clusters also help in minimize source blending issues (further discussion on this selection will be provided in a subsequent section).
\end{itemize}

To establish a robust and uniformly selected spectroscopic member sample, we identify candidate member galaxies satisfying (1) A projected spatial separation: separation $\leq 5 R_{200}$. and (2) A line-of-sight velocity difference condition: $|z_{\rm spec} - z_{\rm cluster}| \cdot c < 3000~\text{km~s}^{-1}$. 
This relatively loose criterion is intended to ensure completeness at the pre-selection stage. A refined member selection is subsequently applied based on the cluster gravitational potential (Section~2.3), which accounts for the radial dependence of the escape velocity. Then we retain only those clusters for which the number of spectroscopic members provided by DESI exceeds the number contributed by the SGA catalog, thereby prioritizing the deeper and more uniform DESI spectroscopy to the low mass galaxies, while still utilizing the SGA as a valuable complementary dataset. We note that this criterion does not explicitly account for spatial variations in spectroscopic completeness within individual clusters, as the current DESI DR1 data may contain inhomogeneities due to survey tiling and fiber assignment. However, since our analysis relies on stacking galaxies in bins of cluster-centric radius and stellar mass, and combines multiple clusters, the impact of such incompleteness is expected to be mitigated in a statistical sense.

We identify that the $5\times R_{200}$ area, specially for some local massive galaxy clusters, can extend to several degrees in diameter, therefore some regions can overlap with each other. In our catalog, cluster AXES\_E\_2416 and AXES\_E\_2420 has a spatial distance about 1.8 deg, sharing a large overlap area; cluster AXES\_E\_2496 and AXES\_C\_2498 are identical but with different ID. So we remove AXES\_C\_2498 and AXES\_E\_2420 in the following analyze. The final sample for H\,{\sc{i}} stacking includes 36 galaxy clusters from AXES-SDSS, and listed in Table \ref{tab1}. 

\begin{figure*}
    \centering
    \includegraphics[width=0.48\linewidth]{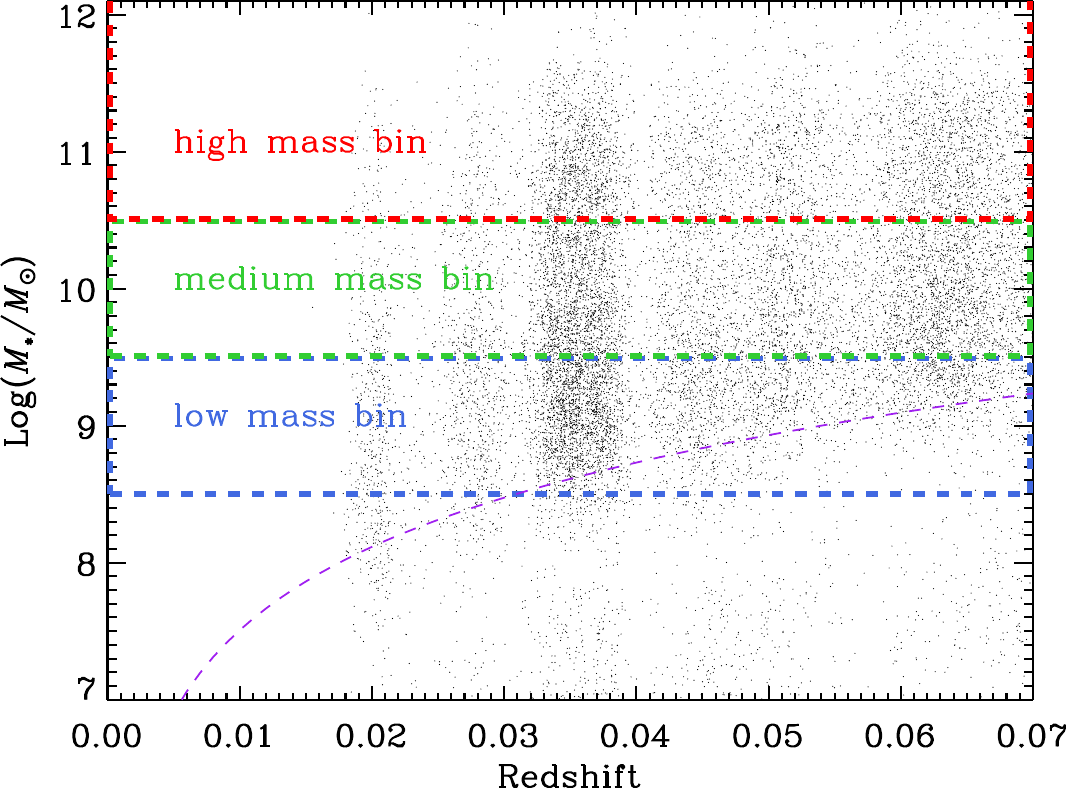}
    \includegraphics[width=0.48\linewidth]{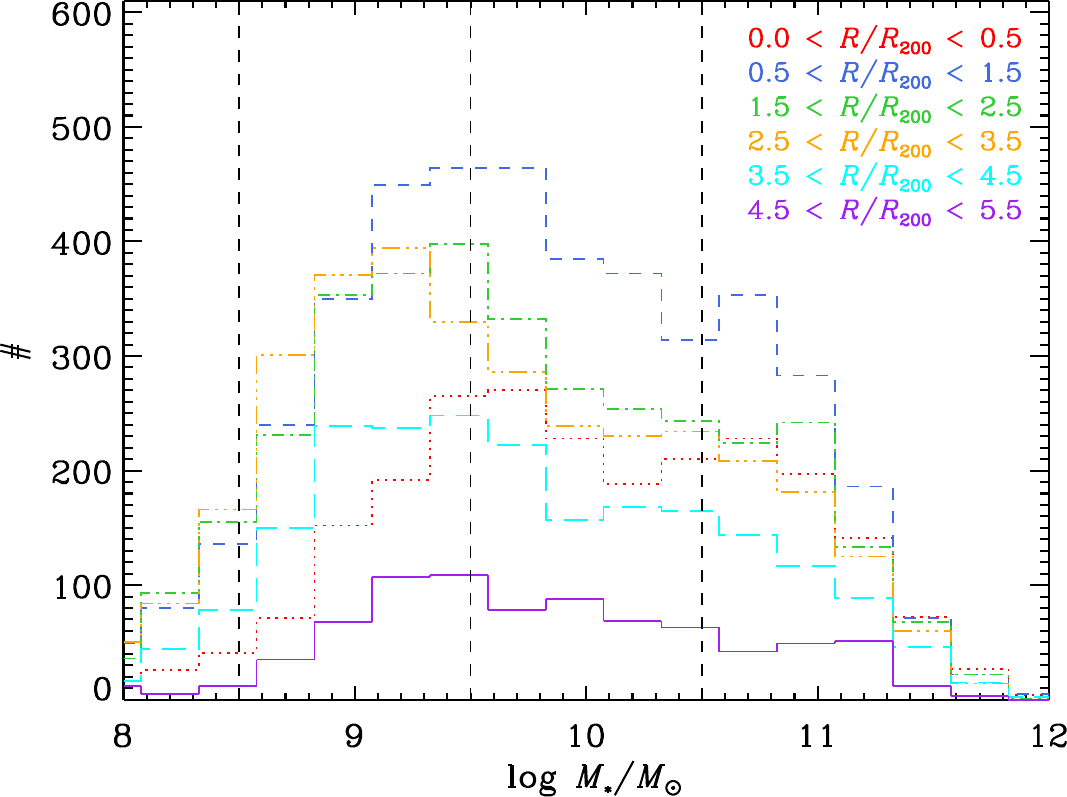}    
    \caption{{\bf Left: }Redshift vs $M_*$ of the member galaxies we selected. The stellar mass lower to $10^9 M_\odot$ is not complete to the redshift 0.06. The dashed purple line shows the stellar mass for a target with $r = 20$ AB mag and $g-r = 0.5$ at each redshift, which is roughly the detection limit of the DESI spectroscopic sample.
    {\bf Right:} Histogram of the stellar mass at different $R/R_{200}$. The stellar mass distribution for galaxies at different radius are similar. 
    }
    \label{masszspec}
\end{figure*}

\subsection{Identification of the member galaxies in clusters}

We make use of velocity-distance distribution from DESI and SGA spec-z catalogs to identify the cluster member galaxies. In projected phase--space (clustercentric radius versus line--of--sight velocity), the dynamical boundary of a cluster can be approximated by the escape velocity profile of a Navarro--Frenk--White (NFW) halo \citep{1997ApJ...490..493N, 2001MNRAS.321..155L}. The three--dimensional escape velocity at radius $r$ is obtained from the NFW potential as $v_{\mathrm{esc}}(r) = \sqrt{2\,|\Phi_{\mathrm{NFW}}(r)|}$. The NFW potential at radius $R$ is
\begin{equation}
    \Phi_{\mathrm{NFW}}(r) = -\,\frac{G M_{200}}{R_{200}} \;
\frac{\ln(1 + c\,x)}{x\, \left(\ln(1+c) - \frac{c}{1+c}\right)},~~ x \equiv \frac{R}{R_{200}},
\end{equation}
where $c = 5.71\,\left(\frac{M_{200}}{2\times10^{12}h^{-1}M_\odot} \right)^{-0.084} \left( 1 + z_{\mathrm{cluster}} \right)^{-0.47}$ is the halo concentration, which is adopted from \citet{2008MNRAS.390L..64D}.

Since only line--of--sight velocities are observed, the envelope must be scaled by a projection factor $f_{\mathrm{los}}$. 
Numerical simulations indicate that the typical ratio between the one--dimensional (line--of--sight) and three--dimensional escape velocities lies in the range $0.6$--$0.8$, depending on orbital anisotropy \citep{1999MNRAS.309..610D}.
Following common practice, we adopt $f_{\mathrm{los}} \approx 0.7$, which provides a good match to the observed ``trumpet--shaped'' distribution of member galaxies in projected phase--space (Figure \ref{phaseplot}).
Galaxies lying within
\begin{equation}\label{eqbound}
\left| \frac{v - v_{\mathrm{cluster}}}{\sigma_v} \right| \;\leq\; 
\frac{f_{\mathrm{los}}\,v_{\mathrm{esc}}(R)}{\sigma_v},
\end{equation}
are considered as cluster members, where $v_{\mathrm{cluster}}$ is adopted from the AXES-SDSS catalog, and the $\sigma_v$ is the velocity dispersion, which is estimated by the scaling relation from \citet{2008ApJ...672..122E}: 
\begin{equation}\label{eqsigma}
    \sigma_v=1082.9\left[\frac{h(z)M_{200}}{10^{15}M_\odot}\right]^{0.33},
\end{equation}
where $h(z) = H(z)/100$. 

We stress that the analytic framework adopted here, including the use of the NFW-based escape-velocity profile as a function of $R/R_{200}$ is intended only to provide an approximate dynamical boundary for selecting cluster member galaxies. The method performs reasonably well within $R \lesssim R_{200}$, where the velocity distribution remains relatively well behaved. However, at larger radii (e.g., $R \gtrsim 2R_{200}$), the uncertainties grow substantially due to the breakdown of spherical symmetry, the increasing contribution from interlopers, and the reduced constraining power of the assumed mass profile. Therefore, membership identification based on this formalism should be interpreted as approximate, especially in the cluster outskirts.

In Figure \ref{phaseplot}, we show the selection of the method based on the velocity-distance distribution. Targets at large $R/R_{200}$ with large offset from $(v_{\rm target} - v_{\rm cluster})/\sigma_v$ are treated as fore/background targets that are not bounded by the cluster. The decrease of the galaxy number within $1\times R/R_{200}$ indicates the virialization of the central galaxies, and the flat distribution of the low stellar mass bin is caused by the incompleteness of the sample in the cluster center, where galaxies can be crowded. In Figure \ref{member_selection1} we show the distribution of the member galaxies  we selected for each cluster. 

At the project distance $> 1\times R/R_{200}$, some cluster members have a small offset of $(v_{\rm target} - v_{\rm cluster})/\sigma \sim 1$, which are caused by the disturbed state of the cluster that the redshift of BCG is offset from the cluster dark matter halo redshift, or the member galaxies in a filament at the outskirts of the clusters are systematically offset from the central galaxies. Since our main results are based on a statistical analysis using stacking, we do not adjust the cluster redshifts based on the results of $(v_{\rm target} - v_{\rm cluster})/\sigma$ distribution.

\subsection{Stellar mass of cluster members}

We estimate the stellar mass ($M_*$) for all cluster members using the mass-to-light ratio ($M/L$) relation from \citet{2003ApJS..149..289B}, applied to the $g$- and $r$-band photometry, and adopt the relation between stellar mass-to-light ratio and optical color. Specifically, we use $\log(M_*/L_r) = a + b \times (g-r)$, where $a = -0.306 -0.15$ and $b = 1.097$, for a Kroupa initial mass function \citep{1993MNRAS.262..545K}.

For the targets selected from DESI, the photometry is performed by using the Tractor framework \citep{2016ascl.soft04008L, 2016AJ....151...36L} to ensure consistent flux measurements across optical and WISE bands. On the other hand, galaxies in the SGA catalog often have large angular sizes and were modeled separately with surface brightness model in $g$ and $r$ bands. We apply the same $M_*$ estimation method in \citet{2003ApJS..149..289B} to maintain uniformity across the entire sample. The typical uncertainty in the stellar mass estimates is approximately 0.3~dex.
    
We show the redshifts vs stellar mass in Figure \ref{masszspec} left panel. Our cluster sample are mainly at $0.02<z<0.07$. To estimate the detection limit of the stellar mass, we show the stellar mass with the observed magnitude $r = 20$ with $g-r = 0.5$ in Figure \ref{masszspec}, which is about $10^8M_\odot$ at $z = 0.02$, and about $10^9M_\odot$ at $z = 0.07$. Therefore, the low mass bin of our sample would suffer from the selection bias.

\section{H\,{\sc{i}} spectrum stacking}
\subsection{Subgroups of Cluster Member Galaxies}\label{subgroup}

The sensitivity of FASHI is insufficient to detect the majority of individual galaxies within each cluster. However, its extensive coverage area still enables a statistical investigation of the global H\,{\sc{i}} properties for the cluster population. Meanwhile, the large number of spec-$z$ identified member galaxies enables us to group the galaxies into different classes (see the histogram in the upper panel of Figure \ref{phaseplot}). Elliptical or central dominated galaxies in clusters are generally HI-deficient, so most H\,{\sc{i}} measurements in dense environments are only upper limits \citep{2012MNRAS.422.1835S, 2014MNRAS.444.3388S, 2024ApJ...963...86L}. As a result, the environmental effect on H\,{\sc{i}} content remains poorly constrained with direct detection. Consequently, we lack robust constraints on the intrinsic HI-to-stellar mass ratio across different environments. 

The large sky coverage of FAST together with the extensive spectroscopic redshift samples from DESI and SGA provide an ideal opportunity to investigate this issue by stacking. We divide our member galaxy sample according to stellar mass and environment. Previous observations show that galaxy evolution stage is different at $M_*\sim 3\times 10^{10} M_\odot$, above which the old stellar population increasing rapidly \citep{2003MNRAS.341...54K}. So we set one stellar mass criterion of $\log(M_*/M_\odot) = 10^{10.5}$. On the other hand, H\,{\sc{i}} observations of the low mass galaxies show that the H\,{\sc{i}} mass would exceed the stellar mass at about $M_*<10^9M_\odot$ \citep{2012AJ....143..133H}. From Figure \ref{masszspec} we can see galaxies with $M_*<10^9M_\odot$ would suffer from serious incompleteness at $z>0.04$, we set the other criterion for dwarf and medium mass bin at $\log(M_*/M_\odot) = 10^{9.5}$. Therefore,  we divide our member galaxy sample into three mass bins of $8.5<\log M_*/M_\odot < 9.5$,  $9.5<\log M_*/M_\odot < 10.5$,  $10.5<\log M_*/M_\odot$ (Figure \ref{masszspec} color boxes). Most importantly, to understand the H\,{\sc{i}} properties in different environment, we divide the sample into six bins ($R/R_{200} \in [0, 0.5], (0.5, 1.5], (1.5, 2.5], (2.5, 3.5], (3.5, 4.5], (4.5, 5.5]$) based on the $R/R_{200}$, where $R$ is the projection distance to the center of the galaxy clusters. We have in total 18 subgroups to cover the galaxies with different stellar mass and projection distances. The stellar mass at each $R/R_{200}$ shows a similar histogram (right panel of Figure \ref{masszspec}), and thus the average stellar masses at different projected radius are also similar.

\begin{figure}
    \centering
    \includegraphics[width=0.98\linewidth]{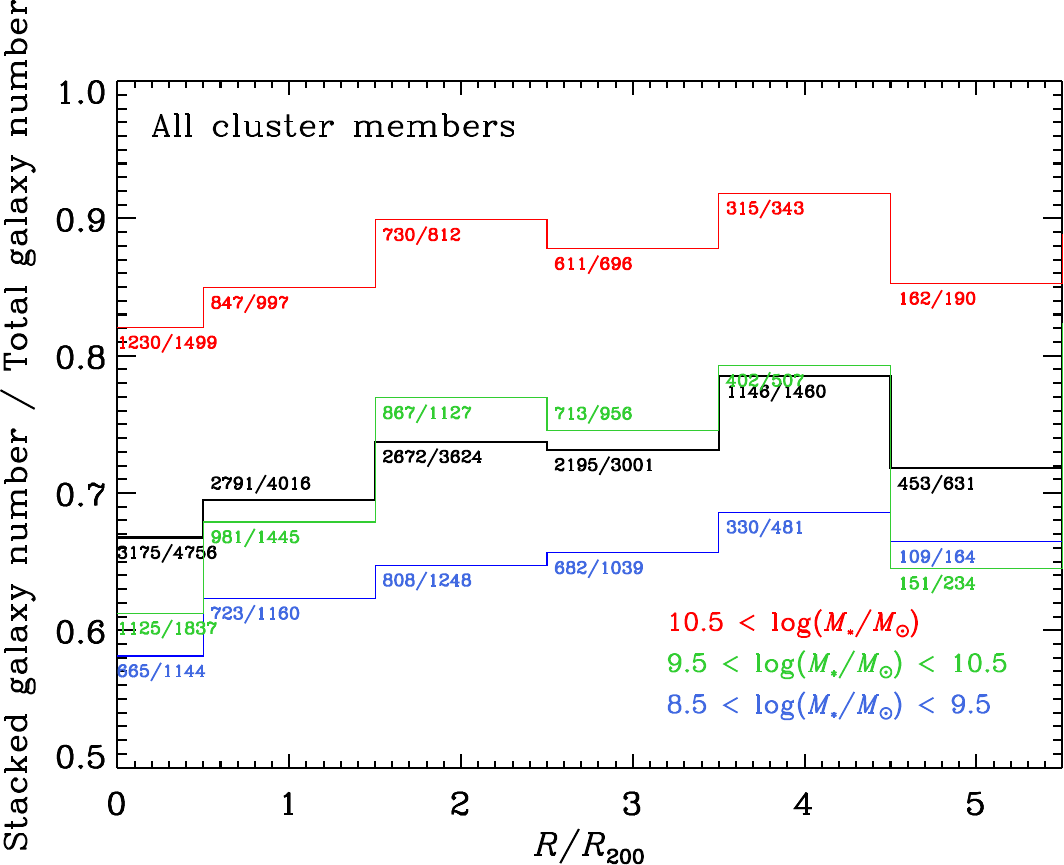}
    \caption{The ratio of the number of targets for stacking and all galaxy numbers. The black line is the total number ratio for all member galaxies, and the blue, green and red lines are the low, medium and high mass bin galaxy members. We denote the number of stacking targets and number of total galaxies in each mass and radius bin near the ratio.  
    }
    \label{NumberRatio}
\end{figure}

\subsection{Identification and Treatment of Blended Sources}\label{blending}

The beam size of the FAST 19-beam receiver is approximately $3'$ at $z < 0.1$, corresponding to about 100 kpc at $z = 0.05$. When two galaxies are close in right ascension, declination, and velocity space, the extracted H\,{\sc{i}} spectra from their positions are nearly identical. If both are included in stacking, this would result in double-counting and lead to an overestimate of the average H\,{\sc{i}} mass. On the other hand, completely removing all blended targets would discard merging systems-which are common in clusters-and bias the sample against such objects. Observations indicate that mergers can expel gas from galaxies during different merging stage \citep{2001A&A...377..812V, 2022Natur.610..461X, 2023A&A...670A..21J, 2023ApJ...954...74C}, resulting in reduced H\,{\sc{i}} mass. Therefore, using only clearly isolated member galaxies would yield an average H\,{\sc{i}} mass higher than the true value, especially for crowded environment such as in clusters.

To address the blending issue, we constructed a catalog with blended systems combined for H\,{\sc{i}} spectral stacking. For each potential blended system, we identify and combine sources within a  radius of $3'$ and a velocity offset $V_{\rm offset} < 300$ km/s into the single massive target, and then removed the blended targets from the original member catalog. This process reduces about 30\% of the total galaxy number. The stellar mass of the removed targets is summed and assigned to this new composite object. The total stellar mass $M_{\rm total}$ is then used to place the system into one of the three stellar mass bins in Figure \ref{masszspec}. In less than 5\% cases, three closest galaxies are combined into one target, and the rest combined targets are the sum of two galaxies. 

Our results show that the stacked galaxy number ($N_{\rm galaxies}^{\rm stacking}$) is about 70\% of the total galaxy number ($N_{\rm galaxies}^{\rm total}$). Then we can correct the $M_{\rm HI}^{\rm avg \, code} (= \Sigma_{i=1}^{N_{\rm galaxies}} M_{\rm HI}^{i} / N_{\rm galaxies}$) into the intrinsic average H\,{\sc{i}} mass $M_{\rm HI}^{\rm avg \, corr} (= M_{\rm HI}^{\rm avg \, code} \times N_{\rm galaxies}^{\rm stacking} / N_{\rm galaxies}^{\rm total})$, which is about 70\% of the averaged H\,{\sc{i}} mass (about 0.15 dex). So the number ratio of $N_{\rm galaxies}^{\rm stacking} / N_{\rm galaxies}^{\rm total}$ can be treated as a correction factor to the H\,{\sc{i}} mass given by H\,{\sc{i}} spectrum stacking code (i.e., $M_{\rm HI}^{\rm avg \, code}$).

We show the target number ratio in each $R/R_{200}$ bin in Figure \ref{NumberRatio}. The low mass bin has the lowest number ratio about 0.6, which will reduce the H\,{\sc{i}} mass from stacking results for 0.2 dex. We note that our treatment of the blending issue will always reduce the number of low mass targets, and the sum of the galaxy mass would increase the stellar mass to less than 0.3 dex. Among the blended systems, approximately 45\% combine galaxies that would individually fall into different stellar mass bins. The net fraction of the total sample affected by such cross-bin blending is ~14\%.

We emphasize that although galaxies appear crowded in projected cluster images and may lead to fiber-collision issues in the DESI spectroscopic catalog, blending is less severe in velocity space due to the high velocity dispersion near cluster centers (e.g., at $R/R_{200} \sim 0$). The spatial limitations discussed here are primarily due to the resolution of FAST. Future high-resolution observations, such as those with MeerKAT or the FAST Core Array \citep{2024AstTI...1...84J}, or wide field interferometry survey such as Widefield ASKAP L-band Legacy All-sky Blind surveY \citep[WALLABY,][]{2020Ap&SS.365..118K}, Apertif imaging survey \citep{2022A&A...667A..38A}
 will help to better resolve and constrain the H\,{\sc{i}} properties in dense environments.

\begin{sidewaysfigure*}
    \centering
    \includegraphics[width=0.17\linewidth]{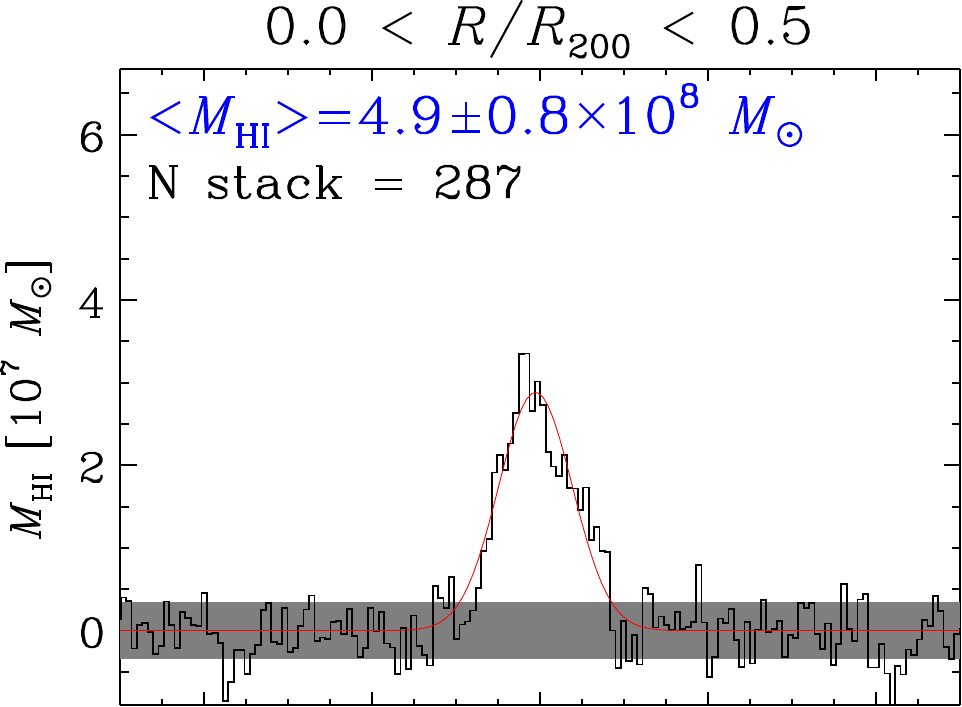}
    \includegraphics[width=0.149\linewidth]{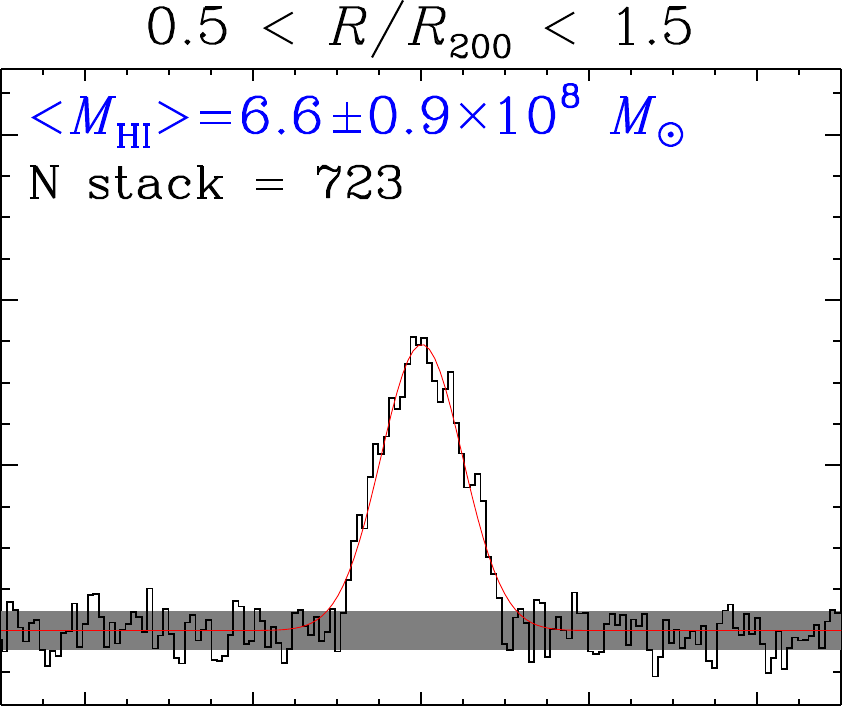}
    \includegraphics[width=0.149\linewidth]{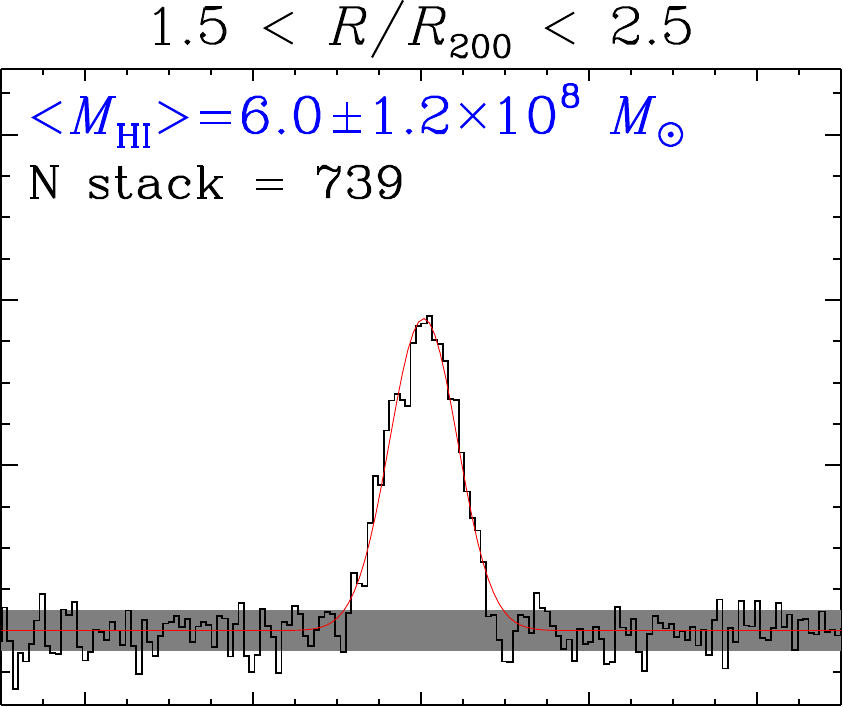}
    \includegraphics[width=0.149\linewidth]{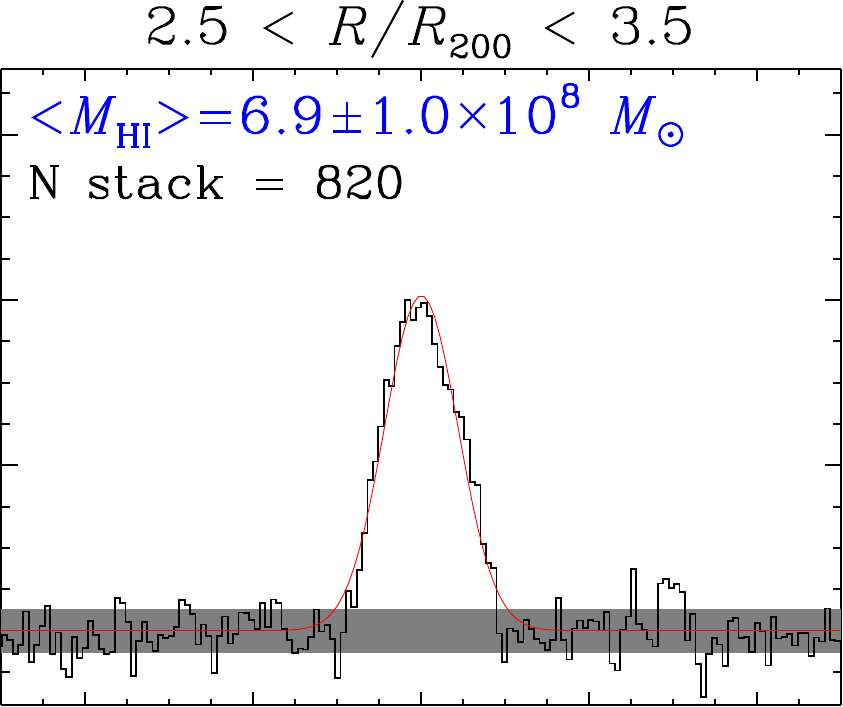}
    \includegraphics[width=0.149\linewidth]{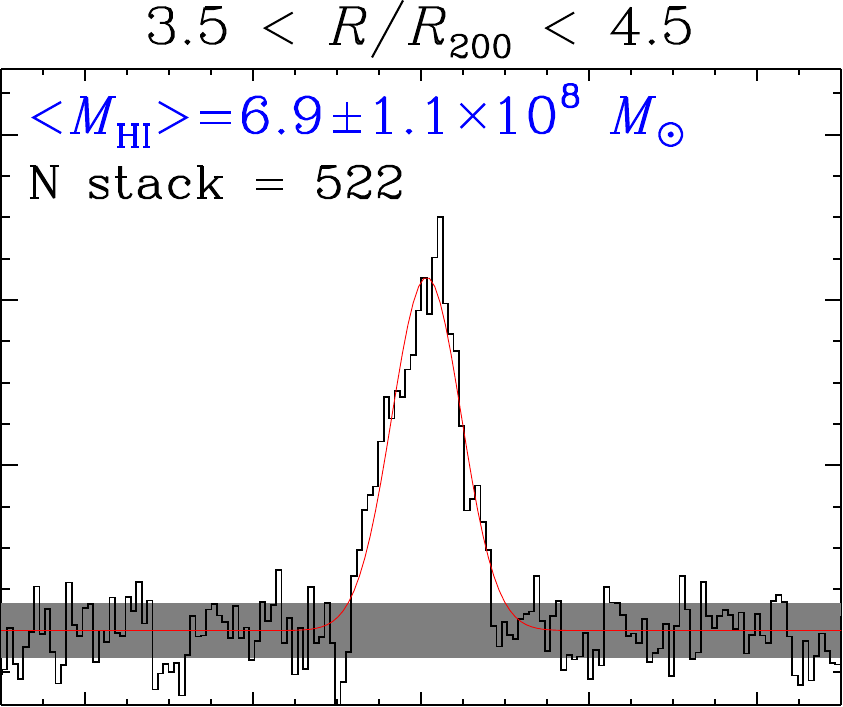}
    \includegraphics[width=0.149\linewidth]{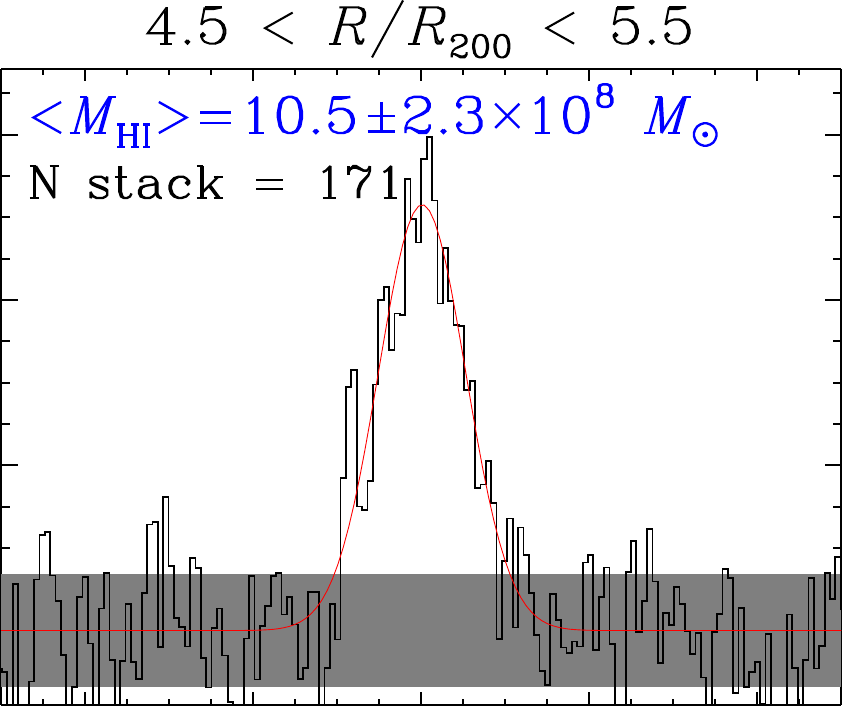}
    \includegraphics[width=0.17\linewidth]{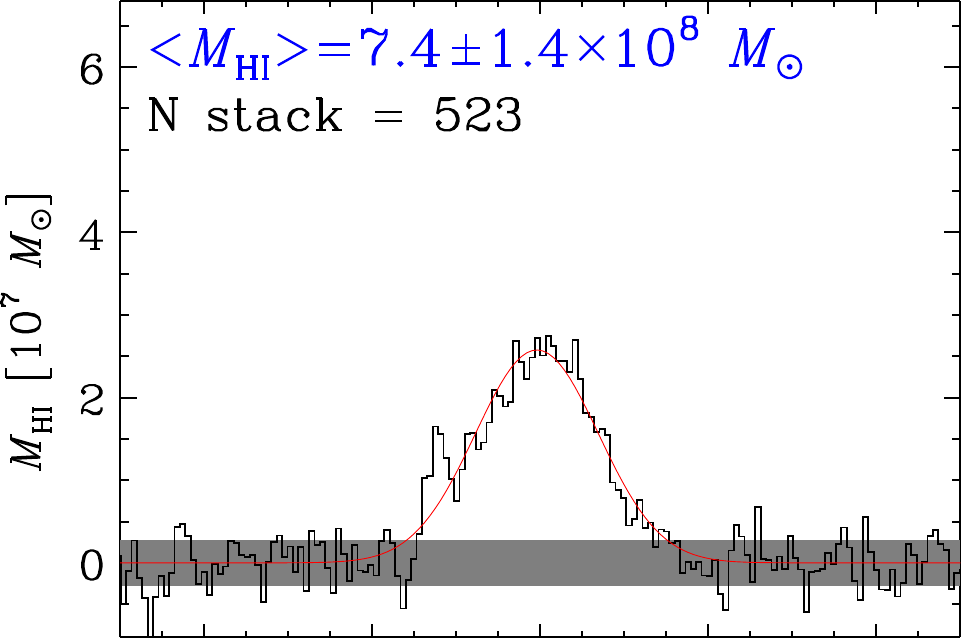}
    \includegraphics[width=0.149\linewidth]{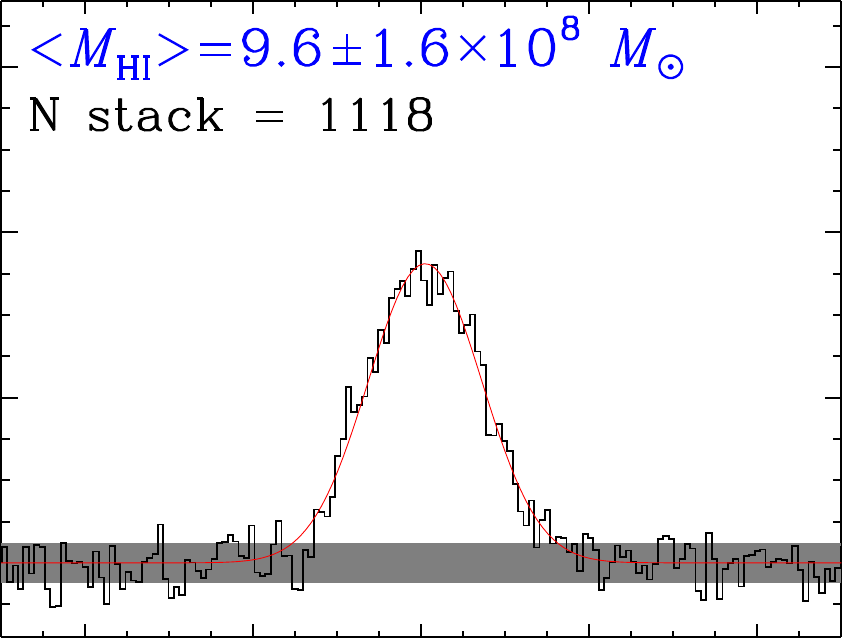}
    \includegraphics[width=0.149\linewidth]{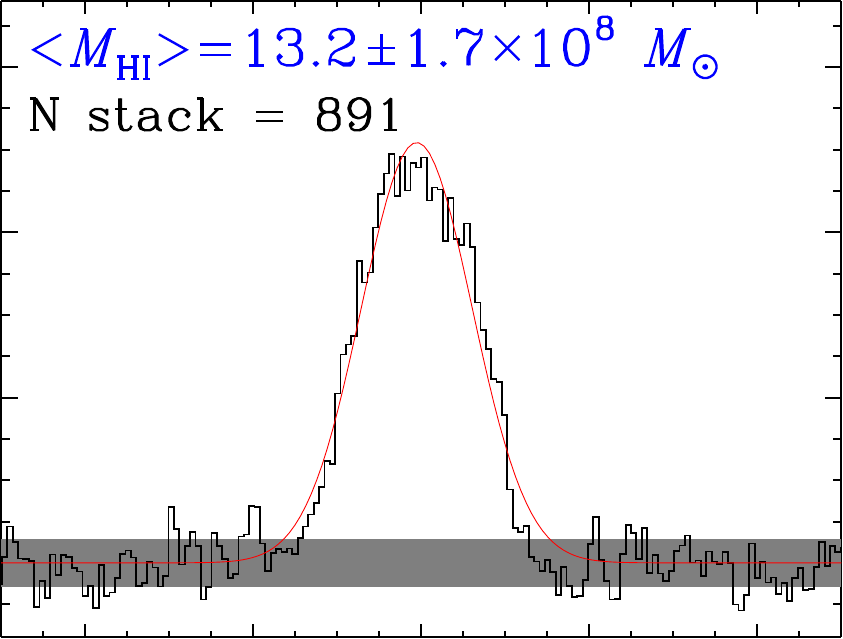}
    \includegraphics[width=0.149\linewidth]{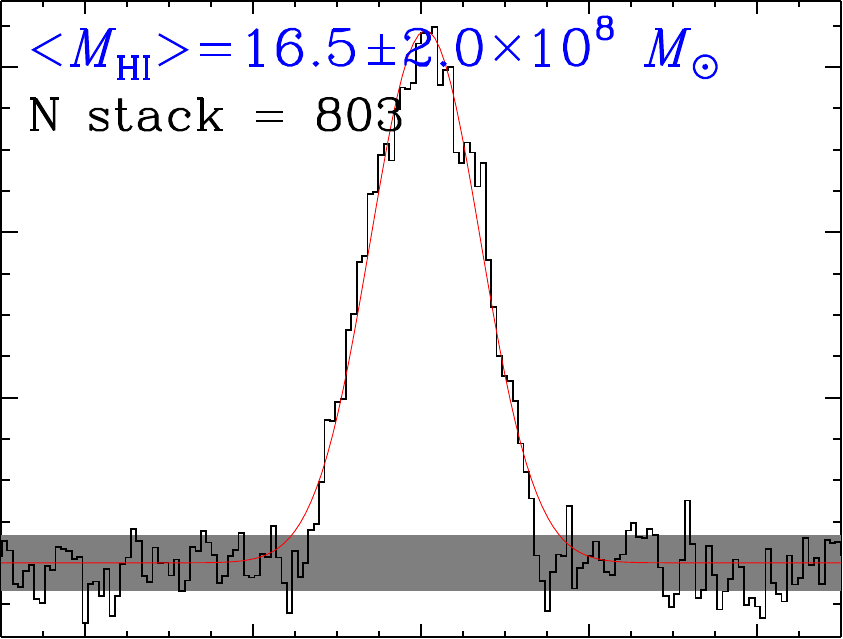}
    \includegraphics[width=0.149\linewidth]{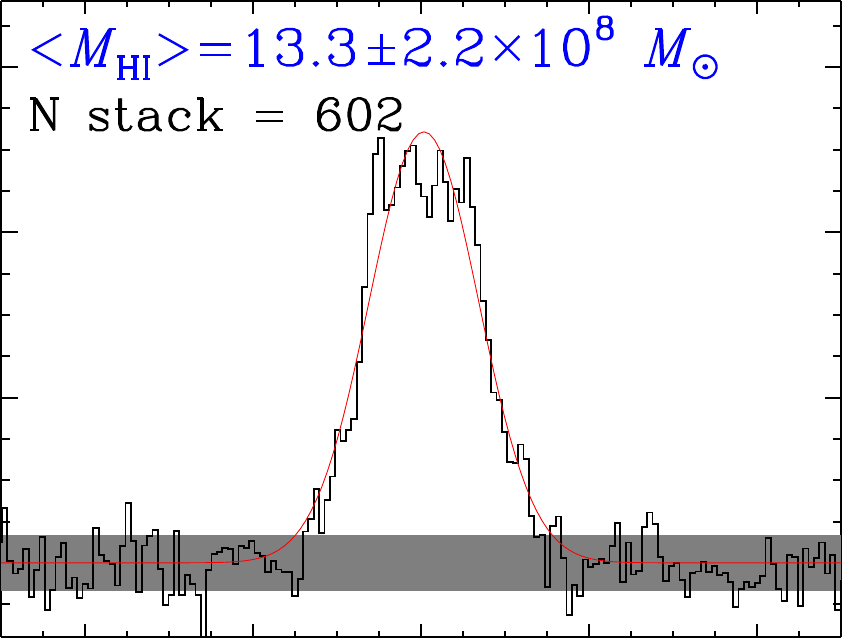}
    \includegraphics[width=0.149\linewidth]{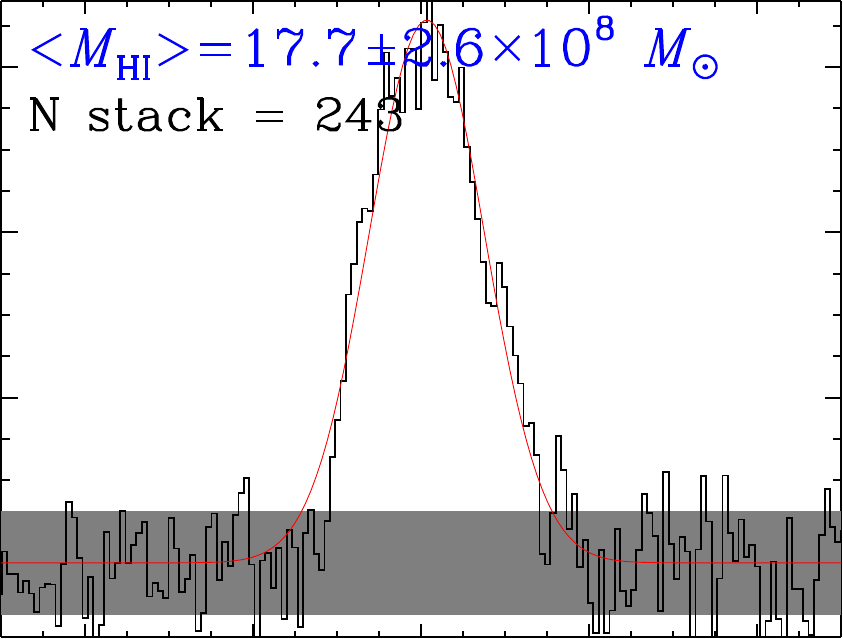}
    \includegraphics[width=0.17\linewidth]{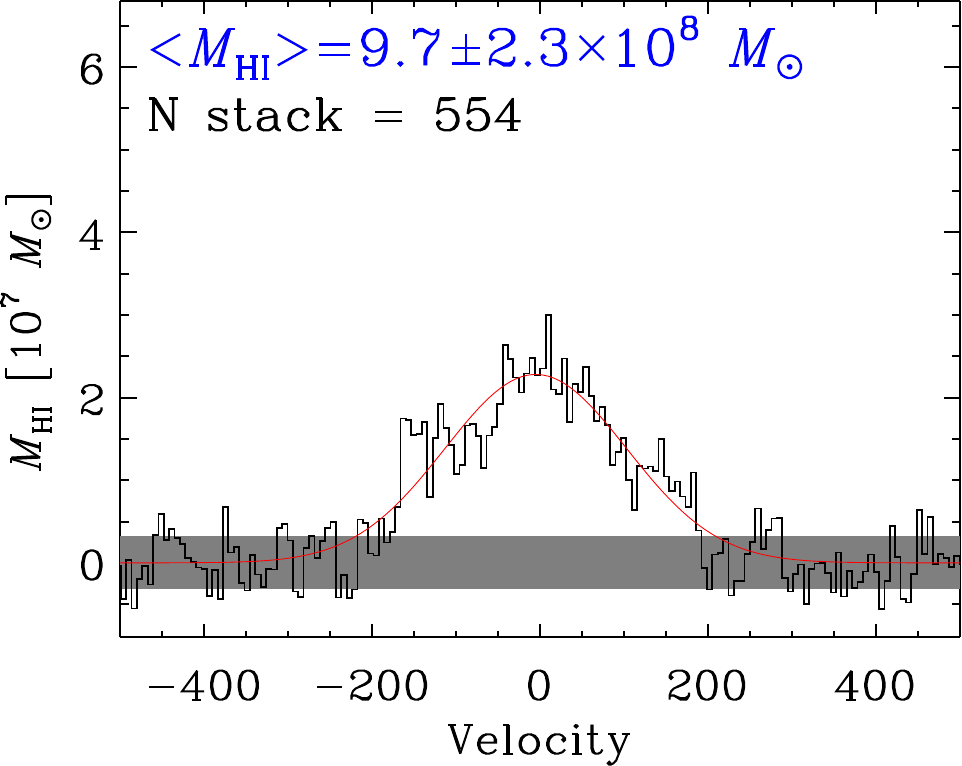}
    \includegraphics[width=0.149\linewidth]{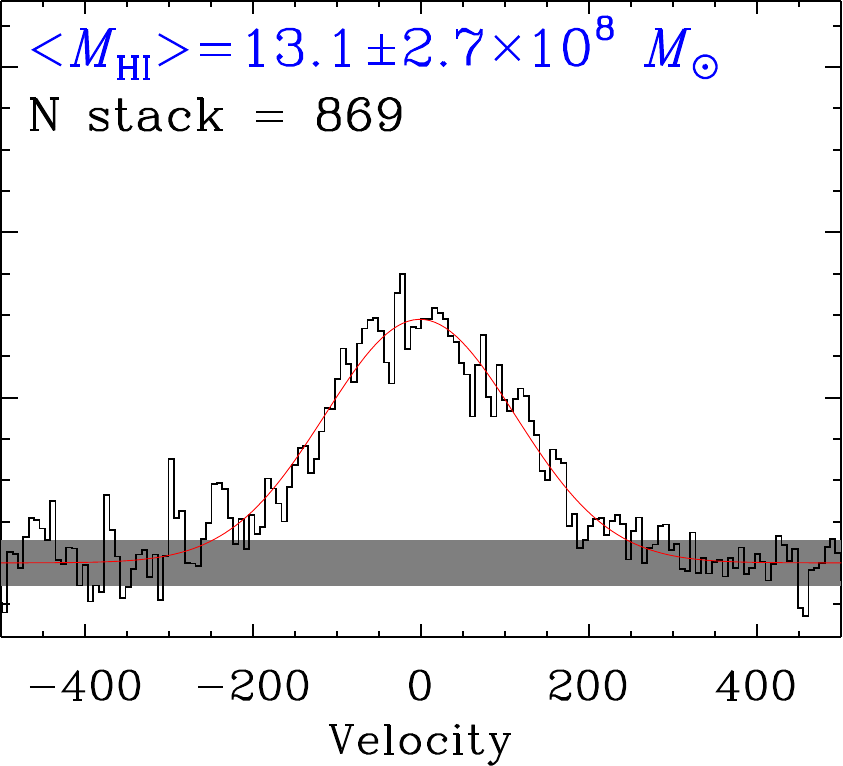}
    \includegraphics[width=0.149\linewidth]{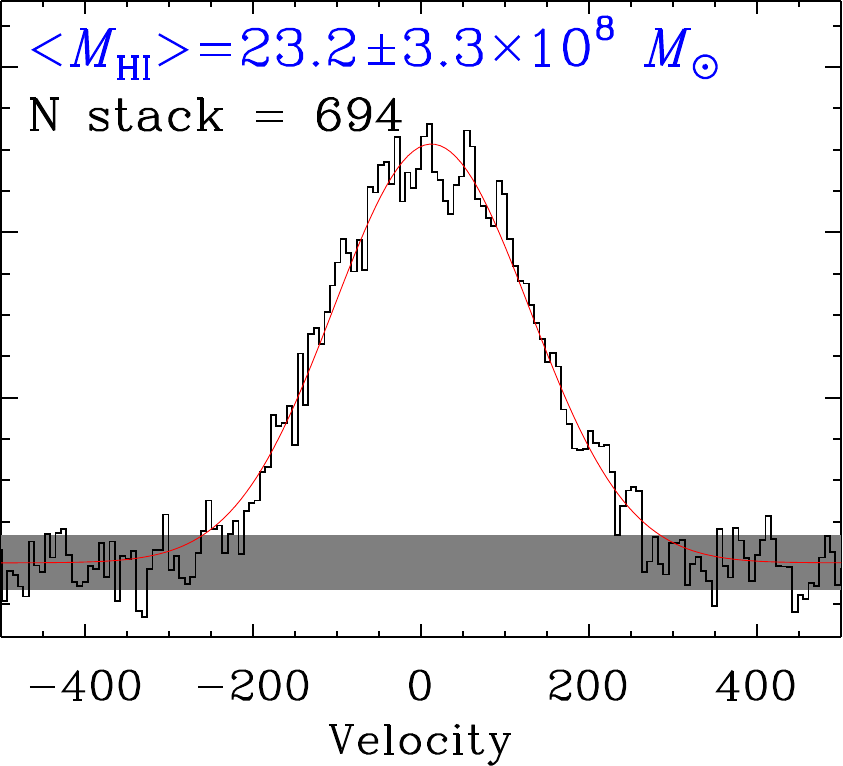}
    \includegraphics[width=0.149\linewidth]{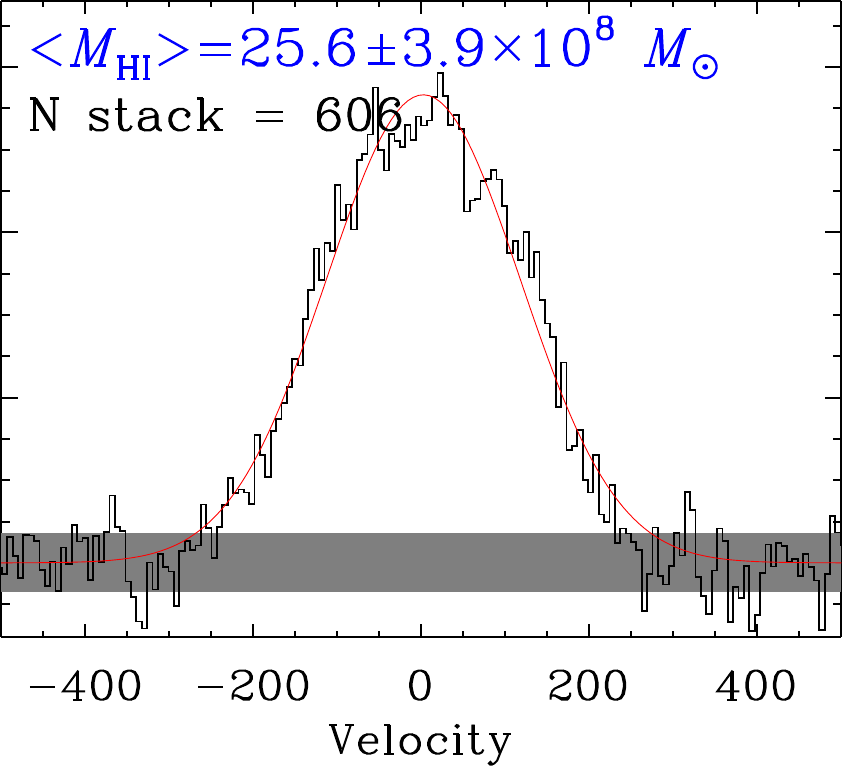}
    \includegraphics[width=0.149\linewidth]{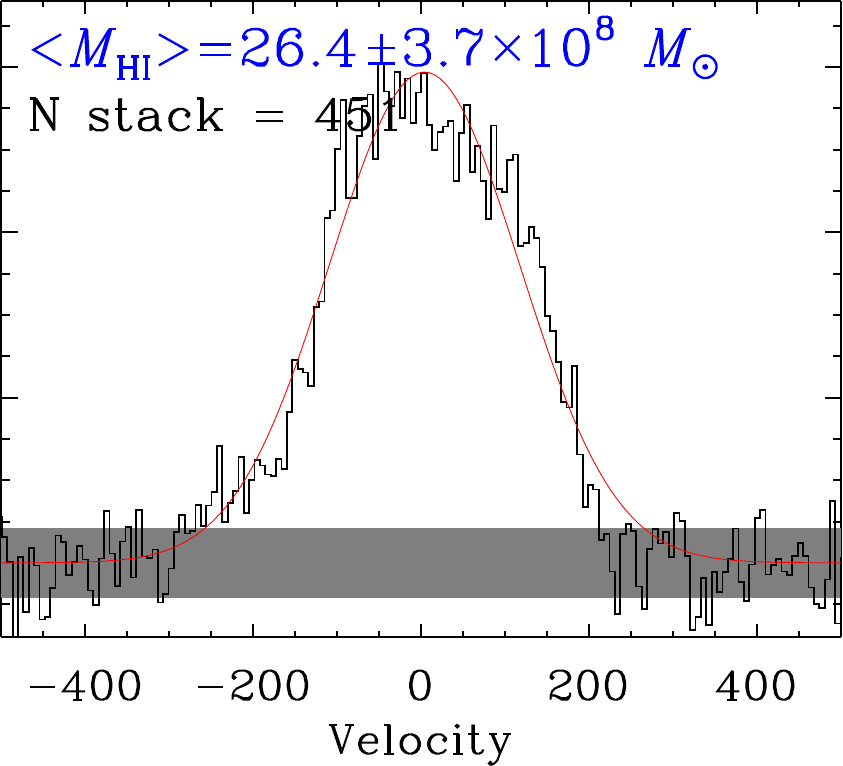}
    \includegraphics[width=0.149\linewidth]{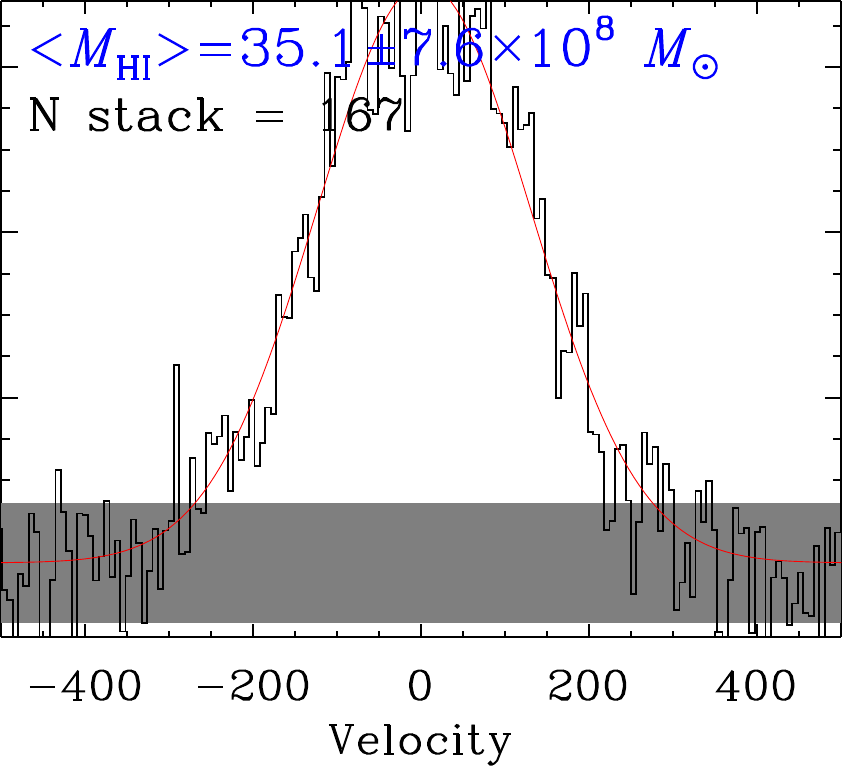}
    \caption{Stacking H\,{\sc{i}} spectra for the 18 bins. The top, middle and bottom panels are the stacking results for low, medium and high mass bins, respectively. The $R/R_{200}$ for each column is denoted on top of the panels. The stacked H\,{\sc{i}} spectra do not show double horn profile, but single-peaked or top-flatted. So we fit the H\,{\sc{i}} flux by gaussian function, and obtain the H\,{\sc{i}} mass by the gaussian fitting. The shaded regions represent 1$\sigma$ noise of each spectrum.}
    \label{HIstack}
\end{sidewaysfigure*}

\subsection{Stacking method of H\,{\sc{i}} spectrum}

To measure the average H\,{\sc{i}} flux of the member galaxies, we first extract a 1D spectrum at the coordinates of each galaxy from the FASHI data cubes (which are in units of mJy/beam) using a 6' diameter aperture, with conversion from mJy/beam to mJy. The extracted spectra are then stacked using the HISS code \citep[][]{2019MNRAS.487.4901H}\footnote{\url{https://github.com/healytwin1/HISS}}. HISS takes as input the ID, redshift, stellar mass, and filename of each extracted spectrum. It converts the flux density into H\,{\sc{i}} mass per spectral bin, shifts each spectrum to the rest frame, and then performs the stacking with a weighting factor of $1/{\rm rms}^2$ for each spectrum.

We include all member galaxies with spectroscopic redshifts in the stacking, regardless of whether they are individually detected in HI. After stacking, all 18 bins (three bins for stellar mass and six bins for $R/R_{200}$) show significant average H\,{\sc{i}} detections (Figure \ref{HIstack}). The stacked spectrum yields the average H\,{\sc{i}} mass $\langle M_{\rm HI} \rangle$ for each bin. We then divide $\langle M_{\rm HI} \rangle$ by the mean stellar mass $\langle M_* \rangle$ of galaxies in that bin to obtain the average HI-to-stellar mass ratio $\langle M_{\rm HI}\rangle/\langle M_* \rangle$.

The $6'$ aperture corresponds to $\sim$140 kpc at $z=0.02$ and $\sim$400 kpc at $z=0.06$, which is sufficiently large to encompass most of the H\,{\sc{i}} associated with individual galaxies.

We adopt a fixed angular aperture ($6'$ diameter) rather than a fixed physical aperture. A fixed physical aperture would require redshift-dependent corrections to account for flux outside the extraction region, introducing additional model-dependent uncertainties. Since our analysis focuses on relative trends across $R/R_{200}$ and stellar mass, rather than absolute H\,{\sc{i}} masses, the choice of aperture does not affect our conclusions as long as it is applied uniformly.

The uncertainty of the stacked H\,{\sc{i}} flux is relatively small, since the large number of targets reduces the rms noise by a factor of $1 / \sqrt{N}$. However, the non-uniform distribution of H\,{\sc{i}} masses within the stacking sample can also affect the averaged result. If only a few targets have bright H\,{\sc{i}} emission while most have very faint or undetected H\,{\sc{i}} signals, the stacked flux would be dominated by those few bright sources, and the resulting average would not represent the typical H\,{\sc{i}} content of the sample. To account for this, we perform 100 times of random resampling, in each of which 50\% of the galaxies in a stacking bin are randomly selected and their H\,{\sc{i}} spectra are re-stacked. The dispersion among the 100 stacked flux measurements is taken as an estimate of the uncertainty of the average H\,{\sc{i}} flux. In all 100 stacking, the stacked H\,{\sc{i}} signals are significantly detected, indicating the robustness of our stacking results. We fit the distribution of the 100 stacked fluxes with a Gaussian function and adopt its standard deviation ($\sigma$) as one component of the total flux uncertainty. In addition, we include a 10\% calibration error, which are added in quadrature to the flux uncertainty.

HISS provides the number of spectra used in the stacking process. By comparing the number of input apertures with the number of successfully extracted spectra, we can roughly estimate the FAST sky coverage to our input sample. Our results show that the minimum value of $N_{\rm input} / N_{\rm output}$ is 97\%, indicating that the FASHI survey has effectively covered almost all the selected clusters.

\begin{figure*}
    \centering
    \includegraphics[width=0.48\linewidth]{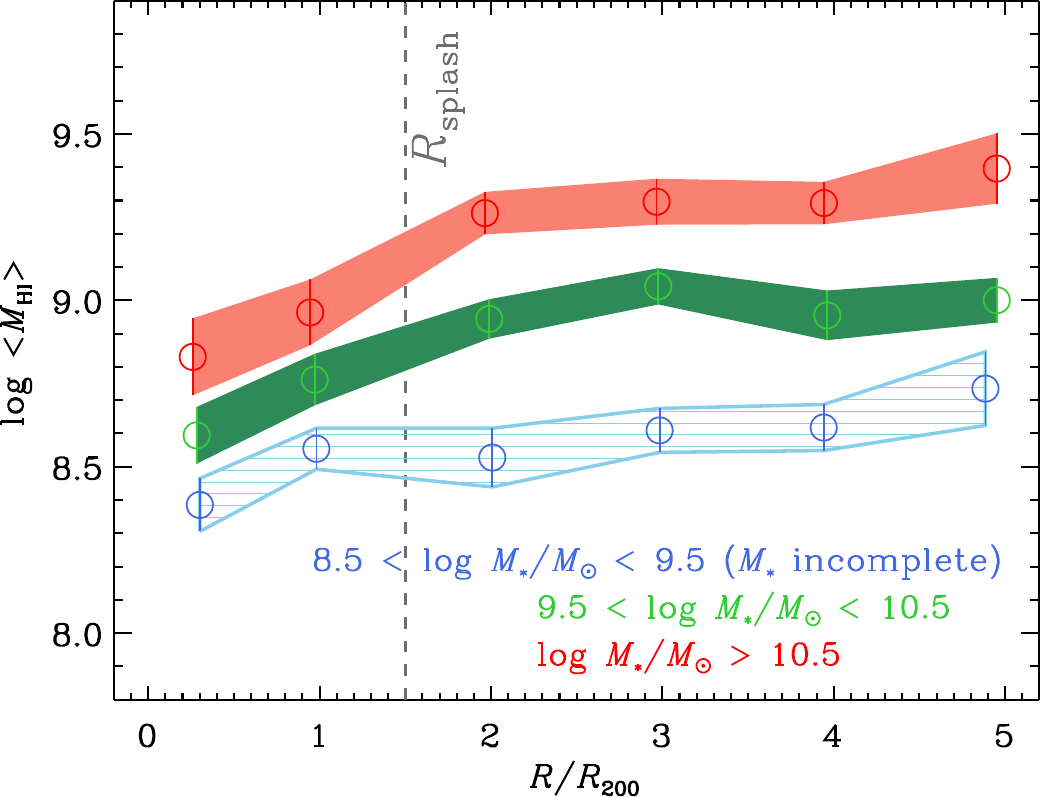}
    \includegraphics[width=0.48\linewidth]{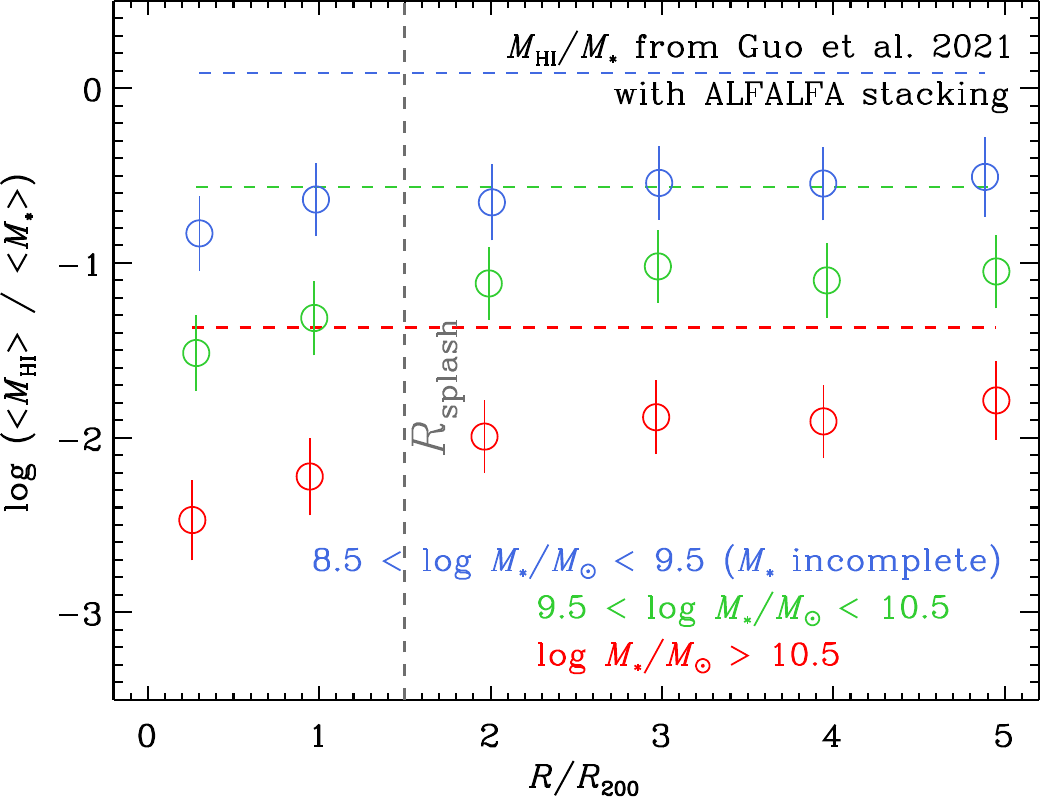}
    \caption{{\bf Left:} Averaged H\,{\sc{i}} mass at different radius for the low mass bin (blue), medium mass bin (green) and high mass bin (red).The shaded regions represent the uncertainties of the stacking results. We use shaded regions with hatching for the low-mass bins to emphasize the stellar-mass incompleteness (Figure~\ref{masszspec}). The vertical dashed line shows the splashback radius of clusters at $\sim 1.5\,R_{200}$ \citep{2019MNRAS.487.2900S}. 
    {\bf right:} Averaged H\,{\sc{i}} mass fraction as a function of radius and stellar mass bins. The dashed lines show the $M_{\rm HI}/M_*$ ratios for low- (blue), intermediate- (green), and high-mass (red) stellar mass bins from \citet{2021ApJ...918...53G}, which are the stacking results from ALFALFA. The values of $M_{\rm HI}/M_*$ are taken at the average stellar masses of the three bins, $\log(M_*/M_\odot) = 9.2$, $10.0$, and $11.2$, and are treated as representative of field galaxies. The $M_{\rm HI}/M_*$ ratios in our sample are lower than those of field galaxies even at $5R_{200}$, implying that a similar pre-processing effect may occur in the cluster outskirts.
    }
    \label{MHI}
\end{figure*}

\begin{figure*}
    \centering
    \includegraphics[width=0.48\linewidth]{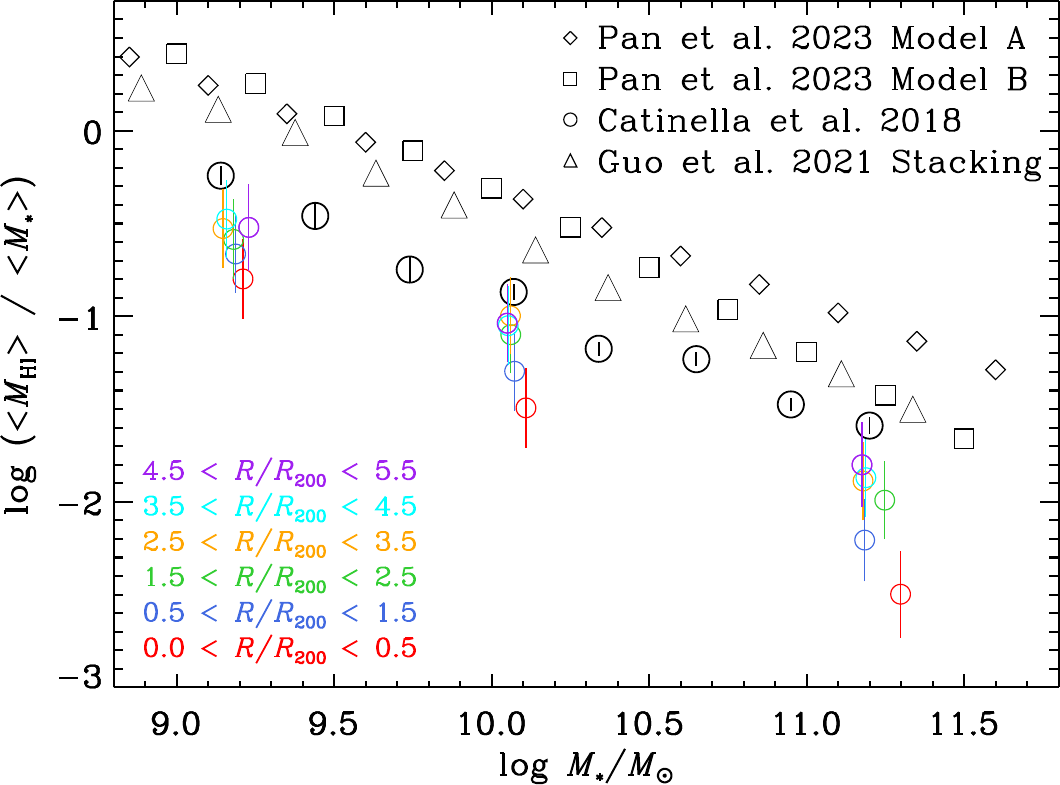}
    \includegraphics[width=0.48\linewidth]{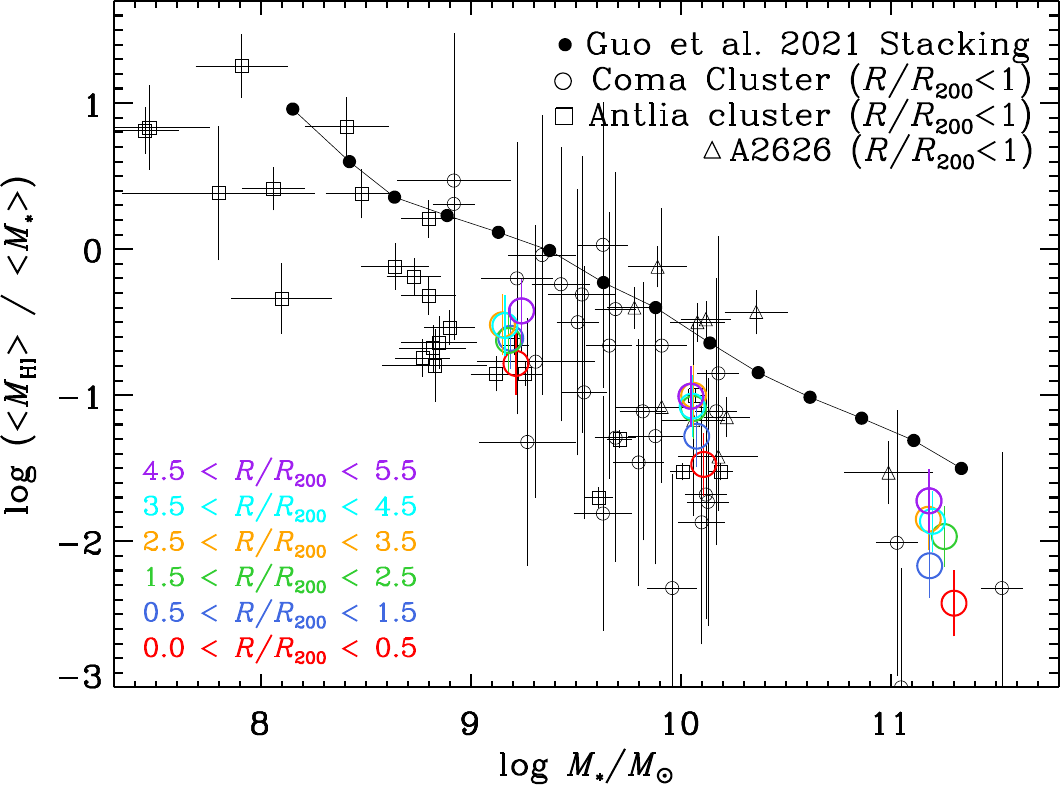}
    \caption{{\bf Left panel:} Averaged H\,{\sc{i}} mass $M_{\rm HI}/M_*$ at different radius (color dots). Results from ALFALFA stacking results \citep[open triangles][]{2021ApJ...918...53G}, xGASS project \citep[open circles][]{2018MNRAS.476..875C} and HI-selected sample from MIGHTEE-HI project \citep{2023MNRAS.525..256P} are shown for comparisons. Our mass-selected sample (open colored circles) shows the lowest $M_{\rm HI}/M_*$ values among the samples. Previous studies typically include statistical uncertainties that scale with the number of stacked sources ($\sim 1/\sqrt{N}$), which are generally smaller than the uncertainties adopted in this work. Our error estimates additionally incorporate the intrinsic scatter of the stacking procedure. Specifically, for each bin we randomly select 50\% of the sample and perform the stacking 100 times, and the dispersion of the resulting stacked measurements is adopted as the uncertainty. As a result, our error bars are generally larger and more conservative.
    {\bf Right panel:} Comparison of our stacked $M_{\rm HI}/M_*$ measurements (colored dots) with individual galaxy H\,{\sc{i}} measurements in three cluster environments: Coma \citep[][]{2021A&A...650A..76H}, Antlia \citep[][]{2015MNRAS.452.1617H}, and Abell 2626 \citep[][]{2025arXiv250515060D}. The field galaxy stacking relation from ALFALFA \citep{2021ApJ...918...53G} is shown for reference. Our stacking results are broadly consistent with H\,{\sc{i}} measurements in other cluster environments, while providing significantly improved precision. The large scatter in individual cluster galaxy measurements reflects the intrinsic diversity of H\,{\sc{i}} content among cluster members at fixed stellar mass.
    }
    \label{MHIMstar_R}
\end{figure*}

\begin{figure*}[ht!]
    \centering
    \includegraphics[width=0.8\linewidth]{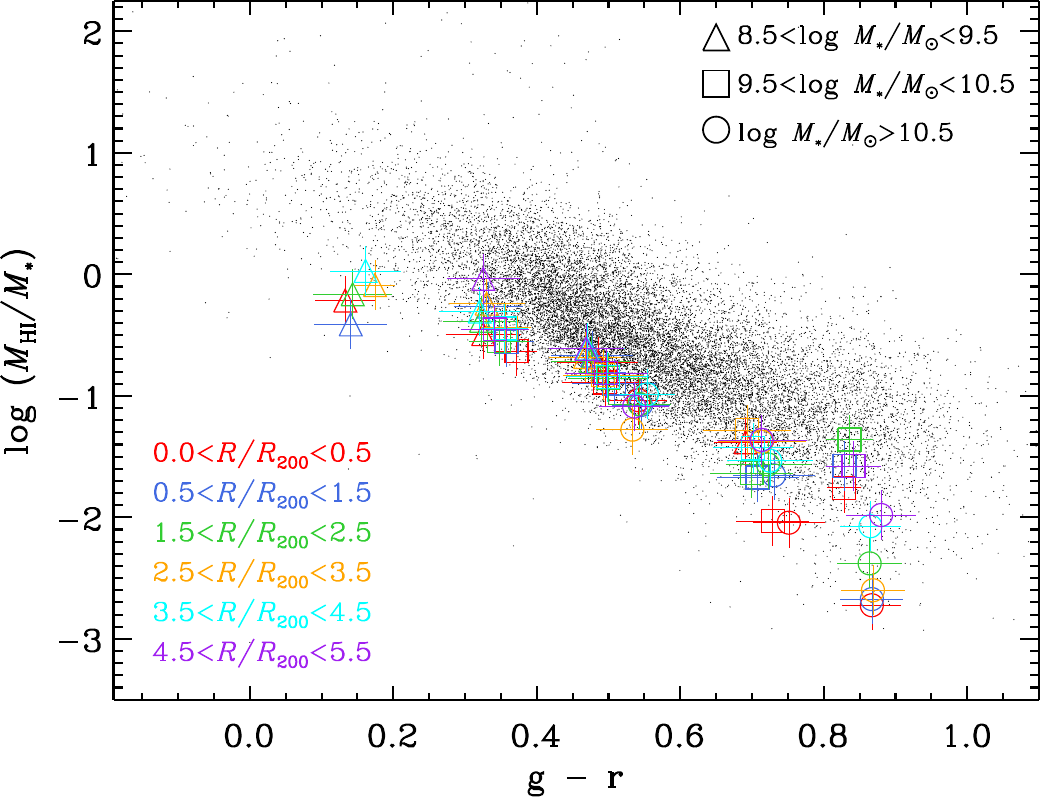}
    \caption{
    Stacked $M_{\rm HI}/M_*$ as a function of $g-r$ color. Small black dots show individual FASHI-detected field galaxies. Colored symbols with error bars denote the stacked $\langle M_{\rm HI}/M_*\rangle$ measurements for cluster member galaxies in different projected cluster-centric distance bins, as indicated in the legend. Different symbols represent the three stellar mass bins: triangles for $8.5<\log M_*/M_\odot<9.5$, squares for $9.5<\log M_*/M_\odot<10.5$, and circles for $\log M_*/M_\odot>10.5$. Error bars include both the stacked spectral rms uncertainty and a 10\% flux calibration error. The cluster stacked measurements lie systematically below the field galaxy distribution at all $g-r$ colors, indicating that H\,{\sc{i}} depletion is already present even among optically blue (or star forming) cluster galaxies.
    }
    \label{ColorHI}
\end{figure*}

\section{H\,{\sc{i}} stacking results}

\subsection{Radial and Mass Dependence of H\,{\sc{i}} Content}

The average H\,{\sc{i}} spectra of the 18 bins (three bins for stellar mass and six bins for $R/R_{200}$) are presented in Figure \ref{HIstack}. Each bin has about 170 to more than thousand targets. There are systematically less targets in the $5R/R_{200}$ bins which lead to a larger uncertainty in the average H\,{\sc{i}} masses. On the other hand, there are also less targets at the cluster centers, partially because of the source blending (see Section \ref{blending}) before stacking. 

We show the H\,{\sc{i}} mass measured in each bin in Figure \ref{MHI} left panel. For each $M_*$ bin, H\,{\sc{i}} mass clearly declines toward the cluster center. The deficiency of H\,{\sc{i}} in the central regions suggests that cluster galaxies have lost their gas reservoir due to environment effects. This decline in H\,{\sc{i}} mass begins at approximately 2$R_{200}$, consistent with the location and concept of the splashback radius \citep{2014ApJ...789....1D, 2019MNRAS.487.2900S} or shock radius \citep{2025PASA...42....8Z}, and with the radial extent of the hot intracluster medium traced by eROSITA X-ray stacking \citep{2025arXiv250925317Z}. From $5R/R_{200}$ to the cluster center, the H\,{\sc{i}} mass drops to about 0.4 dex. Interestingly, the H\,{\sc{i}} mass in the dwarf galaxy bin declines from $5R_{200}$, while the most massive bin keeps their H\,{\sc{i}} mass at $M_{\rm HI} = 10^{9.3}M_\odot$, and starts to lose gas inside of $2R_{200}$. The difference in the trends is consistent with previous results showing that low-mass galaxies lose their gas more easily \citep{2013MNRAS.429.2191Z}.

The right panel of Figure \ref{MHI} shows the H\,{\sc{i}} mass to stellar mass ratio ($M_{\rm HI}/M_*$), which serves as a proxy for the gas mass fraction.
Since gas fractions intrinsically depend on the stellar mass, the $M_{\rm HI}/M_*$ at different radius for each $M_*$ bin would reflect mainly the effect of the environment. The average stellar masses are similar across different $R/R_{200}$ bins (Figure \ref{masszspec}), thus the decline in total H\,{\sc{i}} mass directly translates into a decrease in gas fraction toward the cluster center, which is expected in dense environment \citep[e.g., ][]{1993Natur.366..429W, 1998ApJ...503..569E}. There is a clear trend of declining gas fraction toward the cluster center across all three stellar mass bins. The dwarf galaxy bin exhibits the highest gas fraction, consistent with the known $M_{\rm HI}$–$M_*$ distribution \citep[e.g.,][]{2010MNRAS.403..683C, 2012ApJ...756..113H, 2015MNRAS.447.1610M, 2018MNRAS.476..875C, 2018ApJ...864...40P, 2021ApJ...918...53G, 2023MNRAS.525..256P}.

The most massive galaxies show a drop of about 0.5 dex, while the incomplete dwarf galaxy sample (see Figure~\ref{masszspec}) declines by approximately 0.2 dex. Therefore, we cannot conclude that the H\,{\sc{i}} content in the low-mass bins is truly reduced by a smaller dex value.

We compare our $M_{\rm HI}/M_*$ results with the $M_{\rm HI}/M_* - M_*$ results from \citet{2021ApJ...918...53G} \footnote{The $M_{\rm HI}/M_* - M_*$ results in \citet{2021ApJ...918...53G} is consistent with the stacking results from \citep{2017MNRAS.466.1275B}, and extended the lower end of the stellar mass to $10^8M_\odot$.}, which is the stacking analyzes of the $M_*-M_{\rm HI}/M_*$ relation from ALFALFA data, and more comparable to the average results from HISS stacking code. 
In Figure \ref{MHI}, we can see that at large cluster-centric distances (up to 5$R_{200}$), the average $M_{\rm HI}/M_*$ mass fraction remains $1 \sigma$ lower than that of field galaxies presented by \citet{2021ApJ...918...53G}. This suggests an ongoing pre-processing framework at the very outskirts of clusters, hence a quenching process in groups before falling into the massive cluster.  

In Figure~\ref{MHIMstar_R} left panel, we present the $M_{\rm HI}/M_*$ vs $M_*$ results from our work in comparison with \citet{2021ApJ...918...53G}, as well as MIGHTEE-HI project \citep{2023MNRAS.525..256P}, which represents gas-rich systems, and the result from xGASS project \citep{2018MNRAS.476..875C}, which average the H\,{\sc{i}} mass including non-detections. We find consistently lower $M_{\rm HI}/M_*$ mass fractions in all projected radii.

In the right panel of Figure~\ref{MHIMstar_R}, we compare our stacking results with individual H\,{\sc{i}} measurements in the Coma cluster \citep{2021A&A...650A..76H}, Antlia cluster \citep{2015MNRAS.452.1617H}, and Abell~2626 \citep{2025arXiv250515060D}. Our stacked $M_{\rm HI}/M_*$ values are broadly consistent with H\,{\sc{i}} measurements in these cluster environments, while the individual measurements exhibit significantly larger scatter, reflecting the intrinsic diversity of H\,{\sc{i}} content among cluster members at fixed stellar mass.

\subsection{Color Dependence of the H\,{\sc{i}} Content}
The observed decline of H\,{\sc{i}} content toward cluster centers may partly reflect the well-known color--density relation \citep{1984ApJ...285..426B, 1980ApJ...236..351D, 2000MNRAS.317..782B}, in which cluster environments preferentially host red and passive galaxies. To test whether the observed H\,{\sc{i}} deficiency is primarily driven by the larger fraction of red and passive galaxies in clusters, we further divide the sample into five $g-r$ color bins (0--0.2, 0.2--0.4, 0.4--0.6, 0.6--0.8, and 0.8--1.0), while preserving the stellar mass and radial binning adopted above.

Since massive cluster galaxies are predominantly red, the bluest color bins ($g-r<0.2$) contain few massive systems, while the reddest bins lack low-mass galaxies. Splitting the sample simultaneously by color, stellar mass, and cluster-centric radius therefore significantly reduces the number of galaxies in individual bins. To ensure reliable stacking measurements, we only retain bins containing at least 10 galaxies, with typical numbers of 20--400 galaxies per bin.

We note that when a massive red galaxy is blended with a low-mass blue companion, the flux-combined color is generally dominated by the massive system, assigning the blended target to a red color bin as the massive galaxy. The stacked H\,{\sc{i}} signal, however, includes contributions from both galaxies, which may slightly elevate the measured $\langle M_{\rm HI}/M_*\rangle$ values in the redder bins. This effect is not purely an artifact: in dense environments, massive galaxies readily accrete or merge with low-mass satellites, so the blended systems may genuinely represent a phase of elevated H\,{\sc{i}} content in otherwise gas-poor massive galaxies.

Figure~\ref{ColorHI} presents the stacked HI-to-stellar mass ratios as a function of $g-r$ color. We compare our results with the FASHI field sample \citep{2025ApJS..281...66C}. The FASHI detected galaxies are biased toward gas-rich systems, but this bias is modest in the $g-r$ plane since gas-poor field galaxies are predominantly red and mostly detected blue galaxies are already well-sampled.

We find that across all color bins, cluster galaxies exhibit systematically lower $M_{\rm HI}/M_*$ values than field galaxies at similar colors, with offsets of approximately 0.3--1 dex. Within each color bin, the lowest $\langle M_{\rm HI}/M_*\rangle$ values are found in the innermost radial bins, consistent with the radial decline seen in Section~4.1. The H\,{\sc{i}} deficiency persists even among the bluest galaxies, which would still be classified as star-forming systems based on their optical colors. Interestingly, the bluest color bin (mainly dwarf galaxies) does not show much higher $M_{\rm HI}/M_*$ values than the adjacent blue bin, possibly reflecting the higher susceptibility of low-mass galaxies to environmental gas removal.

The systematically lower H\,{\sc{i}} fractions in clusters are therefore not solely driven by the larger fraction of red and passive galaxies in dense environments. Instead, the results suggest that environmental processes reduce the H\,{\sc{i}} content before galaxies become fully quenched in their stellar populations. The result is consistent with a scenario in which H\,{\sc{i}} stripping and pre-processing occur prior to, or on shorter timescales than, the optical color transformation of cluster galaxies.

\begin{sidewaysfigure*}
    \centering
    \includegraphics[width=0.17\linewidth]{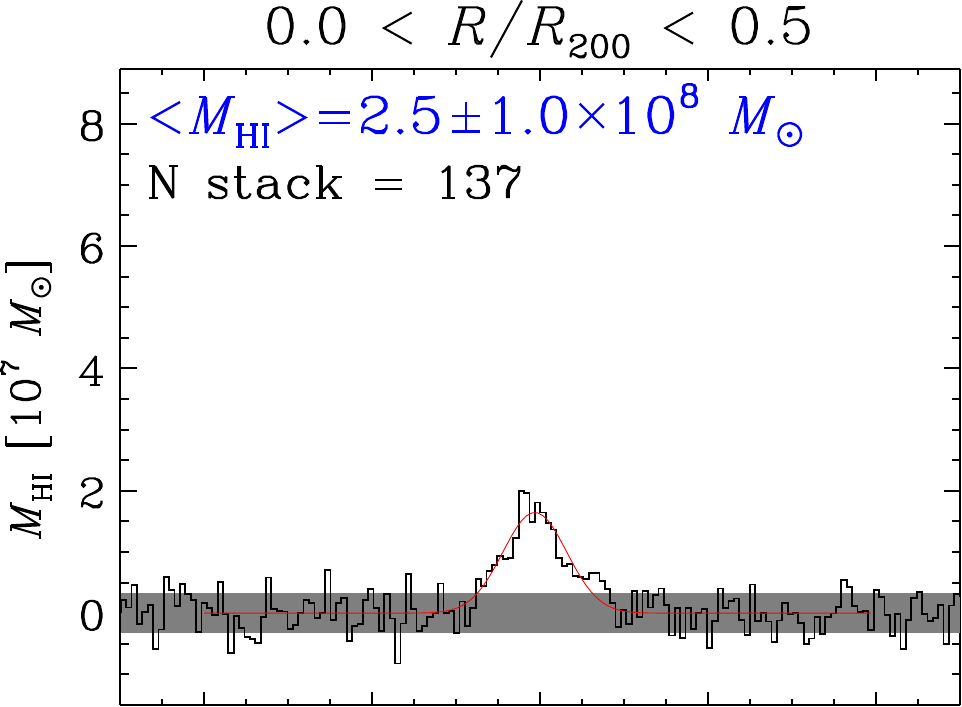}
    \includegraphics[width=0.149\linewidth]{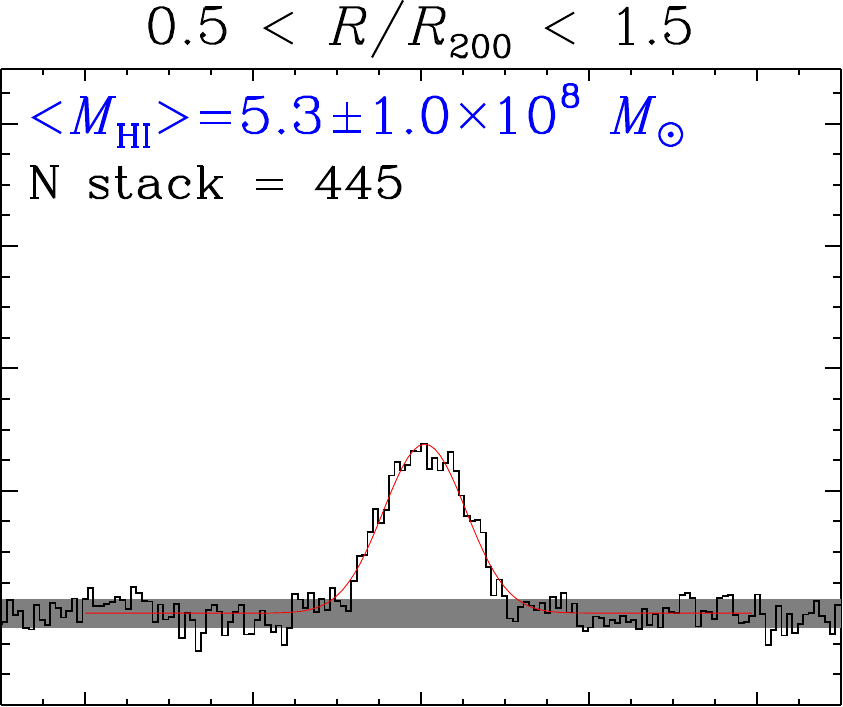}
    \includegraphics[width=0.149\linewidth]{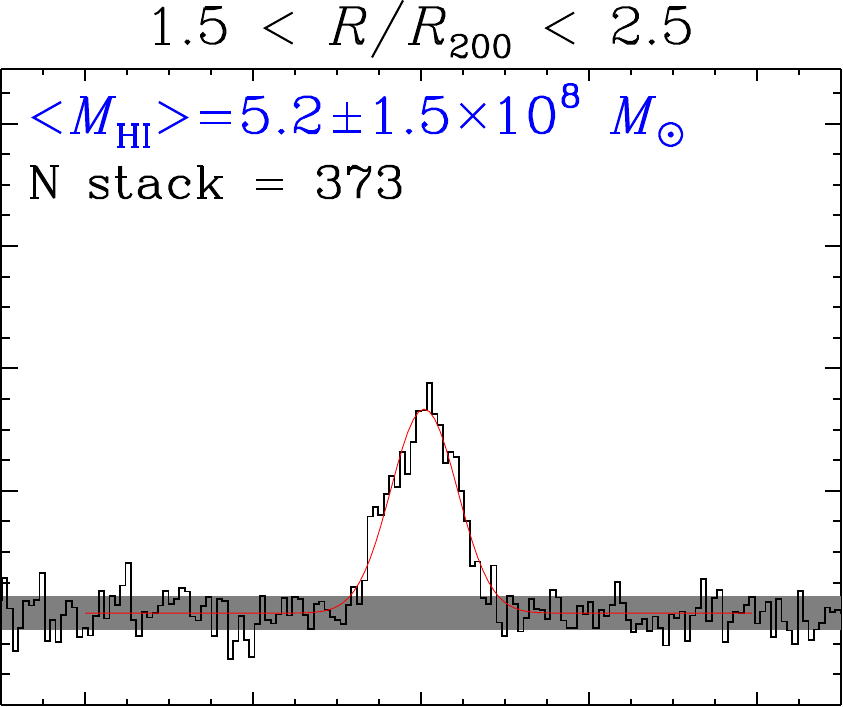}
    \includegraphics[width=0.149\linewidth]{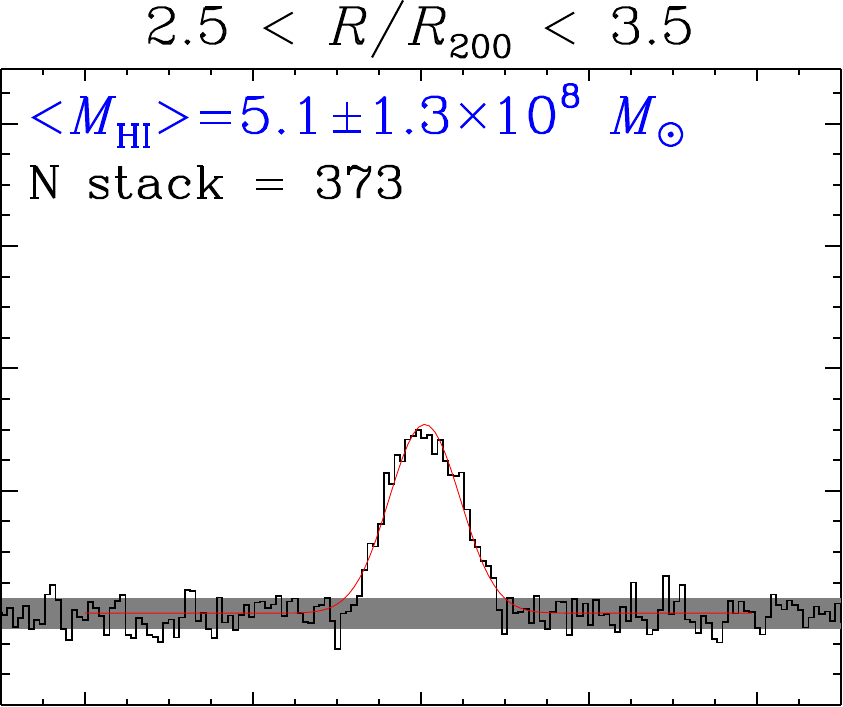}
    \includegraphics[width=0.149\linewidth]{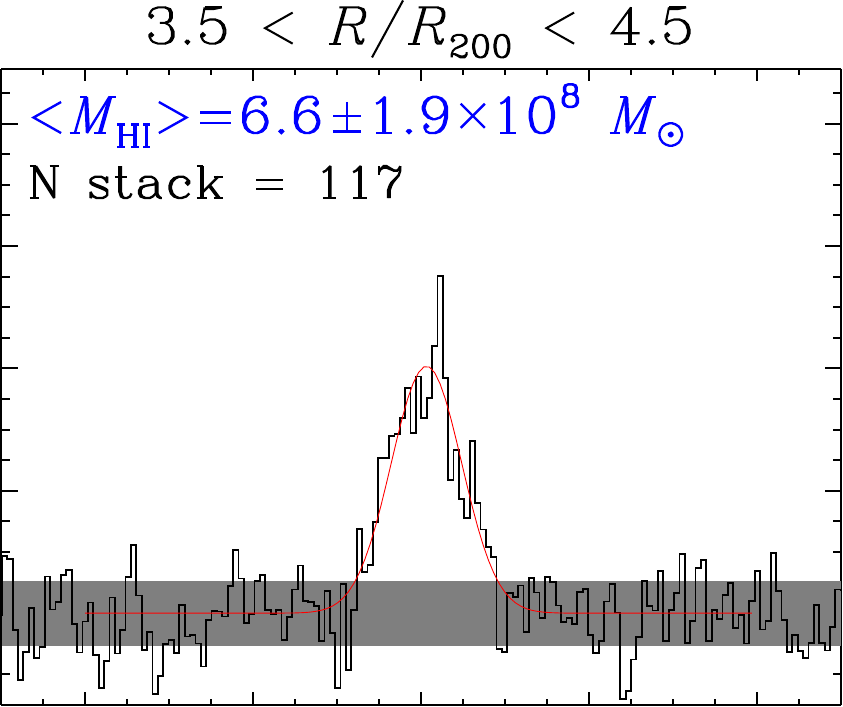}
    \includegraphics[width=0.149\linewidth]{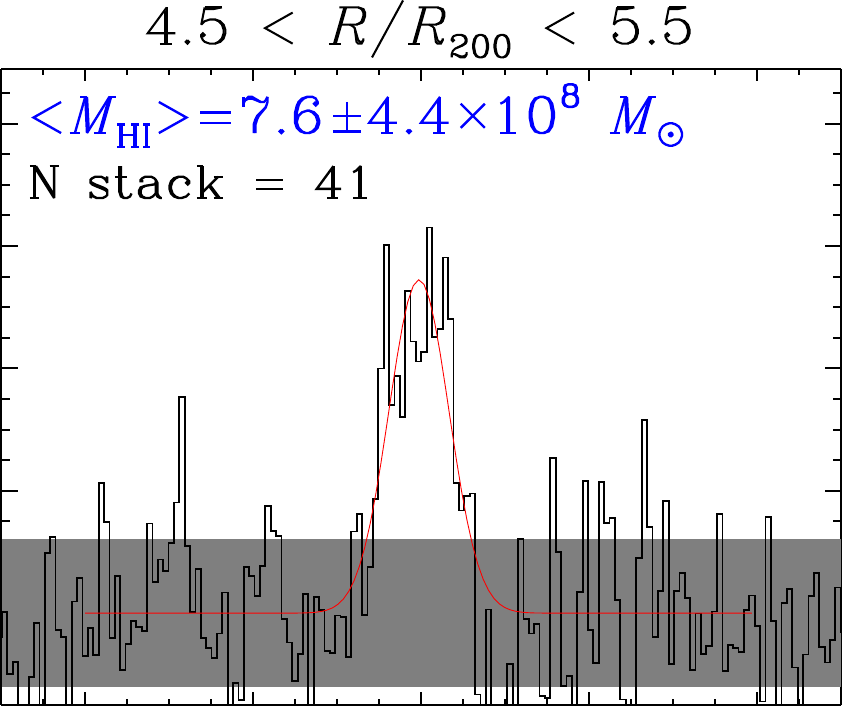}
    \includegraphics[width=0.17\linewidth]{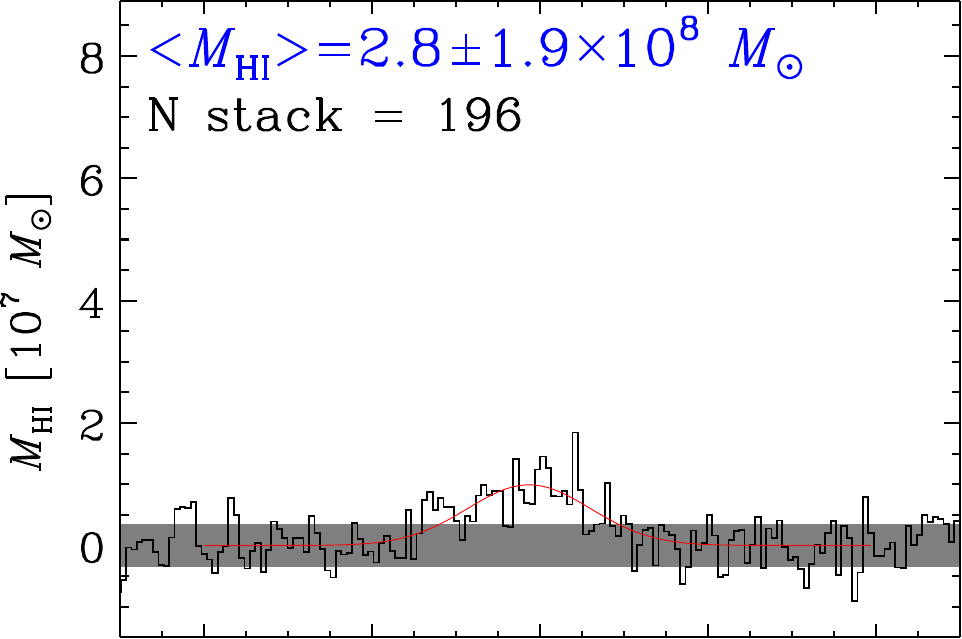}
    \includegraphics[width=0.149\linewidth]{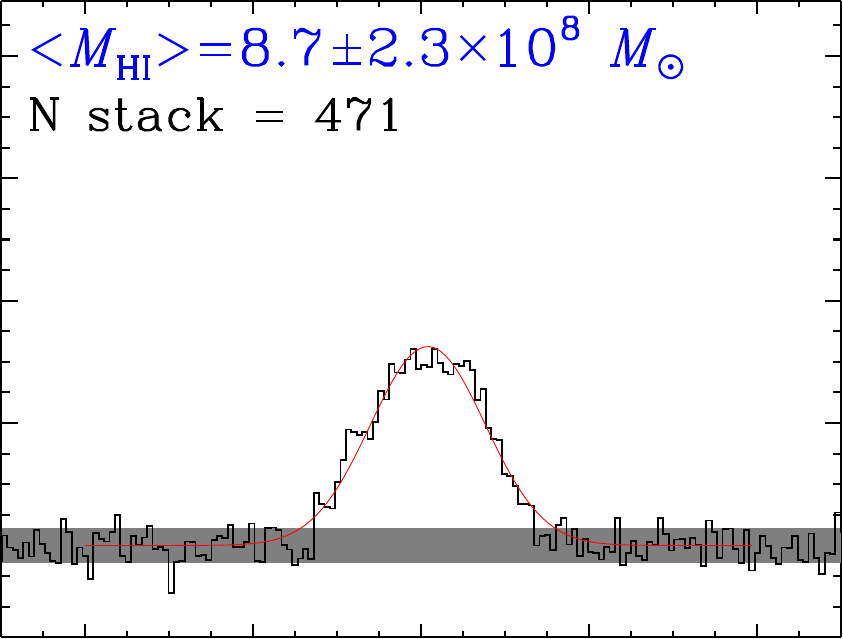}
    \includegraphics[width=0.149\linewidth]{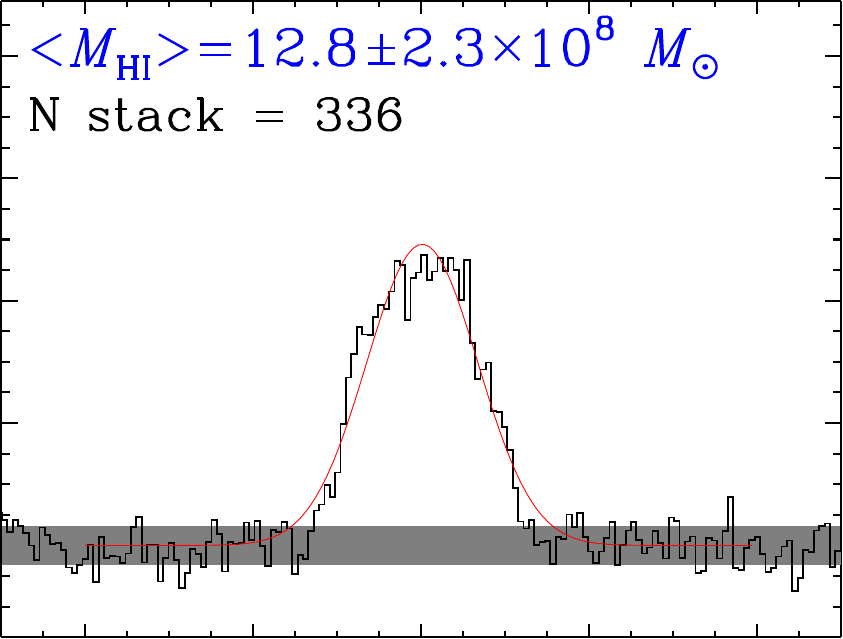}
    \includegraphics[width=0.149\linewidth]{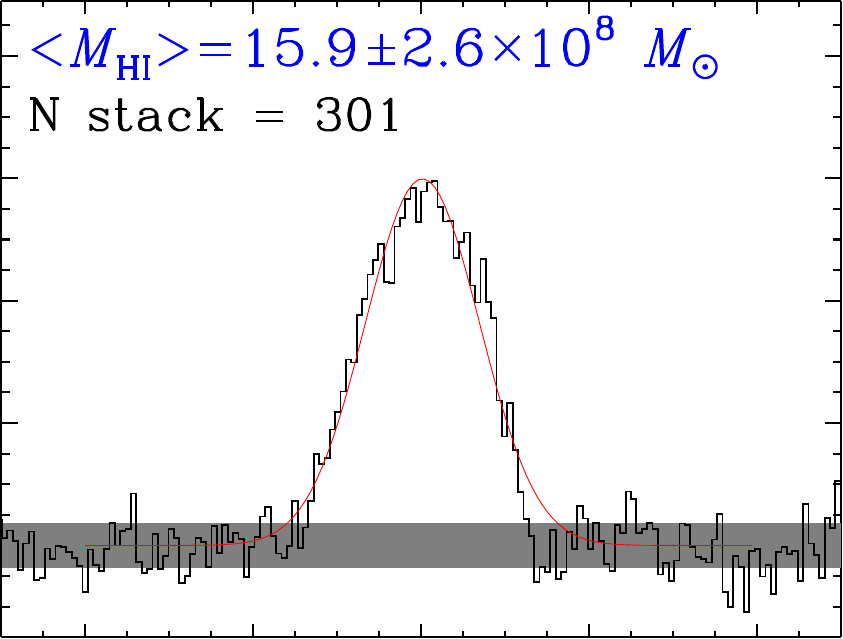}
    \includegraphics[width=0.149\linewidth]{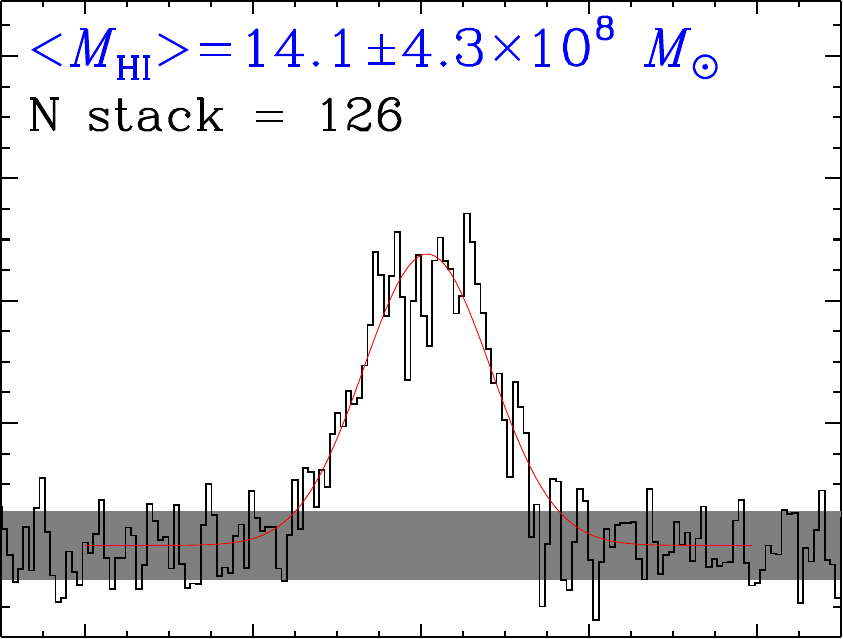}
    \includegraphics[width=0.149\linewidth]{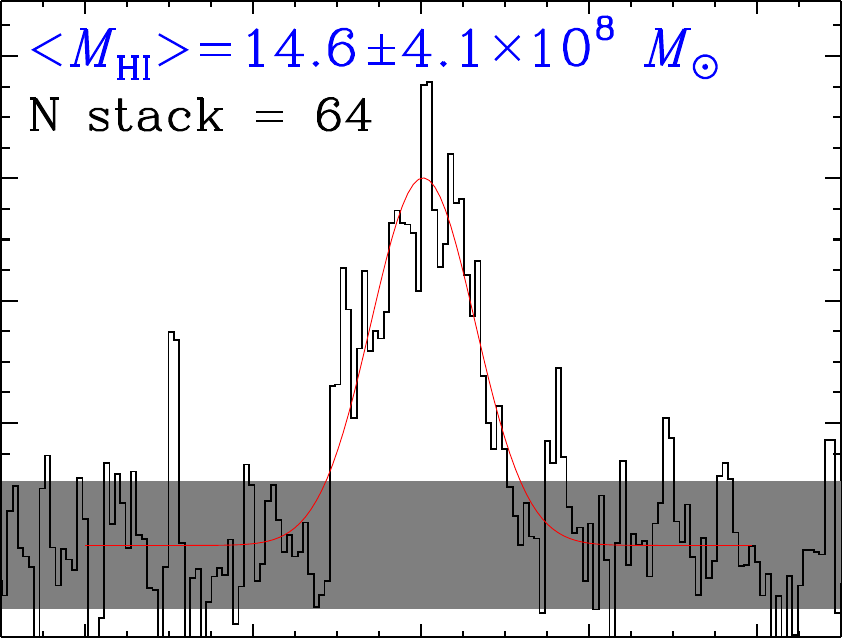}
    \includegraphics[width=0.17\linewidth]{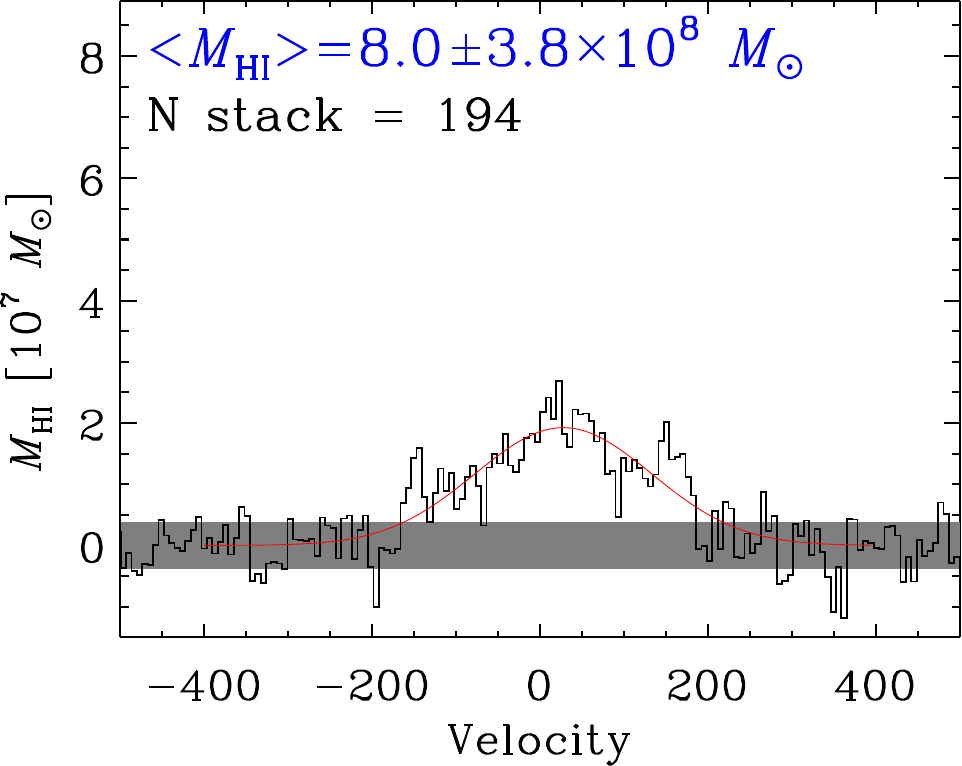}
    \includegraphics[width=0.149\linewidth]{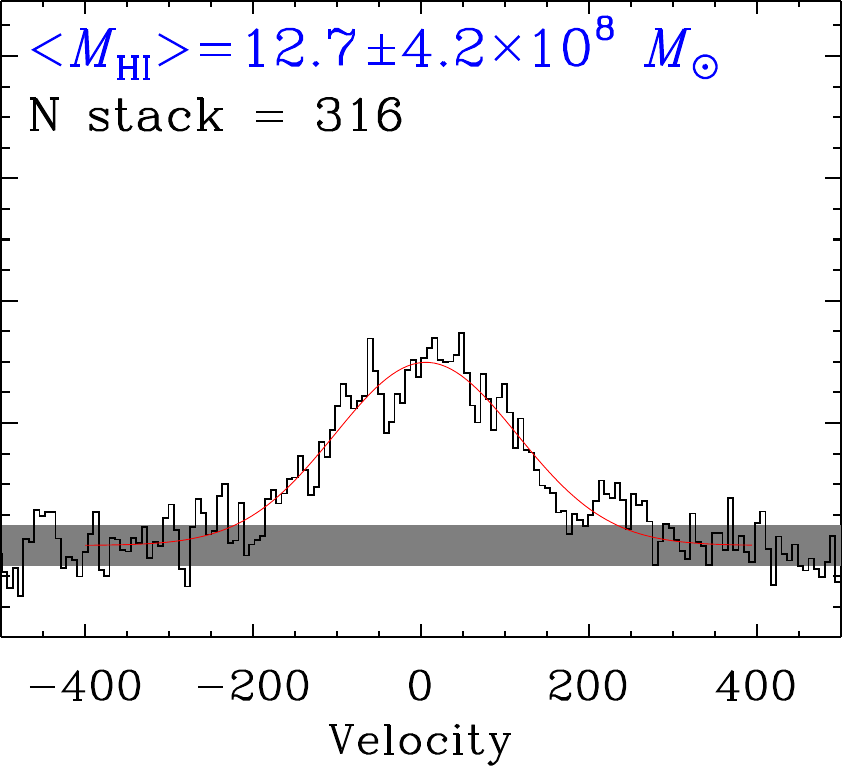}
    \includegraphics[width=0.149\linewidth]{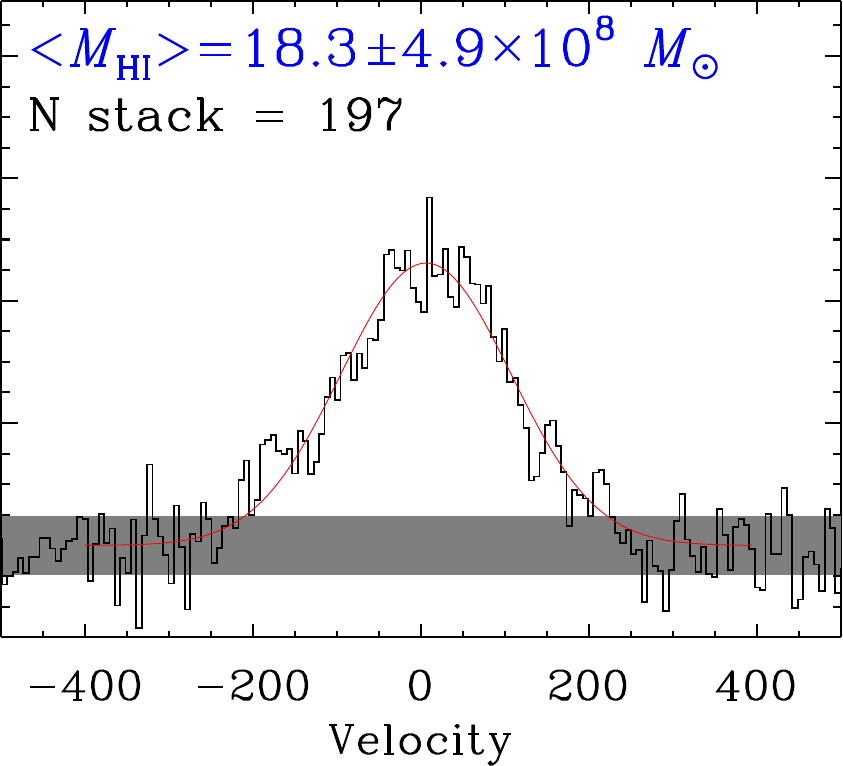}
    \includegraphics[width=0.149\linewidth]{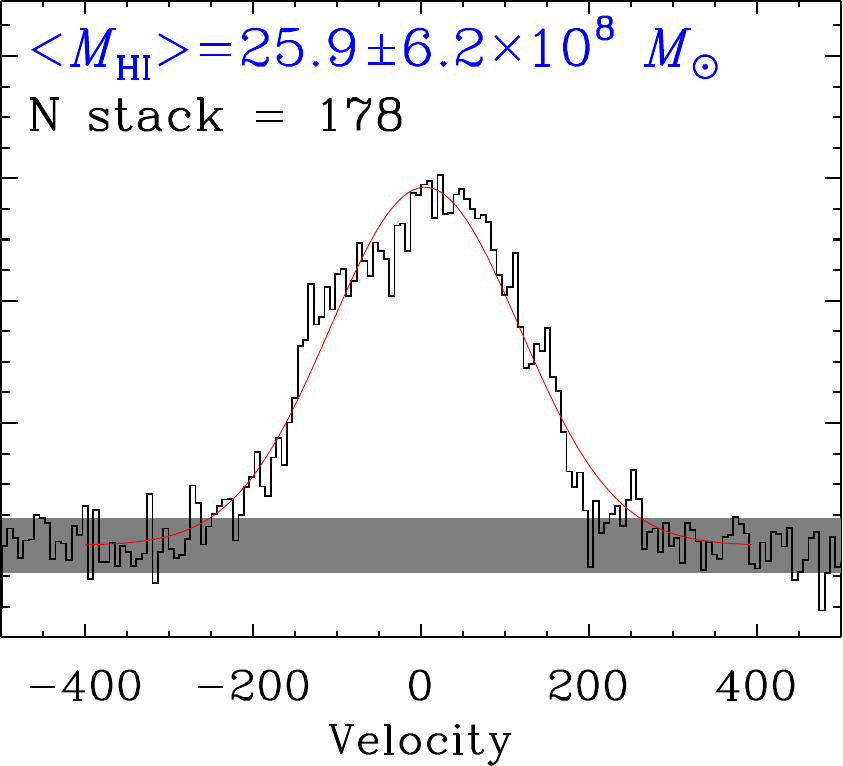}
    \includegraphics[width=0.149\linewidth]{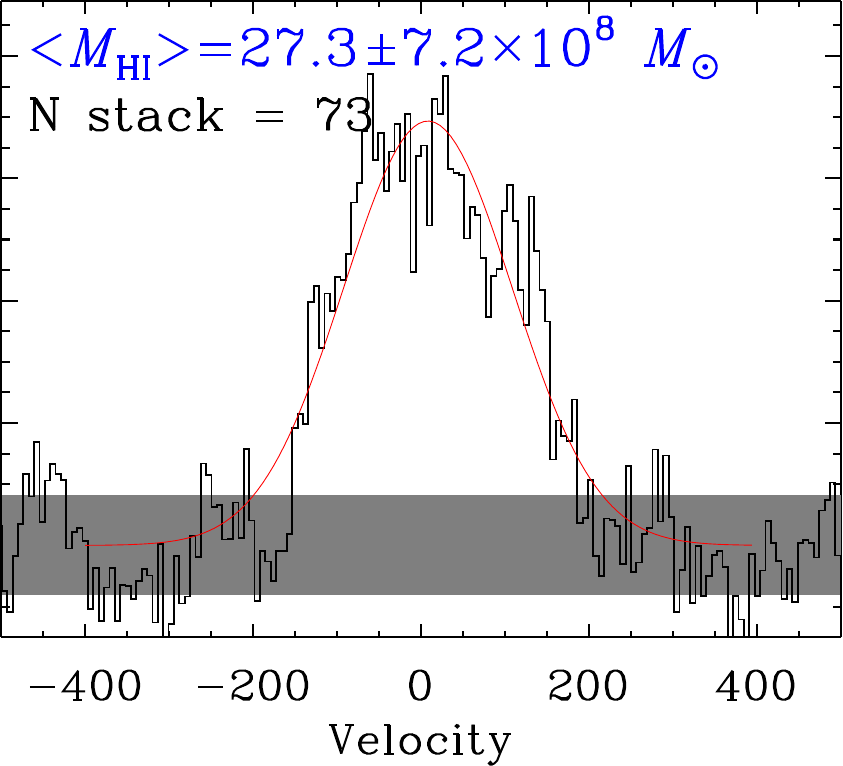}
    \includegraphics[width=0.149\linewidth]{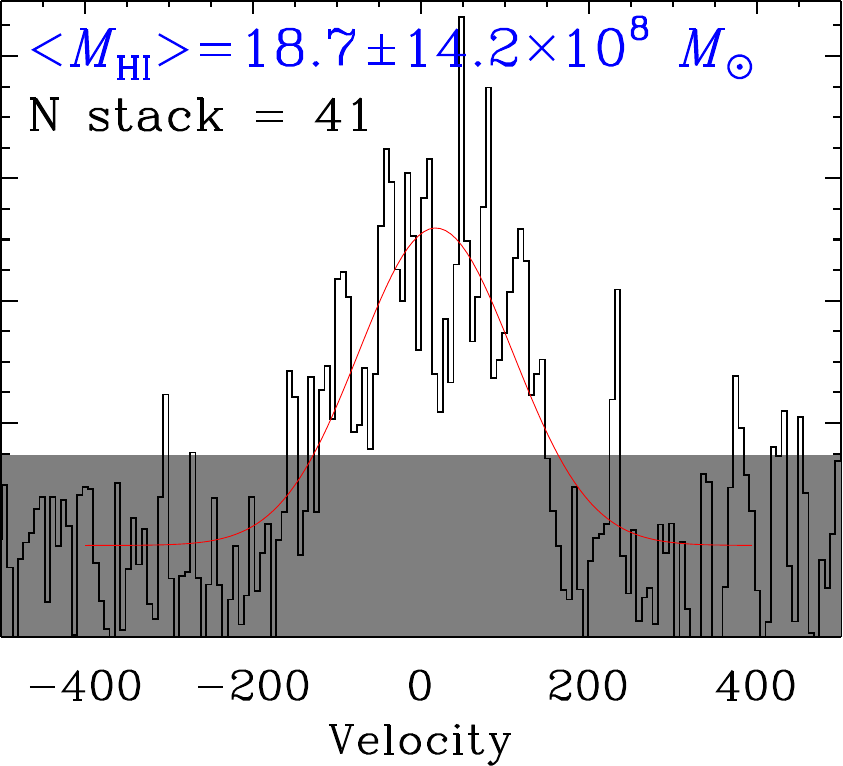}
    \caption{Same as Figure~\ref{HIstack}, but for the relaxed clusters.}
    \label{HIstack_relaxed}
\end{sidewaysfigure*}

\begin{sidewaysfigure*}
    \centering
    \includegraphics[width=0.17\linewidth]{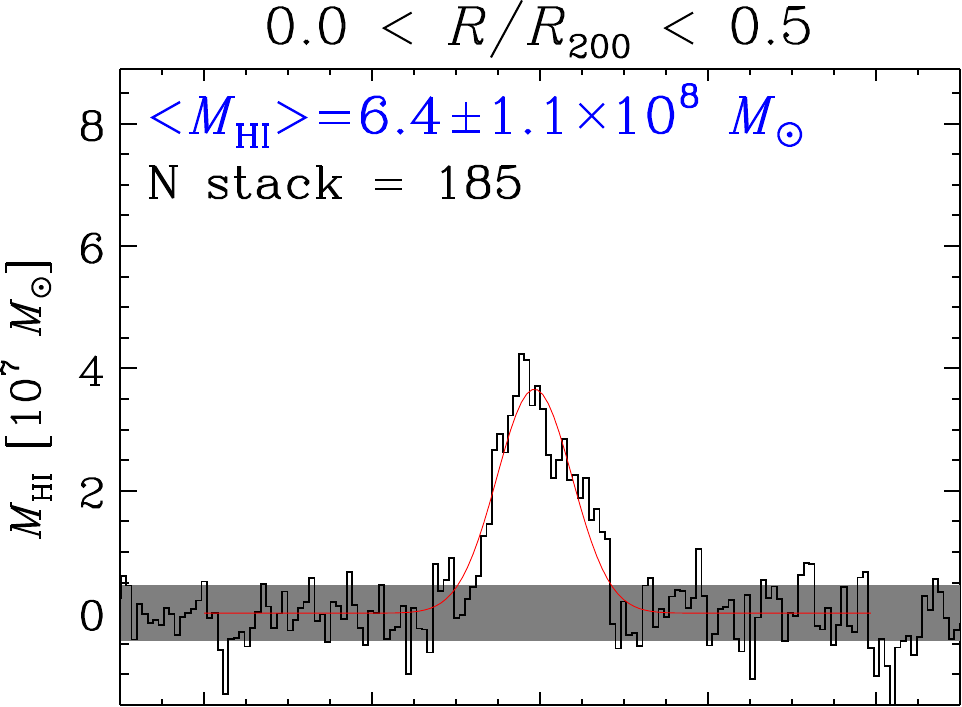}
    \includegraphics[width=0.149\linewidth]{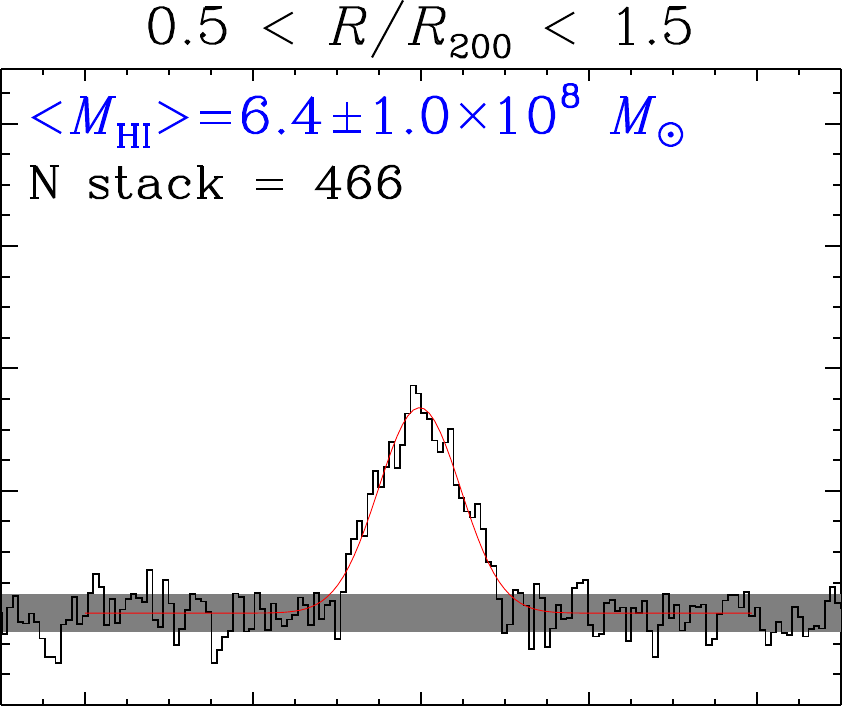}
    \includegraphics[width=0.149\linewidth]{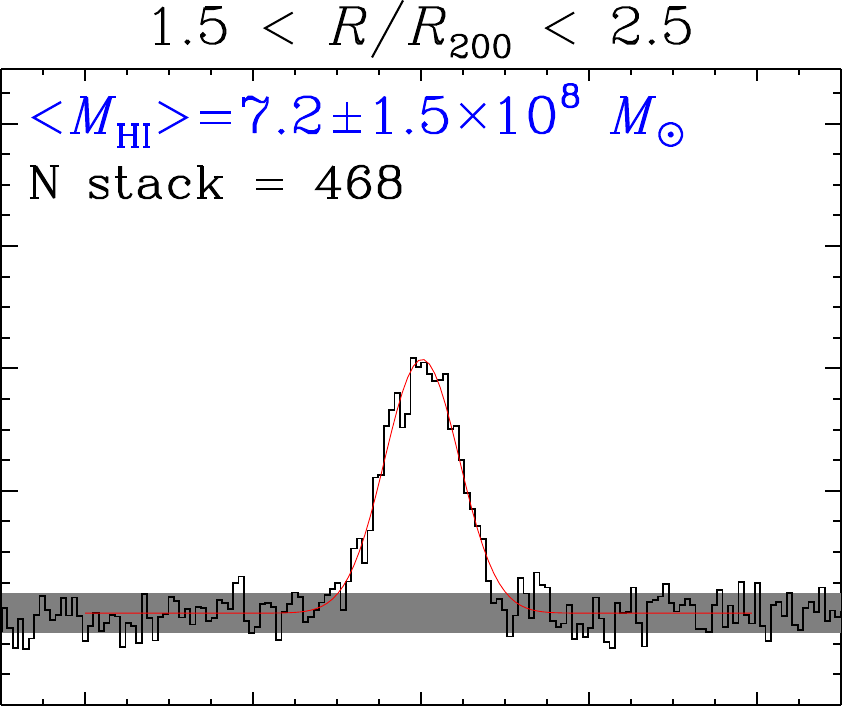}
    \includegraphics[width=0.149\linewidth]{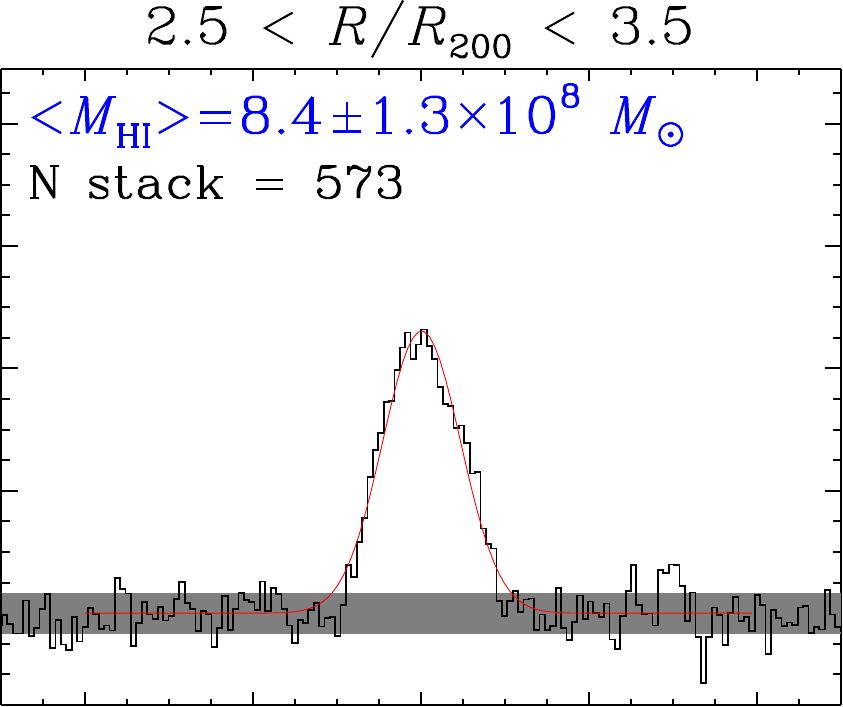}
    \includegraphics[width=0.149\linewidth]{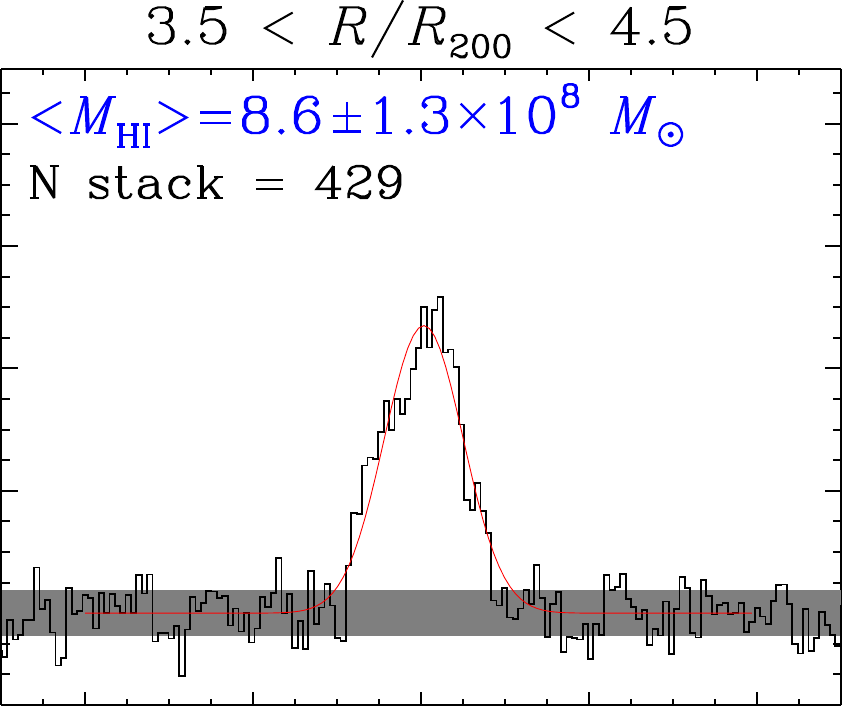}
    \includegraphics[width=0.149\linewidth]{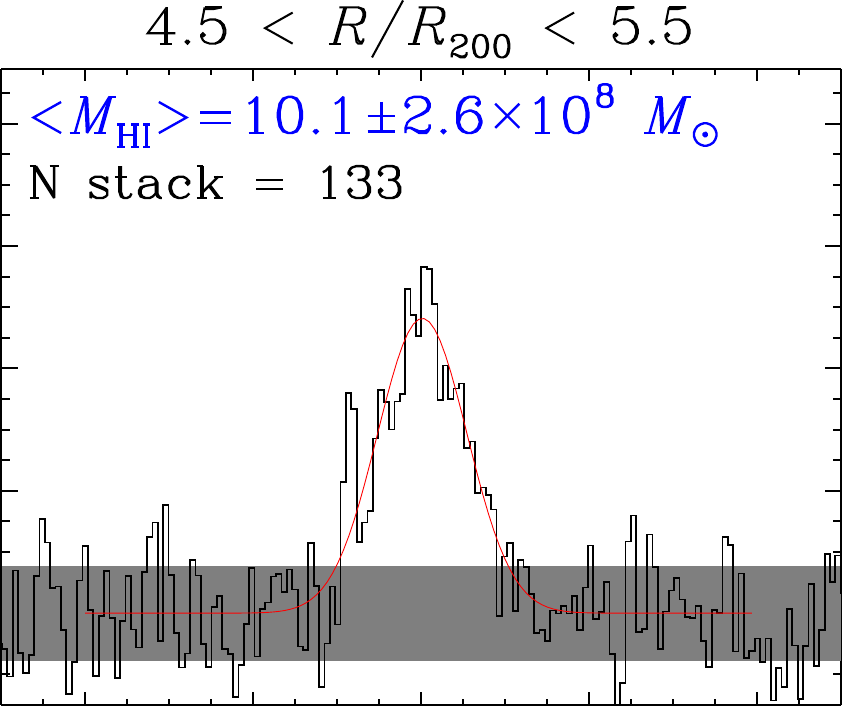}
    \includegraphics[width=0.17\linewidth]{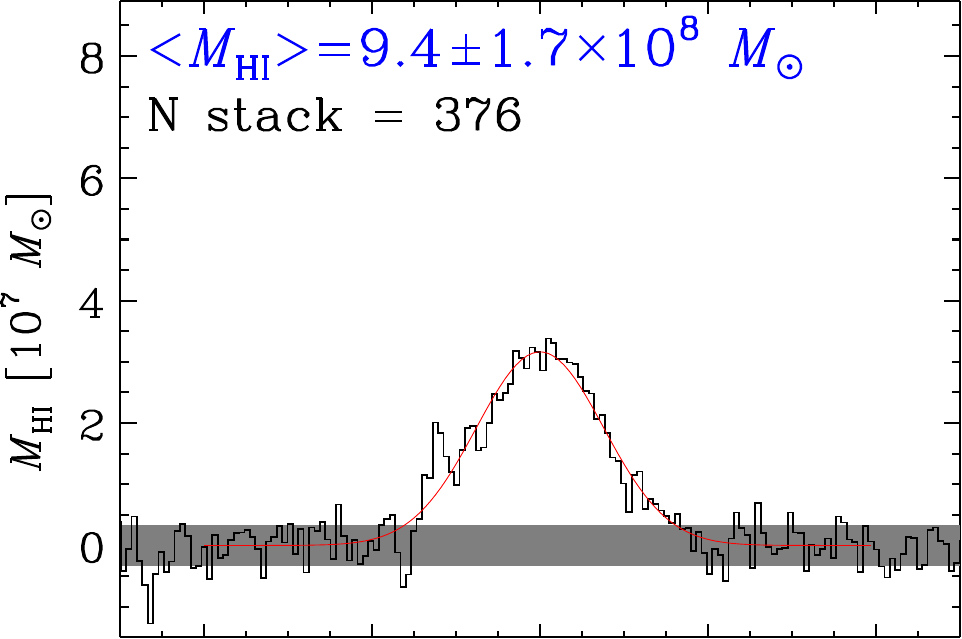}
    \includegraphics[width=0.149\linewidth]{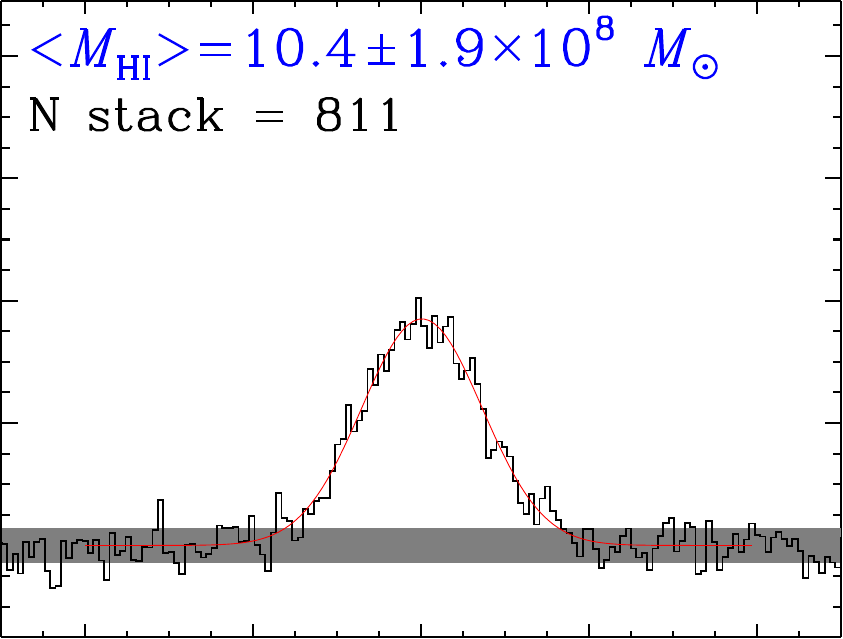}
    \includegraphics[width=0.149\linewidth]{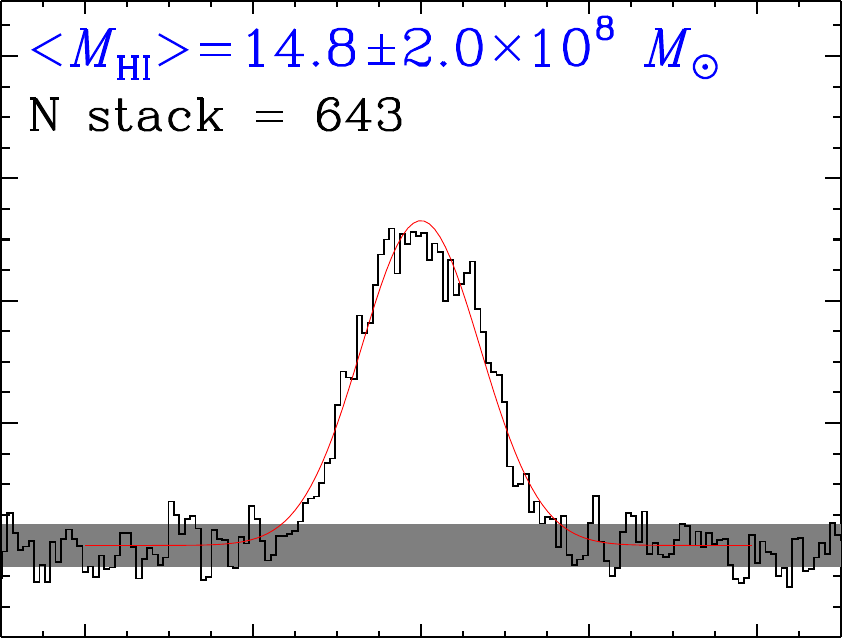}
    \includegraphics[width=0.149\linewidth]{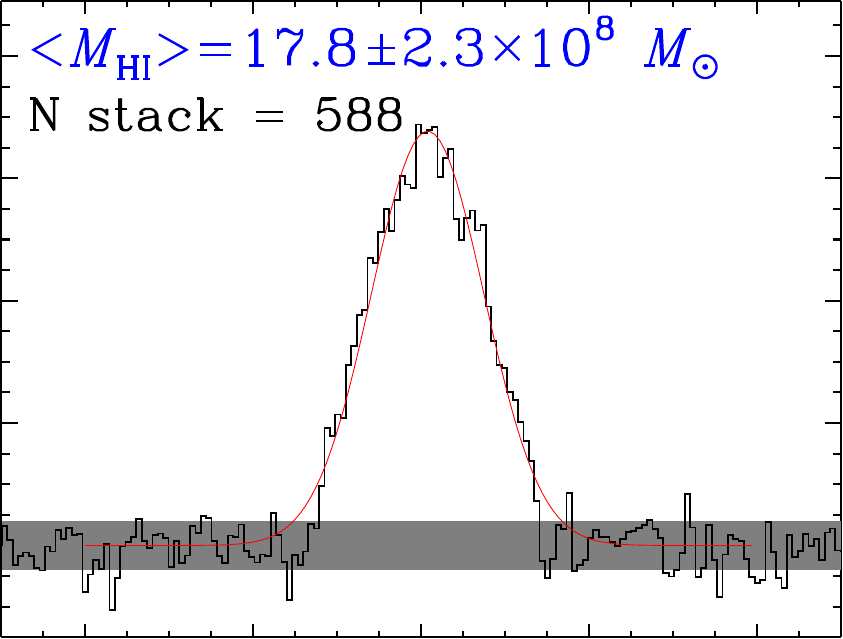}
    \includegraphics[width=0.149\linewidth]{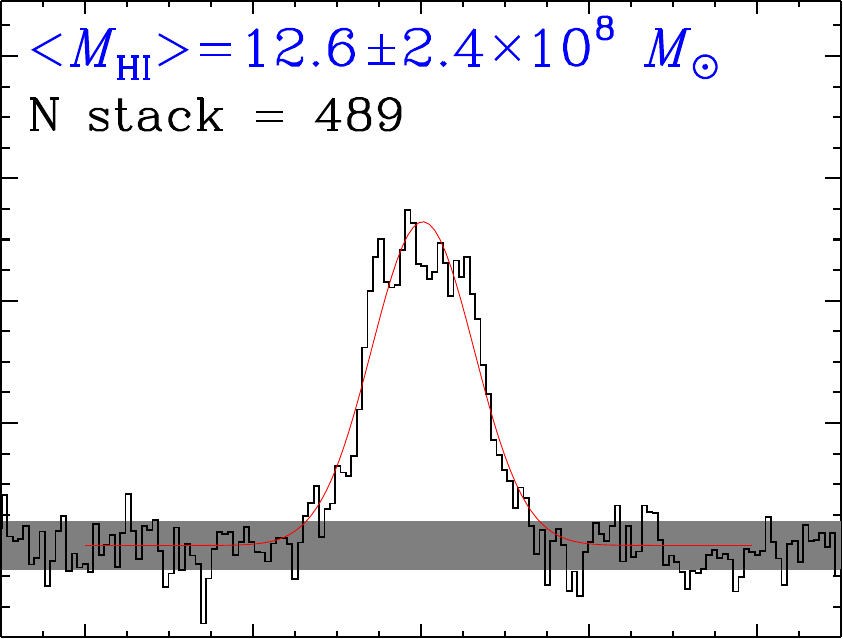}
    \includegraphics[width=0.149\linewidth]{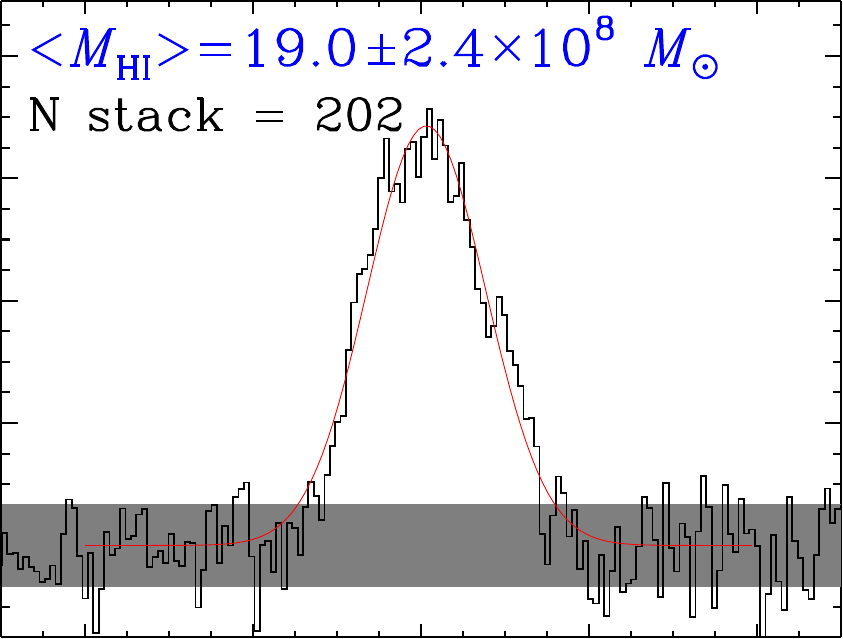}
    \includegraphics[width=0.17\linewidth]{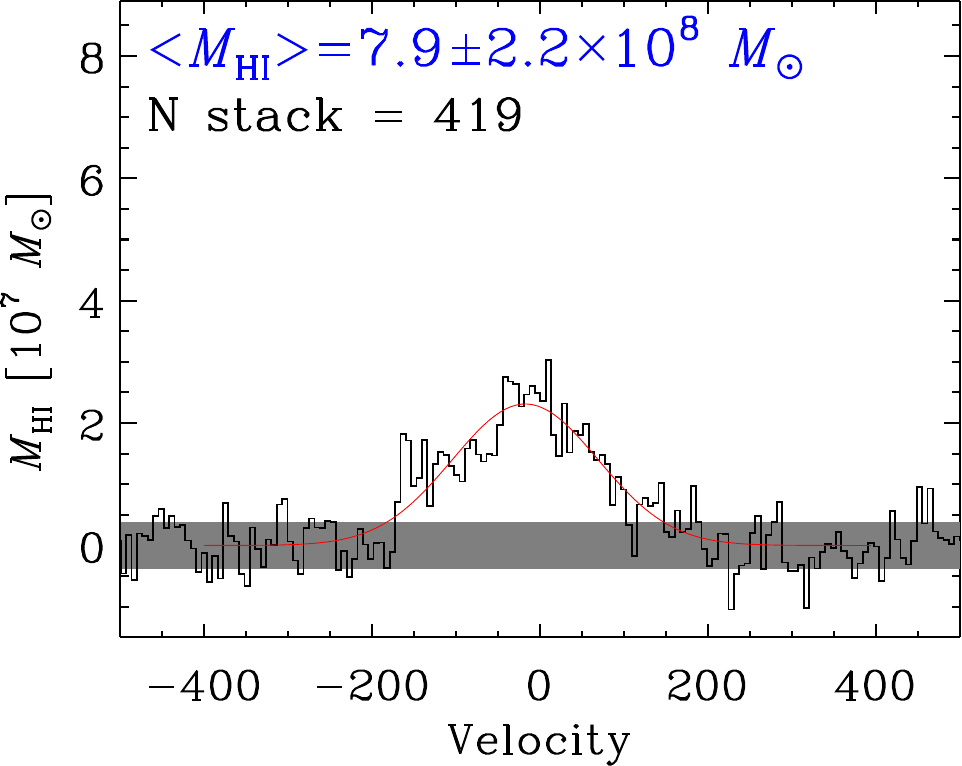}
    \includegraphics[width=0.149\linewidth]{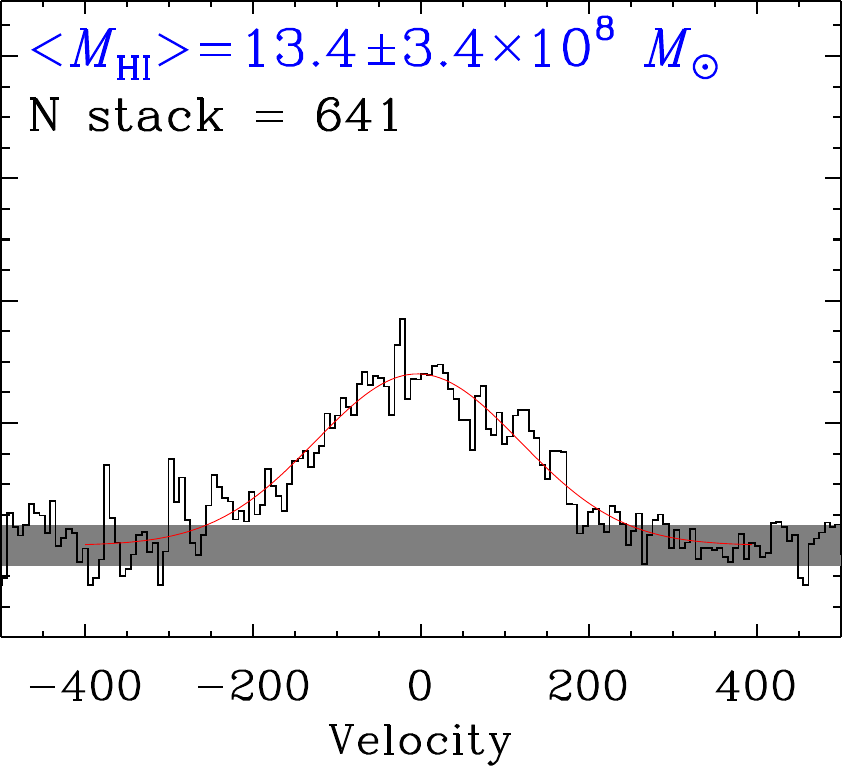}
    \includegraphics[width=0.149\linewidth]{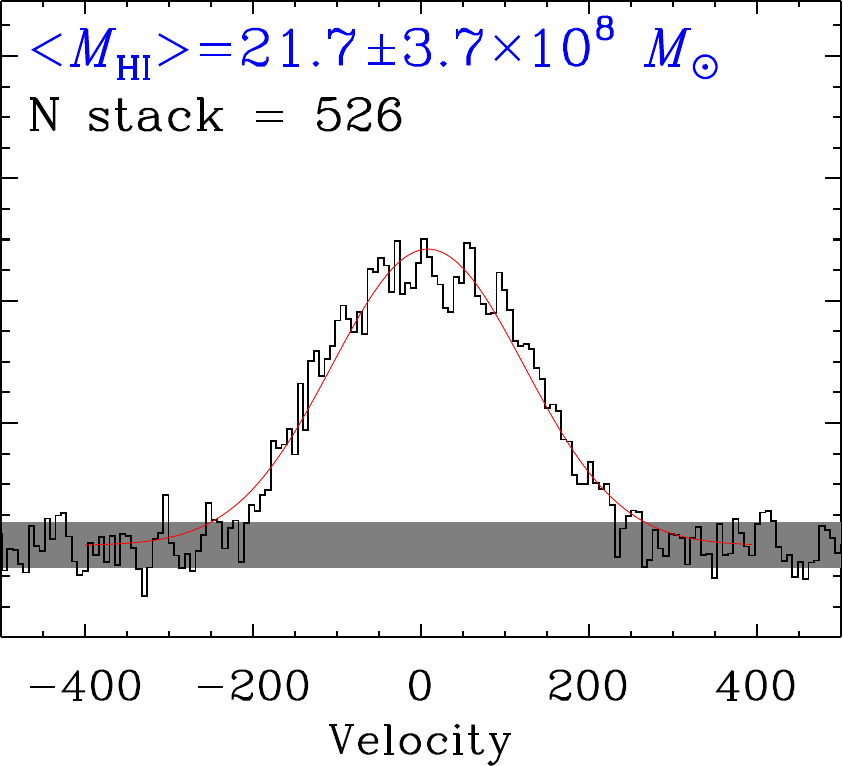}
    \includegraphics[width=0.149\linewidth]{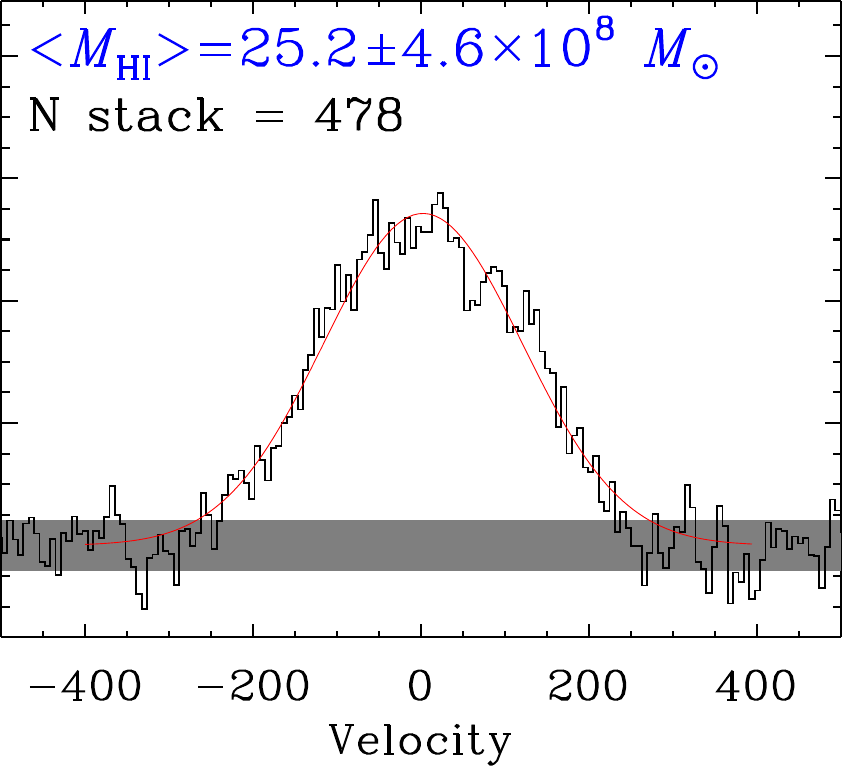}
    \includegraphics[width=0.149\linewidth]{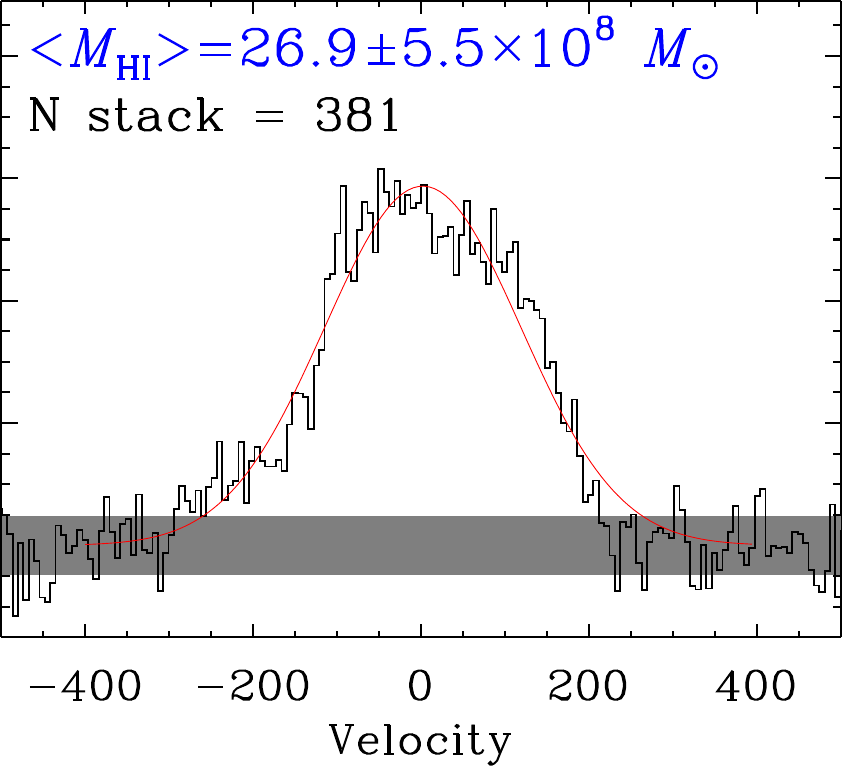}
    \includegraphics[width=0.149\linewidth]{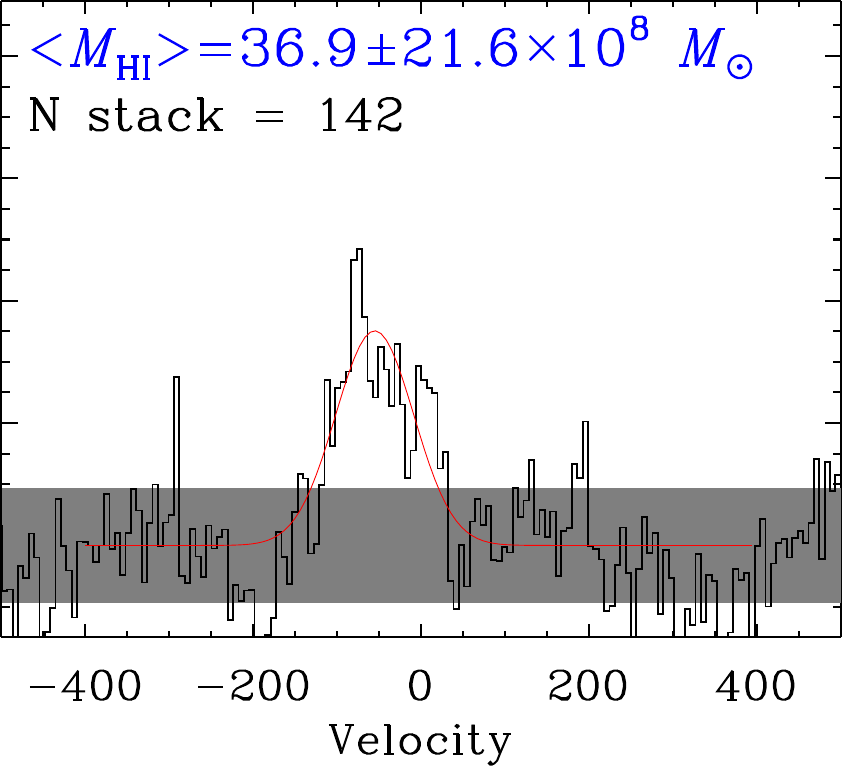}
    \caption{Same as Figure~\ref{HIstack}, but for the disturbed clusters.}
    \label{HIstack_disturbed}
\end{sidewaysfigure*}

\begin{figure}
    \centering
    \includegraphics[width=0.95\linewidth]{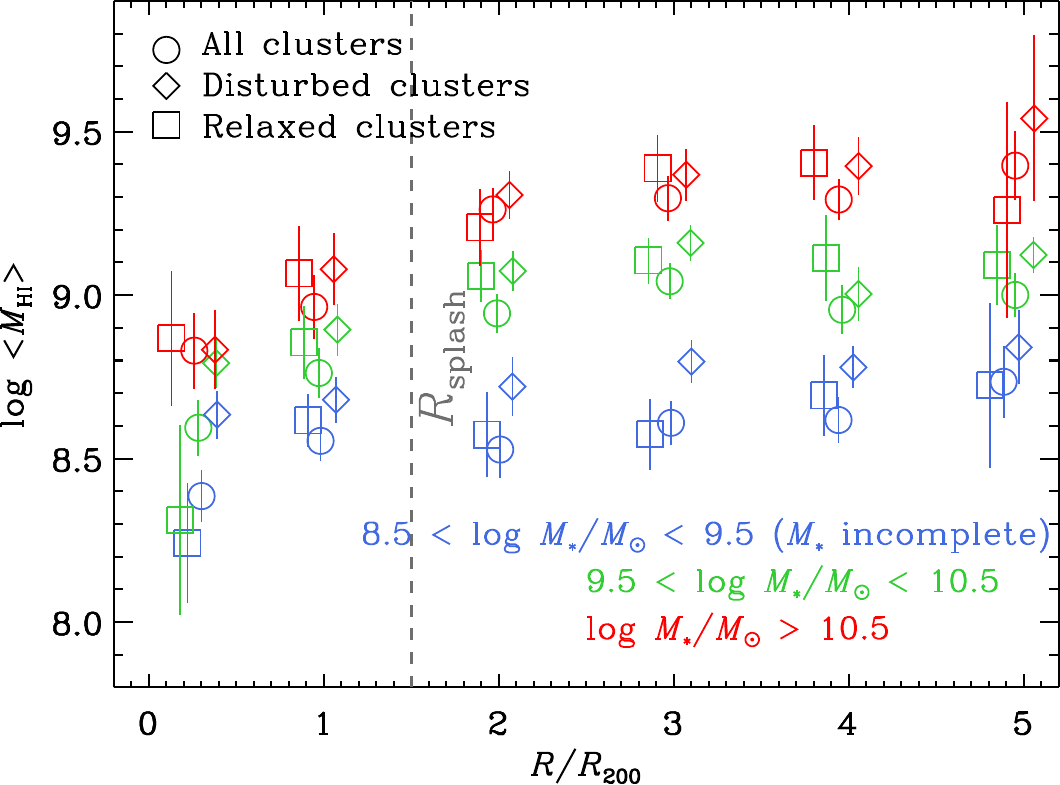}
    \includegraphics[width=0.95\linewidth]{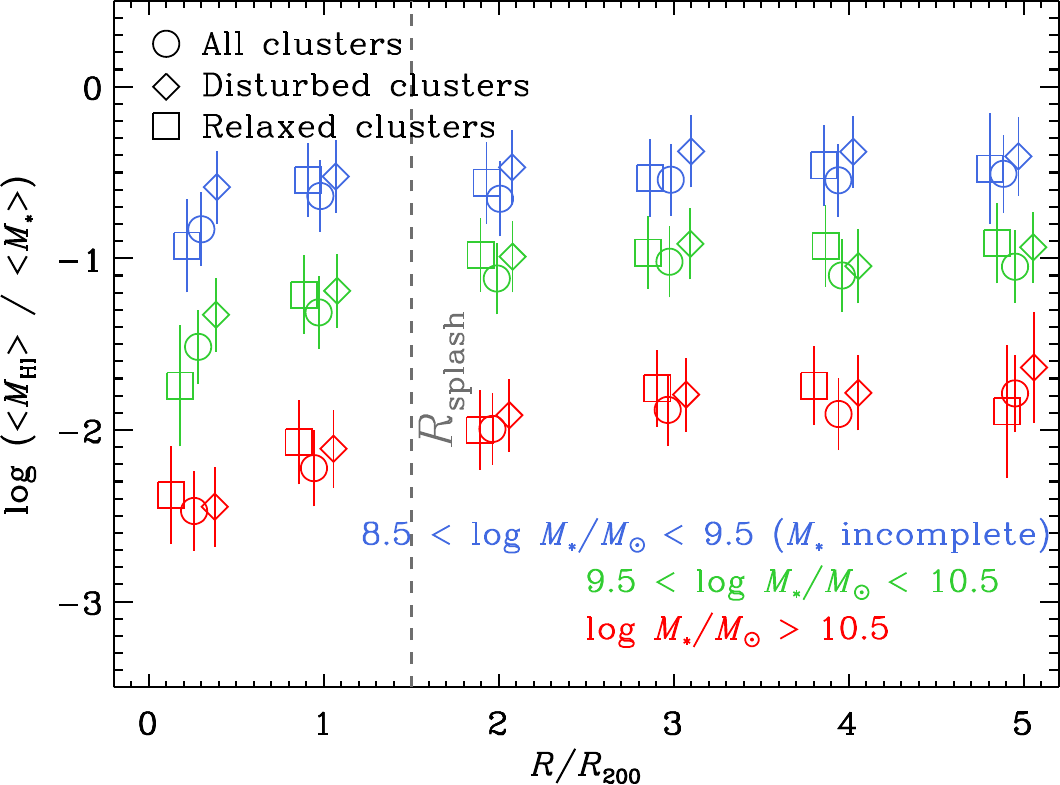}
    \caption{Average H\,{\sc{i}} mass (top panel) and average HI-to-stellar mass ratio (bottom panel) in different radial and mass bins for galaxies in relaxed clusters (open squares) and disturbed clusters (open diamonds), compared with the full cluster sample (open circles). For clarity, data points at the same radial bin are horizontally offset by 0.1~dex to avoid overlapping error bars; the offset has no physical meaning. While the overall trends are consistent with those of the full cluster sample, the total stacked signal may fall outside the relaxed and disturbed values due to baseline subtraction and the exclusion of five galaxy cluster from the stack, although the associated uncertainties remain self-consistent. A roughly $1\sigma$ difference is observed in the innermost radial bin, potentially indicating a mild dependence of the gas content on the cluster dynamical state. The splashback radius of clusters at $\sim 1.5,R_{200}$ is shown by the vertical dashed line \citep{2019MNRAS.487.2900S}. 
    }
    \label{MHI_stage}
\end{figure}

\begin{figure*}
    \centering
    \includegraphics[width=0.95\linewidth]{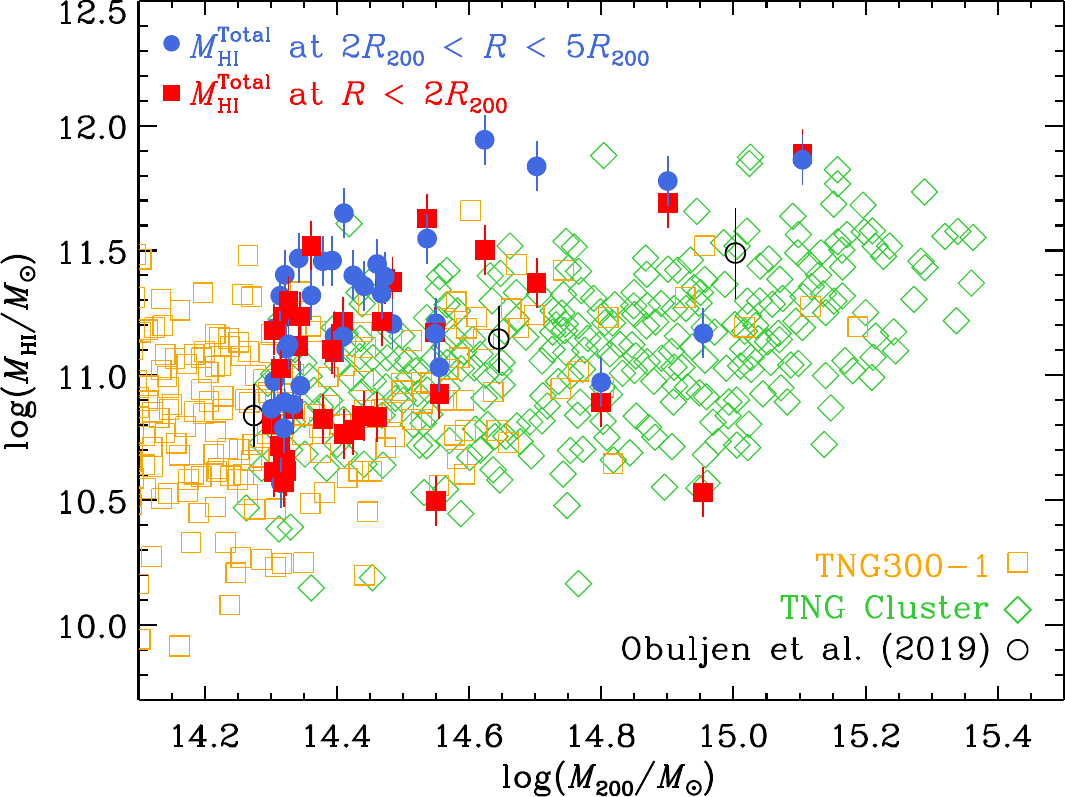}
    \caption{H\,{\sc{i}} total mass in each clusters within $2R_{200}$ (red squares) or $2R_{200}<R<5R_{200}$ (blue solid dots), in comparison with the simulation results from TNG Clusters \citep[green diamonds][]{2024A&A...686A.157N} and TNG300 \citep[coral boxes][]{2018MNRAS.475..648P, 2018MNRAS.475..624N, 2018MNRAS.477.1206N, 2018MNRAS.480.5113M, 2018MNRAS.475..676S}. H\,{\sc{i}} abundances from SDSS data from \citet{2019MNRAS.486.5124O} is shown in black circles.
    }
    \label{MHIMhalo}
\end{figure*}

\begin{figure*}
    \centering
    \includegraphics[width=0.95\linewidth]{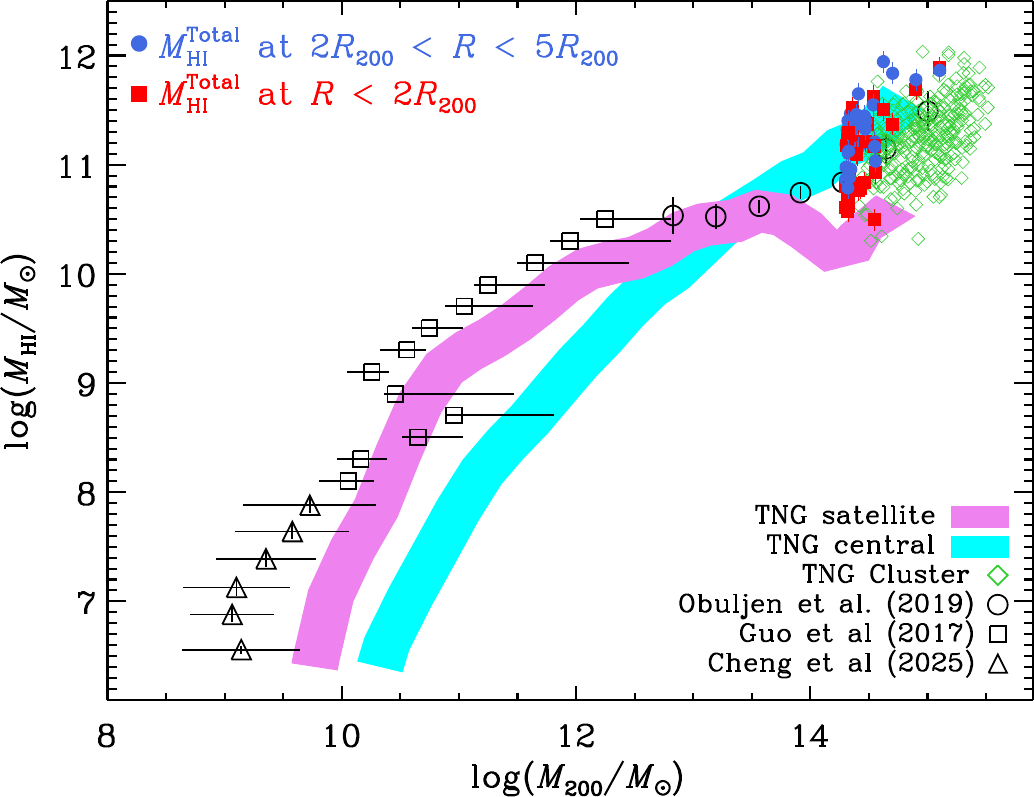}
    \caption{H\,{\sc{i}} mass as a function of halo mass over a wide mass range. The integrated H\,{\sc{i}} mass within $R<2R_{200}$ (red squares) and $2R_{200}<R<5R_{200}$ (blue solid circles) for individual clusters is shown. The $M_{\rm HI}$-$M_{\rm halo}$ relations for group galaxies from \citet{2019MNRAS.486.5124O} and \citet{2017ApJ...846...61G} are shown as black open circles and open squares, respectively. The HI-selected dwarf galaxies ($M_{\rm HI}<10^{8}\,M_\odot$) are shown as open triangles \citep{2025ApJS..281...66C}. We further compare our measurements with the TNG cluster simulation results (green diamonds; \citealt{2024A&A...686A.157N}) and the TNG100 simulation results for central galaxies (cyan shaded region) and satellite galaxies (violet shaded region; \citealt{2018ApJ...866..135V}). Overall, the simulation results are broadly consistent with the observations, but are systematically lower than the observed H\,{\sc{i}} masses over a wide range of halo masses.
    }
    \label{MHIMhaloAll}
\end{figure*}

\subsection{Relaxed and Disturbed clusters}

The morphology of galaxy clusters indicates their evolutionary stage. Dynamically relaxed clusters have more symmetric distribution of ICM around their BCG, and are distributed more smoothly. When galaxy clusters are merging, clusters would be dynamically disturbed, and show irregular distribution of galaxies, and show substructures, shells or tidal streams formed by galaxies \citep[e.g., ][]{1971PASP...83..313R, 1982AJ.....87....7S, 1986RvMP...58....1S}. Identification of a cluster's dynamical state relies on high-resolution images of the ICM together with a suite of morphological indicators (e.g. the distance between the BCG and the X-ray center and measures of symmetry). Combining these metrics is necessary but can produce ambiguous or conflicting signals, so making strict criteria to classify the cluster stage robustly is non-trivial \citep{2024ApJ...967...14C}.

To assess the dynamical state of our clusters, we adopt the multi-indicator framework developed by \citet{2026A&A...708A.262K} for classifying systems as relaxed or disturbed. While earlier approaches often relied on a single diagnostic, this framework combines several independent indicators-including the Sparsity, stellar mass gap, center offset, satellite stellar mass fraction, kuiper's V test, and mirror asymmetry-into a multi-dimensional indicator space. The class-conditional Gaussian mixture model method enables a probabilistic estimate of the dynamical state for each cluster by modeling multi-indicator spaces for different dynamical states.

Using their recipe, we derive for each system the probability of being relaxed or disturbed, and we flag clusters with probabilities $>0.5$ as belonging to the corresponding category. In total, we identify 9 relaxed and 22 disturbed systems and shown in Table \ref{tab1}. It is important to note that these classifications are based solely on galaxies within $R < R_{200}$; therefore, even clusters flagged as relaxed may still host additional substructures or companion systems at larger radii (Figure \ref{member_selection1}, \ref{member_selection2}, \ref{member_selection3}).

The classification results are listed in Table~\ref{tab1}. Among the cluster catalog, we find that five clusters (AXES\_C\_6155, AXES\_C\_39639, AXES\_C\_55129, AXES\_C\_60247, and AXES\_C\_82283) do not have reliable classification results. The first four clusters have fewer than 10 member galaxies with $M_*>10^{10}M_\odot$ within the central region ($R< 0.5R_{200}$), which is significantly lower than the expected number for clusters in our sample \citep[roughly $\sim30$ members for $M_{200}\sim2\times10^{14}\,M_\odot$;][]{2025arXiv251206138N}. AXES\_C\_82283 is affected by a bright foreground star ($V\sim8$~AB mag), which likely reduces the spectroscopic completeness and compromises the dynamical state classification. In addition, the apparent X-ray extent radii of these five clusters are $\sim3$~arcmin, among the smallest size in the AXES cluster sample (typically 5 arcmin). The small X-ray extent and the small number of galaxy members suggest that their X-ray emission may also be affected by AGN contamination. We therefore exclude these five clusters from the relaxed and disturbed cluster stacking analyses. This exclusion does not affect our results, as the analysis is based on stacking and these clusters contribute only a small number of member galaxies.

The member galaxies of these two subsamples (relaxed and disturbed) are subsequently stacked as the same 18 bins for comparative analysis. The stacking results are shown in Figure \ref{HIstack_relaxed} and \ref{HIstack_disturbed}, and compared with the full stacking sample in Figure \ref{MHI_stage}. We find that the H\,{\sc{i}} mass and the HI–to–stellar mass ratio generally follow a similar trend to those of the full cluster sample, with differences less than $1\sigma$ in significance. The similarity of the results across different dynamical states suggests that the average H\,{\sc{i}} environment of clusters is broadly comparable. No statistically significant difference is found between relaxed and disturbed clusters in any radial bin.
The higher H\,{\sc{i}} content observed in disturbed clusters at $R<R_{200}$ is consistent with previous findings for clusters with blue BCGs, which are embedded in bluer, more gas-rich large-scale structures extending to several $R_{200}$ (e.g., Figure 7 of \citet{2025A&A...700A.264D}).

\subsection{Total H\,{\sc{i}} mass in clusters}

The $M_{\rm HI}$–$M_{\rm halo}$ relation bridges the gap between dark matter structure formation and the baryonic processes governing galaxy evolution, making it a fundamental observable for both galaxy physics and cosmology \citep{2017ApJ...846...61G, 2017MNRAS.470..340P, 2019MNRAS.486.5124O,2021MNRAS.506.4893C, 2022ApJ...941...48L, 2023MNRAS.523.2693D, 2023ApJ...952L..41K}. With the large sky coverage of FASHI, we also estimate the total H\,{\sc{i}} mass in each cluster by H\,{\sc{i}} spectrum stacking. The number of member galaxies extending out to $5R_{200}$ ranges from about 500 to 1500. To maintain a reasonable signal-to-noise ratio in the stacked spectra and to explore the total H\,{\sc{i}} content in different cluster regions, we divide the galaxies in each cluster according to their projected cluster-centric distance, showing the total H\,{\sc{i}} mass within projected distance $R < 2R_{200}$ and in the outer region of $2R_{200} < R < 5R_{200}$. The total H\,{\sc{i}} mass is estimated as the product of the average stacked H\,{\sc{i}} mass and the number of spectra used in stacking. The results are presented in Figure \ref{MHIMhalo}.

Our results also show that $M_{\rm HI}^{2<R/R_{200}<5}$ are slightly larger than the $M_{\rm HI}^{R/R_{200}<2}$ by less than 0.5 dex, and both generally consistent with the distribution predicted by the TNG300 and TNG-Cluster simulations \citep{2018MNRAS.475..648P, 2018MNRAS.475..624N, 2018MNRAS.477.1206N, 2018MNRAS.480.5113M, 2018MNRAS.475..676S, 2024A&A...686A.157N}, as well as with observational results \citep{2019MNRAS.486.5124O}. In $\Lambda$CDM-based models, the $M_{\rm HI}$–$M_{\rm halo}$ relation describes how the average neutral hydrogen mass scales with the host halo mass, reflecting the combined effects of gas accretion, star formation, and feedback processes. Hydrodynamical simulations and semi-analytic models both predict that the H\,{\sc{i}} fraction declines toward higher halo masses, primarily due to stronger AGN and environmental feedback in massive systems \citep{2016MNRAS.456.3553V, 2018ApJ...866..135V, 2019MNRAS.483.4922B}. The overall consistency between our stacking results and these predictions supports the scenario that cluster environments efficiently deplete H\,{\sc{i}} gas in the inner regions, while the outskirts retain a higher neutral gas content. We caution, however, that our cluster sample spans a limited range in $M_{200}$ ($\sim$0.3~dex), which precludes a detailed test of the slope of the $M_{\rm HI}$–$M_{\rm halo}$ relation at the cluster scale. The comparison with simulations should therefore be viewed as a consistency check rather than a model discrimination.

We further note that the identification of H\,{\sc{i}} in cosmological hydrodynamic simulations is itself non-trivial. At the $\sim$1~kpc resolution of current simulations, the cold interstellar medium cannot be directly resolved, and each model employs different subgrid prescriptions to partition gas into atomic and molecular phases. These prescriptions differ in their treatment of self-shielding, the metagalactic UV background, and the partitioning between H\,{\sc{i}} and H$_2$, and can introduce systematic offsets of $\sim$0.2--0.5~dex in the predicted $M_{\rm HI}$ at fixed halo mass \citep{2020MNRAS.497..146D}. Therefore, the comparison with any individual simulation should be interpreted with caution.

\section{Discussion}

\subsection{Uncertainty of the Results}\label{sec:uncertainty}

While providing exceptional sensitivity, the $3'$ beam size of FAST introduces significant challenges in resolving individual galaxies within dense cluster environments. This limited angular resolution is the primary source of uncertainty in our analysis. The main effects are twofold: first, beam blending can occur when multiple galaxies, such as galaxy pair, fall within the same beam, leading to an overestimation of the H\,{\sc{i}} mass for individual sources and a potential overestimate of the average H\,{\sc{i}} mass in stacked measurements. Second, the flux from extended low-surface-brightness emission might be resolved out or inaccurately measured, which could lead to an underestimation of the total H\,{\sc{i}} content, particularly in the outer regions of galaxies or in the intra-cluster medium (e.g., diffuse H\,{\sc{i}} in galaxy groups). We correct the blending issue by combining the nearby targets into one aperture for spectrum extraction and stacking, which would bias the results for dwarf galaxies, which are gas rich and more easy to be affected by the environment and ICM. 

Besides the bias introduced by the FAST beam size, we extract the H\,{\sc{i}} spectra using a $6'$ aperture, corresponding to $\sim$200 kpc at $z \sim 0.03$. This aperture is sufficiently large to encompass most of the H\,{\sc{i}} associated with individual galaxies, but it may still miss extended H\,{\sc{i}} components, particularly in interacting or merging systems. For example, previous H\,{\sc{i}} observations of galaxy groups have shown that a significant fraction of the H\,{\sc{i}} gas does not always reside within massive galaxies, but can instead be distributed between group members or stripped into the intragroup medium as diffuse or tidal H\,{\sc{i}} \citep{2023A&A...670A..21J}. As a result, our H\,{\sc{i}} stacking analysis is also expected to underestimate such diffuse H\,{\sc{i}} components.

\citet{2025ApJ...993L..18D} compared extraction apertures of 150--600~kpc in physical diameter for stacking of intermediate-redshift field galaxies (MALS/MeerKAT interferometric data), finding that the stacked H\,{\sc{i}} flux increased by a factor of $\sim 2$ between 150 and 300~kpc apertures due to extended H\,{\sc{i}} reservoirs beyond the stellar disk. Our fixed angular aperture ($\sim$140--400~kpc across $z=0.02$--0.06) covers a comparable physical range, and the aperture-size dependence reported by \citet{2025ApJ...993L..18D} provides an approximate upper bound ($\sim 0.3$~dex) on the systematic uncertainty from our aperture choice. For cluster galaxies, the effect is expected to be smaller because environmental processes preferentially strip the extended, low-column-density H\,{\sc{i}} that dominates the large-aperture signal in field galaxies.

Since a larger aperture would introduce more noise in H\,{\sc{i}} spectrum, we still keep the 6' aperture in this study. Future high-resolution H\,{\sc{i}} observations, for instance with MeerKAT or the planned FAST Core Array \citep{2024AstTI...1...84J}, will be essential to resolve these blended sources, confirm the magnitude of the radial decline, and provide more precise measurements of the H\,{\sc{i}} properties in the challenging central regions of galaxy clusters.

From the optical side, all H\,{\sc{i}} spectra are extracted based on the RA, Dec, and velocity provided by the DESI and SGA catalogs. Therefore, the main uncertainties originate from the input catalogs, particularly for faint targets primarily selected from DESI. As a multi-fiber spectrograph, all targets in the same configuration run share the same exposure time. However, the success rate of redshift determination differs between red and blue galaxies: emission lines in blue galaxies make spectroscopic redshift measurements easier, whereas red galaxies require high S/N continuum to detect absorption lines. This issue is especially severe for dwarf cluster members, which are mostly quenched in dense environments and may lack reliable spectroscopic redshifts \citep{2025arXiv251019958M}. Moreover, the high surface number density of galaxies in cluster centers can lead to fiber collisions in the focal plane, causing some targets to be missed. Consequently, optical catalogs in cluster regions tend to be incomplete for dwarf satellite galaxies, which are typically quenched with low H\,{\sc{i}} content, contributing to the decline in $M_{\rm HI}/M_*$ seen in Figure \ref{MHI}.

Another source of incompleteness arises from the spectroscopic redshift catalogs. Approximately 5\% of targets with spectroscopic redshifts lack reliable photometry in DESI. We removed this small fraction of targets, which is not expected to significantly affect our results. In addition, some bright cluster members (e.g., $r < 19$ mag) are still missing in the DR1 release. As shown in Figure \ref{phaseplot}, the spatial distribution of member galaxies as a function of $R/R_{200}$ appears representative across different projected distances, suggesting that these missing targets are unlikely to introduce a systematic bias.

Other sources of uncertainty include: (1) stellar mass estimation, where the typical scatter is 0.3 dex, propagating into the derived H\,{\sc{i}} mass fractions and related parameters; (2) cluster parameter estimation, including intrinsic scatter in the scaling relations (Equation \ref{eqsigma}) and potential peculiar velocities of clusters, as indicated by velocity offsets in Figure \ref{member_selection1}; (3) cluster dynamical states. Since we separate clusters into relaxed and disturbed to compare their H\,{\sc{i}} properties, while increasing evidence indicates that H\,{\sc{i}} abundance is correlated with filaments \citep{2017MNRAS.466.4692K, 2018ApJ...866...78H, 2023MNRAS.518.1361T, 2024MNRAS.529.2595B}. A detailed filamentary analysis by 1DREAM code \citep{2022A&C....4100658C} will be presented in a forthcoming paper. As a first systematic exploration of H\,{\sc{i}} in galaxy clusters based on FASHI and DESI, we focus here on global population trends, while deferring detailed dynamical and filamentary analyses to future work.

Finally, we compare our $M_{\rm HI}/M_*$ results with previous studies \citep[e.g.,][]{2018MNRAS.476..875C,2021ApJ...918...53G,2023MNRAS.525..256P}. Our stacking approach effectively increases the signal-to-noise ratio, allowing us to recover H\,{\sc{i}} emission from systems that are undetected in individual observations. As a result, the stacked measurements may yield a lower average gas fraction compared to studies that only consider HI-detected galaxies. We note that \citet{2018MNRAS.476..875C} already account for non-detected H\,{\sc{i}} in their average values, which partially mitigates this effect, and would not change the conclusion of our results.

\subsection{H\,{\sc{i}} mass at $<2 R_{200}$: quick gas striping in cluster center}

\citet{2021MNRAS.507.5580H} estimated the H\,{\sc{i}} abundance in dense environments by stacking WSRT data cubes for galaxy groups identified from SDSS. The WSRT beam size of $108'' \times 22''$ is smaller than that of FAST, leading to less severe source blending. Owing to its relatively small sky coverage, WSRT is more suitable for investigating less massive clusters. In contrast, the $3'$ beam of FAST is only adequate for detecting the isolated massive galaxies, which are less affected by blending. The wide coverage and high sensitivity of the FASHI survey thus provide an opportunity to probe the H\,{\sc{i}} content of massive clusters, serving as a complementary dataset to the interferometric studies such as \citet{2021MNRAS.507.5580H} that focus on smaller areas. The two results are quantitatively consistent, while our analysis averages over an order of magnitude more galaxies, yielding statistically more robust conclusions.

Beyond direct H\,{\sc{i}} detections from data cubes, \citet{2013MNRAS.429.2191Z} employed the $M_{\rm HI}/M_*$ scaling relation calibrated by \citet{2012MNRAS.424.1471L}$,$ where the H\,{\sc{i}} gas fraction is expressed as a function of the stellar surface density within the half-light radius, NUV–$r$ color, and the $g–r$ color gradient between $R_{50}$ and $R_{90}$. This empirical relation achieves a $1\sigma$ scatter of about 0.3 dex. Using this calibration, \citet{2013MNRAS.429.2191Z} estimated H\,{\sc{i}} masses for galaxies in 300 clusters spanning halo masses from $2\times10^{12} M_\odot$ up to the most massive systems of $\sim2\times10^{14} M_\odot$. Since the H\,{\sc{i}} mass fraction of red, gas-poor galaxies is often undetected, the calibration may introduce a bias for these extremely gas-deficient systems, which dominate the cluster centers, leading to an overestimation of their H\,{\sc{i}} content \citep{2012MNRAS.422.1835S, 2023arXiv231203601L}. Despite this limitation at the extreme low-HI end, the scaling-relation-based results also show a clear declining trend of H\,{\sc{i}} mass from $\sim$10 $R_{200}$ to the cluster center. More importantly, galaxies with higher stellar surface densities appear less susceptible to gas loss.

All these studies consistently show a decline of about 0.5 dex in both H\,{\sc{i}} mass and $M_{\rm HI}/M_*$ from $\sim$2 $R_{200}$ to the cluster center, indicating that the dense environment starts to strongly affect galaxies around $R_{200}$. Recent X-ray stacking analyses from the SRG/eROSITA All-Sky Survey have revealed that the diffuse hot gas in galaxy clusters can extend out to $2R_{200}$ \citep[e.g.,][]{2025arXiv250925317Z}, suggesting that the intracluster medium occupies a much larger volume than previously assumed. This widespread $\sim 10^7$ K gas may heat or evaporate the neutral hydrogen in infalling galaxies, further contributing to the observed decline of H\,{\sc{i}} content toward the cluster center. This turnover radius is in good agreement with the environmental trends predicted by the EAGLE simulations for cluster galaxies \citep{2016MNRAS.461.2630M}. Simulations also show that ram pressure and tidal interactions efficiently remove diffuse gas, with H\,{\sc{i}} being preferentially stripped. Recent optical studies of jellyfish galaxies reveal that ram pressure stripping can already begin beyond the virial radius ($\sim$1.3 $R_{200}$) and last for $\sim$ 0.6 Gyr after pericentric passage, implying that galaxies undergo significant gas loss even before reaching the cluster core \citep{2023A&A...679A.157R, 2024MNRAS.533..341S}. 

On the other hand, although most galaxies in the cluster center ($<1\times R_{200}$) are quenched, some continue forming stars with rates consistent with the star-forming main sequence \citep[e.g.,][]{2007ApJ...660L..43N, 2014ApJS..214...15S, 2022ApJS..262...31L, 2022ApJS..263...40M, 2025ApJS..279...43C}. This suggests that molecular gas traced by CO is less susceptible to stripping than the more diffuse H\,{\sc{i}} component. Wide-field CO mapping to clusters will be crucial to test this scenario \citep[e.g., ][but with wider coverage]{2022ApJS..262...31L, 2022ApJS..263...40M} .

\subsection{H\,{\sc{i}} mass beyond $2R_{200}$: Pre-Processing of infall galaxies}\label{sec:preprocessing}

At distances beyond $2R_{200}$, we find that the H\,{\sc{i}} mass and $M_{\rm HI}/M_\star$  remain nearly constant, showing little dependence on clustercentric radius. This result implies that the majority of galaxies in the cluster outskirts have not yet averagely experienced strong environmental effects capable of efficiently removing their gas reservoirs. The invariance of $M_{\rm HI}/M_\star$ at large radii may indicate that the pre-processing of galaxies occurs in galaxy groups or along filaments connected out of the cluster, where galaxies can already experience tidal or hydrodynamical interactions. 

Theoretical and simulation works have demonstrated that galaxy evolution within the cosmic web is far from isolated. Galaxies that appear in cluster outskirts today often trace filamentary structures, where environmental influence already becomes significant. Recent simulation-based studies \citep[e.g.,][]{2018ApJ...866...78H, 2023MNRAS.518.1361T, 2024MNRAS.529.2595B, 2025arXiv250922802C} have shown that galaxies residing in filaments experience multiple pre-cluster interactions, such as mild ram-pressure stripping, tidal effects, and starvation, collectively referred to as pre-processing. These processes suppress star formation and reduce the available gas content even before the galaxies cross the cluster’s virial radius. Indeed, as shown in Figures~\ref{member_selection1}--\ref{member_selection3}, the majority of our cluster member galaxies at $R/R_{200}\gtrsim 1$ appear to lie within filamentary structures rather than being isotropically distributed. Their H\,{\sc{i}} content may therefore have been reduced in the overdense regions, regardless of projected cluster-centric distance. This likely implies that member galaxies would continue to show suppressed H\,{\sc{i}} content even when probed out to larger cluster-centric distances (e.g., $\lesssim 10 R_{200}$). Our future approach to separate galaxies by their filamentary directions could help solidify this hypothesis.

It is also possible that orbital mixing contributes to the observed constancy. Cosmological simulations indicate that some galaxies beyond 1–3R$_{200}$ are backsplash systems that have previously passed through the cluster core \citep[e.g., ][]{2025arXiv250917697Z}. These galaxies may have already lost part of their gas content during pericentric passage and now appear in the outskirts, lowering the overall gas fraction at large radii. The combination of newly infalling, gas-rich galaxies and these gas-depleted backsplash systems would naturally smooth the radial gradient of H\,{\sc{i}} fraction.

Observations of nearby clusters also support this scenario of the mixture galaxy populations. \citet{2024MNRAS.528..919P} studied the local galaxy cluster A2670, and found that galaxies at 2–3$R_{200}$ already show disturbed morphology even at $> 4 \times R_{200}$. \citet{2022ApJS..258...32S} showed in the SuperGroup A1882 that star formation in low-mass galaxies is suppressed out to $\sim4$ times the virial radius of major substructures, implying that quenching processes can begin in infall regions before galaxies reach the dense group environment.

In summary, the nearly constant H\,{\sc{i}} fraction beyond $2R_{200}$ likely results from a combination of early pre-processing, the predominance of filamentary environments, and orbital mixing of infall and backsplash populations. Distinguishing these effects requires future studies that include more comprehensive identifications of substructures and subgroups.

\subsection{–Halo Relation Across a Broad Halo Mass Range}

To examine the H\,{\sc{i}} mass within dark matter halos over a broader mass range, we compile and present the $M_{\rm HI}$–$M_{\rm halo}$ results for halos with $10<\log(M_{200}/M_\odot)<14$ from \citet{2019MNRAS.486.5124O} and for halos with $\log(M_{200}/M_\odot) < 10$ from \citet{2025ApJS..281...66C}.
A consistent trend is seen between simulation results and observational constraints \citep{2018ApJ...866..135V, 2024A&A...686A.157N}, although an offset less than 0.5 dex appears (Figure~\ref{MHIMhaloAll}).

The overall $M_{\rm HI}$–$M_{\rm halo}$ relation is not expected to be strictly monotonic. Semi-analytic models such as SHARK \citep{2020MNRAS.498...44C} predict three distinct regimes: a monotonic rise of H\,{\sc{i}} mass with halo mass at the low-mass end ($M_{\rm halo}\lesssim10^{11.8}M_\odot$), where feedback processes are relatively weak and cold gas can be efficiently retained; a decline or flattening in the transition regime ($10^{11.8}\lesssim M_{\rm halo}/M_\odot\lesssim10^{13}$), driven by the increasing importance of AGN feedback and virial heating that suppress gas cooling; and a subsequent upturn at the most massive end ($M_{\rm halo}\gtrsim10^{13}M_\odot$). 

In this high-mass regime, massive halos typically correspond to galaxy groups or clusters that host multiple gas-rich satellite galaxies, whose combined contribution elevates the total H\,{\sc{i}} content. Moreover, in dynamically young or disturbed clusters where the hot intracluster medium and virial shocks have not yet been fully established, a larger fraction of H\,{\sc{i}} may survive either within satellite galaxies or in the intragroup medium, further enhancing the integrated H\,{\sc{i}} mass. The comparison between the relaxed and disturbed clusters in Figure \ref{MHI_stage} only shows a 
tendency of higher H\,{\sc{i}} mass and $M_{\rm HI}/M_*$ for low-mass galaxies in disturbed clusters. A larger cluster sample with additional spectroscopic redshifts from future DESI data releases or the ongoing 4MOST survey will be required to robustly quantify any potential difference.

In the future, more detailed comparisons-where simulations not only reproduce the total H\,{\sc{i}} mass but also the radial distribution and mass-dependent deficiency of H\,{\sc{i}} within clusters-will be highly valuable. Such efforts, especially those that trace the physical processes governing the interaction between galaxies and the intracluster medium (such as ram-pressure stripping, tidal interactions) as well as the star formation activity, will greatly improve our understanding of the environmental mechanisms responsible for gas loss and quenching. Direct, statistically robust comparisons between observations and simulations will help to validate and refine the subgrid physics models that encode these complex environmental processes.

\subsection{Gas Consumption in Clusters}

The downsizing effect indicates that massive galaxies form their stars earlier and evolve more rapidly than lower-mass systems \citep{1996AJ....112..839C}. As a result, much of the neutral hydrogen gas in massive galaxies was consumed through star formation in the early Universe, leading to a lower $M_{\rm HI}/M_*$ ratio. Our findings further reveal that massive galaxies in cluster centers exhibit an even greater deficiency in atomic gas, suggesting that the cluster environment plays an additional role in gas removal. Environmental mechanisms such as ram-pressure stripping may efficiently deplete the remaining gas reservoirs, thereby accelerating the evolution of these galaxies.

However, the influence of the large-scale cosmic web-particularly the filaments that funnel galaxies into clusters-appears to be more complex. \citet{2017MNRAS.466.4692K} present a seemingly counterintuitive result in the nearby Universe: galaxies with stellar masses $\log(M_*/M_\odot)>11$ located near the spines of filaments possess systematically higher H\,{\sc{i}} fractions compared to a control sample in the field. They interpret this as evidence that massive galaxies may accrete cold gas from the intrafilament medium, replenishing their H\,{\sc{i}} reservoirs via cold-mode accretion. This finding seems to contradict the severe gas deficiency we observe in the cores of clusters.

Therefore, discussions of environmental effects must extend beyond a galaxy’s current location to encompass its full environmental history. Recent work by \citet{2025arXiv250917697Z} reveals that filaments regulate galaxy evolution in a mass-dependent manner. For low-mass galaxies, filaments primarily channel them into group environments, where pre-processing (e.g., tidal stripping and galaxy harassment) quenches their star formation. In contrast, massive central galaxies that entered filaments roughly 9 Gyr ago can continue to grow efficiently within the filaments themselves. Thus, the star formation histories of galaxies in dense environments are shaped by a sequence of environmental stages, leading to the observed diversity in gas fractions.

\subsection{Dwarf Galaxies Lost Gas Earlier? Massive Galaxies Lost Gas Quicker?}

Our results suggest a potential mass-dependent trend in H\,{\sc{i}} depletion: the H\,{\sc{i}} mass in massive galaxies appears to drop more sharply within the inner cluster regions ($R < 2 R_{200}$), while low-mass galaxies exhibit a more gradual decline starting from as far out as $5 R_{200}$. However, this interpretation must be treated with caution due to the significant impact of our blending correction. As detailed in Section \ref{subgroup}, the process of combining blended sources disproportionately removes low-mass galaxies from the stacking sample and moves them into higher mass bins. The resulting correction factor is the largest for the dwarf galaxy bin (reducing the inferred H\,{\sc{i}} mass by $\sim$0.2 dex), and the stellar mass distribution itself is altered. Consequently, the apparent early-onset H\,{\sc{i}} loss in dwarfs could be, at least in part, a systematic effect of our methodology rather than a pure physical signal.

Bearing this major caveat in mind, the differing slopes could hypothetically point to distinct dominant gas-removal mechanisms. Massive galaxies, with their deeper potential wells, may only be effectively stripped by the strong ram pressure encountered in the cluster core. In contrast, the shallow potentials of dwarf galaxies make them vulnerable to a broader range of environmental effects \citep{2022NatAs...6..647C}, such as tidal interactions and starvation, which can operate efficiently in the cluster outskirts during pre-processing. The nearly flat H\,{\sc{i}} distribution for massive galaxies beyond $2 R_{200}$ suggests that their primary gas consumption mechanisms are less influenced by the cluster environment until they venture closer to the center. A definitive conclusion on the mass dependence of H\,{\sc{i}} loss radii must await observations capable of resolving these populations without the confounding effects of beam blending.

We also note that our sample consists exclusively of massive clusters ($M_{200} > 10^{14}\,M_\odot$), and that the timescale of gas stripping and quenching may differ in lower-mass groups \citep{2021MNRAS.501.5073O}. Our conclusions therefore apply to the massive cluster regime.

\subsection{Star-Forming Galaxy Fraction in Galaxy Clusters}

Previous observations, utilizing various methodological approaches, have consistently confirmed that the fraction of star-forming galaxies declines significantly toward the centers of galaxy clusters \citep{2015ApJ...806..101H}. However, a key finding is that even at large cluster-centric distances, out to approximately $5R_{200}$, the star-forming fraction remains lower than that observed in field galaxies. 

From the perspective of cold gas content, our results provide a crucial physical basis for this phenomenon: the average H\,{\sc{i}} mass fraction in these outer regions is also systematically lower than the typical values from H\,{\sc{i}} stacking results. This deficiency in the neutral gas reservoir, essential for future star formation, strongly suggests that environmental processes act over very large scales to pre-process galaxies even before they enter the dense cluster core ($<1 R/R_{200}$). Our results thereby bridge the gap between the observed star formation quenching and the removal of its fundamental fuel, highlighting the role of the cluster environment in shaping galactic evolution across a wide range of densities.

The lower H\,{\sc{i}} mass fraction out to $5R_{200}$ also raises the question of how {\it field galaxies} should be defined. Galaxies are known to reside within a cosmic web of filaments, groups, compact groups, and clusters \citep{1989Sci...246..897G}, and H\,{\sc{i}} gas is likewise distributed coherently along these structures, connecting galaxies \citep[e.g.,][]{2009A&A...504...15P}. 
As shown in Figures~\ref{member_selection1}--\ref{member_selection3} and discussed in Section~\ref{sec:preprocessing}, our cluster members at $R/R_{200}\gtrsim1$ predominantly trace filamentary structures, further blurring the boundary between `cluster' and the surrounding large-scale environment.

Denser environments can accrete more cold gas (e.g., \citealt{2017MNRAS.466.4692K}), enhancing both star formation and feedback, and thus inevitably influencing galaxy properties. Conversely, galaxies in cosmic voids \citep{2021dcv..book.....T} - true isolated systems - tend to host smaller dark matter halos, contain less H\,{\sc{i}} gas, and may exhibit distinct feedback mechanisms. Therefore, galaxies outside clusters, even at 5–10 $R_{200}$, are not necessarily isolated but instead occupy environments of varying density.
Rather than being separated by a sharp boundary, the cluster and field environments are interwoven through the filamentary network. Comparisons between `cluster' and `field' galaxies thus effectively probe galaxy properties across a continuum of environmental densities. Future work quantifying galaxy populations across this gradient will help clarify how the star-forming fraction and H\,{\sc{i}} content evolve with environment, revealing the smooth progression of pre-processing effects.

\section{Conclusion}
In this study, we have conducted a systematic investigation of the neutral hydrogen content within galaxy clusters by leveraging the unique combination of the FASHI blind H\,{\sc{i}} survey from FAST and the extensive spectroscopic member catalog from DESI. Our stacking analysis, which carefully accounts for the challenges of source blending in the FAST beam, reveals a clear and consistent picture of environmental impact on galaxy evolution.

Our main results are as follows:

\begin{itemize}
\item We find a pronounced radial decline in both the average H\,{\sc{i}} mass and the HI-to-stellar mass ratio from the cluster outskirts ($\sim5R_{200}$) inward to the core. This provides direct, statistical evidence that the cluster environment effectively depletes the cold gas reservoirs of its member galaxies.

\item Both the extended hot gas traced by eROSITA X-ray stacking and the cold H\,{\sc{i}} depletion traced by FASHI identify $\sim 2R_{200}$ as a critical transition radius, suggesting this is where infalling galaxies first encounter a dense intracluster medium capable of stripping their gas reservoirs.

\item Crucially, the detected H\,{\sc{i}} deficiency persists even at large cluster-centric distances ($\sim5R_{200}$), where the density of the intracluster medium is low. This strongly indicates that pre-processing in the surrounding large-scale structure, such as filaments, groups, and maybe accretion shocks plays a vital role in quenching galaxies before they enter the cluster core proper.

\item By splitting the sample by $g-r$ color, we find that cluster galaxies show systematically lower $M_{\rm HI}/M_*$ than field galaxies at all colors, with the deficit reaching $\sim 0.5$~dex in all color bins. This indicates that the environmental H\,{\sc{i}} deficiency is not solely due to the larger fraction of quenched red galaxies in clusters; even blue, star-forming cluster members have lost a significant fraction of their cold gas.

\item By dividing the cluster sample into relaxed and disturbed systems, we do not find a significant difference in the H\,{\sc{i}} mass, 
although a weak trend is seen that disturbed clusters exhibit higher $M_{\rm HI}$ and $M_{\rm HI}/M_*$ at inner radii.

\item By stacking all H\,{\sc{i}} emission, we derive the total H\,{\sc{i}} content as a function of halo mass and recover the global $M_{\rm HI}$–$M_{\rm halo}$ relation. Across a wide halo mass range, the observed H\,{\sc{i}} masses are systematically higher than those predicted by the simulations, by up to $\sim0.3$ dex (Figure~\ref{MHIMhaloAll}).

\end{itemize}

Our findings bridge the gap between the well-established decline in star-forming galaxy fractions and the physical removal of the star-forming fuel. The good agreement of our overall H\,{\sc{i}} mass scale with the TNG-Cluster simulation provides encouraging validation for modern models of galaxy evolution. Future high-resolution H\,{\sc{i}} observations with facilities like MeerKAT and the FAST Core Array, as well as the future data release of DESI, CHANCES and eROSITA will be essential to resolve blended sources in cluster cores, and further constrain the complex interplay between galaxies and their environment. 


\begin{acknowledgments}
We thank the anonymous referee for the constructive and detailed suggestions that have significantly improved the quality of this work. We thank Hong Guo, Yara L. Jaff\'e and Amelie Saintonge for helpful discussions.
This work is supported by National SKA Program of China No. 2025SKA0150101. This work is sponsored (in part) by the Chinese Academy of Sciences (CAS) through a grant to the CAS South America Center for Astronomy. C.C. acknowledges NSFC grant No. 11803044 and 12173045. This work is supported by the China Manned Space Program with grant no. CMS-CSST-2025-A07. C.C. is supported by Chinese Academy of Sciences South America Center for Astronomy (CASSACA) Key Research Project E52H540101 and E52H540301.

E.I. acknowledge support from ANID MILENIO NCN2024\_112 and ANID FONDECYT Regular 1221846. J.M. gratefully acknowledge financial support from ANID- MILENIO - NCN2024\_112. HMH acknowledges support from Fondo Rubin/Chile 2024, DIA2322, Agencia Nacional de Investigaci\'on y Desarrollo (ANID) through Fondecyt project 3230176, Millennium Science Initiative Program NCN2024\_112 and ANID BASAL project FB210003.

WX thanks the support of National Nature Science Foundation of China (Nos 11988101, 12022306, 12203063), the support by National Key R$\&$D Program of China No. 2022YFF0503403, the support from the Ministry of Science and Technology of China (Nos. 2020SKA0110100), the science research grants from the China Manned Space Project (Nos. CMS-CSST-2025-A03, CMS-CSST-2021-B01, CMS-CSST-2021-A01), CAS Project for Young Scientists in Basic Research (No. YSBR-062), and the support from K.C.Wong Education Foundation.

The DESI Legacy Imaging Surveys consist of three individual and complementary projects: the Dark Energy Camera Legacy Survey (DECaLS), the Beijing-Arizona Sky Survey (BASS), and the Mayall z-band Legacy Survey (MzLS). DECaLS, BASS and MzLS together include data obtained, respectively, at the Blanco telescope, Cerro Tololo Inter-American Observatory, NSF’s NOIRLab; the Bok telescope, Steward Observatory, University of Arizona; and the Mayall telescope, Kitt Peak National Observatory, NOIRLab. NOIRLab is operated by the Association of Universities for Research in Astronomy (AURA) under a cooperative agreement with the National Science Foundation. Pipeline processing and analyses of the data were supported by NOIRLab and the Lawrence Berkeley National Laboratory (LBNL). Legacy Surveys also uses data products from the Near-Earth Object Wide-field Infrared Survey Explorer (NEOWISE), a project of the Jet Propulsion Laboratory/California Institute of Technology, funded by the National Aeronautics and Space Administration. Legacy Surveys was supported by: the Director, Office of Science, Office of High Energy Physics of the U.S. Department of Energy; the National Energy Research Scientific Computing Center, a DOE Office of Science User Facility; the U.S. National Science Foundation, Division of Astronomical Sciences; the National Astronomical Observatories of China, the Chinese Academy of Sciences and the Chinese National Natural Science Foundation. LBNL is managed by the Regents of the University of California under contract to the U.S. Department of Energy. The complete acknowledgments can be found at https://www.legacysurvey.org/acknowledgment/.

The Siena Galaxy Atlas was made possible by funding support from the U.S. Department of Energy, Office of Science, Office of High Energy Physics under Award Number DE-SC0020086 and from the National Science Foundation under grant AST-1616414.

DESI construction and operations is managed by the Lawrence Berkeley National Laboratory. This research is supported by the U.S. Department of Energy, Office of Science, Office of High-Energy Physics, under Contract No. DE–AC02–05CH11231, and by the National Energy Research Scientific Computing Center, a DOE Office of Science User Facility under the same contract. Additional support for DESI is provided by the U.S. National Science Foundation, Division of Astronomical Sciences under Contract No. AST-0950945 to the NSF’s National Optical-Infrared Astronomy Research Laboratory; the Science and Technology Facilities Council of the United Kingdom; the Gordon and Betty Moore Foundation; the Heising-Simons Foundation; the French Alternative Energies and Atomic Energy Commission (CEA); the National Council of Science and Technology of Mexico (CONACYT); the Ministry of Science and Innovation of Spain, and by the DESI Member Institutions. The DESI collaboration is honored to be permitted to conduct astronomical research on Iolkam Du’ag (Kitt Peak), a mountain with particular significance to the Tohono O’odham Nation.

\end{acknowledgments}

\begin{contribution}


Dr. Cheng Cheng and Edo Ibar conceived the initial research idea. 
Dr. Chuan-Peng Zhang provided the H\,{\sc{i}} stacking results. 
Dr. Hyowon Kim contributed to the classification of relaxed and disturbed clusters. 
Dr. Cheng Cheng prepared the first draft of the manuscript. 
All authors contributed to the analysis and interpretation of the results and 
to the writing of the manuscript.


\end{contribution}

%
\facilities{FAST, DESI, KPNO:Mayall (Mosaic-3), Steward:Bok (90Prime), CTIO:Blanco (DECam)}

\software{astropy \citep{2013A&A...558A..33A,2018AJ....156..123A,2022ApJ...935..167A},  HISS\footnote{\url{https://github.com/healytwin1/HISS}} \citep[][]{2019MNRAS.487.4901H}  
          }


\appendix

\section{Ra Dec distribution of the galaxies in our cluster sample}

In Figure \ref{phaseplot} we show the phase plot for all clusters together. Here we show the spatial distribution of the galaxies in all clusters analyses in this work.

\setcounter{figure}{0}
\renewcommand{\thefigure}{A-\arabic{figure}}

\begin{figure*}
    \centering
    \includegraphics[width=0.32\linewidth]{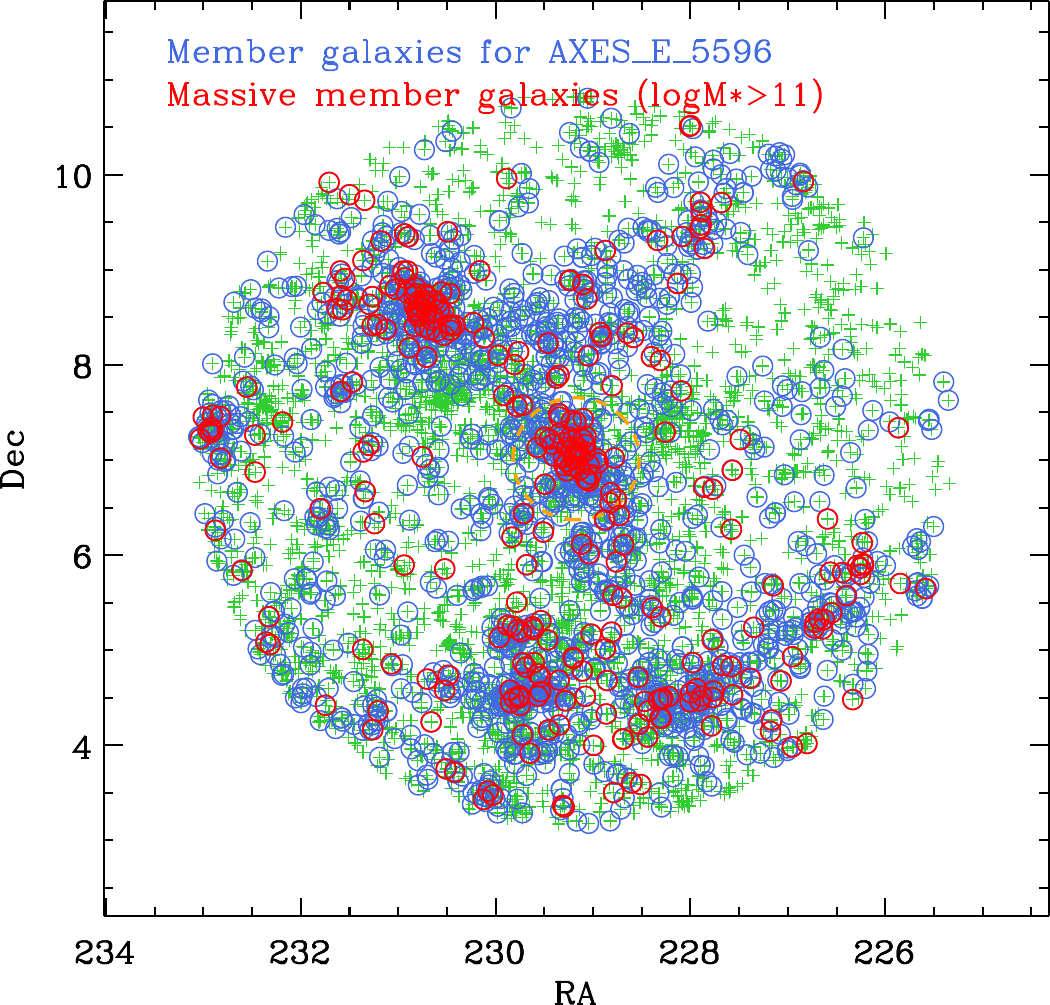}  
    \includegraphics[width=0.32\linewidth]{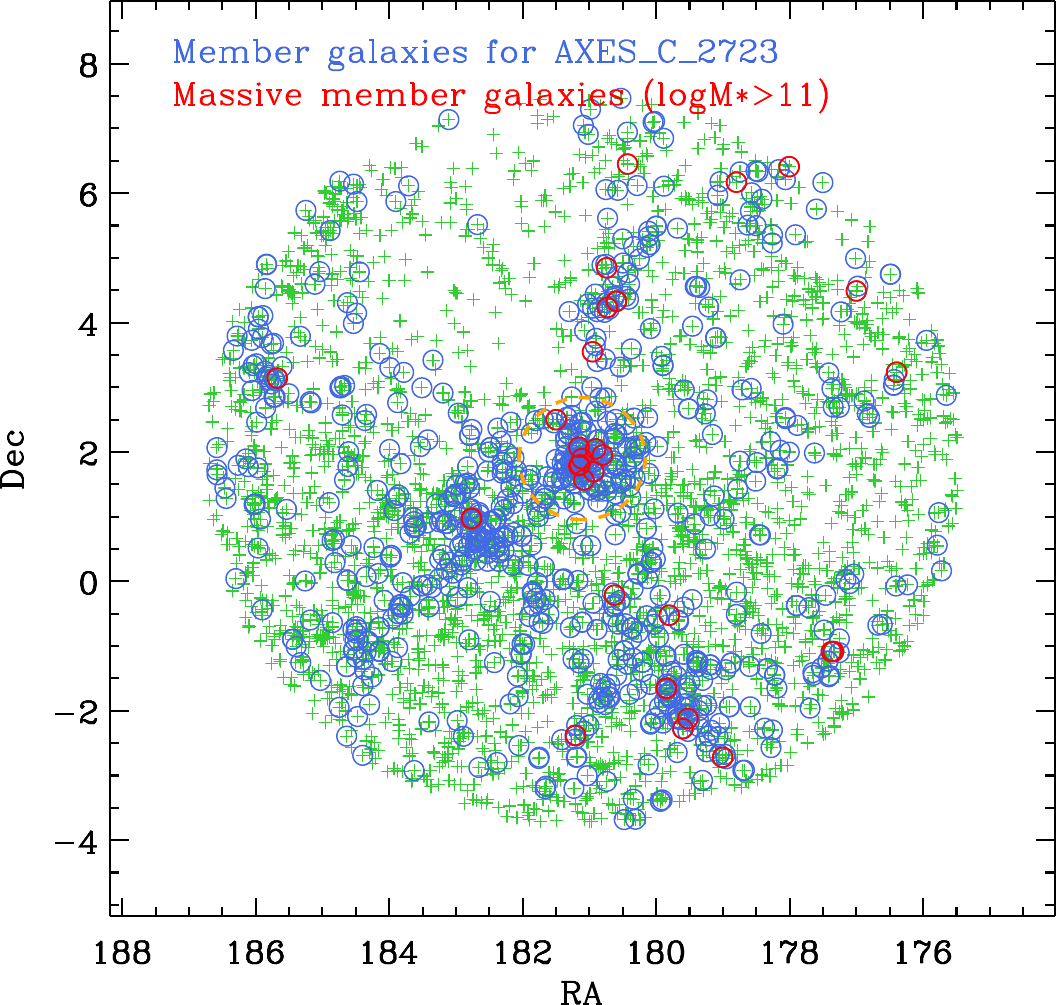}  
    \includegraphics[width=0.32\linewidth]{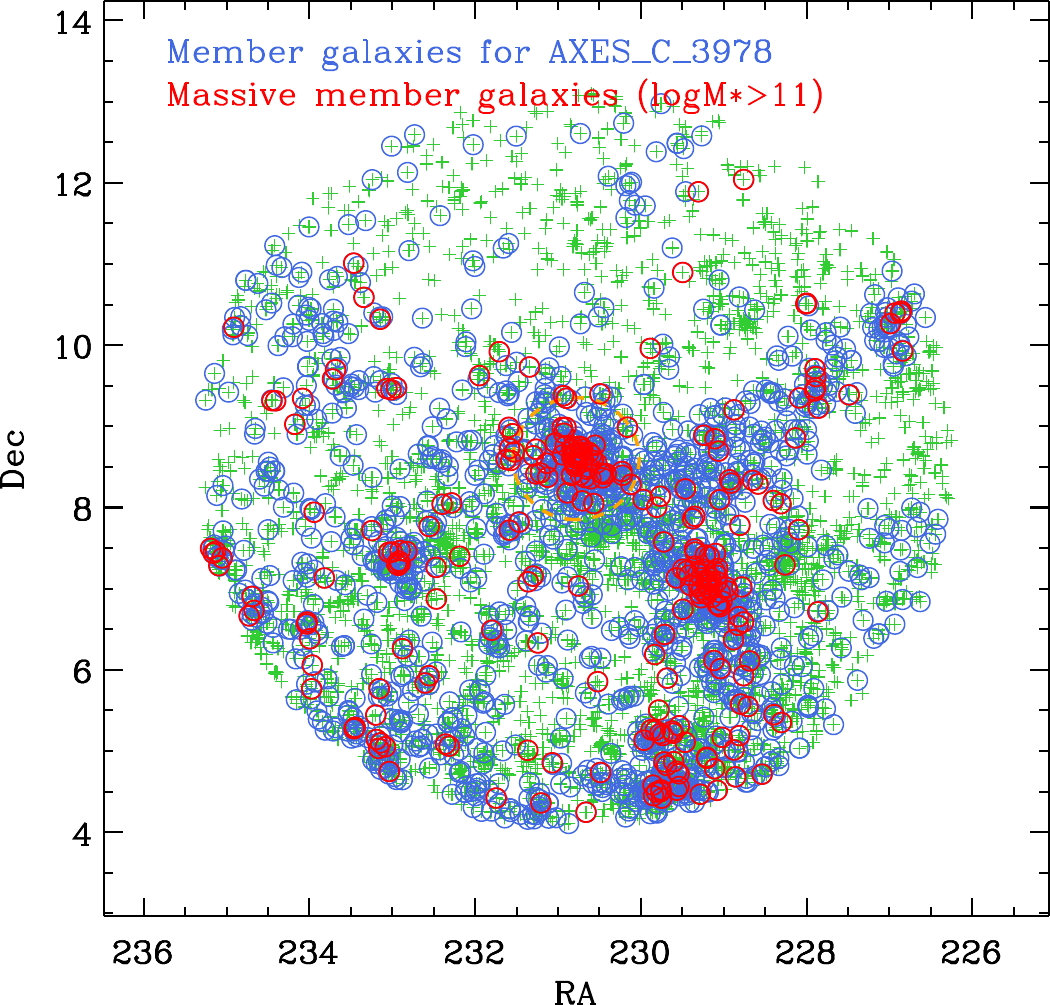}  
    \includegraphics[width=0.32\linewidth]{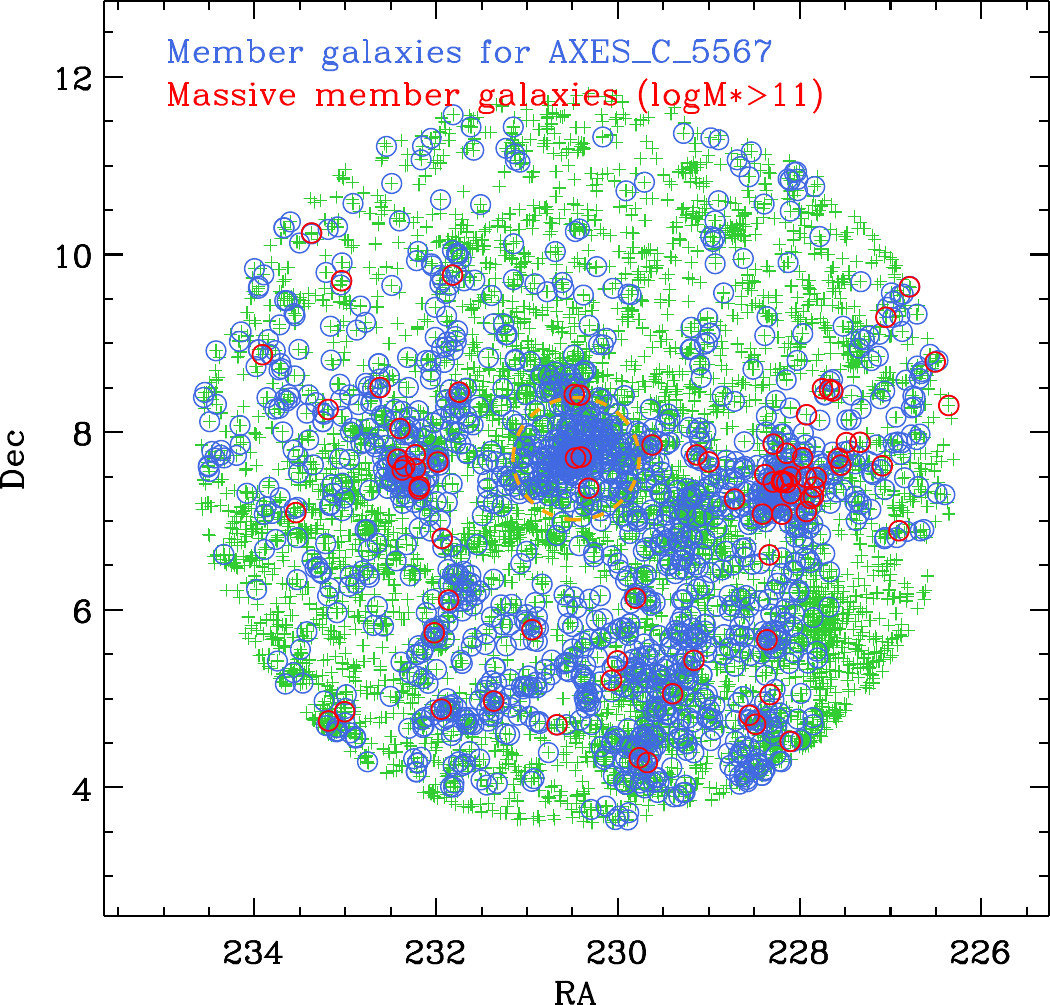}  
    \includegraphics[width=0.32\linewidth]{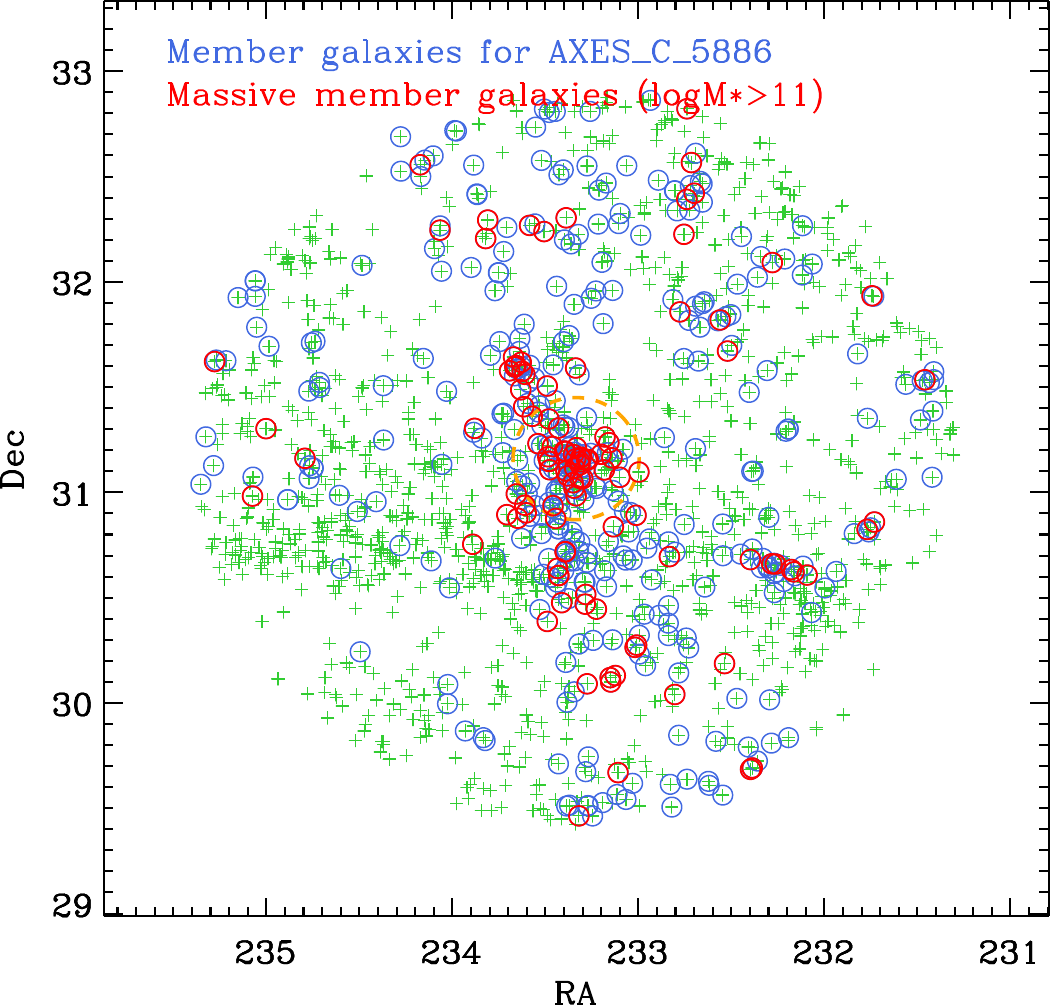}  
    \includegraphics[width=0.32\linewidth]{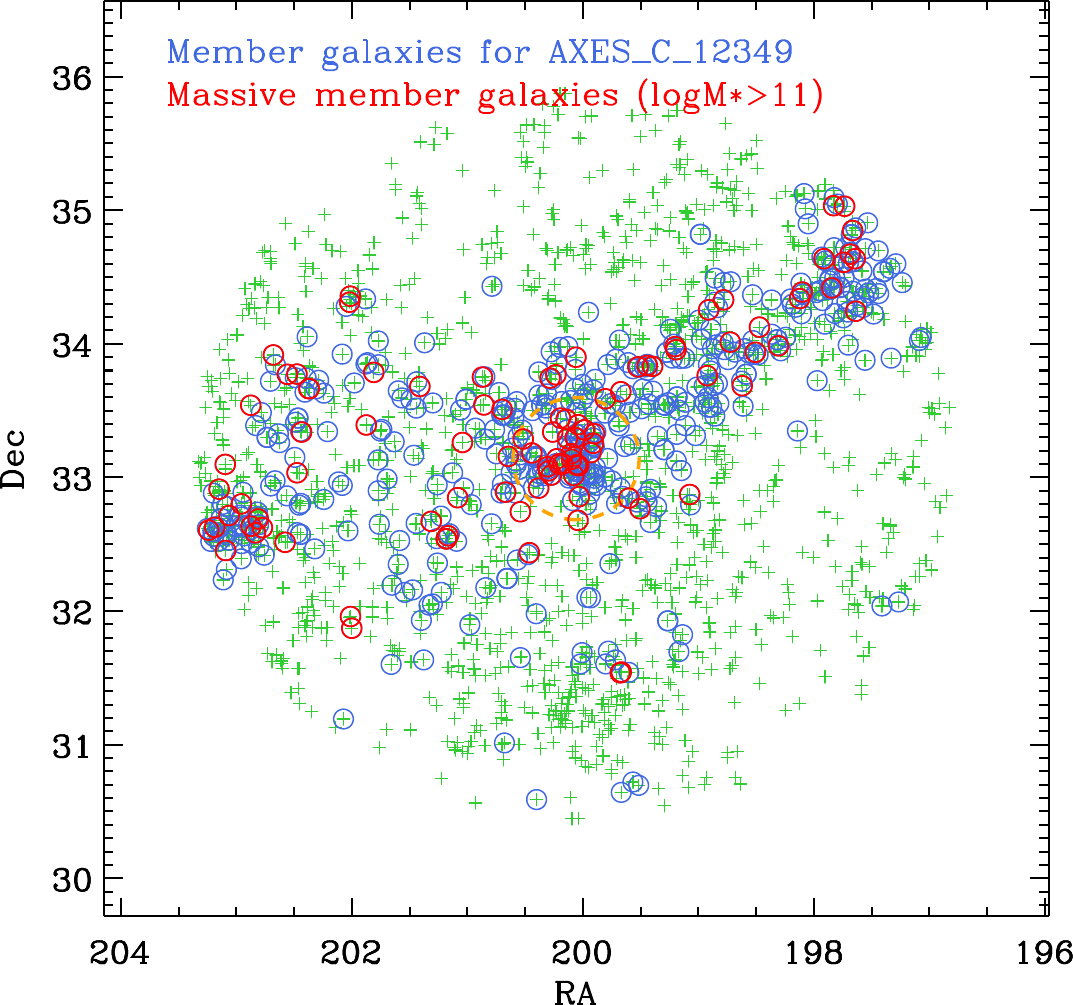}
    \includegraphics[width=0.32\linewidth]{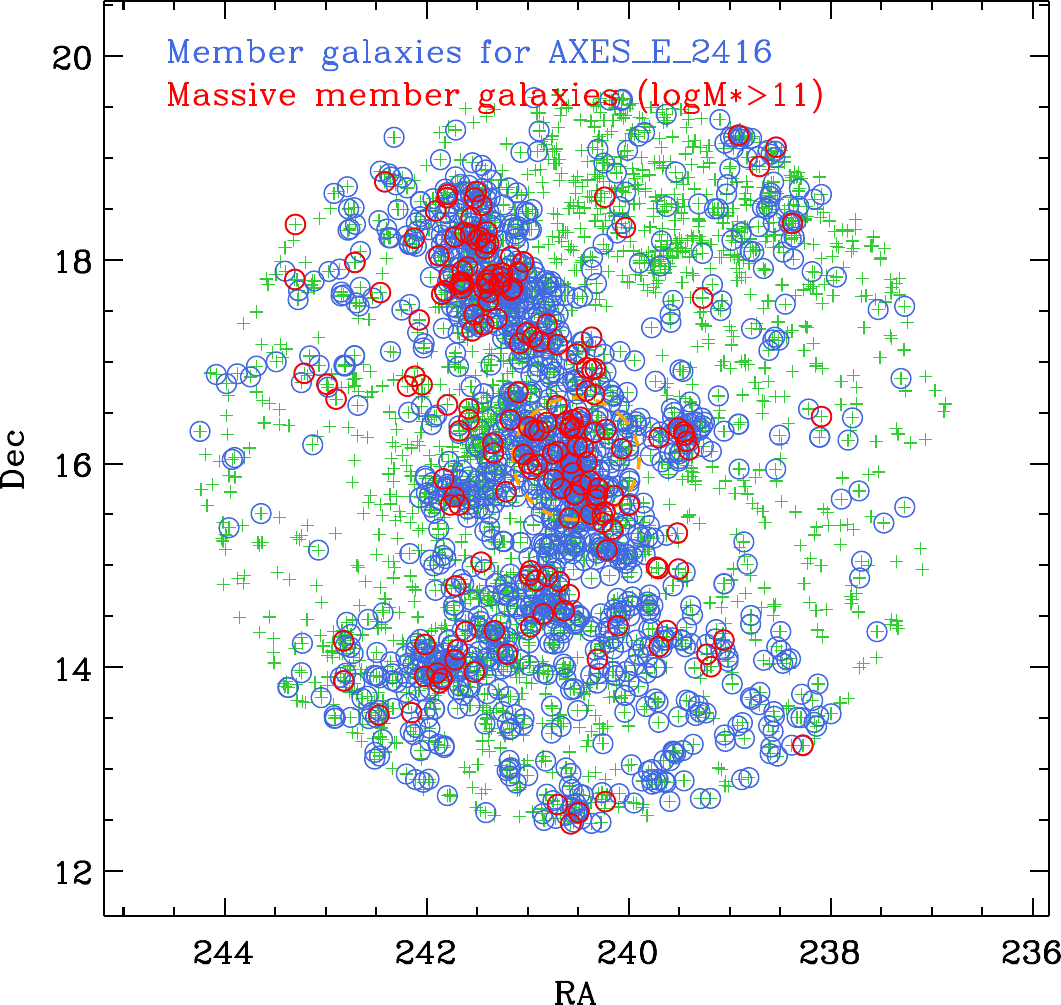}
    \includegraphics[width=0.32\linewidth]{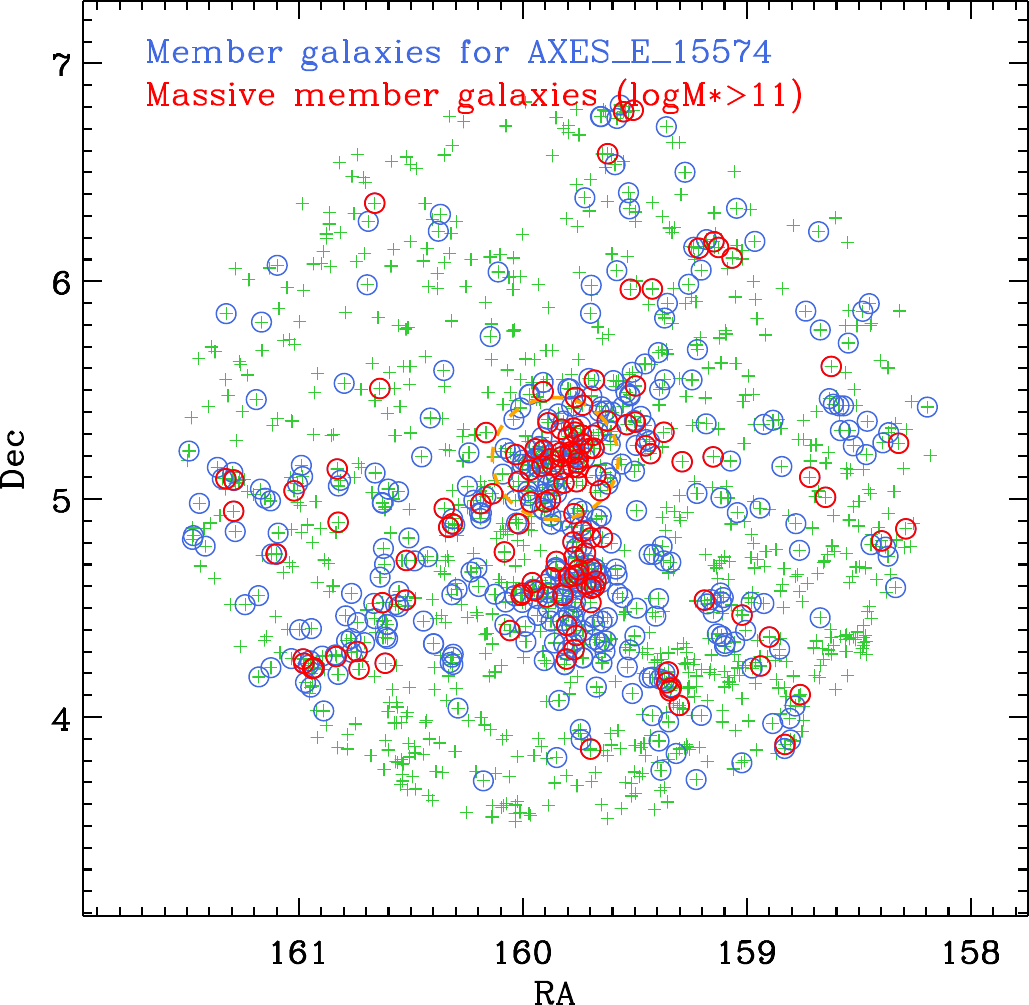}
    \includegraphics[width=0.32\linewidth]{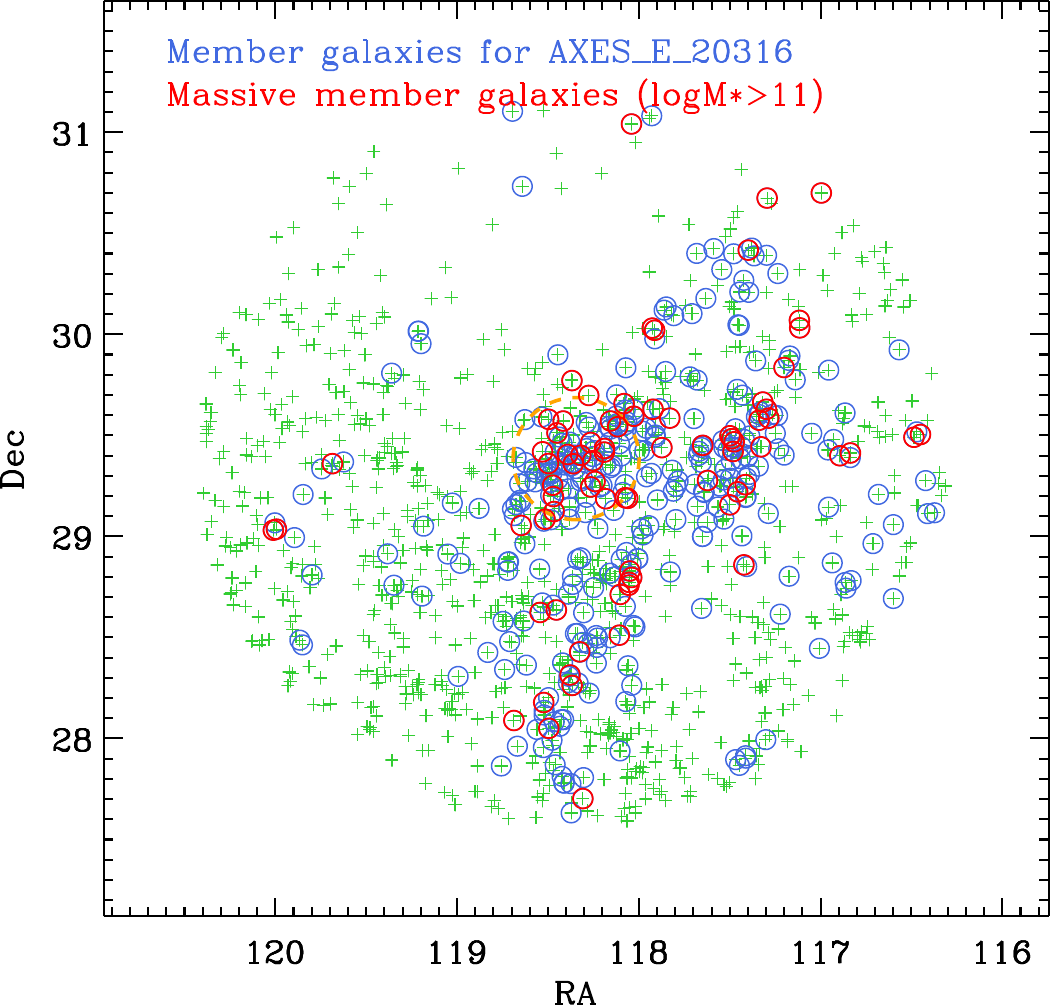}
    \includegraphics[width=0.32\linewidth]{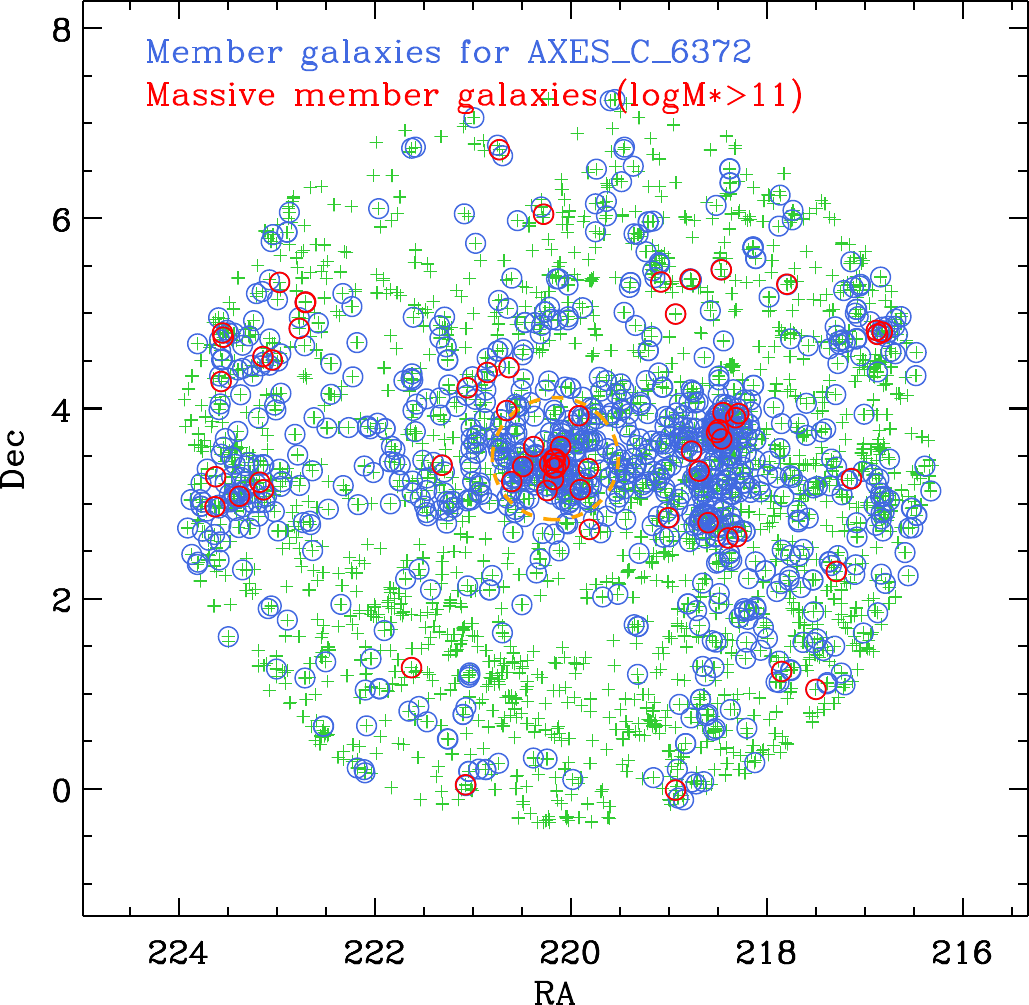} 
    \includegraphics[width=0.32\linewidth]{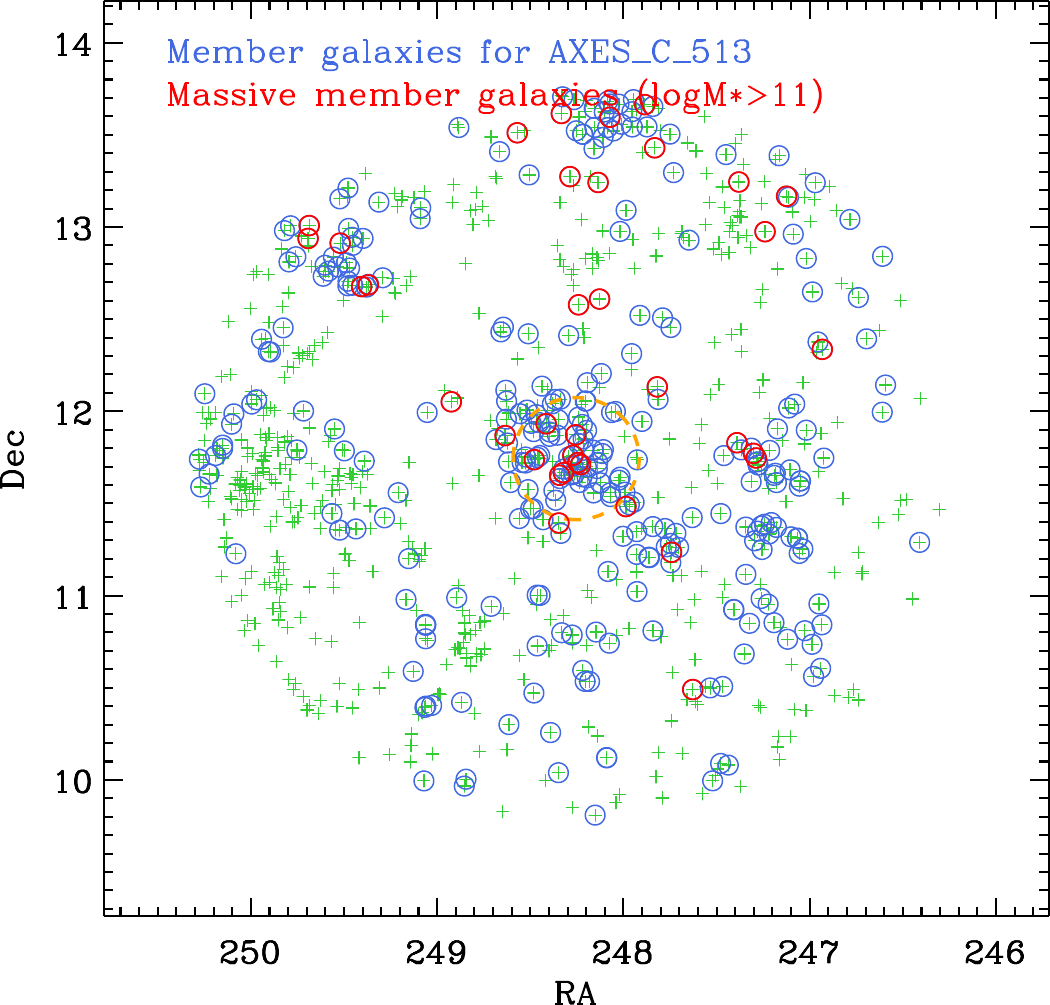}
    \includegraphics[width=0.32\linewidth]{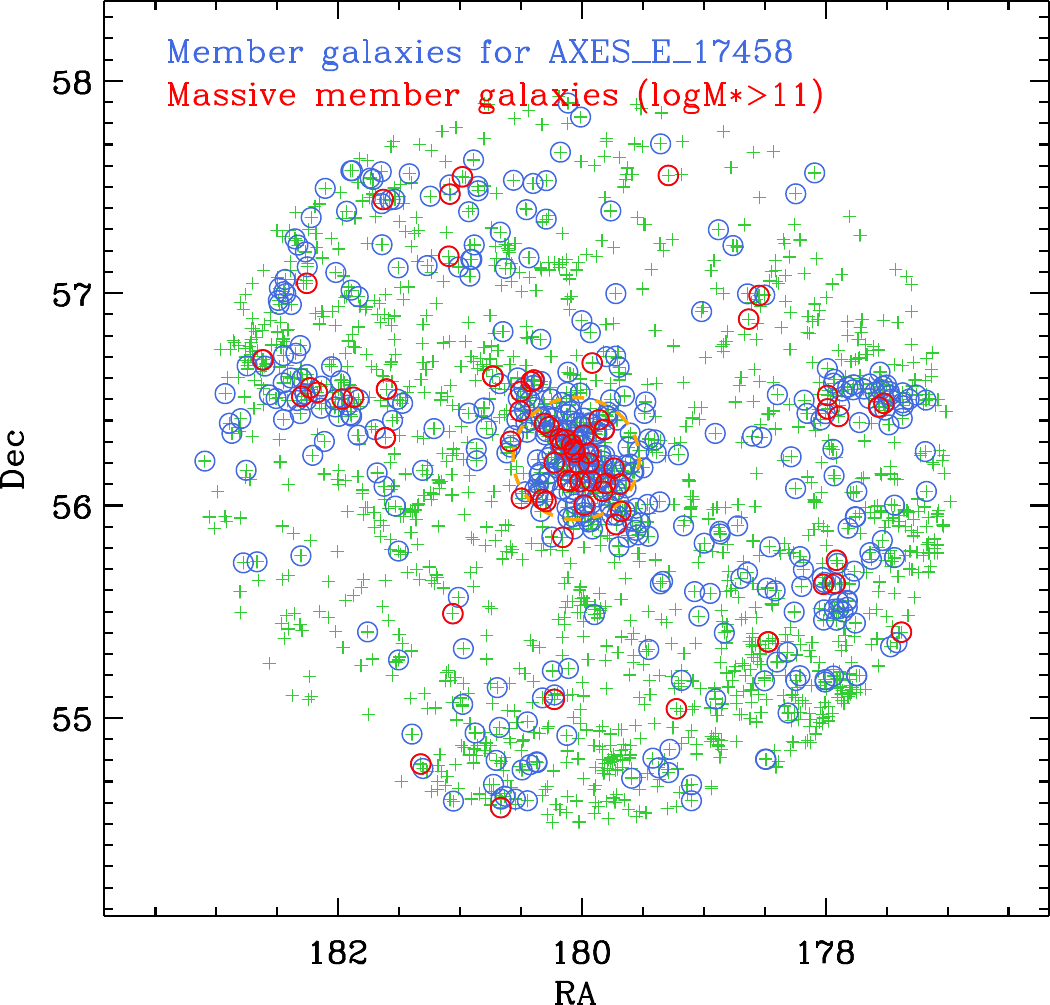}
    \caption{The R.A.--Decl. distribution of the sample clusters. The green plus symbols denote all the galaxies selected from the DESI and SGA catalogs with $|z_{\rm spec} - z_{\rm cluster}| \cdot c < 3000~\text{km~s}^{-1}$ and projected spatial separation $\leq 5 R_{200}$. The blue circles indicate cluster member galaxies selected using the phase-space diagram shown in Figure~\ref{phaseplot}. The dashed orange circle at the center of each panel indicates $1\times R_{200}$. The red circles highlight the locations of massive galaxies ($\log(M_*/M_\odot) > 11$).}
    \label{member_selection1}
\end{figure*}
\begin{figure*}
    \centering
    \includegraphics[width=0.32\linewidth]{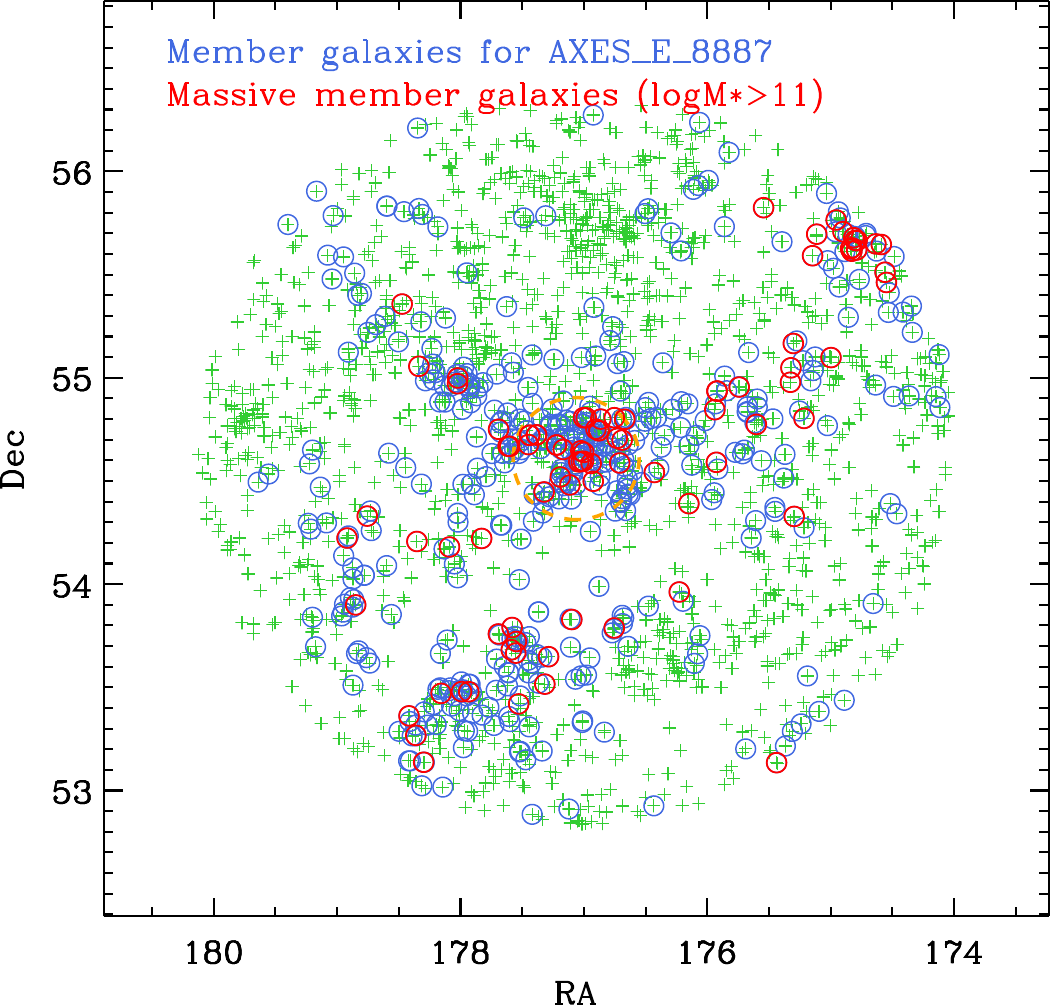}
    \includegraphics[width=0.32\linewidth]{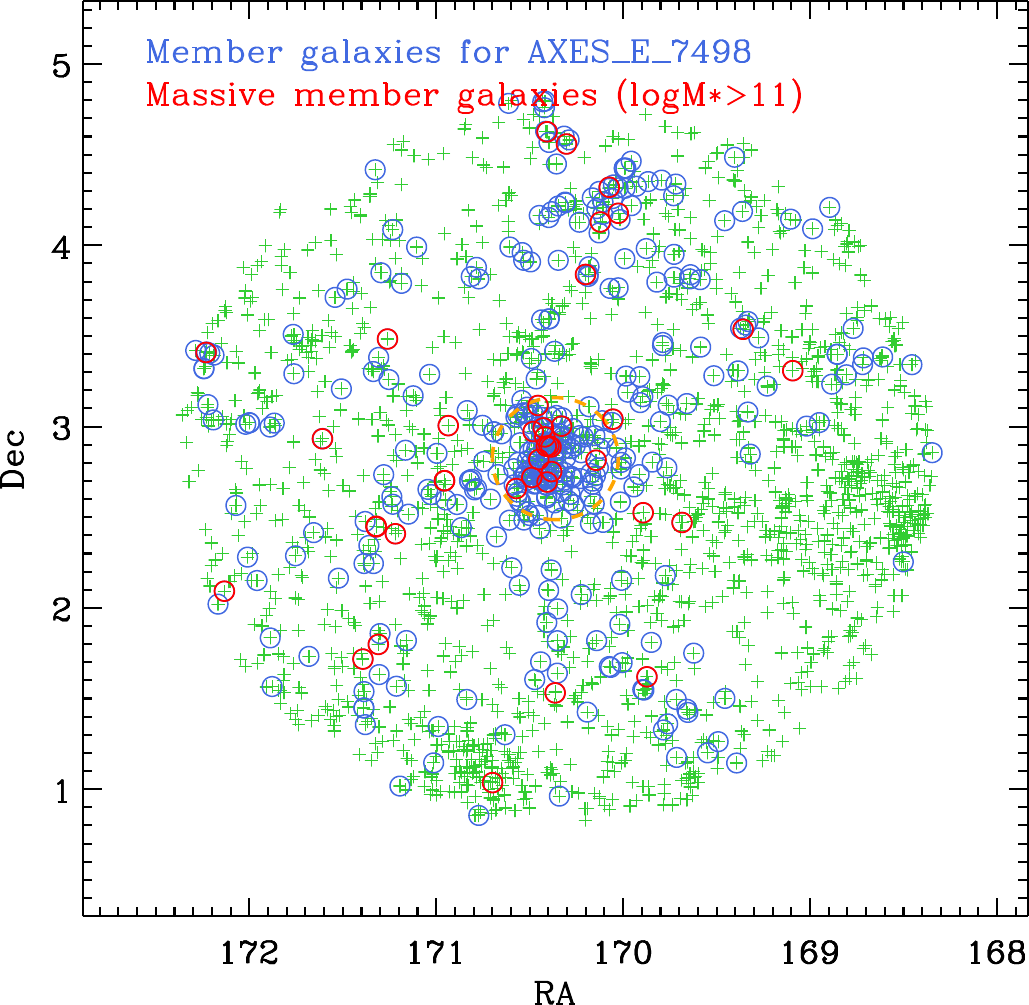}
    \includegraphics[width=0.32\linewidth]{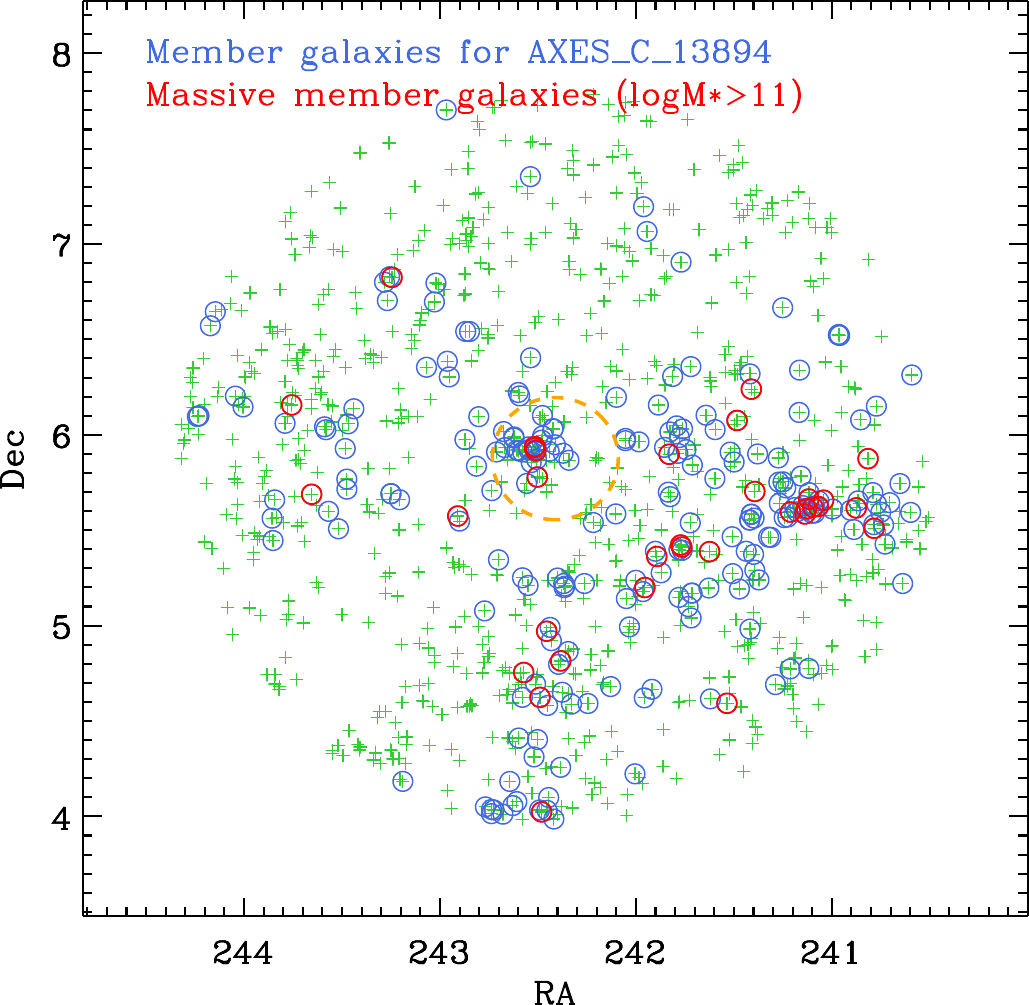}
    \includegraphics[width=0.32\linewidth]{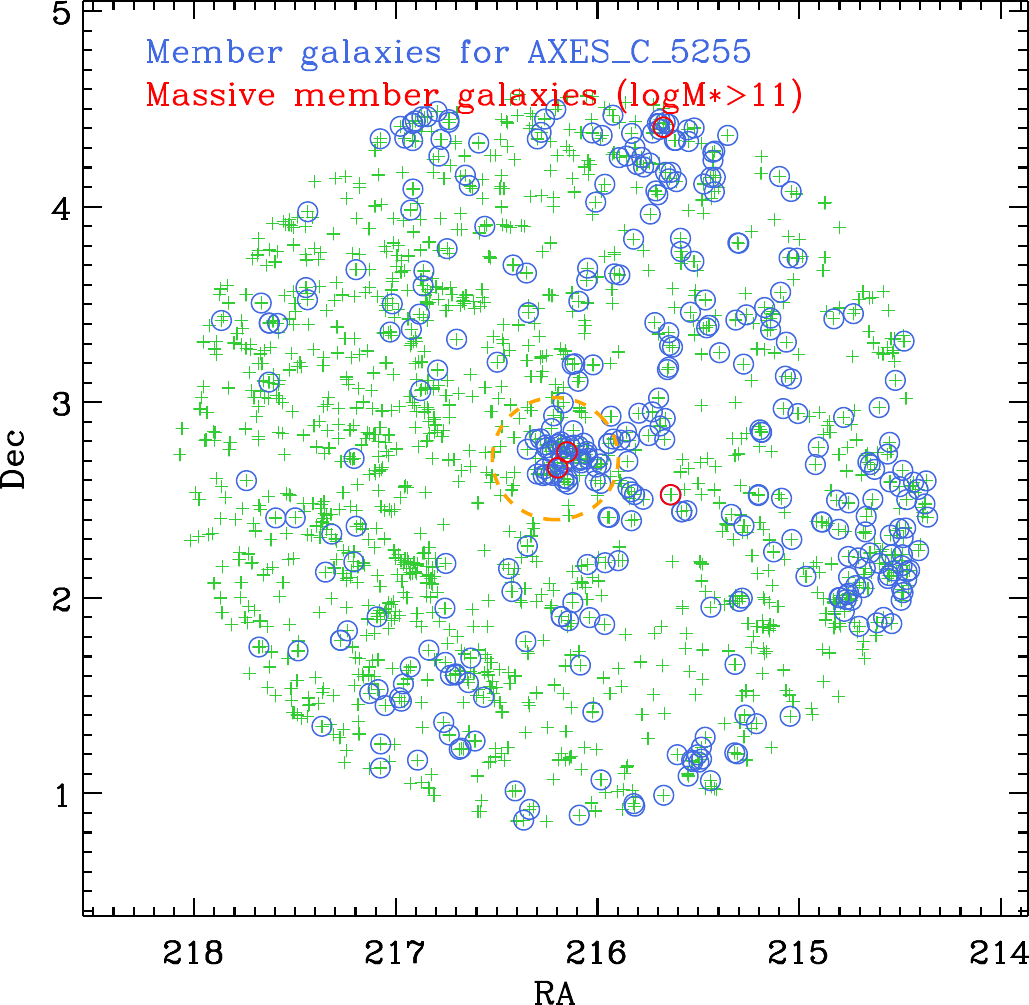}  
    \includegraphics[width=0.32\linewidth]{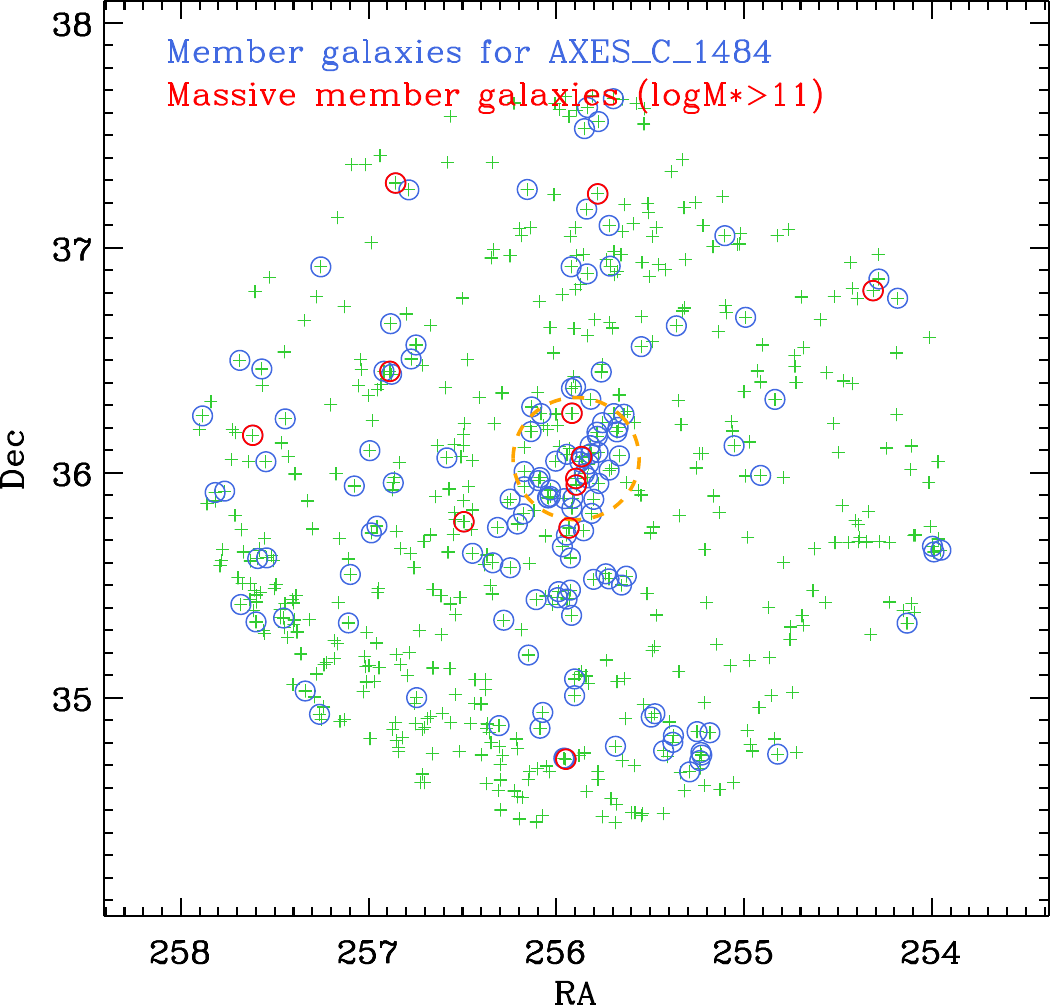}  
    \includegraphics[width=0.32\linewidth]{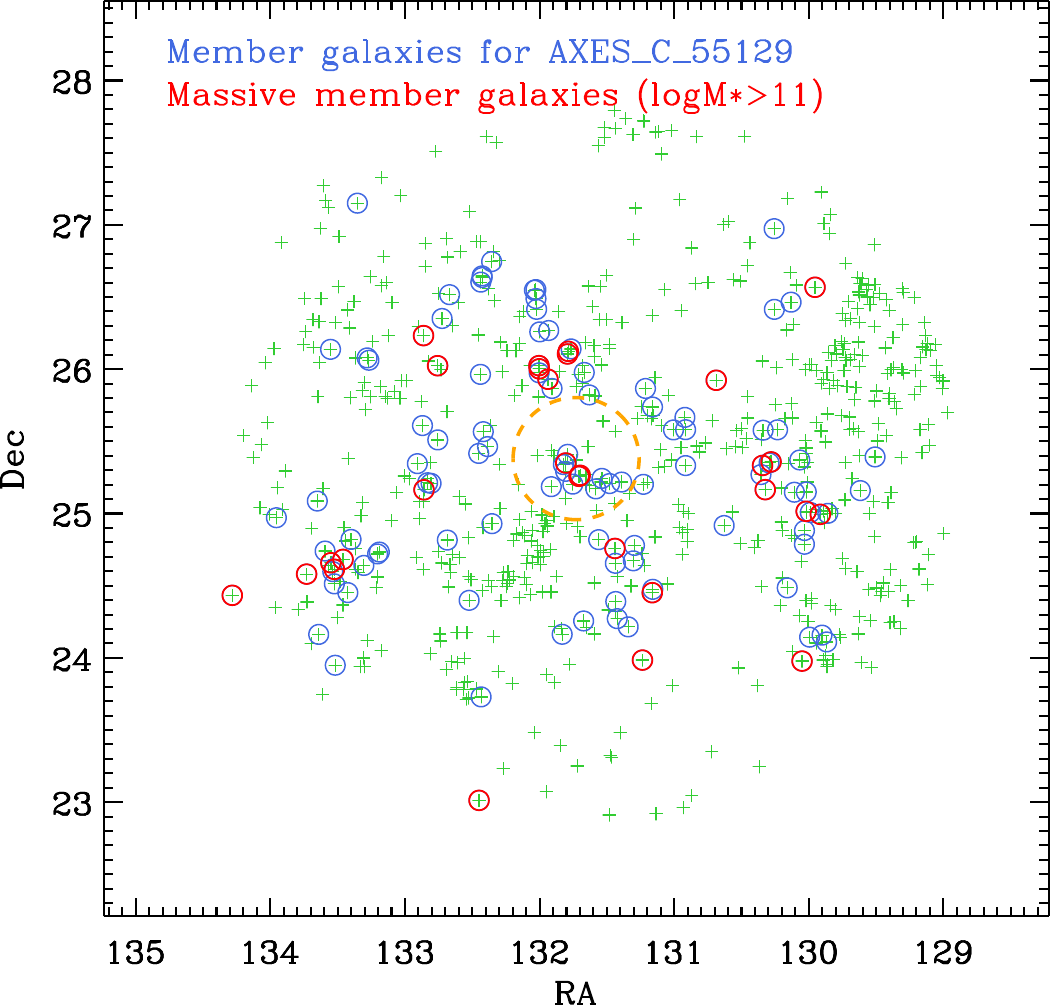}  
    \includegraphics[width=0.32\linewidth]{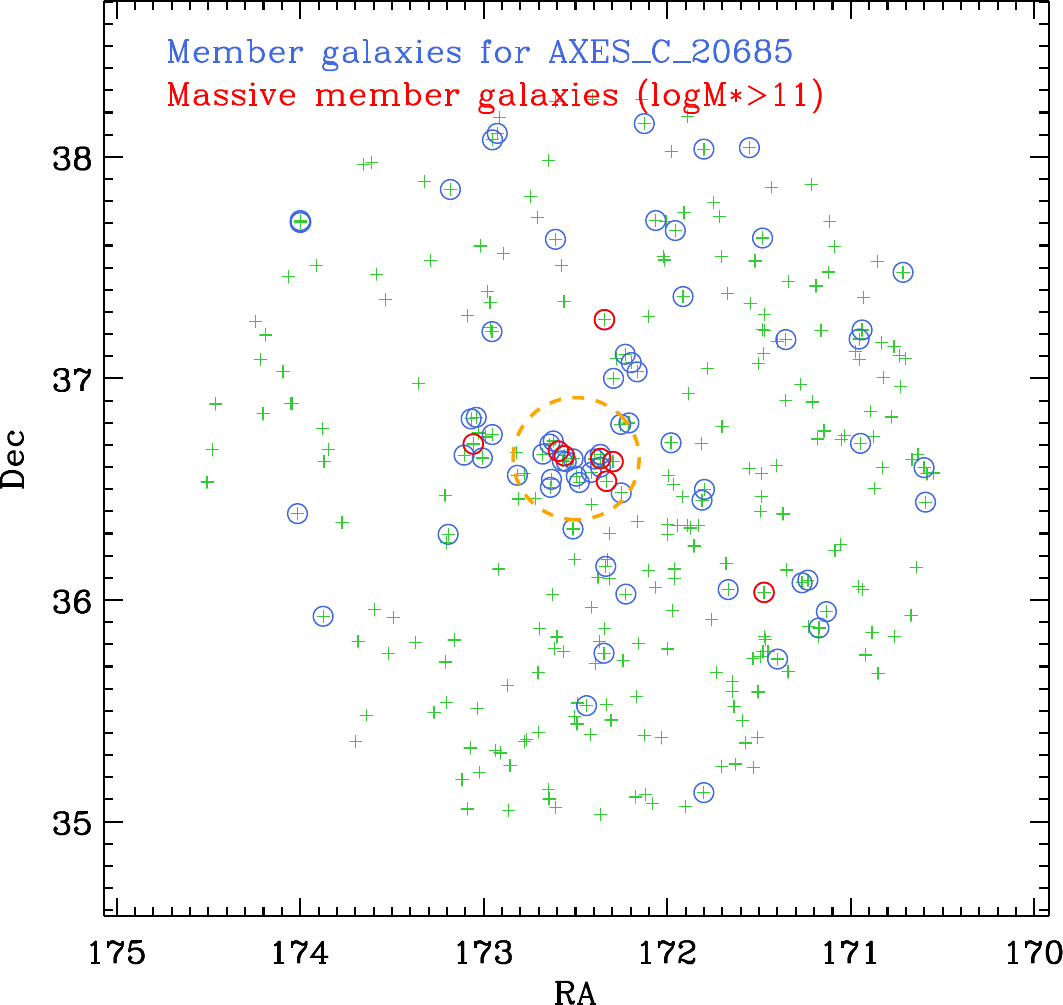}  
    \includegraphics[width=0.32\linewidth]{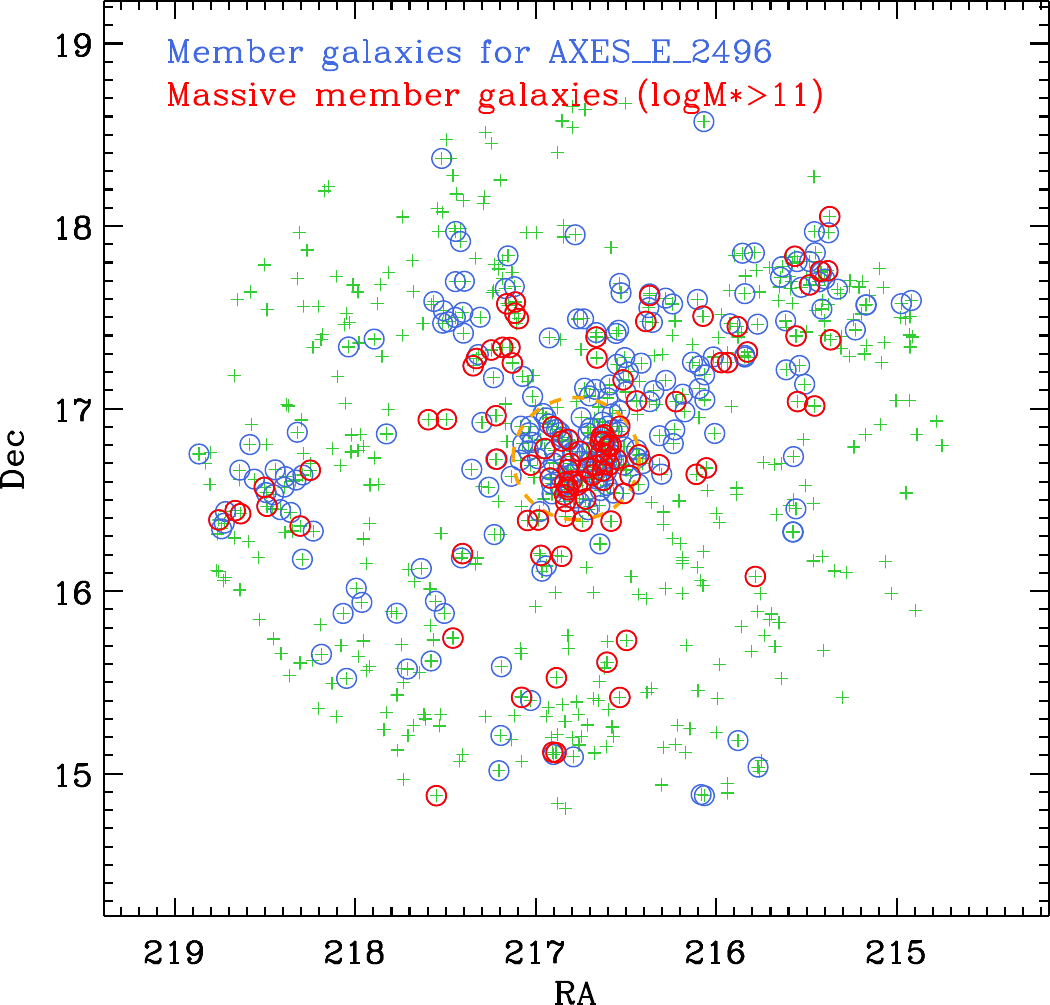}
    \includegraphics[width=0.32\linewidth]{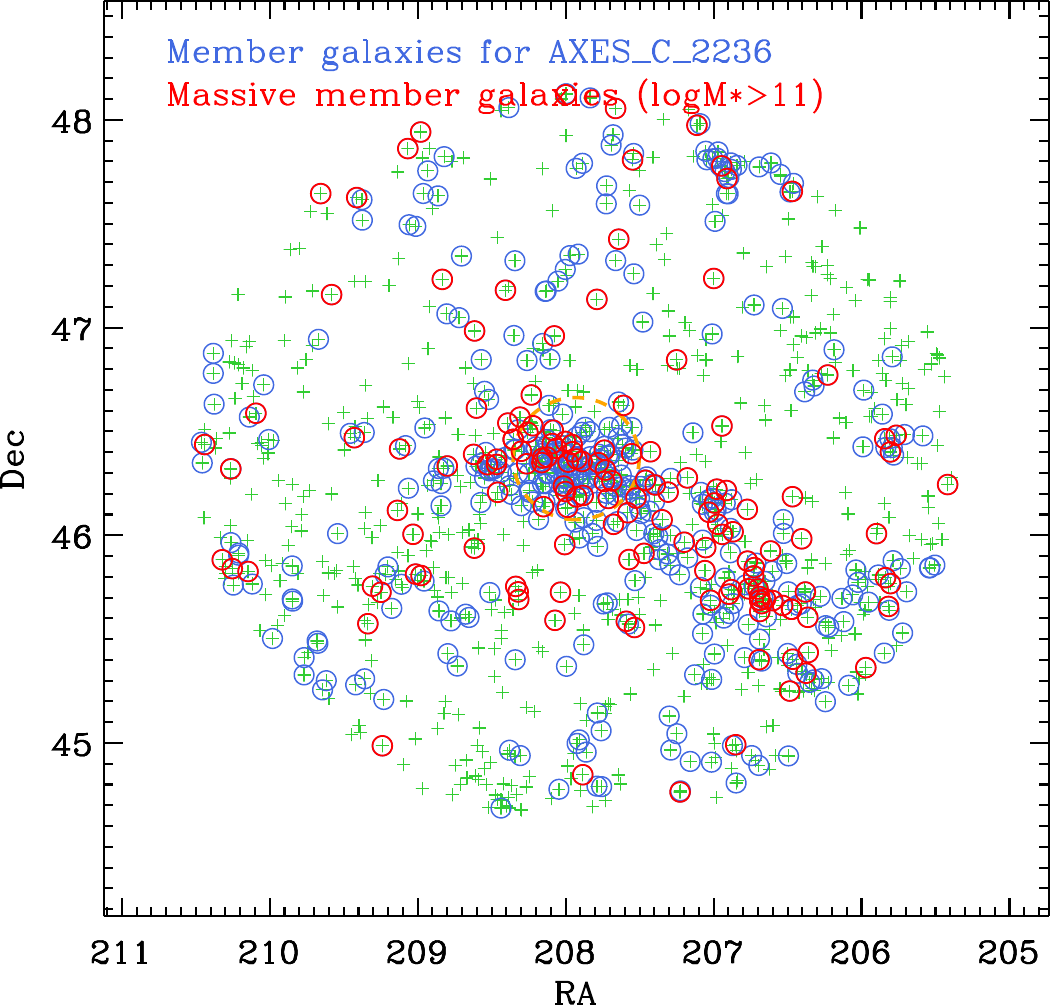}
    \includegraphics[width=0.32\linewidth]{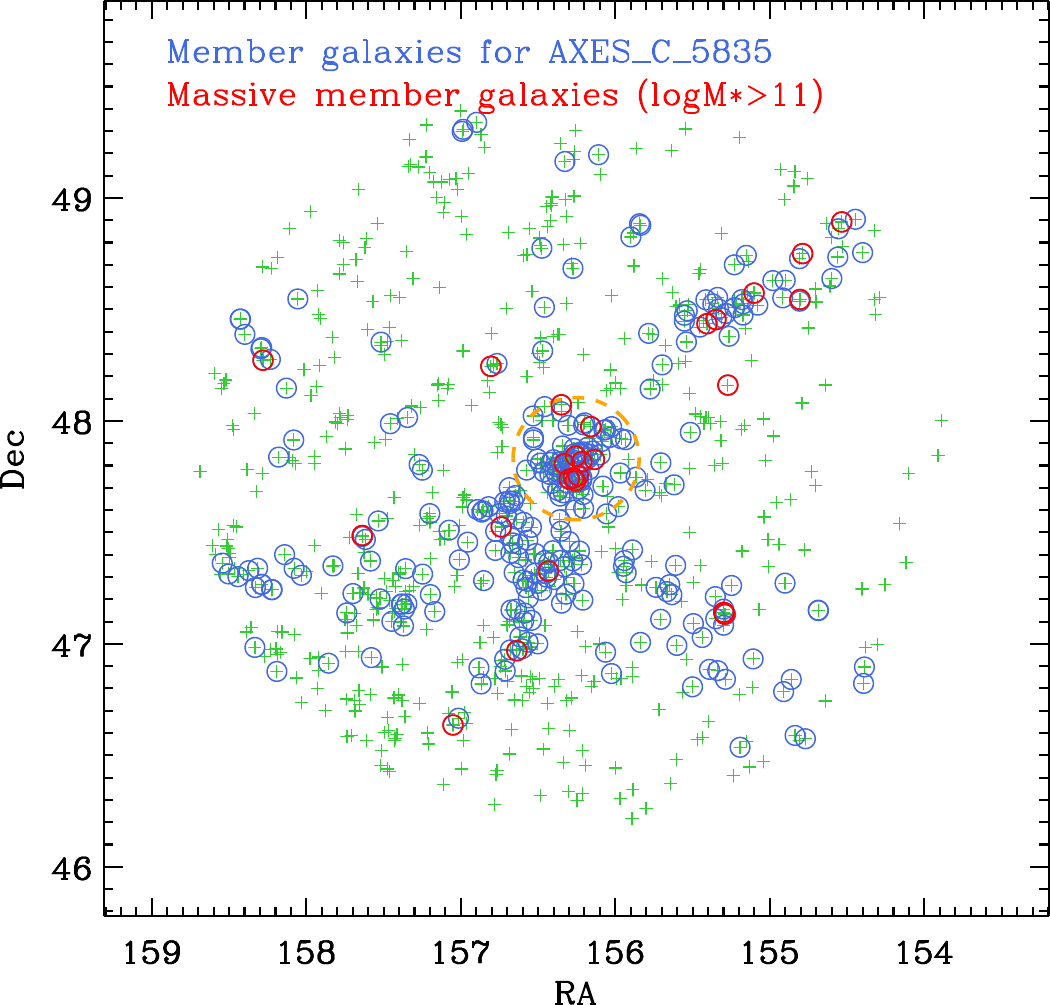}
    \includegraphics[width=0.32\linewidth]{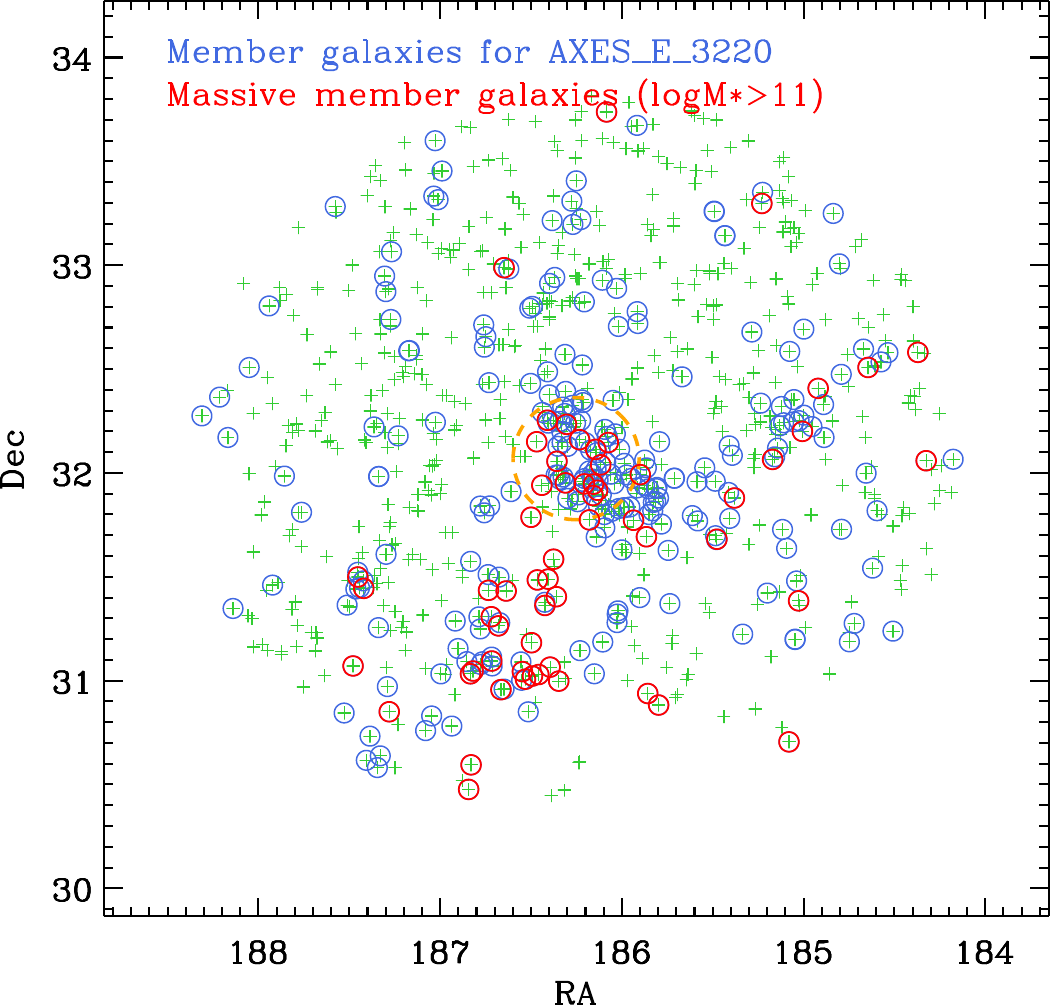}
    \includegraphics[width=0.32\linewidth]{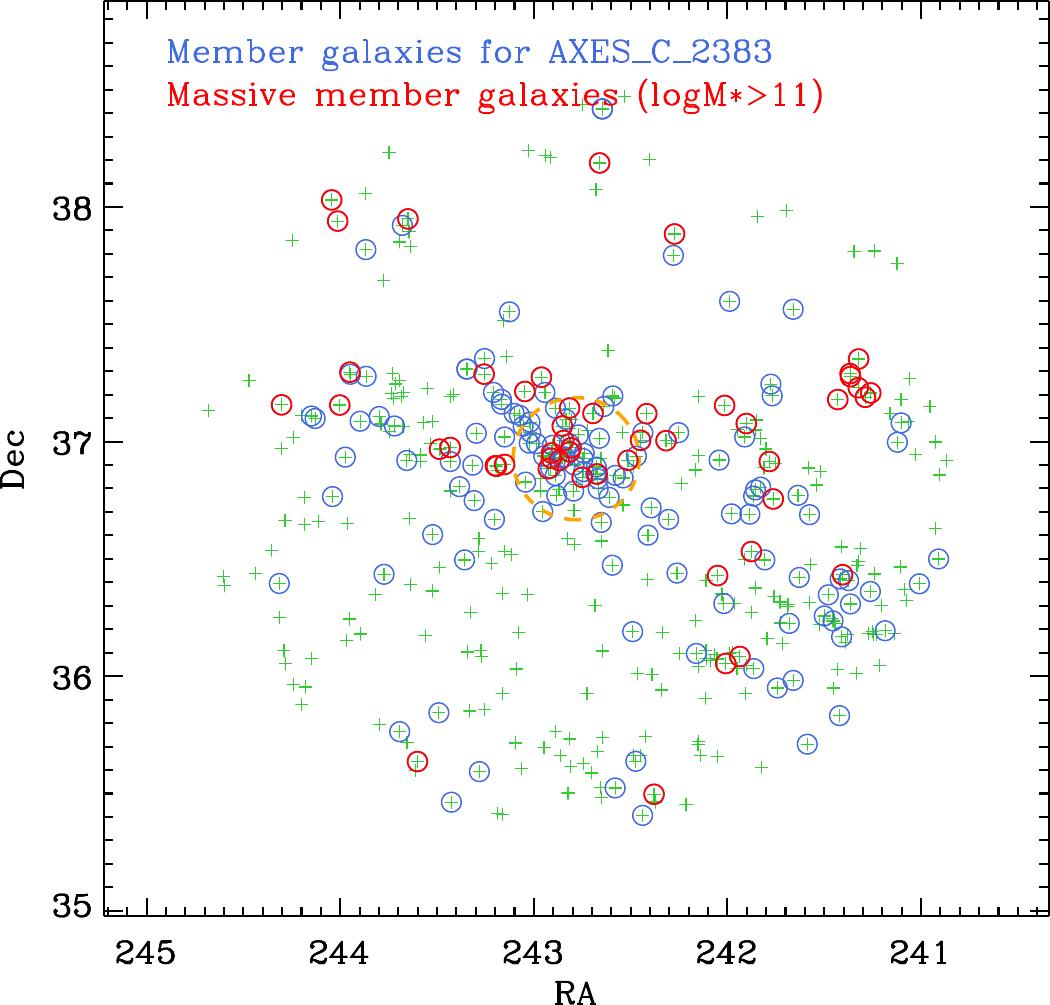} 
    \caption{Same as Figure \ref{member_selection1}.
    }
    \label{member_selection2}
\end{figure*}
\begin{figure*}
    \centering
    \includegraphics[width=0.32\linewidth]{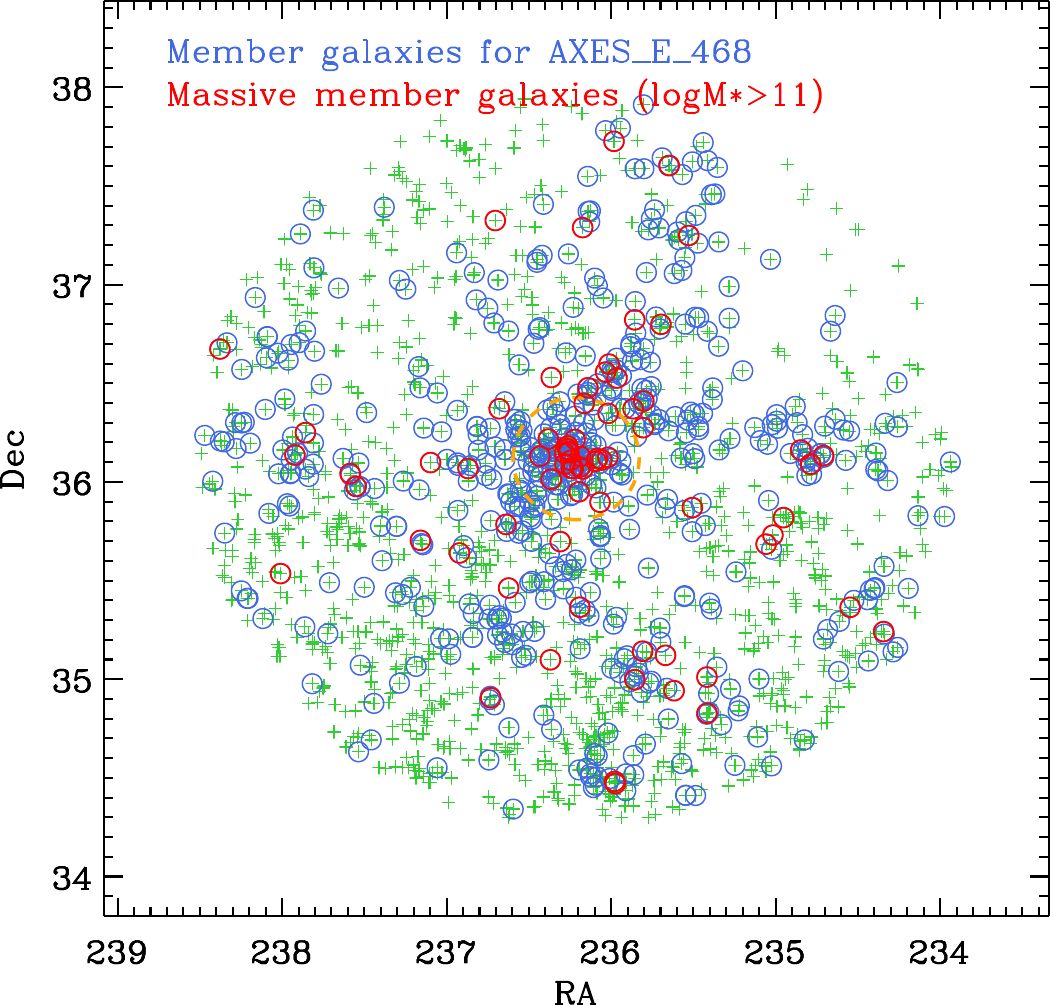}
    \includegraphics[width=0.32\linewidth]{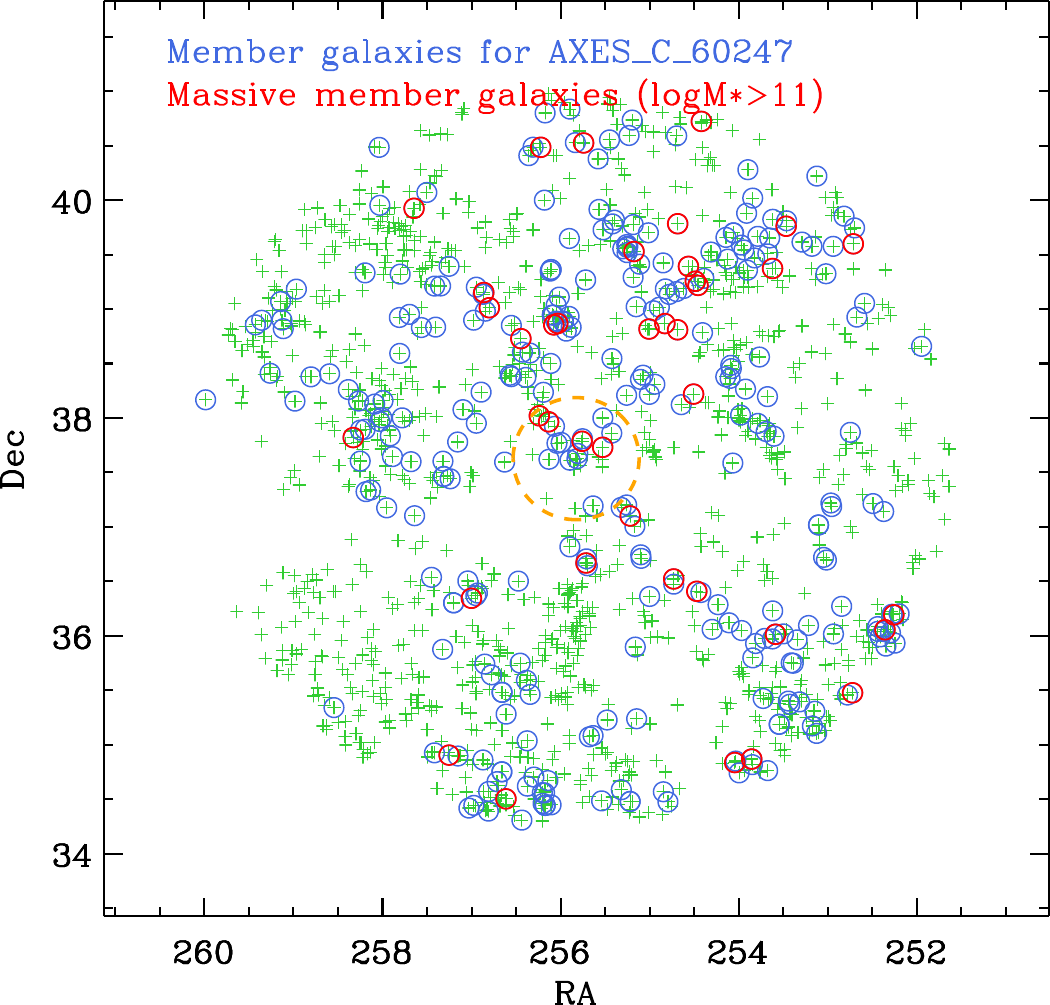}
    \includegraphics[width=0.32\linewidth]{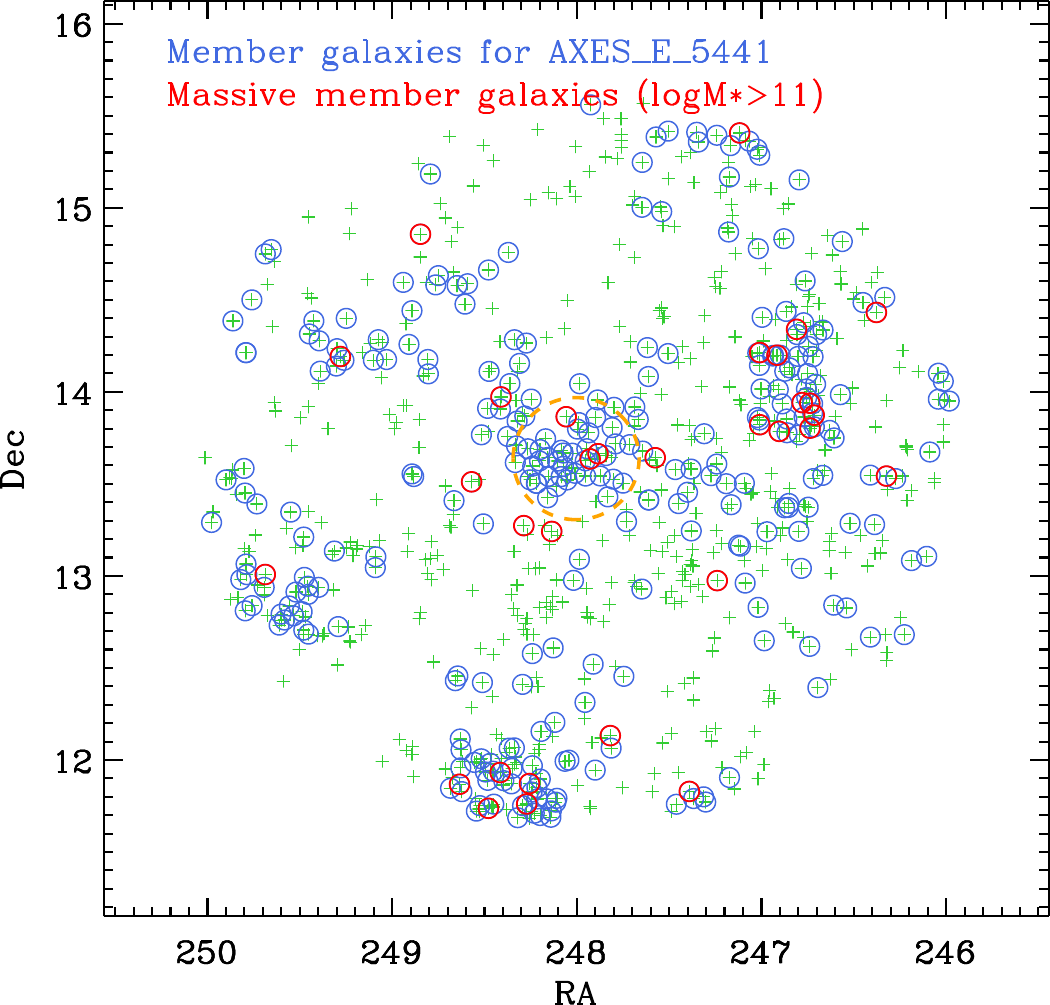}
    \includegraphics[width=0.32\linewidth]{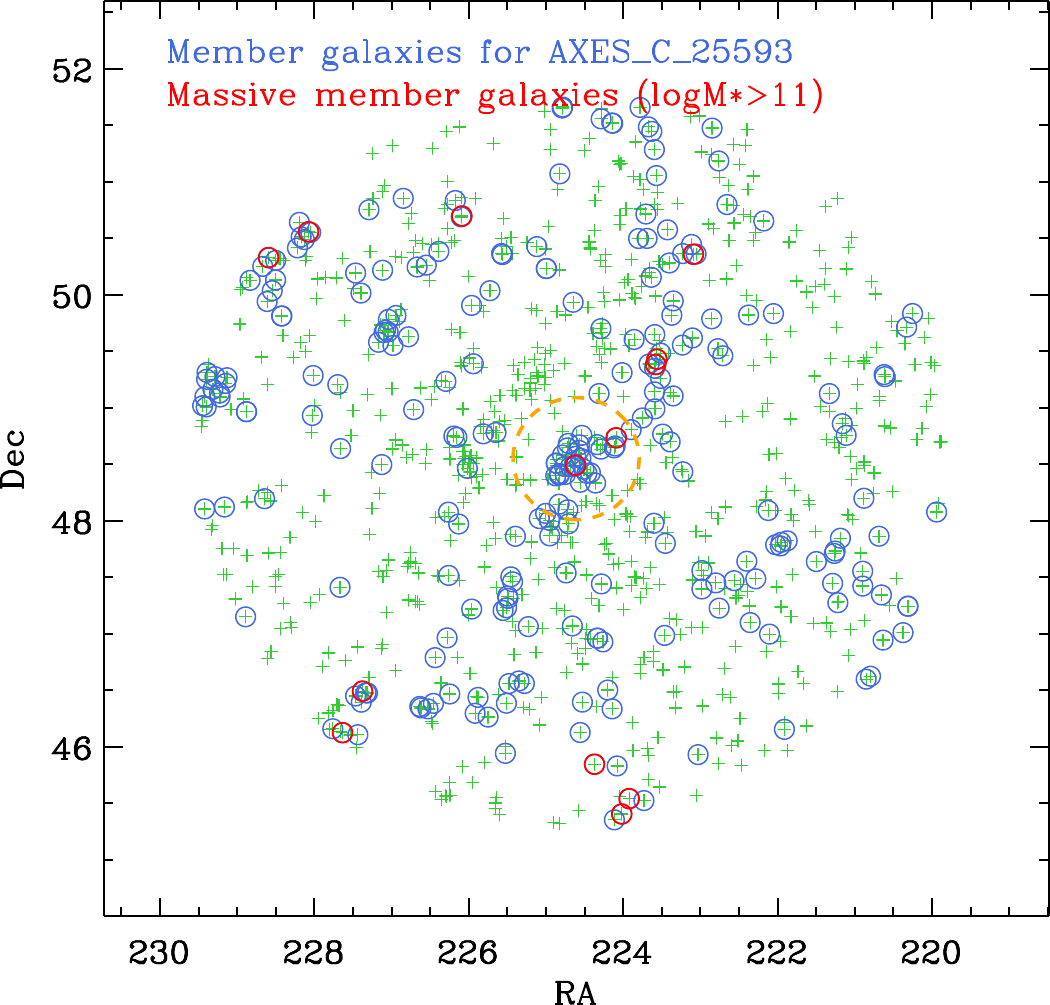}
    \includegraphics[width=0.32\linewidth]{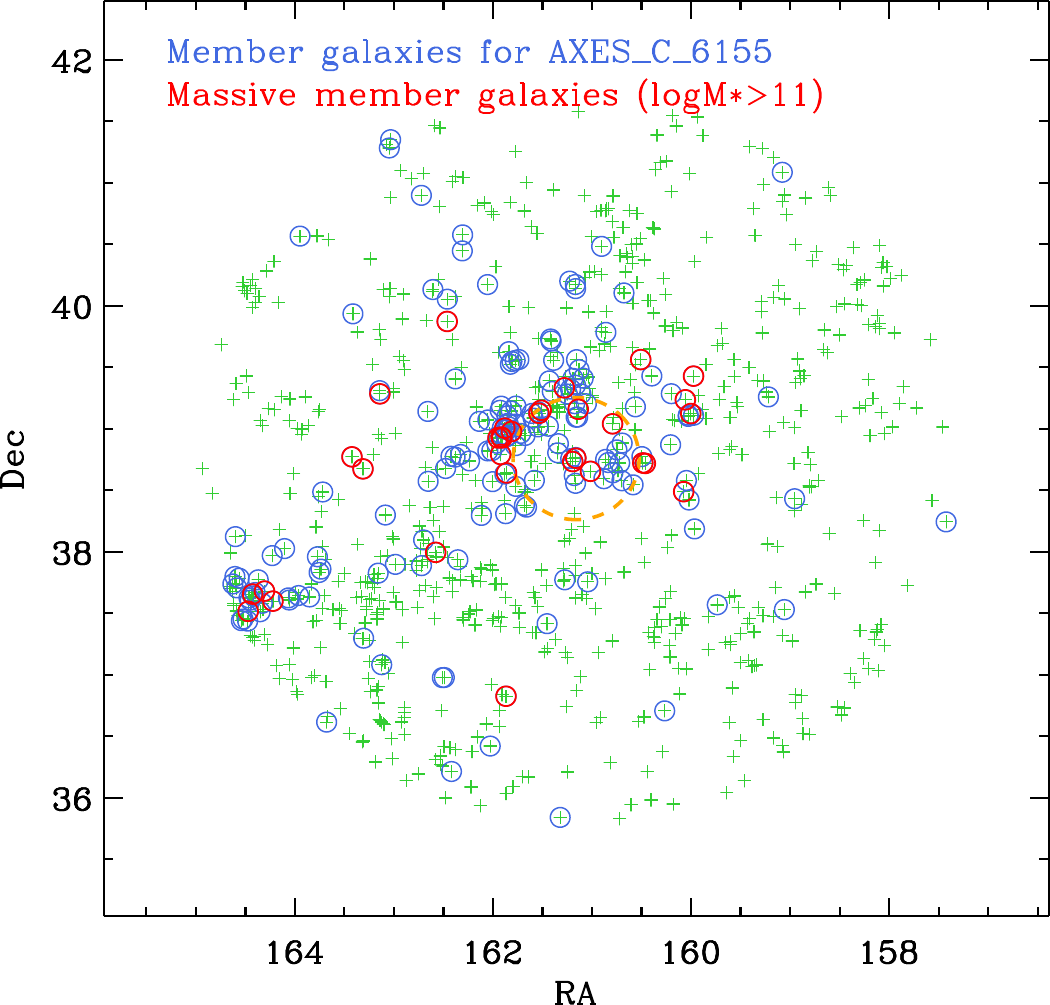}  
    \includegraphics[width=0.32\linewidth]{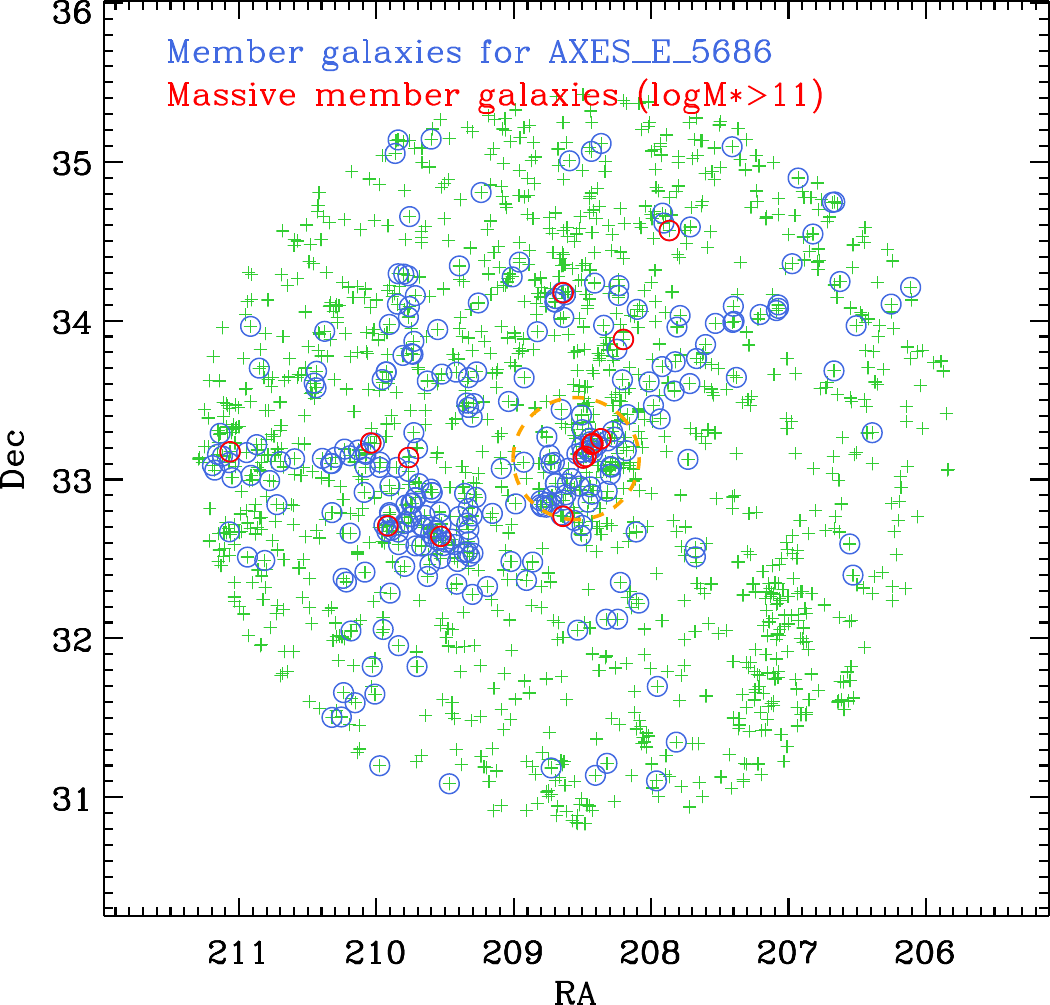}  
    \includegraphics[width=0.32\linewidth]{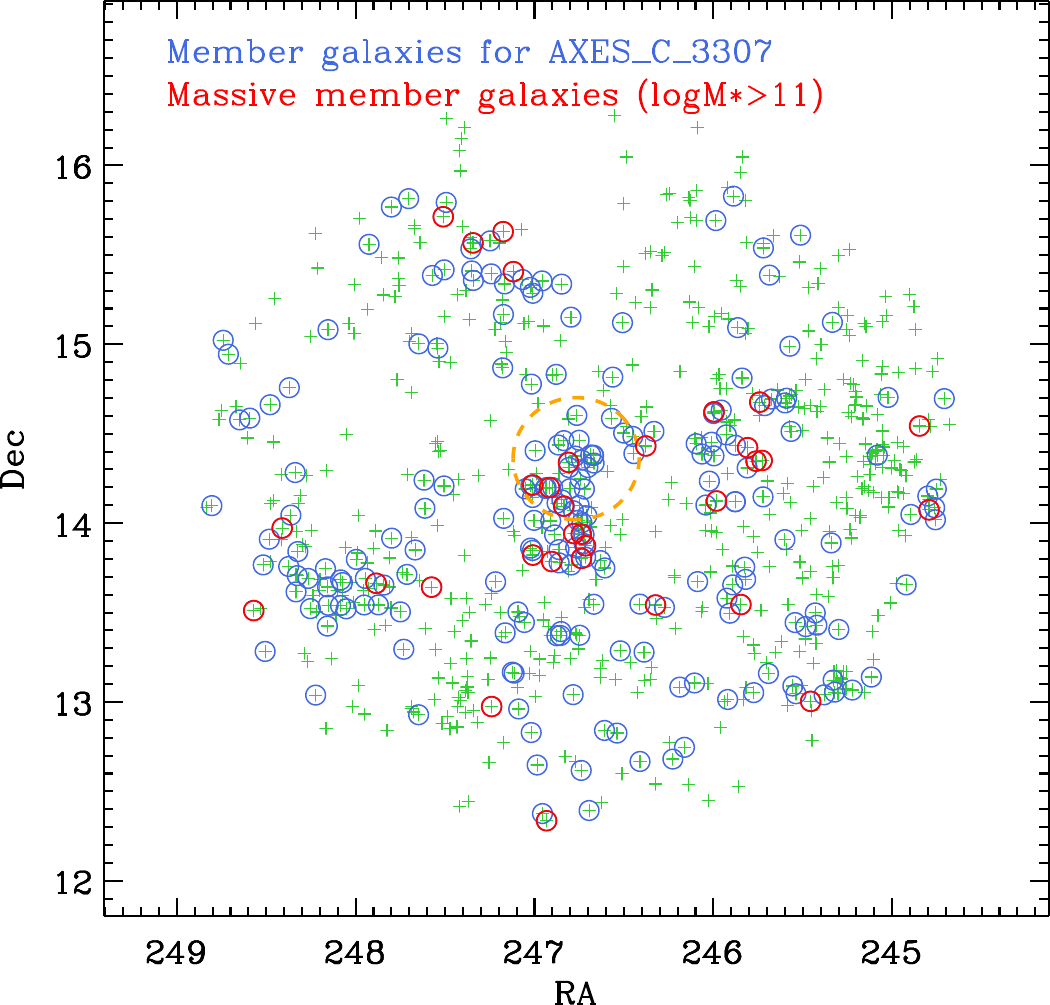}  
    \includegraphics[width=0.32\linewidth]{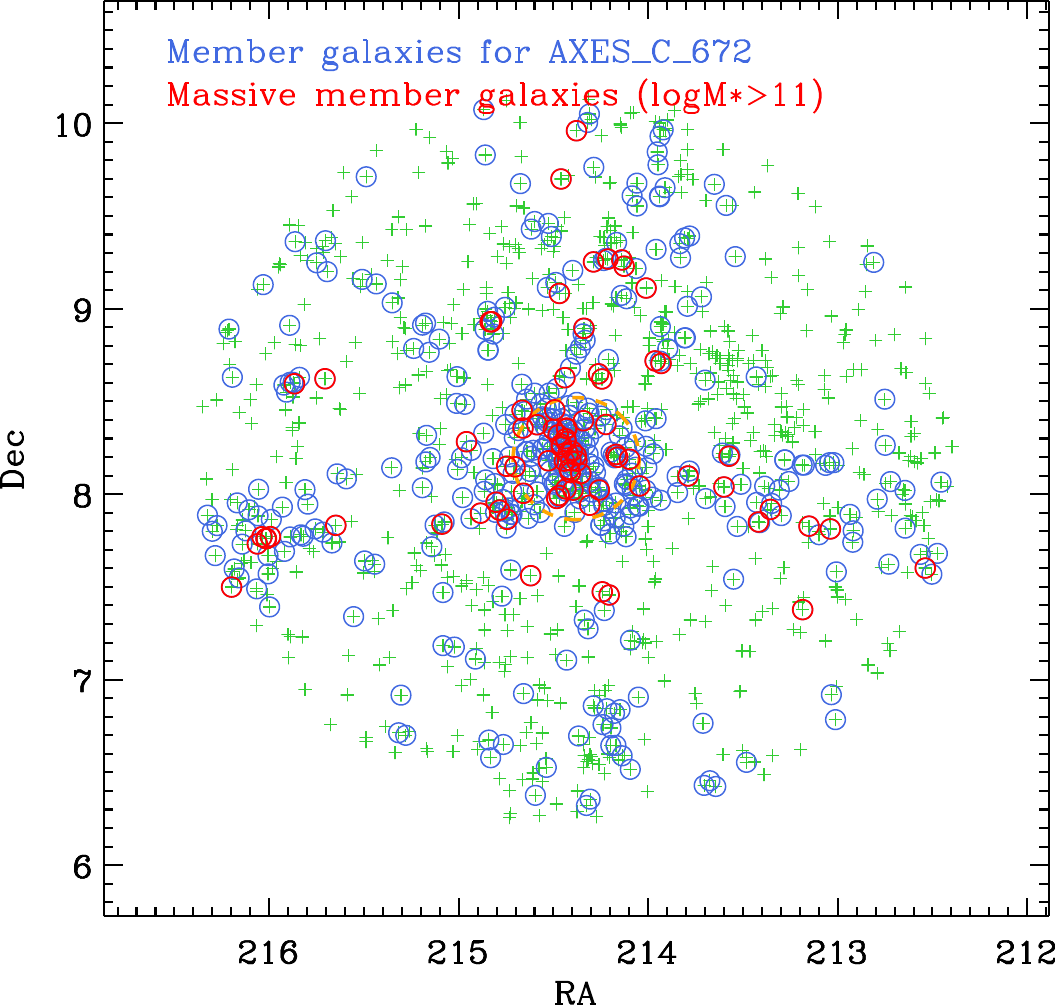}  
    \includegraphics[width=0.32\linewidth]{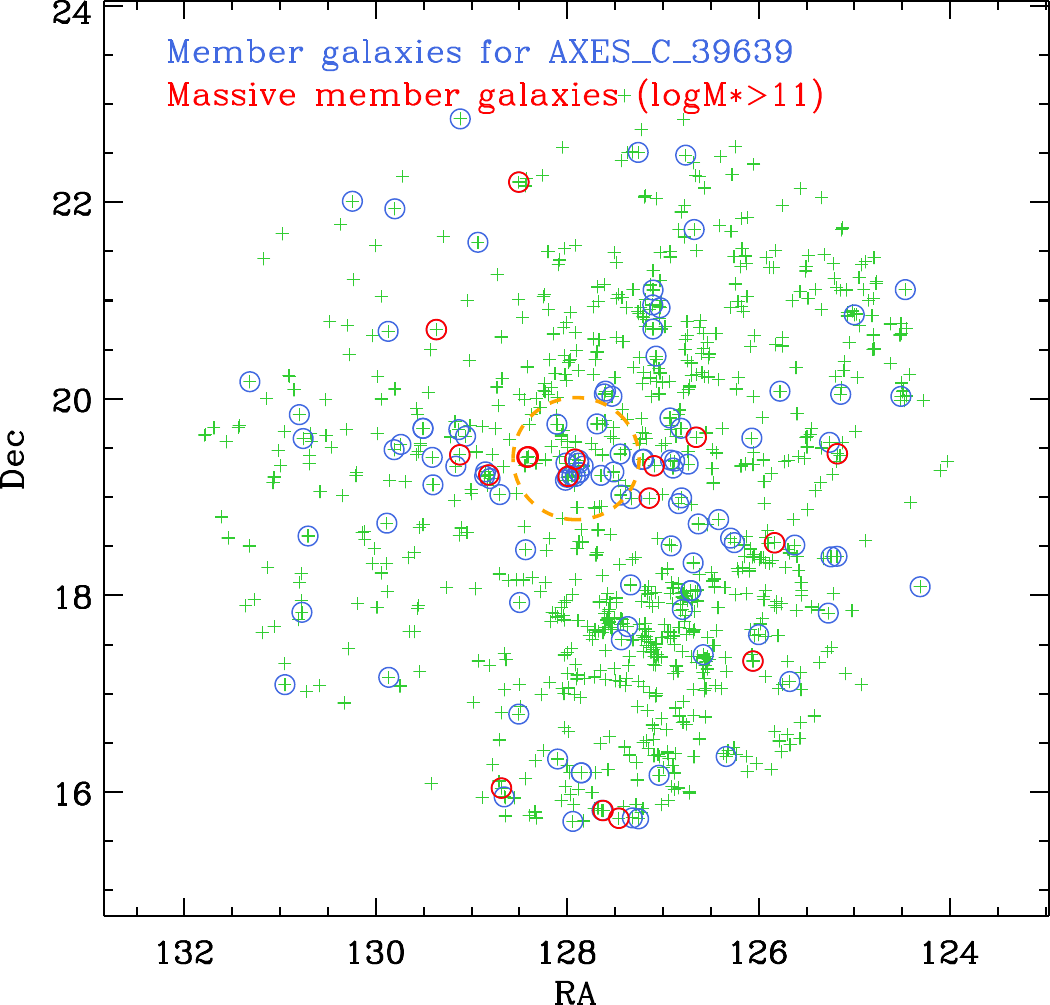}  
    \includegraphics[width=0.32\linewidth]{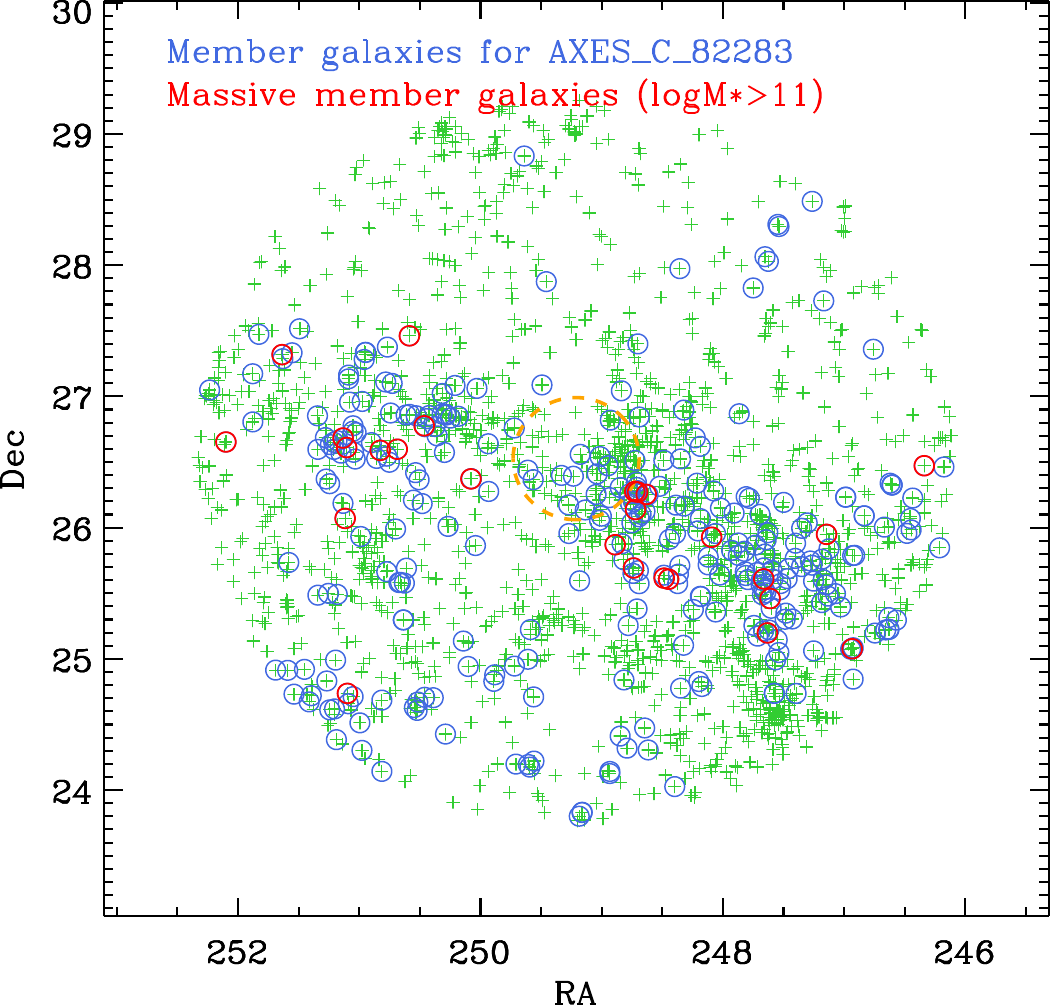}
    \includegraphics[width=0.32\linewidth]{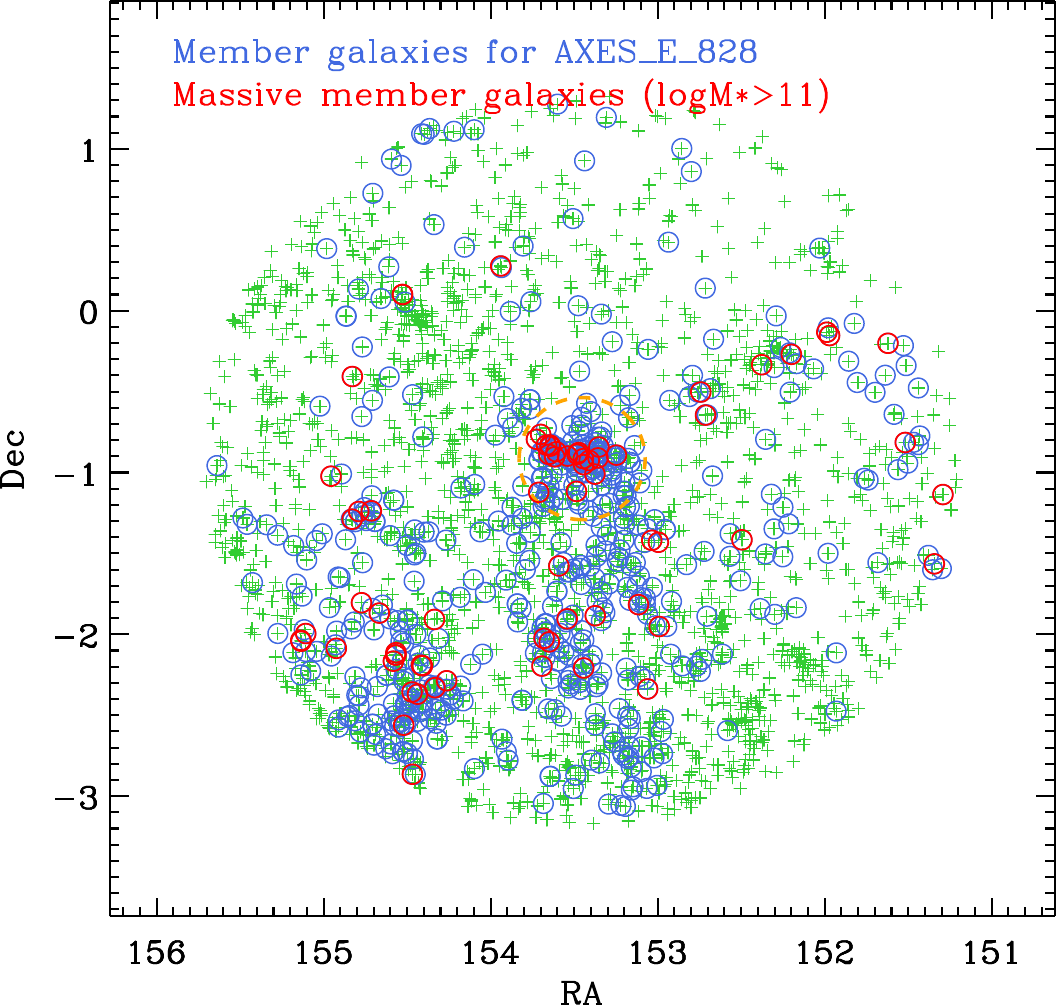}
    \includegraphics[width=0.32\linewidth]{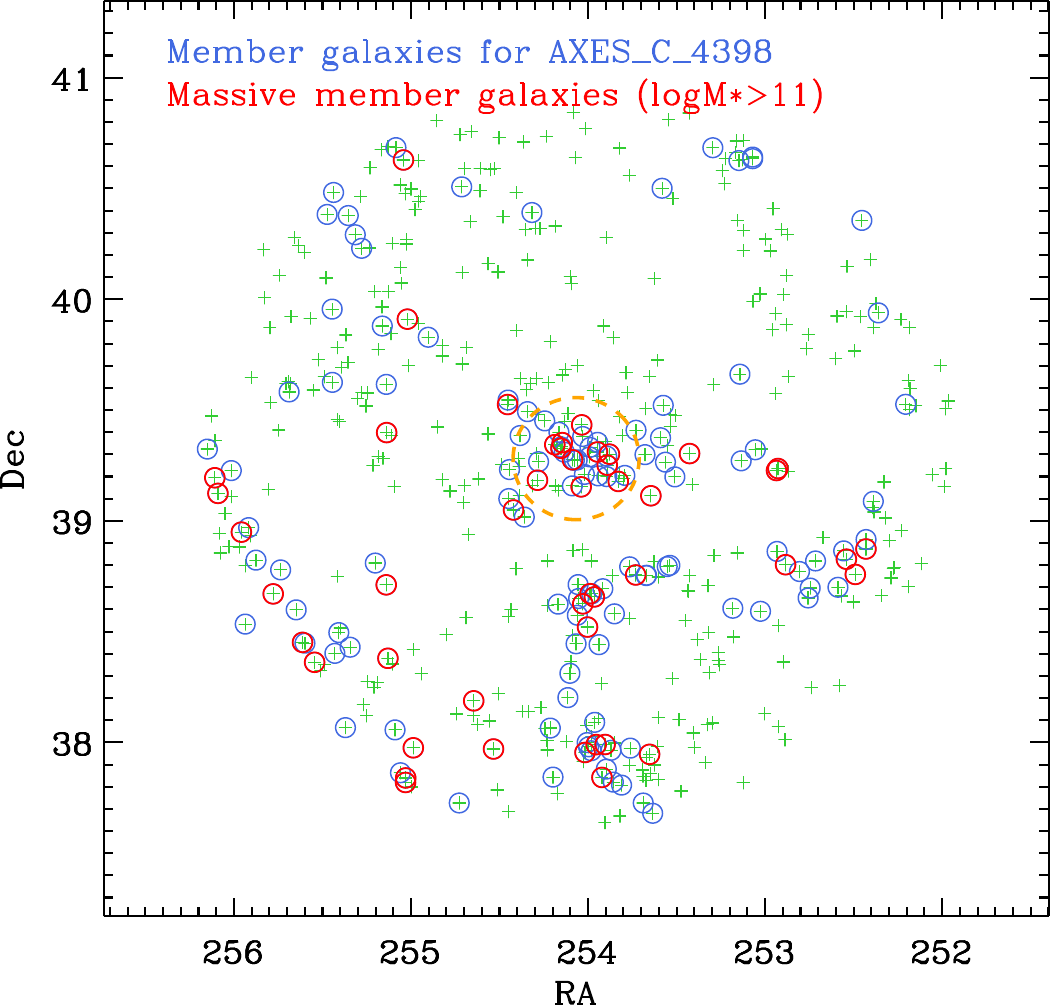}
    \caption{Same as Figure \ref{member_selection1}.
    }
    \label{member_selection3}
\end{figure*}

%





\begin{thebibliography}{}
\expandafter\ifx\csname natexlab\endcsname\relax\def\natexlab#1{#1}\fi
\providecommand{\url}[1]{\href{#1}{#1}}
\providecommand{\dodoi}[1]{doi:~\href{http://doi.org/#1}{\nolinkurl{#1}}}
\providecommand{\doeprint}[1]{\href{http://ascl.net/#1}{\nolinkurl{http://ascl.net/#1}}}
\providecommand{\doarXiv}[1]{\href{https://arxiv.org/abs/#1}{\nolinkurl{https://arxiv.org/abs/#1}}}

\bibitem[{E.~A.~K. {Adams} {et~al.}(2022){Adams}, {Adebahr}, {de Blok}, {D{\'e}nes}, {Hess}, {van der Hulst}, {Kutkin}, {Lucero}, {Morganti}, {Moss}, {Oosterloo}, {Orr{\'u}}, {Schulz}, {van Amesfoort}, {Berger}, {Boersma}, {Bouwhuis}, {van den Brink}, {van Cappellen}, {Connor}, {Coolen}, {Damstra}, {van Diepen}, {Dijkema}, {Ebbendorf}, {Grange}, {de Goei}, {Gunst}, {Holties}, {Hut}, {Ivashina}, {J{\'o}zsa}, {van Leeuwen}, {Loose}, {Maan}, {Mancini}, {Mika}, {Mulder}, {Norden}, {Offringa}, {Oostrum}, {Pastor-Marazuela}, {Pisano}, {Ponomareva}, {Romein}, {Ruiter}, {Schoenmakers}, {van der Schuur}, {Sluman}, {Smits}, {Stuurwold}, {Verstappen}, {Vilchez}, {Vohl}, {Wierenga}, {Wijnholds}, {Woestenburg}, {Zanting}, \& {Ziemke}}]{2022A&A...667A..38A}
{Adams}, E.~A.~K., {Adebahr}, B., {de Blok}, W.~J.~G., {et~al.} 2022, \bibinfo{title}{{First release of Apertif imaging survey data},} \aap, 667, A38, \dodoi{10.1051/0004-6361/202244007}

\bibitem[{ {Astropy Collaboration} {et~al.}(2013){Astropy Collaboration}, {Robitaille}, {Tollerud}, {Greenfield}, {Droettboom}, {Bray}, {Aldcroft}, {Davis}, {Ginsburg}, {Price-Whelan}, {Kerzendorf}, {Conley}, {Crighton}, {Barbary}, {Muna}, {Ferguson}, {Grollier}, {Parikh}, {Nair}, {Unther}, {Deil}, {Woillez}, {Conseil}, {Kramer}, {Turner}, {Singer}, {Fox}, {Weaver}, {Zabalza}, {Edwards}, {Azalee Bostroem}, {Burke}, {Casey}, {Crawford}, {Dencheva}, {Ely}, {Jenness}, {Labrie}, {Lim}, {Pierfederici}, {Pontzen}, {Ptak}, {Refsdal}, {Servillat}, \& {Streicher}}]{2013A&A...558A..33A}
{Astropy Collaboration}, {Robitaille}, T.~P., {Tollerud}, E.~J., {et~al.} 2013, \bibinfo{title}{{Astropy: A community Python package for astronomy},} \aap, 558, A33, \dodoi{10.1051/0004-6361/201322068}

\bibitem[{ {Astropy Collaboration} {et~al.}(2018){Astropy Collaboration}, {Price-Whelan}, {Sip{\H{o}}cz}, {G{\"u}nther}, {Lim}, {Crawford}, {Conseil}, {Shupe}, {Craig}, {Dencheva}, {Ginsburg}, {VanderPlas}, {Bradley}, {P{\'e}rez-Su{\'a}rez}, {de Val-Borro}, {Aldcroft}, {Cruz}, {Robitaille}, {Tollerud}, {Ardelean}, {Babej}, {Bach}, {Bachetti}, {Bakanov}, {Bamford}, {Barentsen}, {Barmby}, {Baumbach}, {Berry}, {Biscani}, {Boquien}, {Bostroem}, {Bouma}, {Brammer}, {Bray}, {Breytenbach}, {Buddelmeijer}, {Burke}, {Calderone}, {Cano Rodr{\'\i}guez}, {Cara}, {Cardoso}, {Cheedella}, {Copin}, {Corrales}, {Crichton}, {D'Avella}, {Deil}, {Depagne}, {Dietrich}, {Donath}, {Droettboom}, {Earl}, {Erben}, {Fabbro}, {Ferreira}, {Finethy}, {Fox}, {Garrison}, {Gibbons}, {Goldstein}, {Gommers}, {Greco}, {Greenfield}, {Groener}, {Grollier}, {Hagen}, {Hirst}, {Homeier}, {Horton}, {Hosseinzadeh}, {Hu}, {Hunkeler}, {Ivezi{\'c}}, {Jain}, {Jenness}, {Kanarek}, {Kendrew}, {Kern}, {Kerzendorf}, {Khvalko}, {King}, {Kirkby}, {Kulkarni},
  {Kumar}, {Lee}, {Lenz}, {Littlefair}, {Ma}, {Macleod}, {Mastropietro}, {McCully}, {Montagnac}, {Morris}, {Mueller}, {Mumford}, {Muna}, {Murphy}, {Nelson}, {Nguyen}, {Ninan}, {N{\"o}the}, {Ogaz}, {Oh}, {Parejko}, {Parley}, {Pascual}, {Patil}, {Patil}, {Plunkett}, {Prochaska}, {Rastogi}, {Reddy Janga}, {Sabater}, {Sakurikar}, {Seifert}, {Sherbert}, {Sherwood-Taylor}, {Shih}, {Sick}, {Silbiger}, {Singanamalla}, {Singer}, {Sladen}, {Sooley}, {Sornarajah}, {Streicher}, {Teuben}, {Thomas}, {Tremblay}, {Turner}, {Terr{\'o}n}, {van Kerkwijk}, {de la Vega}, {Watkins}, {Weaver}, {Whitmore}, {Woillez}, {Zabalza}, \& {Astropy Contributors}}]{2018AJ....156..123A}
{Astropy Collaboration}, {Price-Whelan}, A.~M., {Sip{\H{o}}cz}, B.~M., {et~al.} 2018, \bibinfo{title}{{The Astropy Project: Building an Open-science Project and Status of the v2.0 Core Package},} \aj, 156, 123, \dodoi{10.3847/1538-3881/aabc4f}

\bibitem[{ {Astropy Collaboration} {et~al.}(2022){Astropy Collaboration}, {Price-Whelan}, {Lim}, {Earl}, {Starkman}, {Bradley}, {Shupe}, {Patil}, {Corrales}, {Brasseur}, {N{\"o}the}, {Donath}, {Tollerud}, {Morris}, {Ginsburg}, {Vaher}, {Weaver}, {Tocknell}, {Jamieson}, {van Kerkwijk}, {Robitaille}, {Merry}, {Bachetti}, {G{\"u}nther}, {Aldcroft}, {Alvarado-Montes}, {Archibald}, {B{\'o}di}, {Bapat}, {Barentsen}, {Baz{\'a}n}, {Biswas}, {Boquien}, {Burke}, {Cara}, {Cara}, {Conroy}, {Conseil}, {Craig}, {Cross}, {Cruz}, {D'Eugenio}, {Dencheva}, {Devillepoix}, {Dietrich}, {Eigenbrot}, {Erben}, {Ferreira}, {Foreman-Mackey}, {Fox}, {Freij}, {Garg}, {Geda}, {Glattly}, {Gondhalekar}, {Gordon}, {Grant}, {Greenfield}, {Groener}, {Guest}, {Gurovich}, {Handberg}, {Hart}, {Hatfield-Dodds}, {Homeier}, {Hosseinzadeh}, {Jenness}, {Jones}, {Joseph}, {Kalmbach}, {Karamehmetoglu}, {Ka{\l}uszy{\'n}ski}, {Kelley}, {Kern}, {Kerzendorf}, {Koch}, {Kulumani}, {Lee}, {Ly}, {Ma}, {MacBride}, {Maljaars}, {Muna}, {Murphy}, {Norman},
  {O'Steen}, {Oman}, {Pacifici}, {Pascual}, {Pascual-Granado}, {Patil}, {Perren}, {Pickering}, {Rastogi}, {Roulston}, {Ryan}, {Rykoff}, {Sabater}, {Sakurikar}, {Salgado}, {Sanghi}, {Saunders}, {Savchenko}, {Schwardt}, {Seifert-Eckert}, {Shih}, {Jain}, {Shukla}, {Sick}, {Simpson}, {Singanamalla}, {Singer}, {Singhal}, {Sinha}, {Sip{\H{o}}cz}, {Spitler}, {Stansby}, {Streicher}, {{\v{S}}umak}, {Swinbank}, {Taranu}, {Tewary}, {Tremblay}, {de Val-Borro}, {Van Kooten}, {Vasovi{\'c}}, {Verma}, {de Miranda Cardoso}, {Williams}, {Wilson}, {Winkel}, {Wood-Vasey}, {Xue}, {Yoachim}, {Zhang}, {Zonca}, \& {Astropy Project Contributors}}]{2022ApJ...935..167A}
{Astropy Collaboration}, {Price-Whelan}, A.~M., {Lim}, P.~L., {et~al.} 2022, \bibinfo{title}{{The Astropy Project: Sustaining and Growing a Community-oriented Open-source Project and the Latest Major Release (v5.0) of the Core Package},} \apj, 935, 167, \dodoi{10.3847/1538-4357/ac7c74}

\bibitem[{C.~M. {Baugh} {et~al.}(2019){Baugh}, {Gonzalez-Perez}, {Lagos}, {Lacey}, {Helly}, {Jenkins}, {Frenk}, {Benson}, {Bower}, \& {Cole}}]{2019MNRAS.483.4922B}
{Baugh}, C.~M., {Gonzalez-Perez}, V., {Lagos}, C. D.~P., {et~al.} 2019, \bibinfo{title}{{Galaxy formation in the Planck Millennium: the atomic hydrogen content of dark matter haloes},} \mnras, 483, 4922, \dodoi{10.1093/mnras/sty3427}

\bibitem[{E.~F. {Bell} {et~al.}(2003){Bell}, {McIntosh}, {Katz}, \& {Weinberg}}]{2003ApJS..149..289B}
{Bell}, E.~F., {McIntosh}, D.~H., {Katz}, N., \& {Weinberg}, M.~D. 2003, \bibinfo{title}{{The Optical and Near-Infrared Properties of Galaxies. I. Luminosity and Stellar Mass Functions},} \apjs, 149, 289, \dodoi{10.1086/378847}

\bibitem[{M.~J.~I. {Brown} {et~al.}(2000){Brown}, {Webster}, \& {Boyle}}]{2000MNRAS.317..782B}
{Brown}, M.~J.~I., {Webster}, R.~L., \& {Boyle}, B.~J. 2000, \bibinfo{title}{{The clustering of colour-selected galaxies},} \mnras, 317, 782, \dodoi{10.1046/j.1365-8711.2000.03688.x}

\bibitem[{T. {Brown} {et~al.}(2017){Brown}, {Catinella}, {Cortese}, {Lagos}, {Dav{\'e}}, {Kilborn}, {Haynes}, {Giovanelli}, \& {Rafieferantsoa}}]{2017MNRAS.466.1275B}
{Brown}, T., {Catinella}, B., {Cortese}, L., {et~al.} 2017, \bibinfo{title}{{Cold gas stripping in satellite galaxies: from pairs to clusters},} \mnras, 466, 1275, \dodoi{10.1093/mnras/stw2991}

\bibitem[{T.-E. {Bulichi} {et~al.}(2024){Bulichi}, {Dav{\'e}}, \& {Kraljic}}]{2024MNRAS.529.2595B}
{Bulichi}, T.-E., {Dav{\'e}}, R., \& {Kraljic}, K. 2024, \bibinfo{title}{{How galaxy properties vary with filament proximity in the SIMBA simulations},} \mnras, 529, 2595, \dodoi{10.1093/mnras/stae667}

\bibitem[{H. {Butcher} \& A. {Oemler}(1984){Butcher} \& {Oemler}}]{1984ApJ...285..426B}
{Butcher}, H., \& {Oemler}, Jr., A. 1984, \bibinfo{title}{{The evolution of galaxies in clusters. V. A study of populations since Z 0.5.},} \apj, 285, 426, \dodoi{10.1086/162519}

\bibitem[{M. {Canducci} {et~al.}(2022){Canducci}, {Awad}, {Taghribi}, {Mohammadi}, {Mastropietro}, {De Rijcke}, {Peletier}, {Smith}, {Bunte}, \& {Ti{\v{n}}o}}]{2022A&C....4100658C}
{Canducci}, M., {Awad}, P., {Taghribi}, A., {et~al.} 2022, \bibinfo{title}{{1-DREAM: 1D Recovery, Extraction and Analysis of Manifolds in noisy environments},} Astronomy and Computing, 41, 100658, \dodoi{10.1016/j.ascom.2022.100658}

\bibitem[{M.~C. {Casas} {et~al.}(2024){Casas}, {Putnam}, {Mantz}, {Allen}, \& {Somboonpanyakul}}]{2024ApJ...967...14C}
{Casas}, M.~C., {Putnam}, K., {Mantz}, A.~B., {Allen}, S.~W., \& {Somboonpanyakul}, T. 2024, \bibinfo{title}{{Optical Photometric Indicators of Galaxy Cluster Relaxation},} \apj, 967, 14, \dodoi{10.3847/1538-4357/ad41de}

\bibitem[{B. {Catinella} {et~al.}(2010){Catinella}, {Schiminovich}, {Kauffmann}, {Fabello}, {Wang}, {Hummels}, {Lemonias}, {Moran}, {Wu}, {Giovanelli}, {Haynes}, {Heckman}, {Basu-Zych}, {Blanton}, {Brinchmann}, {Budav{\'a}ri}, {Gon{\c{c}}alves}, {Johnson}, {Kennicutt}, {Madore}, {Martin}, {Rich}, {Tacconi}, {Thilker}, {Wild}, \& {Wyder}}]{2010MNRAS.403..683C}
{Catinella}, B., {Schiminovich}, D., {Kauffmann}, G., {et~al.} 2010, \bibinfo{title}{{The GALEX Arecibo SDSS Survey - I. Gas fraction scaling relations of massive galaxies and first data release},} \mnras, 403, 683, \dodoi{10.1111/j.1365-2966.2009.16180.x}

\bibitem[{B. {Catinella} {et~al.}(2018){Catinella}, {Saintonge}, {Janowiecki}, {Cortese}, {Dav{\'e}}, {Lemonias}, {Cooper}, {Schiminovich}, {Hummels}, {Fabello}, {Ger{\'e}b}, {Kilborn}, \& {Wang}}]{2018MNRAS.476..875C}
{Catinella}, B., {Saintonge}, A., {Janowiecki}, S., {et~al.} 2018, \bibinfo{title}{{xGASS: total cold gas scaling relations and molecular-to-atomic gas ratios of galaxies in the local Universe},} \mnras, 476, 875, \dodoi{10.1093/mnras/sty089}

\bibitem[{G. {Chauhan} {et~al.}(2021){Chauhan}, {Lagos}, {Stevens}, {Bravo}, {Rhee}, {Power}, {Obreschkow}, \& {Meyer}}]{2021MNRAS.506.4893C}
{Chauhan}, G., {Lagos}, C. d.~P., {Stevens}, A. R.~H., {et~al.} 2021, \bibinfo{title}{{Unveiling the atomic hydrogen-halo mass relation via spectral stacking},} \mnras, 506, 4893, \dodoi{10.1093/mnras/stab1925}

\bibitem[{G. {Chauhan} {et~al.}(2020){Chauhan}, {Lagos}, {Stevens}, {Obreschkow}, {Power}, \& {Meyer}}]{2020MNRAS.498...44C}
{Chauhan}, G., {Lagos}, C. d.~P., {Stevens}, A. R.~H., {et~al.} 2020, \bibinfo{title}{{The physical drivers of the atomic hydrogen-halo mass relation},} \mnras, 498, 44, \dodoi{10.1093/mnras/staa2251}

\bibitem[{C. {Cheng} {et~al.}(2025{\natexlab{a}}){Cheng}, {Huang}, {Du}, {Zhang}, {Zhang}, {Zhu}, \& {Orellana}}]{2025ApJS..281...66C}
{Cheng}, C., {Huang}, J.-S., {Du}, W., {et~al.} 2025{\natexlab{a}}, \bibinfo{title}{{H I-detected Dwarf Galaxies in the FASHI Survey: Insights from Single- and Double-peaked Emission-line Samples},} \apjs, 281, 66, \dodoi{10.3847/1538-4365/ae1690}

\bibitem[{C. {Cheng} {et~al.}(2023){Cheng}, {Xu}, {Appleton}, {Duc}, {Tang}, {Dai}, {Huang}, {Lisenfeld}, {Renaud}, {He}, \& {Feng}}]{2023ApJ...954...74C}
{Cheng}, C., {Xu}, C.~K., {Appleton}, P.~N., {et~al.} 2023, \bibinfo{title}{{Deep H I Mapping of Stephan's Quintet and Its Neighborhood},} \apj, 954, 74, \dodoi{10.3847/1538-4357/ace03e}

\bibitem[{C. {Cheng} {et~al.}(2025{\natexlab{b}}){Cheng}, {Wang}, {Liang}, {Sun}, {Ibar}, {Brinch}, {Yan}, {Huang}, {Li}, \& {Molina}}]{2025ApJS..279...43C}
{Cheng}, C., {Wang}, X., {Liang}, P., {et~al.} 2025{\natexlab{b}}, \bibinfo{title}{{Probing Obscured Star Formation in Galaxy Clusters Using JWST Medium-band Images: 3.3 {\ensuremath{\mu}}m PAH Emitter Sample in A2744},} \apjs, 279, 43, \dodoi{10.3847/1538-4365/ade147}

\bibitem[{N. {Choque-Challapa} {et~al.}(2025){Choque-Challapa}, {Smith}, {Lacerna}, {Aguerri}, \& {Palma}}]{2025arXiv251023260C}
{Choque-Challapa}, N., {Smith}, R., {Lacerna}, I., {Aguerri}, J. A.~L., \& {Palma}, D. 2025, \bibinfo{title}{{The spatial distribution of dwarf and giant galaxies in and around Virgo cluster},} arXiv e-prints, arXiv:2510.23260, \dodoi{10.48550/arXiv.2510.23260}

\bibitem[{K. {Chun} {et~al.}(2025){Chun}, {Shin}, {Ko}, {Smith}, {Park}, \& {Nam}}]{2025arXiv250922802C}
{Chun}, K., {Shin}, J., {Ko}, J., {et~al.} 2025, \bibinfo{title}{{The Role of Pre-Processing in Tidal Feature Formation within Galaxy Clusters},} arXiv e-prints, arXiv:2509.22802, \dodoi{10.48550/arXiv.2509.22802}

\bibitem[{J. {Chung} {et~al.}(2021){Chung}, {Kim}, {Rey}, \& {Lee}}]{2021ApJ...923..235C}
{Chung}, J., {Kim}, S., {Rey}, S.-C., \& {Lee}, Y. 2021, \bibinfo{title}{{Star-forming Dwarf Galaxies in Filamentary Structures around the Virgo Cluster: Probing Chemical Pre-processing in Filament Environments},} \apj, 923, 235, \dodoi{10.3847/1538-4357/ac3002}

\bibitem[{M.~L.~M. {Collins} \& J.~I. {Read}(2022){Collins} \& {Read}}]{2022NatAs...6..647C}
{Collins}, M. L.~M., \& {Read}, J.~I. 2022, \bibinfo{title}{{Observational constraints on stellar feedback in dwarf galaxies},} Nature Astronomy, 6, 647, \dodoi{10.1038/s41550-022-01657-4}

\bibitem[{L. {Cortese} {et~al.}(2021){Cortese}, {Catinella}, \& {Smith}}]{2021PASA...38...35C}
{Cortese}, L., {Catinella}, B., \& {Smith}, R. 2021, \bibinfo{title}{{The Dawes Review 9: The role of cold gas stripping on the star formation quenching of satellite galaxies},} \pasa, 38, e035, \dodoi{10.1017/pasa.2021.18}

\bibitem[{L.~L. {Cowie} {et~al.}(1996){Cowie}, {Songaila}, {Hu}, \& {Cohen}}]{1996AJ....112..839C}
{Cowie}, L.~L., {Songaila}, A., {Hu}, E.~M., \& {Cohen}, J.~G. 1996, \bibinfo{title}{{New Insight on Galaxy Formation and Evolution From Keck Spectroscopy of the Hawaii Deep Fields},} \aj, 112, 839, \dodoi{10.1086/118058}

\bibitem[{M. {Crone Odekon} {et~al.}(2018){Crone Odekon}, {Hallenbeck}, {Haynes}, {Koopmann}, {Phi}, \& {Wolfe}}]{2018ApJ...852..142C}
{Crone Odekon}, M., {Hallenbeck}, G., {Haynes}, M.~P., {et~al.} 2018, \bibinfo{title}{{The Effect of Filaments and Tendrils on the H I Content of Galaxies},} \apj, 852, 142, \dodoi{10.3847/1538-4357/aaa1e8}

\bibitem[{S. {Damsted} {et~al.}(2024){Damsted}, {Finoguenov}, {Lietzen}, {Mamon}, {Comparat}, {Tempel}, {Dmitrieva}, {Clerc}, {Collins}, {Gozaliasl}, \& {Eckert}}]{2024A&A...690A..52D}
{Damsted}, S., {Finoguenov}, A., {Lietzen}, H., {et~al.} 2024, \bibinfo{title}{{AXES-SDSS: Comparison of SDSS galaxy groups with all-sky X-ray extended sources},} \aap, 690, A52, \dodoi{10.1051/0004-6361/202449591}

\bibitem[{R. {Dav{\'e}} {et~al.}(2020){Dav{\'e}}, {Crain}, {Stevens}, {Narayanan}, {Saintonge}, {Catinella}, \& {Cortese}}]{2020MNRAS.497..146D}
{Dav{\'e}}, R., {Crain}, R.~A., {Stevens}, A. R.~H., {et~al.} 2020, \bibinfo{title}{{Galaxy cold gas contents in modern cosmological hydrodynamic simulations},} \mnras, 497, 146, \dodoi{10.1093/mnras/staa1894}

\bibitem[{T. {Deb} {et~al.}(2025){Deb}, {Keating}, {Zabel}, {Moretti}, {Bacchini}, {Davis}, {Poggianti}, {Gullieuszik}, {Vulcani}, {Jeff{\'e}}, {Tomicic}, \& {Brown}}]{2025arXiv250515060D}
{Deb}, T., {Keating}, G.~K., {Zabel}, N., {et~al.} 2025, \bibinfo{title}{{SYMPHANY- SYnergy of Molecular PHase And Neutral hYdrogen in galaxies in A2626},} arXiv e-prints, arXiv:2505.15060, \dodoi{10.48550/arXiv.2505.15060}

\bibitem[{A. {Dekel} \& Y. {Birnboim}(2006){Dekel} \& {Birnboim}}]{2006MNRAS.368....2D}
{Dekel}, A., \& {Birnboim}, Y. 2006, \bibinfo{title}{{Galaxy bimodality due to cold flows and shock heating},} \mnras, 368, 2, \dodoi{10.1111/j.1365-2966.2006.10145.x}

\bibitem[{D. {DePalma} {et~al.}(2025){DePalma}, {Gupta}, {Chen}, {Simcoe}, {Balashev}, {Boettcher}, {Cantalupo}, {Chen}, {Combes}, {Faucher-Gigu{\`e}re}, {Johnson}, {Kl{\"o}ckner}, {Krogager}, {Li}, {L{\'o}pez}, {Noterdaeme}, {Petitjean}, {Qu}, {Rudie}, {Schaye}, \& {Zahedy}}]{2025ApJ...993L..18D}
{DePalma}, D., {Gupta}, N., {Chen}, H.-W., {et~al.} 2025, \bibinfo{title}{{H I Properties of Field Galaxies at z ≍ 0.2─0.6: Insights into Declining Cosmic Star Formation},} \apjl, 993, L18, \dodoi{10.3847/2041-8213/ae0d8b}

\bibitem[{ {DESI Collaboration} {et~al.}(2022){DESI Collaboration}, {Abareshi}, {Aguilar}, {Ahlen}, {Alam}, {Alexander}, {Alfarsy}, {Allen}, {Allende Prieto}, {Alves}, {Ameel}, {Armengaud}, {Asorey}, {Aviles}, {Bailey}, {Balaguera-Antol{\'\i}nez}, {Ballester}, {Baltay}, {Bault}, {Beltran}, {Benavides}, {BenZvi}, {Berti}, {Besuner}, {Beutler}, {Bianchi}, {Blake}, {Blanc}, {Blum}, {Bolton}, {Bose}, {Bramall}, {Brieden}, {Brodzeller}, {Brooks}, {Brownewell}, {Buckley-Geer}, {Cahn}, {Cai}, {Canning}, {Capasso}, {Carnero Rosell}, {Carton}, {Casas}, {Castander}, {Cervantes-Cota}, {Chabanier}, {Chaussidon}, {Chuang}, {Circosta}, {Cole}, {Cooper}, {da Costa}, {Cousinou}, {Cuceu}, {Davis}, {Dawson}, {de la Cruz-Noriega}, {de la Macorra}, {de Mattia}, {Della Costa}, {Demmer}, {Derwent}, {Dey}, {Dey}, {Dhungana}, {Ding}, {Dobson}, {Doel}, {Donald-McCann}, {Donaldson}, {Douglass}, {Duan}, {Dunlop}, {Edelstein}, {Eftekharzadeh}, {Eisenstein}, {Enriquez-Vargas}, {Escoffier}, {Evatt}, {Fagrelius}, {Fan}, {Fanning},
  {Fawcett}, {Ferraro}, {Ereza}, {Flaugher}, {Font-Ribera}, {Forero-Romero}, {Frenk}, {Fromenteau}, {G{\"a}nsicke}, {Garcia-Quintero}, {Garrison}, {Gazta{\~n}aga}, {Gerardi}, {Gil-Mar{\'\i}n}, {Gontcho A Gontcho}, {Gonzalez-Morales}, {Gonzalez-de-Rivera}, {Gonzalez-Perez}, {Gordon}, {Graur}, {Green}, {Grove}, {Gruen}, {Gutierrez}, {Guy}, {Hahn}, {Harris}, {Herrera}, {Herrera-Alcantar}, {Honscheid}, {Howlett}, {Huterer}, {Ir{\v{s}}i{\v{c}}}, {Ishak}, {Jelinsky}, {Jiang}, {Jimenez}, {Jing}, {Joyce}, {Jullo}, {Juneau}, {Kara{\c{c}}ayl{\i}}, {Karamanis}, {Karcher}, {Karim}, {Kehoe}, {Kent}, {Kirkby}, {Kisner}, {Kitaura}, {Koposov}, {Kov{\'a}cs}, {Kremin}, {Krolewski}, {L'Huillier}, {Lahav}, {Lambert}, {Lamman}, {Lan}, {Landriau}, {Lane}, {Lang}, {Lange}, {Lasker}, {Le Guillou}, {Leauthaud}, {Le Van Suu}, {Levi}, {Li}, {Magneville}, {Manera}, {Manser}, {Marshall}, {Martini}, {McCollam}, {McDonald}, {Meisner}, {Mena-Fern{\'a}ndez}, {Meneses-Rizo}, {Mezcua}, {Miller}, {Miquel}, {Montero-Camacho}, {Moon},
  {Moustakas}, {Mueller}, {Mu{\~n}oz-Guti{\'e}rrez}, {Myers}, {Nadathur}, {Najita}, {Napolitano}, {Neilsen}, {Newman}, {Nie}, {Ning}, {Niz}, {Norberg}, {Noriega}, {O'Brien}, {Obuljen}, {Palanque-Delabrouille}, {Palmese}, {Zhiwei}, {Pappalardo}, {PENG}, {Percival}, {Perruchot}, {Pogge}, {Poppett}, {Porredon}, {Prada}, {Prochaska}, {Pucha}, {P{\'e}rez-Fern{\'a}ndez}, {P{\'e}rez-R{\`a}fols}, {Rabinowitz}, \& {Raichoor}}]{2022AJ....164..207D}
{DESI Collaboration}, {Abareshi}, B., {Aguilar}, J., {et~al.} 2022, \bibinfo{title}{{Overview of the Instrumentation for the Dark Energy Spectroscopic Instrument},} \aj, 164, 207, \dodoi{10.3847/1538-3881/ac882b}

\bibitem[{ {DESI Collaboration} {et~al.}(2025){DESI Collaboration}, {Abdul-Karim}, {Adame}, {Aguado}, {Aguilar}, {Ahlen}, {Alam}, {Aldering}, {Alexander}, {Alfarsy}, {Allen}, {Allende Prieto}, {Alves}, {Anand}, {Andrade}, {Armengaud}, {Avila}, {Aviles}, {Awan}, {Bailey}, {Baleato Lizancos}, {Ballester}, {Bault}, {Bautista}, {BenZvi}, {Beraldo e Silva}, {Bermejo-Climent}, {Beutler}, {Bianchi}, {Blake}, {Blum}, {Bolton}, {Bonici}, {Brieden}, {Brodzeller}, {Brooks}, {Buckley-Geer}, {Burtin}, {Canning}, {Carnero Rosell}, {Carr}, {Carrilho}, {Casas}, {Castander}, {Cereskaite}, {Cervantes-Cota}, {Chaussidon}, {Chaves-Montero}, {Chen}, {Chen}, {Claybaugh}, {Cole}, {Cooper}, {Cousinou}, {Cuceu}, {Davis}, {Dawson}, {de Belsunce}, {de la Cruz}, {de la Macorra}, {de Mattia}, {Deiosso}, {Della Costa}, {Demina}, {Demirbozan}, {DeRose}, {Dey}, {Dey}, {Ding}, {Ding}, {Doel}, {Douglass}, {Dowicz}, {Ebina}, {Edelstein}, {Eisenstein}, {Elbers}, {Emas}, {Escoffier}, {Fagrelius}, {Fan}, {Fanning}, {Fawcett},
  {Fern\'andez-Garc\'ia}, {Ferraro}, {Findlay}, {Font-Ribera}, {Forero-Romero}, {Forero-S\'anchez}, {Frenk}, {G\''ansicke}, {Galbany}, {Garc\'ia-Bellido}, {Garcia-Quintero}, {Garrison}, {Gazta\~naga}, {Gil-Mar\'in}, {Gnedin}, {Gontcho}, {Gonzalez-Morales}, {Gonzalez-Perez}, {Gordon}, {Graur}, {Green}, {Gruen}, {Gsponer}, {Guandalin}, {Gutierrez}, {Guy}, {Hahn}, {Han}, {Han}, {He}, {Herrera-Alcantar}, {Honscheid}, {Hou}, {Howlett}, {Huterer}, {Ir\v\{s\}i\v\{c\}}, {Ishak}, {Jacques}, {Jimenez}, {Jing}, {Joachimi}, {Joudaki}, {Joyce}, {Jullo}, {Juneau}, {Kara\c\{c\}ayl\{\'i\}}, {Karim}, {Kehoe}, {Kent}, {Khederlarian}, {Kirkby}, {Kisner}, {Kitaura}, {Kizhuprakkat}, {Kong}, {Koposov}, {Kremin}, {Krolewski}, {Lahav}, {Lai}, {Lamman}, {Lan}, {Landriau}, {Lang}, {Lange}, {Lasker}, {Le Goff}, {Le Guillou}, {Leauthaud}, {Levi}, {Li}, {Li}, {Lodha}, {Lokken}, {Luo}, {Magneville}, {Manera}, {Manser}, {Margala}, {Martini}, {Maus}, {McCullough}, {McDonald}, {Medina}, {Medina-Varela}, {Meisner}, {Mena-Fern\'andez},
  {Menegas}, {Mezcua}, {Miquel}, {Montero-Camacho}, {Moon}, {Moustakas}, {Mu\~{n}oz-Guti\'errez}, {Mu\~{n}oz-Santos}, {Myers}, {Myles}, {Nadathur}, {Najita}, {Napolitano}, {Newman}, {Nikakhtar}, {Nikutta}, {Niz}, {Noriega}, {Padmanabhan}, {Paillas}, {Palanque-Delabrouille}, {Palmese}, {Pan}, {Pan}, {Parkinson}, {Peacock}, {Percival}, {P\'erez-Fern\'andez}, {P\'erez-R\`afols}, \& {Peterson}}]{2025arXiv250314745D}
{DESI Collaboration}, {Abdul-Karim}, M., {Adame}, A.~G., {et~al.} 2025, \bibinfo{title}{{Data Release 1 of the Dark Energy Spectroscopic Instrument},} arXiv e-prints, arXiv:2503.14745, \dodoi{10.48550/arXiv.2503.14745}

\bibitem[{A. {Dev} {et~al.}(2023){Dev}, {Driver}, {Meyer}, {Roychowdhury}, {Rhee}, {Stevens}, {Lagos}, {Bland-Hawthorn}, {Catinella}, {Hopkins}, {Loveday}, {Obreschkow}, {Phillipps}, \& {Robotham}}]{2023MNRAS.523.2693D}
{Dev}, A., {Driver}, S.~P., {Meyer}, M., {et~al.} 2023, \bibinfo{title}{{Galaxy And Mass Assembly (GAMA): The group H I mass as a function of halo mass},} \mnras, 523, 2693, \dodoi{10.1093/mnras/stad1575}

\bibitem[{A. {Diaferio}(1999){Diaferio}}]{1999MNRAS.309..610D}
{Diaferio}, A. 1999, \bibinfo{title}{{Mass estimation in the outer regions of galaxy clusters},} \mnras, 309, 610, \dodoi{10.1046/j.1365-8711.1999.02864.x}

\bibitem[{B. {Diemer} \& A.~V. {Kravtsov}(2014){Diemer} \& {Kravtsov}}]{2014ApJ...789....1D}
{Diemer}, B., \& {Kravtsov}, A.~V. 2014, \bibinfo{title}{{Dependence of the Outer Density Profiles of Halos on Their Mass Accretion Rate},} \apj, 789, 1, \dodoi{10.1088/0004-637X/789/1/1}

\bibitem[{L. {Doubrawa} {et~al.}(2025){Doubrawa}, {Smith}, {Mendes de Oliveira}, {O'Mill}, {Nakazono}, {Herpich}, {Gon{\c{c}}alves}, {Kanaan}, {Ribeiro}, \& {Schoenell}}]{2025A&A...700A.264D}
{Doubrawa}, L., {Smith}, R., {Mendes de Oliveira}, C., {et~al.} 2025, \bibinfo{title}{{SCALE: Using S-PLUS photometric information to analyse the brightest cluster galaxy alignment with satellite galaxies},} \aap, 700, A264, \dodoi{10.1051/0004-6361/202556043}

\bibitem[{A. {Dressler}(1980){Dressler}}]{1980ApJ...236..351D}
{Dressler}, A. 1980, \bibinfo{title}{{Galaxy morphology in rich clusters: implications for the formation and evolution of galaxies.},} \apj, 236, 351, \dodoi{10.1086/157753}

\bibitem[{A.~R. {Duffy} {et~al.}(2008){Duffy}, {Schaye}, {Kay}, \& {Dalla Vecchia}}]{2008MNRAS.390L..64D}
{Duffy}, A.~R., {Schaye}, J., {Kay}, S.~T., \& {Dalla Vecchia}, C. 2008, \bibinfo{title}{{Dark matter halo concentrations in the Wilkinson Microwave Anisotropy Probe year 5 cosmology},} \mnras, 390, L64, \dodoi{10.1111/j.1745-3933.2008.00537.x}

\bibitem[{V.~R. {Eke} {et~al.}(1998){Eke}, {Navarro}, \& {Frenk}}]{1998ApJ...503..569E}
{Eke}, V.~R., {Navarro}, J.~F., \& {Frenk}, C.~S. 1998, \bibinfo{title}{{The Evolution of X-Ray Clusters in a Low-Density Universe},} \apj, 503, 569, \dodoi{10.1086/306008}

\bibitem[{A.~E. {Evrard} {et~al.}(2008){Evrard}, {Bialek}, {Busha}, {White}, {Habib}, {Heitmann}, {Warren}, {Rasia}, {Tormen}, {Moscardini}, {Power}, {Jenkins}, {Gao}, {Frenk}, {Springel}, {White}, \& {Diemand}}]{2008ApJ...672..122E}
{Evrard}, A.~E., {Bialek}, J., {Busha}, M., {et~al.} 2008, \bibinfo{title}{{Virial Scaling of Massive Dark Matter Halos: Why Clusters Prefer a High Normalization Cosmology},} \apj, 672, 122, \dodoi{10.1086/521616}

\bibitem[{Y. {Fujita}(2004){Fujita}}]{2004PASJ...56...29F}
{Fujita}, Y. 2004, \bibinfo{title}{{Pre-Processing of Galaxies before Entering a Cluster},} \pasj, 56, 29, \dodoi{10.1093/pasj/56.1.29}

\bibitem[{M.~J. {Geller} \& J.~P. {Huchra}(1989){Geller} \& {Huchra}}]{1989Sci...246..897G}
{Geller}, M.~J., \& {Huchra}, J.~P. 1989, \bibinfo{title}{{Mapping the Universe},} Science, 246, 897, \dodoi{10.1126/science.246.4932.897}

\bibitem[{H. {Guo} {et~al.}(2021){Guo}, {Jones}, {Wang}, \& {Lin}}]{2021ApJ...918...53G}
{Guo}, H., {Jones}, M.~G., {Wang}, J., \& {Lin}, L. 2021, \bibinfo{title}{{Star Formation and Quenching of Central Galaxies from Stacked HI Measurements},} \apj, 918, 53, \dodoi{10.3847/1538-4357/ac062e}

\bibitem[{H. {Guo} {et~al.}(2017){Guo}, {Li}, {Zheng}, {Mo}, {Jing}, {Zu}, {Lim}, \& {Xu}}]{2017ApJ...846...61G}
{Guo}, H., {Li}, C., {Zheng}, Z., {et~al.} 2017, \bibinfo{title}{{Constraining the H I-Halo Mass Relation from Galaxy Clustering},} \apj, 846, 61, \dodoi{10.3847/1538-4357/aa85e7}

\bibitem[{C.~P. {Haines} {et~al.}(2015){Haines}, {Pereira}, {Smith}, {Egami}, {Babul}, {Finoguenov}, {Ziparo}, {McGee}, {Rawle}, {Okabe}, \& {Moran}}]{2015ApJ...806..101H}
{Haines}, C.~P., {Pereira}, M.~J., {Smith}, G.~P., {et~al.} 2015, \bibinfo{title}{{LoCuSS: The Slow Quenching of Star Formation in Cluster Galaxies and the Need for Pre-processing},} \apj, 806, 101, \dodoi{10.1088/0004-637X/806/1/101}

\bibitem[{C.~P. {Haines} {et~al.}(2018){Haines}, {Finoguenov}, {Smith}, {Babul}, {Egami}, {Mazzotta}, {Okabe}, {Pereira}, {Bianconi}, {McGee}, {Ziparo}, {Campusano}, \& {Loyola}}]{2018MNRAS.477.4931H}
{Haines}, C.~P., {Finoguenov}, A., {Smith}, G.~P., {et~al.} 2018, \bibinfo{title}{{LoCuSS: The infall of X-ray groups on to massive clusters},} \mnras, 477, 4931, \dodoi{10.1093/mnras/sty651}

\bibitem[{E.~J. {Hallman} \& M. {Markevitch}(2004){Hallman} \& {Markevitch}}]{2004ApJ...610L..81H}
{Hallman}, E.~J., \& {Markevitch}, M. 2004, \bibinfo{title}{{Chandra Observation of the Merging Cluster A168: A Late Stage in the Evolution of a Cold Front},} \apjl, 610, L81, \dodoi{10.1086/423449}

\bibitem[{S. {Han} {et~al.}(2018){Han}, {Smith}, {Choi}, {Cortese}, {Catinella}, {Contini}, \& {Yi}}]{2018ApJ...866...78H}
{Han}, S., {Smith}, R., {Choi}, H., {et~al.} 2018, \bibinfo{title}{{YZiCS: Preprocessing of Dark Halos in the Hydrodynamic Zoom-in Simulation of Clusters},} \apj, 866, 78, \dodoi{10.3847/1538-4357/aadfe2}

\bibitem[{M.~P. {Haynes} \& R. {Giovanelli}(1984){Haynes} \& {Giovanelli}}]{1984AJ.....89..758H}
{Haynes}, M.~P., \& {Giovanelli}, R. 1984, \bibinfo{title}{{Neutral hydrogen in isolated galaxies. IV. Results for the Arecibo sample.},} \aj, 89, 758, \dodoi{10.1086/113573}

\bibitem[{J. {Healy} {et~al.}(2019){Healy}, {Blyth}, {Elson}, {van Driel}, {Butcher}, {Schneider}, {Lehnert}, \& {Minchin}}]{2019MNRAS.487.4901H}
{Healy}, J., {Blyth}, S.~L., {Elson}, E., {et~al.} 2019, \bibinfo{title}{{HISS, a new tool for H I stacking: application to NIBLES spectra},} \mnras, 487, 4901, \dodoi{10.1093/mnras/stz1555}

\bibitem[{J. {Healy} {et~al.}(2021){Healy}, {Blyth}, {Verheijen}, {Hess}, {Serra}, {van der Hulst}, {Jarrett}, {Yim}, \& {J{\'o}zsa}}]{2021A&A...650A..76H}
{Healy}, J., {Blyth}, S.-L., {Verheijen}, M.~A.~W., {et~al.} 2021, \bibinfo{title}{{H I content in Coma cluster substructure},} \aap, 650, A76, \dodoi{10.1051/0004-6361/202038738}

\bibitem[{K.~M. {Hess} {et~al.}(2015){Hess}, {Jarrett}, {Carignan}, {Passmoor}, \& {Goedhart}}]{2015MNRAS.452.1617H}
{Hess}, K.~M., {Jarrett}, T.~H., {Carignan}, C., {Passmoor}, S.~S., \& {Goedhart}, S. 2015, \bibinfo{title}{{KAT-7 science verification: cold gas, star formation, and substructure in the nearby Antlia Cluster},} \mnras, 452, 1617, \dodoi{10.1093/mnras/stv1372}

\bibitem[{W. {Hu} {et~al.}(2021){Hu}, {Cortese}, {Staveley-Smith}, {Catinella}, {Chauhan}, {Lagos}, {Oosterloo}, \& {Chen}}]{2021MNRAS.507.5580H}
{Hu}, W., {Cortese}, L., {Staveley-Smith}, L., {et~al.} 2021, \bibinfo{title}{{The atomic hydrogen content of galaxies as a function of group-centric radius},} \mnras, 507, 5580, \dodoi{10.1093/mnras/stab2431}

\bibitem[{S. {Huang} {et~al.}(2012{\natexlab{a}}){Huang}, {Haynes}, {Giovanelli}, \& {Brinchmann}}]{2012ApJ...756..113H}
{Huang}, S., {Haynes}, M.~P., {Giovanelli}, R., \& {Brinchmann}, J. 2012{\natexlab{a}}, \bibinfo{title}{{The Arecibo Legacy Fast ALFA Survey: The Galaxy Population Detected by ALFALFA},} \apj, 756, 113, \dodoi{10.1088/0004-637X/756/2/113}

\bibitem[{S. {Huang} {et~al.}(2012{\natexlab{b}}){Huang}, {Haynes}, {Giovanelli}, {Brinchmann}, {Stierwalt}, \& {Neff}}]{2012AJ....143..133H}
{Huang}, S., {Haynes}, M.~P., {Giovanelli}, R., {et~al.} 2012{\natexlab{b}}, \bibinfo{title}{{Gas, Stars, and Star Formation in ALFALFA Dwarf Galaxies},} \aj, 143, 133, \dodoi{10.1088/0004-6256/143/6/133}

\bibitem[{Y.~L. {Jaff{\'e}} {et~al.}(2013){Jaff{\'e}}, {Poggianti}, {Verheijen}, {Deshev}, \& {van Gorkom}}]{2013MNRAS.431.2111J}
{Jaff{\'e}}, Y.~L., {Poggianti}, B.~M., {Verheijen}, M. A.~W., {Deshev}, B.~Z., \& {van Gorkom}, J.~H. 2013, \bibinfo{title}{{BUDHIES I: characterizing the environments in and around two clusters at z⋍0.2},} \mnras, 431, 2111, \dodoi{10.1093/mnras/stt250}

\bibitem[{P. {Jiang} {et~al.}(2024){Jiang}, {Chen}, {Gan}, {Sun}, {Zhu}, {Li}, {Zhu}, {Wu}, {Chen}, {Zhang}, \& {An}}]{2024AstTI...1...84J}
{Jiang}, P., {Chen}, R., {Gan}, H., {et~al.} 2024, \bibinfo{title}{{The FAST Core Array},} Astronomical Techniques and Instruments, 1, 84, \dodoi{10.61977/ati2024012}

\bibitem[{Y. {Jing} {et~al.}(2024){Jing}, {Wang}, {Xu}, {Liu}, {Chen}, {Liang}, {Xu}, {Cao}, {Wang}, {Hu}, {Zhang}, {Guo}, {Gao}, {Ai}, {Gan}, {Gao}, {Han}, {Hou}, {Hou}, {Jiang}, {Kong}, {Li}, {Liu}, {Shao}, {Pan}, {Pan}, {Qian}, {Sun}, {Tang}, {Yang}, {Zhang}, {Zhang}, \& {Zhu}}]{2024SCPMA..6759514J}
{Jing}, Y., {Wang}, J., {Xu}, C., {et~al.} 2024, \bibinfo{title}{{HiFAST: An HI data calibration and imaging pipeline for FAST},} Science China Physics, Mechanics, and Astronomy, 67, 259514, \dodoi{10.1007/s11433-023-2333-8}

\bibitem[{M.~G. {Jones} {et~al.}(2023){Jones}, {Verdes-Montenegro}, {Moldon}, {Damas Segovia}, {Borthakur}, {Luna}, {Yun}, {del Olmo}, {Perea}, {Cannon}, {Lopez Gutierrez}, {Cluver}, {Garrido}, \& {Sanchez}}]{2023A&A...670A..21J}
{Jones}, M.~G., {Verdes-Montenegro}, L., {Moldon}, J., {et~al.} 2023, \bibinfo{title}{{Disturbed, diffuse, or just missing? A global study of the H I content of Hickson compact groups},} \aap, 670, A21, \dodoi{10.1051/0004-6361/202244622}

\bibitem[{G. {Kauffmann} {et~al.}(2003){Kauffmann}, {Heckman}, {White}, {Charlot}, {Tremonti}, {Peng}, {Seibert}, {Brinkmann}, {Nichol}, {SubbaRao}, \& {York}}]{2003MNRAS.341...54K}
{Kauffmann}, G., {Heckman}, T.~M., {White}, S. D.~M., {et~al.} 2003, \bibinfo{title}{{The dependence of star formation history and internal structure on stellar mass for {}10$^{5}$ low-redshift galaxies},} \mnras, 341, 54, \dodoi{10.1046/j.1365-8711.2003.06292.x}

\bibitem[{V.~A. {Kilborn} {et~al.}(2009){Kilborn}, {Forbes}, {Barnes}, {Koribalski}, {Brough}, \& {Kern}}]{2009MNRAS.400.1962K}
{Kilborn}, V.~A., {Forbes}, D.~A., {Barnes}, D.~G., {et~al.} 2009, \bibinfo{title}{{Southern GEMS groups - II. HI distribution, mass functions and HI deficient galaxies},} \mnras, 400, 1962, \dodoi{10.1111/j.1365-2966.2009.15587.x}

\bibitem[{H. {Kim} {et~al.}(2026){Kim}, {Canducci}, {Smith}, {Tino}, {Jaffe}, {Seong Hwang}, {Shin}, \& {Chun}}]{2026A&A...708A.262K}
{Kim}, H., {Canducci}, M., {Smith}, R., {et~al.} 2026, \bibinfo{title}{{New classification method for the dynamical state of galaxy clusters with a Gaussian mixture model},} \aap, 708, A262, \dodoi{10.1051/0004-6361/202557129}

\bibitem[{S. {Kim} {et~al.}(2016){Kim}, {Rey}, {Bureau}, {Yoon}, {Chung}, {Jerjen}, {Lisker}, {Jeong}, {Sung}, {Lee}, {Lee}, \& {Chung}}]{2016ApJ...833..207K}
{Kim}, S., {Rey}, S.-C., {Bureau}, M., {et~al.} 2016, \bibinfo{title}{{Large-scale Filamentary Structures around the Virgo Cluster Revisited},} \apj, 833, 207, \dodoi{10.3847/1538-4357/833/2/207}

\bibitem[{D. {Kleiner} {et~al.}(2017){Kleiner}, {Pimbblet}, {Jones}, {Koribalski}, \& {Serra}}]{2017MNRAS.466.4692K}
{Kleiner}, D., {Pimbblet}, K.~A., {Jones}, D.~H., {Koribalski}, B.~S., \& {Serra}, P. 2017, \bibinfo{title}{{Evidence for H I replenishment in massive galaxies through gas accretion from the cosmic web},} \mnras, 466, 4692, \dodoi{10.1093/mnras/stw3328}

\bibitem[{B.~S. {Koribalski} {et~al.}(2020){Koribalski}, {Staveley-Smith}, {Westmeier}, {Serra}, {Spekkens}, {Wong}, {Lee-Waddell}, {Lagos}, {Obreschkow}, {Ryan-Weber}, {Zwaan}, {Kilborn}, {Bekiaris}, {Bekki}, {Bigiel}, {Boselli}, {Bosma}, {Catinella}, {Chauhan}, {Cluver}, {Colless}, {Courtois}, {Crain}, {de Blok}, {D{\'e}nes}, {Duffy}, {Elagali}, {Fluke}, {For}, {Heald}, {Henning}, {Hess}, {Holwerda}, {Howlett}, {Jarrett}, {Jones}, {Jones}, {J{\'o}zsa}, {Jurek}, {J{\"u}tte}, {Kamphuis}, {Karachentsev}, {Kerp}, {Kleiner}, {Kraan-Korteweg}, {L{\'o}pez-S{\'a}nchez}, {Madrid}, {Meyer}, {Mould}, {Murugeshan}, {Norris}, {Oh}, {Oosterloo}, {Popping}, {Putman}, {Reynolds}, {Rhee}, {Robotham}, {Ryder}, {Schr{\"o}der}, {Shao}, {Stevens}, {Taylor}, {van{\^A} der Hulst}, {Verdes-Montenegro}, {Wakker}, {Wang}, {Whiting}, {Winkel}, \& {Wolf}}]{2020Ap&SS.365..118K}
{Koribalski}, B.~S., {Staveley-Smith}, L., {Westmeier}, T., {et~al.} 2020, \bibinfo{title}{{WALLABY {\textendash} an SKA Pathfinder H I survey},} \apss, 365, 118, \dodoi{10.1007/s10509-020-03831-4}

\bibitem[{M. {Korsaga} {et~al.}(2023){Korsaga}, {Famaey}, {Freundlich}, {Posti}, {Ibata}, {Boily}, {Kraljic}, {Esparza-Arredondo}, {Ramos Almeida}, \& {Koulidiati}}]{2023ApJ...952L..41K}
{Korsaga}, M., {Famaey}, B., {Freundlich}, J., {et~al.} 2023, \bibinfo{title}{{Disk Galaxies Are Self-similar: The Universality of the H I-to-Halo Mass Ratio for Isolated Disks},} \apjl, 952, L41, \dodoi{10.3847/2041-8213/ace364}

\bibitem[{P. {Kroupa} {et~al.}(1993){Kroupa}, {Tout}, \& {Gilmore}}]{1993MNRAS.262..545K}
{Kroupa}, P., {Tout}, C.~A., \& {Gilmore}, G. 1993, \bibinfo{title}{{The Distribution of Low-Mass Stars in the Galactic Disc},} \mnras, 262, 545, \dodoi{10.1093/mnras/262.3.545}

\bibitem[{D. {Lang} {et~al.}(2016{\natexlab{a}}){Lang}, {Hogg}, \& {Mykytyn}}]{2016ascl.soft04008L}
{Lang}, D., {Hogg}, D.~W., \& {Mykytyn}, D. 2016{\natexlab{a}}, {The Tractor: Probabilistic astronomical source detection and measurement},, Astrophysics Source Code Library, record ascl:1604.008 \doeprint{1604.008}

\bibitem[{D. {Lang} {et~al.}(2016{\natexlab{b}}){Lang}, {Hogg}, \& {Schlegel}}]{2016AJ....151...36L}
{Lang}, D., {Hogg}, D.~W., \& {Schlegel}, D.~J. 2016{\natexlab{b}}, \bibinfo{title}{{WISE Photometry for 400 Million SDSS Sources},} \aj, 151, 36, \dodoi{10.3847/0004-6256/151/2/36}

\bibitem[{B. {Lee} {et~al.}(2022){Lee}, {Wang}, {Chung}, {Ho}, {Wang}, {Michiyama}, {Molina}, {Kim}, {Shao}, {Kilborn}, {Wang}, {Lin}, {Kim}, {Catinella}, {Cortese}, {Deg}, {Denes}, {Elagali}, {For}, {Kleiner}, {Koribalski}, {Lee-Waddell}, {Rhee}, {Spekkens}, {Westmeier}, {Wong}, {Bigiel}, {Bosma}, {Holwerda}, {van der Hulst}, {Roychowdhury}, {Verdes-Montenegro}, \& {Zwaan}}]{2022ApJS..262...31L}
{Lee}, B., {Wang}, J., {Chung}, A., {et~al.} 2022, \bibinfo{title}{{ALMA/ACA CO Survey of the IC 1459 and NGC 4636 Groups: Environmental Effects on the Molecular Gas of Group Galaxies},} \apjs, 262, 31, \dodoi{10.3847/1538-4365/ac7eba}

\bibitem[{C. {Li} {et~al.}(2012){Li}, {Kauffmann}, {Fu}, {Wang}, {Catinella}, {Fabello}, {Schiminovich}, \& {Zhang}}]{2012MNRAS.424.1471L}
{Li}, C., {Kauffmann}, G., {Fu}, J., {et~al.} 2012, \bibinfo{title}{{The clustering of galaxies as a function of their photometrically estimated atomic gas content},} \mnras, 424, 1471, \dodoi{10.1111/j.1365-2966.2012.21337.x}

\bibitem[{X. {Li} {et~al.}(2023){Li}, {Li}, {Mo}, {Hu}, {Wang}, \& {Xiao}}]{2023arXiv231203601L}
{Li}, X., {Li}, C., {Mo}, H.~J., {et~al.} 2023, \bibinfo{title}{{On the existence, rareness and uniqueness of quenched HI-rich galaxies in the local Universe},} arXiv e-prints, arXiv:2312.03601, \dodoi{10.48550/arXiv.2312.03601}

\bibitem[{X. {Li} {et~al.}(2024){Li}, {Li}, {Mo}, {Hu}, {Wang}, \& {Xiao}}]{2024ApJ...963...86L}
{Li}, X., {Li}, C., {Mo}, H.~J., {et~al.} 2024, \bibinfo{title}{{On the Existence, Rareness, and Uniqueness of Quenched H I-rich Galaxies in the Local Universe},} \apj, 963, 86, \dodoi{10.3847/1538-4357/ad1ce3}

\bibitem[{X. {Li} {et~al.}(2022){Li}, {Li}, {Mo}, {Xiao}, \& {Wang}}]{2022ApJ...941...48L}
{Li}, X., {Li}, C., {Mo}, H.~J., {Xiao}, T., \& {Wang}, J. 2022, \bibinfo{title}{{Conditional H I Mass Functions and the H I-to-halo Mass Relation in the Local Universe},} \apj, 941, 48, \dodoi{10.3847/1538-4357/ac9ccb}

\bibitem[{E.~L. {{\L}okas} \& G.~A. {Mamon}(2001){{\L}okas} \& {Mamon}}]{2001MNRAS.321..155L}
{{\L}okas}, E.~L., \& {Mamon}, G.~A. 2001, \bibinfo{title}{{Properties of spherical galaxies and clusters with an NFW density profile},} \mnras, 321, 155, \dodoi{10.1046/j.1365-8711.2001.04007.x}

\bibitem[{P.~A.~A. {Lopes} {et~al.}(2024){Lopes}, {Ribeiro}, \& {Brambila}}]{2024MNRAS.527L..19L}
{Lopes}, P. A.~A., {Ribeiro}, A. L.~B., \& {Brambila}, D. 2024, \bibinfo{title}{{The role of groups in galaxy evolution: compelling evidence of pre-processing out to the turnaround radius of clusters},} \mnras, 527, L19, \dodoi{10.1093/mnrasl/slad134}

\bibitem[{N. {Maddox} {et~al.}(2015){Maddox}, {Hess}, {Obreschkow}, {Jarvis}, \& {Blyth}}]{2015MNRAS.447.1610M}
{Maddox}, N., {Hess}, K.~M., {Obreschkow}, D., {Jarvis}, M.~J., \& {Blyth}, S.~L. 2015, \bibinfo{title}{{Variation of galactic cold gas reservoirs with stellar mass},} \mnras, 447, 1610, \dodoi{10.1093/mnras/stu2532}

\bibitem[{A. {Marasco} {et~al.}(2016){Marasco}, {Crain}, {Schaye}, {Bah{\'e}}, {van der Hulst}, {Theuns}, \& {Bower}}]{2016MNRAS.461.2630M}
{Marasco}, A., {Crain}, R.~A., {Schaye}, J., {et~al.} 2016, \bibinfo{title}{{The environmental dependence of H I in galaxies in the EAGLE simulations},} \mnras, 461, 2630, \dodoi{10.1093/mnras/stw1498}

\bibitem[{F. {Marinacci} {et~al.}(2018){Marinacci}, {Vogelsberger}, {Pakmor}, {Torrey}, {Springel}, {Hernquist}, {Nelson}, {Weinberger}, {Pillepich}, {Naiman}, \& {Genel}}]{2018MNRAS.480.5113M}
{Marinacci}, F., {Vogelsberger}, M., {Pakmor}, R., {et~al.} 2018, \bibinfo{title}{{First results from the IllustrisTNG simulations: radio haloes and magnetic fields},} \mnras, 480, 5113, \dodoi{10.1093/mnras/sty2206}

\bibitem[{H. {M{\'e}ndez-Hern{\'a}ndez} {et~al.}(2025){M{\'e}ndez-Hern{\'a}ndez}, {Lima-Dias}, {Monachesi}, {Jaff{\'e}}, {Haines}, {Teixeira}, {L{\"o}sch}, {Baier-Soto}, {Lima}, {Amrutha B.}, {Bom}, {D'Ago}, {Demarco}, {Finoguenov}, {Haack}, {Lopes}, {Mendes de Oliveira}, {Merluzzi}, {Piraino-Cerda}, {Smith Castelli}, {Sif'on}, {Sodr{\'e}}, {Tejos}, {Torres-Flores}, {Argudo-Fern{\'a}ndez}, {Crossett}, {Ibar}, {Kuchner}, {Lacerna}, {Lopes-Silva}, {Lopez}, {McGee}, {Morelli}, {Nantais}, {Olivares V.}, {Pallero}, {Poggianti}, {Pompei}, {Sampaio}, {Vulcani}, {Zenteno}, {Almeida-Fernandes}, {Bilicki}, {Carvalho}, {Cheng}, {Figueiredo}, {Guti{\'e}rrez-Soto}, {Herpich}, {Kanaan}, {Lacerda}, {Nakazono}, {Oliveira Schwarz}, {Ribeiro}, {Roukema}, {Sartori}, {Santos-Silva}, \& {Schoenell}}]{2025arXiv251019958M}
{M{\'e}ndez-Hern{\'a}ndez}, H., {Lima-Dias}, C., {Monachesi}, A., {et~al.} 2025, \bibinfo{title}{{Targeting cluster galaxies for the 4MOST CHANCES Low-z sub-survey with photometric redshifts},} arXiv e-prints, arXiv:2510.19958, \dodoi{10.48550/arXiv.2510.19958}

\bibitem[{D.~C. {Moln{\'a}r} {et~al.}(2022){Moln{\'a}r}, {Serra}, {van der Hulst}, {Jarrett}, {Boselli}, {Cortese}, {Healy}, {de Blok}, {Cappellari}, {Hess}, {J{\'o}zsa}, {McDermid}, {Oosterloo}, \& {Verheijen}}]{2022A&A...659A..94M}
{Moln{\'a}r}, D.~C., {Serra}, P., {van der Hulst}, T., {et~al.} 2022, \bibinfo{title}{{The Westerbork Coma Survey. A blind, deep, high-resolution H I survey of the Coma cluster},} \aap, 659, A94, \dodoi{10.1051/0004-6361/202142614}

\bibitem[{K. {Morokuma-Matsui} {et~al.}(2022){Morokuma-Matsui}, {Bekki}, {Wang}, {Serra}, {Koyama}, {Morokuma}, {Egusa}, {For}, {Nakanishi}, {Koribalski}, {Okamoto}, {Kodama}, {Lee}, {Maccagni}, {Miura}, {Espada}, {Takeuchi}, {Yang}, {Lee}, {Ueda}, \& {Matsushita}}]{2022ApJS..263...40M}
{Morokuma-Matsui}, K., {Bekki}, K., {Wang}, J., {et~al.} 2022, \bibinfo{title}{{CO(J = 1-0) Mapping Survey of 64 Galaxies in the Fornax Cluster with the ALMA Morita Array},} \apjs, 263, 40, \dodoi{10.3847/1538-4365/ac983b}

\bibitem[{J. {Moustakas} {et~al.}(2023){Moustakas}, {Lang}, {Dey}, {Juneau}, {Meisner}, {Myers}, {Schlafly}, {Schlegel}, {Valdes}, {Weaver}, \& {Zhou}}]{2023ApJS..269....3M}
{Moustakas}, J., {Lang}, D., {Dey}, A., {et~al.} 2023, \bibinfo{title}{{Siena Galaxy Atlas 2020},} \apjs, 269, 3, \dodoi{10.3847/1538-4365/acfaa2}

\bibitem[{J.~P. {Naiman} {et~al.}(2018){Naiman}, {Pillepich}, {Springel}, {Ramirez-Ruiz}, {Torrey}, {Vogelsberger}, {Pakmor}, {Nelson}, {Marinacci}, {Hernquist}, {Weinberger}, \& {Genel}}]{2018MNRAS.477.1206N}
{Naiman}, J.~P., {Pillepich}, A., {Springel}, V., {et~al.} 2018, \bibinfo{title}{{First results from the IllustrisTNG simulations: a tale of two elements - chemical evolution of magnesium and europium},} \mnras, 477, 1206, \dodoi{10.1093/mnras/sty618}

\bibitem[{J.~F. {Navarro} {et~al.}(1997){Navarro}, {Frenk}, \& {White}}]{1997ApJ...490..493N}
{Navarro}, J.~F., {Frenk}, C.~S., \& {White}, S. D.~M. 1997, \bibinfo{title}{{A Universal Density Profile from Hierarchical Clustering},} \apj, 490, 493, \dodoi{10.1086/304888}

\bibitem[{D. {Nelson} {et~al.}(2024){Nelson}, {Pillepich}, {Ayromlou}, {Lee}, {Lehle}, {Rohr}, \& {Truong}}]{2024A&A...686A.157N}
{Nelson}, D., {Pillepich}, A., {Ayromlou}, M., {et~al.} 2024, \bibinfo{title}{{Introducing the TNG-Cluster simulation: Overview and the physical properties of the gaseous intracluster medium},} \aap, 686, A157, \dodoi{10.1051/0004-6361/202348608}

\bibitem[{D. {Nelson} {et~al.}(2018){Nelson}, {Pillepich}, {Springel}, {Weinberger}, {Hernquist}, {Pakmor}, {Genel}, {Torrey}, {Vogelsberger}, {Kauffmann}, {Marinacci}, \& {Naiman}}]{2018MNRAS.475..624N}
{Nelson}, D., {Pillepich}, A., {Springel}, V., {et~al.} 2018, \bibinfo{title}{{First results from the IllustrisTNG simulations: the galaxy colour bimodality},} \mnras, 475, 624, \dodoi{10.1093/mnras/stx3040}

\bibitem[{N.~T. {Nguyen-Dang} {et~al.}(2025){Nguyen-Dang}, {Ota}, {Okabe}, {Oguri}, {Mitsuishi}, {Reiprich}, {Pacaud}, {Bulbul}, {Sanders}, {Br{\"u}ggen}, {Liu}, {Tsujita}, {Chiu}, {Ghirardini}, {Grandis}, {Klein}, {Migkas}, {Miyatake}, {Miyazaki}, \& {Ramos-Ceja}}]{2025arXiv251206138N}
{Nguyen-Dang}, N.~T., {Ota}, N., {Okabe}, N., {et~al.} 2025, \bibinfo{title}{{The eROSITA Final Equatorial-Depth Survey (eFEDS): X-ray stacking analysis of Subaru's optically selected clusters spanning low richness regime},} arXiv e-prints, arXiv:2512.06138.
\newblock \doarXiv{2512.06138}

\bibitem[{K.~G. {Noeske} {et~al.}(2007){Noeske}, {Weiner}, {Faber}, {Papovich}, {Koo}, {Somerville}, {Bundy}, {Conselice}, {Newman}, {Schiminovich}, {Le Floc'h}, {Coil}, {Rieke}, {Lotz}, {Primack}, {Barmby}, {Cooper}, {Davis}, {Ellis}, {Fazio}, {Guhathakurta}, {Huang}, {Kassin}, {Martin}, {Phillips}, {Rich}, {Small}, {Willmer}, \& {Wilson}}]{2007ApJ...660L..43N}
{Noeske}, K.~G., {Weiner}, B.~J., {Faber}, S.~M., {et~al.} 2007, \bibinfo{title}{{Star Formation in AEGIS Field Galaxies since z=1.1: The Dominance of Gradually Declining Star Formation, and the Main Sequence of Star-forming Galaxies},} \apjl, 660, L43, \dodoi{10.1086/517926}

\bibitem[{A. {Obuljen} {et~al.}(2019){Obuljen}, {Alonso}, {Villaescusa-Navarro}, {Yoon}, \& {Jones}}]{2019MNRAS.486.5124O}
{Obuljen}, A., {Alonso}, D., {Villaescusa-Navarro}, F., {Yoon}, I., \& {Jones}, M. 2019, \bibinfo{title}{{The H I content of dark matter haloes at z {\ensuremath{\approx}} 0 from ALFALFA},} \mnras, 486, 5124, \dodoi{10.1093/mnras/stz1118}

\bibitem[{J.~B. {Oke} \& J.~E. {Gunn}(1983){Oke} \& {Gunn}}]{1983ApJ...266..713O}
{Oke}, J.~B., \& {Gunn}, J.~E. 1983, \bibinfo{title}{{Secondary standard stars for absolute spectrophotometry.},} \apj, 266, 713, \dodoi{10.1086/160817}

\bibitem[{K.~A. {Oman} {et~al.}(2021){Oman}, {Bah{\'e}}, {Healy}, {Hess}, {Hudson}, \& {Verheijen}}]{2021MNRAS.501.5073O}
{Oman}, K.~A., {Bah{\'e}}, Y.~M., {Healy}, J., {et~al.} 2021, \bibinfo{title}{{A homogeneous measurement of the delay between the onsets of gas stripping and star formation quenching in satellite galaxies of groups and clusters},} \mnras, 501, 5073, \dodoi{10.1093/mnras/staa3845}

\bibitem[{E. {O'Sullivan} {et~al.}(2018){O'Sullivan}, {Combes}, {Salom{\'e}}, {David}, {Babul}, {Vrtilek}, {Lim}, {Olivares}, {Raychaudhury}, \& {Schellenberger}}]{2018A&A...618A.126O}
{O'Sullivan}, E., {Combes}, F., {Salom{\'e}}, P., {et~al.} 2018, \bibinfo{title}{{Cold gas in a complete sample of group-dominant early-type galaxies},} \aap, 618, A126, \dodoi{10.1051/0004-6361/201833580}

\bibitem[{H. {Padmanabhan} \& G. {Kulkarni}(2017){Padmanabhan} \& {Kulkarni}}]{2017MNRAS.470..340P}
{Padmanabhan}, H., \& {Kulkarni}, G. 2017, \bibinfo{title}{{Constraints on the evolution of the relationship between H I mass and halo mass in the last 12 Gyr},} \mnras, 470, 340, \dodoi{10.1093/mnras/stx1178}

\bibitem[{H. {Pan} {et~al.}(2023){Pan}, {Jarvis}, {Santos}, {Maddox}, {Frank}, {Ponomareva}, {Prandoni}, {Kurapati}, {Baes}, {Mancera Pi{\~n}a}, {Rodighiero}, {Meyer}, {Dav{\'e}}, {Sharma}, {Rajohnson}, {Adams}, {Bowler}, {Sinigaglia}, {van der Hulst}, {Hatfield}, {Sekhar}, \& {Collier}}]{2023MNRAS.525..256P}
{Pan}, H., {Jarvis}, M.~J., {Santos}, M.~G., {et~al.} 2023, \bibinfo{title}{{MIGHTEE-H I: the M$_{H I}$ - M$_{*}$ relation over the last billion years},} \mnras, 525, 256, \dodoi{10.1093/mnras/stad2343}

\bibitem[{V. {Parkash} {et~al.}(2018){Parkash}, {Brown}, {Jarrett}, \& {Bonne}}]{2018ApJ...864...40P}
{Parkash}, V., {Brown}, M. J.~I., {Jarrett}, T.~H., \& {Bonne}, N.~J. 2018, \bibinfo{title}{{Relationships between HI Gas Mass, Stellar Mass, and the Star Formation Rate of HICAT+WISE (H I-WISE) Galaxies},} \apj, 864, 40, \dodoi{10.3847/1538-4357/aad3b9}

\bibitem[{A. {Pillepich} {et~al.}(2018){Pillepich}, {Nelson}, {Hernquist}, {Springel}, {Pakmor}, {Torrey}, {Weinberger}, {Genel}, {Naiman}, {Marinacci}, \& {Vogelsberger}}]{2018MNRAS.475..648P}
{Pillepich}, A., {Nelson}, D., {Hernquist}, L., {et~al.} 2018, \bibinfo{title}{{First results from the IllustrisTNG simulations: the stellar mass content of groups and clusters of galaxies},} \mnras, 475, 648, \dodoi{10.1093/mnras/stx3112}

\bibitem[{F. {Piraino-Cerda} {et~al.}(2024){Piraino-Cerda}, {Jaff{\'e}}, {Louren{\c{c}}o}, {Crossett}, {Salinas}, {Kim}, {Sheen}, {Kelkar}, {Pallero}, \& {Bravo-Alfaro}}]{2024MNRAS.528..919P}
{Piraino-Cerda}, F., {Jaff{\'e}}, Y.~L., {Louren{\c{c}}o}, A.~C., {et~al.} 2024, \bibinfo{title}{{Pre- and post-processing of cluster galaxies out to 5 {\texttimes} R$_{200}$: the extreme case of A2670},} \mnras, 528, 919, \dodoi{10.1093/mnras/stad3957}

\bibitem[{A. {Popping} {et~al.}(2009){Popping}, {Dav{\'e}}, {Braun}, \& {Oppenheimer}}]{2009A&A...504...15P}
{Popping}, A., {Dav{\'e}}, R., {Braun}, R., \& {Oppenheimer}, B.~D. 2009, \bibinfo{title}{{The simulated H I sky at low redshift},} \aap, 504, 15, \dodoi{10.1051/0004-6361/200911811}

\bibitem[{T.~N. {Reynolds} {et~al.}(2021){Reynolds}, {Westmeier}, {Elagali}, {Catinella}, {Cortese}, {Deg}, {For}, {Kamphuis}, {Kleiner}, {Koribalski}, {Lee-Waddell}, {Oh}, {Rhee}, {Serra}, {Spekkens}, {Staveley-Smith}, {Stevens}, {Taylor}, {Wang}, \& {Wong}}]{2021MNRAS.505.1891R}
{Reynolds}, T.~N., {Westmeier}, T., {Elagali}, A., {et~al.} 2021, \bibinfo{title}{{WALLABY pilot survey: first look at the Hydra I cluster and ram pressure stripping of ESO 501-G075},} \mnras, 505, 1891, \dodoi{10.1093/mnras/stab1371}

\bibitem[{J. {Rom{\'a}n} {et~al.}(2023){Rom{\'a}n}, {Rich}, {Ahvazi}, {Sales}, {Li}, {Golini}, {Trujillo}, {Knapen}, {Peletier}, \& {S{\'a}nchez-Alarc{\'o}n}}]{2023A&A...679A.157R}
{Rom{\'a}n}, J., {Rich}, R.~M., {Ahvazi}, N., {et~al.} 2023, \bibinfo{title}{{A giant thin stellar stream in the Coma Galaxy Cluster},} \aap, 679, A157, \dodoi{10.1051/0004-6361/202346780}

\bibitem[{H.~J. {Rood} \& G.~N. {Sastry}(1971){Rood} \& {Sastry}}]{1971PASP...83..313R}
{Rood}, H.~J., \& {Sastry}, G.~N. 1971, \bibinfo{title}{{``Tuning Fork'' Classification of Rich Clusters of Galaxies},} \pasp, 83, 313, \dodoi{10.1086/129128}

\bibitem[{V. {Salinas} {et~al.}(2024){Salinas}, {Jaff{\'e}}, {Smith}, {Shinn}, {Crossett}, {Gullieuszik}, {Gonz{\'a}lez-Tor{\`a}}, {Piraino-Cerda}, {Poggianti}, {Vulcani}, {Biviano}, {Louren{\c{c}}o}, {Bilton}, {Kelkar}, \& {Calder{\'o}n-Castillo}}]{2024MNRAS.533..341S}
{Salinas}, V., {Jaff{\'e}}, Y.~L., {Smith}, R., {et~al.} 2024, \bibinfo{title}{{Constraining the duration of ram pressure stripping features in the optical from the direction of jellyfish galaxy tails},} \mnras, 533, 341, \dodoi{10.1093/mnras/stae1784}

\bibitem[{C.~L. {Sarazin}(1986){Sarazin}}]{1986RvMP...58....1S}
{Sarazin}, C.~L. 1986, \bibinfo{title}{{X-ray emission from clusters of galaxies},} Reviews of Modern Physics, 58, 1, \dodoi{10.1103/RevModPhys.58.1}

\bibitem[{A. {Sengupta} {et~al.}(2022){Sengupta}, {Keel}, {Morrison}, {Windhorst}, {Miller}, \& {Smith}}]{2022ApJS..258...32S}
{Sengupta}, A., {Keel}, W.~C., {Morrison}, G., {et~al.} 2022, \bibinfo{title}{{The Preprocessing of Galaxies in the Early Stages of Cluster Formation in Abell 1882 at z = 0.139},} \apjs, 258, 32, \dodoi{10.3847/1538-4365/ac3761}

\bibitem[{P. {Serra} {et~al.}(2012){Serra}, {Oosterloo}, {Morganti}, {Alatalo}, {Blitz}, {Bois}, {Bournaud}, {Bureau}, {Cappellari}, {Crocker}, {Davies}, {Davis}, {de Zeeuw}, {Duc}, {Emsellem}, {Khochfar}, {Krajnovi{\'c}}, {Kuntschner}, {Lablanche}, {McDermid}, {Naab}, {Sarzi}, {Scott}, {Trager}, {Weijmans}, \& {Young}}]{2012MNRAS.422.1835S}
{Serra}, P., {Oosterloo}, T., {Morganti}, R., {et~al.} 2012, \bibinfo{title}{{The ATLAS$^{3D}$ project - XIII. Mass and morphology of H I in early-type galaxies as a function of environment},} \mnras, 422, 1835, \dodoi{10.1111/j.1365-2966.2012.20219.x}

\bibitem[{P. {Serra} {et~al.}(2014){Serra}, {Oser}, {Krajnovi{\'c}}, {Naab}, {Oosterloo}, {Morganti}, {Cappellari}, {Emsellem}, {Young}, {Blitz}, {Davis}, {Duc}, {Hirschmann}, {Weijmans}, {Alatalo}, {Bayet}, {Bois}, {Bournaud}, {Bureau}, {Crocker}, {Davies}, {de Zeeuw}, {Khochfar}, {Kuntschner}, {Lablanche}, {McDermid}, {Sarzi}, \& {Scott}}]{2014MNRAS.444.3388S}
{Serra}, P., {Oser}, L., {Krajnovi{\'c}}, D., {et~al.} 2014, \bibinfo{title}{{The ATLAS$^{3D}$ project - XXVI. H I discs in real and simulated fast and slow rotators},} \mnras, 444, 3388, \dodoi{10.1093/mnras/stt2496}

\bibitem[{T. {Shin} {et~al.}(2019){Shin}, {Adhikari}, {Baxter}, {Chang}, {Jain}, {Battaglia}, {Bleem}, {Bocquet}, {DeRose}, {Gruen}, {Hilton}, {Kravtsov}, {McClintock}, {Rozo}, {Rykoff}, {Varga}, {Wechsler}, {Wu}, {Zhang}, {Aiola}, {Allam}, {Bechtol}, {Benson}, {Bertin}, {Bond}, {Brodwin}, {Brooks}, {Buckley-Geer}, {Burke}, {Carlstrom}, {Carnero Rosell}, {Carrasco Kind}, {Carretero}, {Castander}, {Choi}, {Cunha}, {Crawford}, {da Costa}, {De Vicente}, {Desai}, {Devlin}, {Dietrich}, {Doel}, {Dunkley}, {Eifler}, {Evrard}, {Flaugher}, {Fosalba}, {Gallardo}, {Garc{\'\i}a-Bellido}, {Gaztanaga}, {Gerdes}, {Gralla}, {Gruendl}, {Gschwend}, {Gupta}, {Gutierrez}, {Hartley}, {Hill}, {Ho}, {Hollowood}, {Honscheid}, {Hoyle}, {Huffenberger}, {Hughes}, {James}, {Jeltema}, {Kim}, {Krause}, {Kuehn}, {Lahav}, {Lima}, {Madhavacheril}, {Maia}, {Marshall}, {Maurin}, {McMahon}, {Menanteau}, {Miller}, {Miquel}, {Mohr}, {Naess}, {Nati}, {Newburgh}, {Niemack}, {Ogando}, {Page}, {Partridge}, {Patil}, {Plazas}, {Rapetti}, {Reichardt},
  {Romer}, {Sanchez}, {Scarpine}, {Schindler}, {Serrano}, {Smith}, {Smith}, {Soares-Santos}, {Sobreira}, {Staggs}, {Stark}, {Stein}, {Suchyta}, {Swanson}, {Tarle}, {Thomas}, {van Engelen}, {Wollack}, \& {Xu}}]{2019MNRAS.487.2900S}
{Shin}, T., {Adhikari}, S., {Baxter}, E.~J., {et~al.} 2019, \bibinfo{title}{{Measurement of the splashback feature around SZ-selected Galaxy clusters with DES, SPT, and ACT},} \mnras, 487, 2900, \dodoi{10.1093/mnras/stz1434}

\bibitem[{J.~S. {Speagle} {et~al.}(2014){Speagle}, {Steinhardt}, {Capak}, \& {Silverman}}]{2014ApJS..214...15S}
{Speagle}, J.~S., {Steinhardt}, C.~L., {Capak}, P.~L., \& {Silverman}, J.~D. 2014, \bibinfo{title}{{A Highly Consistent Framework for the Evolution of the Star-Forming ``Main Sequence'' from z \raisebox{-0.5ex}\textasciitilde 0-6},} \apjs, 214, 15, \dodoi{10.1088/0067-0049/214/2/15}

\bibitem[{V. {Springel} {et~al.}(2018){Springel}, {Pakmor}, {Pillepich}, {Weinberger}, {Nelson}, {Hernquist}, {Vogelsberger}, {Genel}, {Torrey}, {Marinacci}, \& {Naiman}}]{2018MNRAS.475..676S}
{Springel}, V., {Pakmor}, R., {Pillepich}, A., {et~al.} 2018, \bibinfo{title}{{First results from the IllustrisTNG simulations: matter and galaxy clustering},} \mnras, 475, 676, \dodoi{10.1093/mnras/stx3304}

\bibitem[{M.~F. {Struble} \& H.~J. {Rood}(1982){Struble} \& {Rood}}]{1982AJ.....87....7S}
{Struble}, M.~F., \& {Rood}, H.~J. 1982, \bibinfo{title}{{Morphological classification (revised RS) of Abell clusters in D<4 and an analysis of observed correlations.},} \aj, 87, 7, \dodoi{10.1086/113081}

\bibitem[{R. {Taylor} {et~al.}(2012){Taylor}, {Davies}, {Auld}, \& {Minchin}}]{2012MNRAS.423..787T}
{Taylor}, R., {Davies}, J.~I., {Auld}, R., \& {Minchin}, R.~F. 2012, \bibinfo{title}{{The Arecibo Galaxy Environment Survey - V. The Virgo cluster (I)},} \mnras, 423, 787, \dodoi{10.1111/j.1365-2966.2012.20914.x}

\bibitem[{B.~B. {Thompson} {et~al.}(2023){Thompson}, {Smith}, \& {Kraljic}}]{2023MNRAS.518.1361T}
{Thompson}, B.~B., {Smith}, R., \& {Kraljic}, K. 2023, \bibinfo{title}{{Gas accretion and ram pressure stripping of haloes in void walls},} \mnras, 518, 1361, \dodoi{10.1093/mnras/stac2963}

\bibitem[{L.~A. {Thompson}(2021){Thompson}}]{2021dcv..book.....T}
{Thompson}, L.~A. 2021, {The Discovery of Cosmic Voids} (Cambridge: Cambridge University Press), \dodoi{10.1017/9781108867504}

\bibitem[{O. {Urban} {et~al.}(2011){Urban}, {Werner}, {Simionescu}, {Allen}, \& {B{\"o}hringer}}]{2011MNRAS.414.2101U}
{Urban}, O., {Werner}, N., {Simionescu}, A., {Allen}, S.~W., \& {B{\"o}hringer}, H. 2011, \bibinfo{title}{{X-ray spectroscopy of the Virgo Cluster out to the virial radius},} \mnras, 414, 2101, \dodoi{10.1111/j.1365-2966.2011.18526.x}

\bibitem[{L. {Verdes-Montenegro} {et~al.}(2001){Verdes-Montenegro}, {Yun}, {Williams}, {Huchtmeier}, {Del Olmo}, \& {Perea}}]{2001A&A...377..812V}
{Verdes-Montenegro}, L., {Yun}, M.~S., {Williams}, B.~A., {et~al.} 2001, \bibinfo{title}{{Where is the neutral atomic gas in Hickson groups?},} \aap, 377, 812, \dodoi{10.1051/0004-6361:20011127}

\bibitem[{F. {Villaescusa-Navarro} {et~al.}(2016){Villaescusa-Navarro}, {Planelles}, {Borgani}, {Viel}, {Rasia}, {Murante}, {Dolag}, {Steinborn}, {Biffi}, {Beck}, \& {Ragone-Figueroa}}]{2016MNRAS.456.3553V}
{Villaescusa-Navarro}, F., {Planelles}, S., {Borgani}, S., {et~al.} 2016, \bibinfo{title}{{Neutral hydrogen in galaxy clusters: impact of AGN feedback and implications for intensity mapping},} \mnras, 456, 3553, \dodoi{10.1093/mnras/stv2904}

\bibitem[{F. {Villaescusa-Navarro} {et~al.}(2018){Villaescusa-Navarro}, {Genel}, {Castorina}, {Obuljen}, {Spergel}, {Hernquist}, {Nelson}, {Carucci}, {Pillepich}, {Marinacci}, {Diemer}, {Vogelsberger}, {Weinberger}, \& {Pakmor}}]{2018ApJ...866..135V}
{Villaescusa-Navarro}, F., {Genel}, S., {Castorina}, E., {et~al.} 2018, \bibinfo{title}{{Ingredients for 21 cm Intensity Mapping},} \apj, 866, 135, \dodoi{10.3847/1538-4357/aadba0}

\bibitem[{J. {Wang} {et~al.}(2020){Wang}, {Xu}, {Lee}, {Du}, {Overzier}, \& {Shao}}]{2020ApJ...903..103W}
{Wang}, J., {Xu}, W., {Lee}, B., {et~al.} 2020, \bibinfo{title}{{Ram Pressure Stripping of HI-rich Galaxies Infalling into Massive Clusters},} \apj, 903, 103, \dodoi{10.3847/1538-4357/abb9aa}

\bibitem[{J. {Wang} {et~al.}(2021){Wang}, {Staveley-Smith}, {Westmeier}, {Catinella}, {Shao}, {Reynolds}, {For}, {Lee}, {Liang}, {Wang}, {Elagali}, {D{\'e}nes}, {Kleiner}, {Koribalski}, {Lee-Waddell}, {Oh}, {Rhee}, {Serra}, {Spekkens}, {Wong}, {Bekki}, {Bigiel}, {Courtois}, {Hess}, {Holwerda}, {McQuinn}, {Pandey-Pommier}, {van der Hulst}, \& {Verdes-Montenegro}}]{2021ApJ...915...70W}
{Wang}, J., {Staveley-Smith}, L., {Westmeier}, T., {et~al.} 2021, \bibinfo{title}{{WALLABY Pilot Survey: The Diversity of Ram Pressure Stripping of the Galactic H I Gas in the Hydra Cluster},} \apj, 915, 70, \dodoi{10.3847/1538-4357/abfc52}

\bibitem[{S.~D.~M. {White} \& C.~S. {Frenk}(1991){White} \& {Frenk}}]{1991ApJ...379...52W}
{White}, S. D.~M., \& {Frenk}, C.~S. 1991, \bibinfo{title}{{Galaxy Formation through Hierarchical Clustering},} \apj, 379, 52, \dodoi{10.1086/170483}

\bibitem[{S.~D.~M. {White} {et~al.}(1993){White}, {Navarro}, {Evrard}, \& {Frenk}}]{1993Natur.366..429W}
{White}, S. D.~M., {Navarro}, J.~F., {Evrard}, A.~E., \& {Frenk}, C.~S. 1993, \bibinfo{title}{{The baryon content of galaxy clusters: a challenge to cosmological orthodoxy},} \nat, 366, 429, \dodoi{10.1038/366429a0}

\bibitem[{C.~K. {Xu} {et~al.}(2022){Xu}, {Cheng}, {Appleton}, {Duc}, {Gao}, {Tang}, {Yun}, {Dai}, {Huang}, {Lisenfeld}, \& {Renaud}}]{2022Natur.610..461X}
{Xu}, C.~K., {Cheng}, C., {Appleton}, P.~N., {et~al.} 2022, \bibinfo{title}{{A 0.6 Mpc H I structure associated with Stephan's Quintet},} \nat, 610, 461, \dodoi{10.1038/s41586-022-05206-x}

\bibitem[{X. {Yang} {et~al.}(2007){Yang}, {Mo}, {van den Bosch}, {Pasquali}, {Li}, \& {Barden}}]{2007ApJ...671..153Y}
{Yang}, X., {Mo}, H.~J., {van den Bosch}, F.~C., {et~al.} 2007, \bibinfo{title}{{Galaxy Groups in the SDSS DR4. I. The Catalog and Basic Properties},} \apj, 671, 153, \dodoi{10.1086/522027}

\bibitem[{X. {Yang} {et~al.}(2012){Yang}, {Mo}, {van den Bosch}, {Zhang}, \& {Han}}]{2012ApJ...752...41Y}
{Yang}, X., {Mo}, H.~J., {van den Bosch}, F.~C., {Zhang}, Y., \& {Han}, J. 2012, \bibinfo{title}{{Evolution of the Galaxy-Dark Matter Connection and the Assembly of Galaxies in Dark Matter Halos},} \apj, 752, 41, \dodoi{10.1088/0004-637X/752/1/41}

\bibitem[{D. {Zakharova} {et~al.}(2025){Zakharova}, {De Lucia}, {Vulcani}, {Fontanot}, \& {Xie}}]{2025arXiv250917697Z}
{Zakharova}, D., {De Lucia}, G., {Vulcani}, B., {Fontanot}, F., \& {Xie}, L. 2025, \bibinfo{title}{{Environmental history of filament galaxies: stellar mass assembly and star-formation of filament galaxies},} arXiv e-prints, arXiv:2509.17697, \dodoi{10.48550/arXiv.2509.17697}

\bibitem[{C.-P. {Zhang} {et~al.}(2024){Zhang}, {Zhu}, {Jiang}, {Cheng}, {Wang}, {Wang}, {Xu}, {Liu}, {Yu}, {Qian}, {Yu}, {Ai}, {Jing}, {Xu}, {Liu}, {Guan}, {Sun}, {Yang}, {Huang}, {Hao}, \& {FAST Collaboration}}]{2024SCPMA..6719511Z}
{Zhang}, C.-P., {Zhu}, M., {Jiang}, P., {et~al.} 2024, \bibinfo{title}{{The FAST all sky H I survey (FASHI): The first release of catalog},} Science China Physics, Mechanics, and Astronomy, 67, 219511, \dodoi{10.1007/s11433-023-2219-7}

\bibitem[{C.-P. {Zhang} {et~al.}(2026){Zhang}, {Zhu}, {Jiang}, {Guo}, {Xu}, {Liu}, {Yu}, {Cheng}, {Wang}, {Wang}, \& {FAST Collaboration}}]{2026arXiv260631539Z}
{Zhang}, C.-P., {Zhu}, M., {Jiang}, P., {et~al.} 2026, \bibinfo{title}{{The FAST All Sky HI Survey DR2: the FASHI Catalog and the HI Mass Function},} arXiv e-prints, arXiv:2606.31539, \dodoi{10.48550/arXiv.2606.31539}

\bibitem[{M. {Zhang} {et~al.}(2025){Zhang}, {Walker}, {Sullivan}, {Power}, {Cui}, {Li}, \& {Zhang}}]{2025PASA...42....8Z}
{Zhang}, M., {Walker}, K., {Sullivan}, A., {et~al.} 2025, \bibinfo{title}{{The Three Hundred project: The relationship between the shock and splashback radii of simulated galaxy clusters},} \pasa, 42, e008, \dodoi{10.1017/pasa.2024.132}

\bibitem[{W. {Zhang} {et~al.}(2013){Zhang}, {Li}, {Kauffmann}, \& {Xiao}}]{2013MNRAS.429.2191Z}
{Zhang}, W., {Li}, C., {Kauffmann}, G., \& {Xiao}, T. 2013, \bibinfo{title}{{Gas depletion in cluster galaxies depends strongly on their internal structure},} \mnras, 429, 2191, \dodoi{10.1093/mnras/sts490}

\bibitem[{X. {Zhang} {et~al.}(2025){Zhang}, {Bulbul}, {Diemer}, {Bahar}, {Comparat}, {Ghirardini}, {Liu}, {Malavasi}, {Mistele}, {Ramos-Ceja}, {Sanders}, {Zhang}, {Artis}, {Ding}, {Fiorino}, {Kluge}, {Merloni}, {Nandra}, \& {Zelmer}}]{2025arXiv250925317Z}
{Zhang}, X., {Bulbul}, E., {Diemer}, B., {et~al.} 2025, \bibinfo{title}{{The SRG/eROSITA All-Sky Survey. Detection of shock-heated gas beyond the halo boundary into the accretion region},} arXiv e-prints, arXiv:2509.25317, \dodoi{10.48550/arXiv.2509.25317}

\end{thebibliography}
\end{document}